\documentclass[dutch,english]{book}

\input{epsf}
\usepackage{epsfig}

\usepackage{babel}
\catcode`@=11
\@addtoreset{equation}{section}
\def\theequation{\thesection.\arabic{equation}}

\def\lhead{\@ifnextchar[{\@xlhead}{\@ylhead}}
\def\@xlhead[#1]#2{\gdef\@elhead{#1}\gdef\@olhead{#2}}
\def\@ylhead#1{\gdef\@elhead{#1}\gdef\@olhead{#1}}

\def\chead{\@ifnextchar[{\@xchead}{\@ychead}}
\def\@xchead[#1]#2{\gdef\@echead{#1}\gdef\@ochead{#2}}
\def\@ychead#1{\gdef\@echead{#1}\gdef\@ochead{#1}}

\def\rhead{\@ifnextchar[{\@xrhead}{\@yrhead}}
\def\@xrhead[#1]#2{\gdef\@erhead{#1}\gdef\@orhead{#2}}
\def\@yrhead#1{\gdef\@erhead{#1}\gdef\@orhead{#1}}

\def\lfoot{\@ifnextchar[{\@xlfoot}{\@ylfoot}}
\def\@xlfoot[#1]#2{\gdef\@elfoot{#1}\gdef\@olfoot{#2}}
\def\@ylfoot#1{\gdef\@elfoot{#1}\gdef\@olfoot{#1}}

\def\cfoot{\@ifnextchar[{\@xcfoot}{\@ycfoot}}
\def\@xcfoot[#1]#2{\gdef\@ecfoot{#1}\gdef\@ocfoot{#2}}
\def\@ycfoot#1{\gdef\@ecfoot{#1}\gdef\@ocfoot{#1}}

\def\rfoot{\@ifnextchar[{\@xrfoot}{\@yrfoot}}
\def\@xrfoot[#1]#2{\gdef\@erfoot{#1}\gdef\@orfoot{#2}}
\def\@yrfoot#1{\gdef\@erfoot{#1}\gdef\@orfoot{#1}}

\newdimen\headrulewidth
\newdimen\footrulewidth
\newdimen\plainheadrulewidth
\newdimen\plainfootrulewidth
\newdimen\headwidth
\newif\if@fancyplain \@fancyplainfalse
\def\fancyplain#1#2{\if@fancyplain#1\else#2\fi}

\footrulewidth\z@
\plainheadrulewidth\z@
\plainfootrulewidth\z@

\def\@fancyhead#1#2#3#4#5{#1\hbox to\headwidth{\vbox{\hbox
{\rlap{\parbox[b]{\headwidth}{\raggedright#2\strut}}\hfill
\parbox[b]{\headwidth}{\centering#3\strut}\hfill
\llap{\parbox[b]{\headwidth}{\raggedleft#4\strut}}}\headrule}}#5}

\def\@fancyfoot#1#2#3#4#5{#1\hbox to\headwidth{\vbox{\footrule
\hbox{\rlap{\parbox[t]{\headwidth}{\raggedright#2\strut}}\hfill
\parbox[t]{\headwidth}{\centering#3\strut}\hfill
\llap{\parbox[t]{\headwidth}{\raggedleft#4\strut}}}}}#5}

\def\headrule{{\if@fancyplain\headrulewidth\plainheadrulewidth\fi
\hrule\@height\headrulewidth\@width\headwidth \vskip-\headrulewidth}}

\def\footrule{{\if@fancyplain\footrulewidth\plainfootrulewidth\fi
\vskip-0.3\normalbaselineskip\vskip-\footrulewidth
\hrule\@width\headwidth\@height\footrulewidth\vskip0.3\normalbaselineskip}}

\def\ps@fancy{
\let\@mkboth\markboth
\@ifundefined{chapter}{\def\sectionmark##1{\markboth
{\ifnum \c@secnumdepth>\z@
 \thesection\hskip 1em\relax \fi ##1}{}}
\def\subsectionmark##1{\markright {\ifnum \c@secnumdepth >\@ne
 \thesubsection\hskip 1em\relax \fi ##1}}}
{\def\chaptermark##1{\markboth {\ifnum \c@secnumdepth>\m@ne
 \@chapapp\ \thechapter. \ \fi ##1}{}}
\def\sectionmark##1{\markright{\ifnum \c@secnumdepth >\z@
 \thesection. \ \fi ##1}}}
\def\@oddhead{\@fancyhead\relax\@olhead\@ochead\@orhead\hss}
\def\@oddfoot{\@fancyfoot\relax\@olfoot\@ocfoot\@orfoot\hss}
\def\@evenhead{\@fancyhead\hss\@elhead\@echead\@erhead\relax}
\def\@evenfoot{\@fancyfoot\hss\@elfoot\@ecfoot\@erfoot\relax}
\headwidth\textwidth}
\def\ps@fancyplain{\ps@fancy \let\ps@plain\ps@plain@fancy}
\def\ps@plain@fancy{\@fancyplaintrue\ps@fancy}

\catcode`\@=12

\def\a{\alpha}
\def\b{\beta}

\def\d{\delta}
\def\e{\varepsilon}

\def\g{\gamma}

\def\k{\kappa}                  % Also, \varkappa
\def\l{\lambda}
\def\m{\mu}
\def\n{\nu}

\def\r{\rho}                    %       \varrho
\def\s{\sigma}                  %       \varsigma
\def\t{\tau}

\def\z{\zeta}
\def\D{\Delta}
\def\F{\Phi}
\def\G{\Gamma}

\def\L{\Lambda}
\def\O{\Omega}

\newcommand{\extraspace}{\addtolength{\abovedisplayskip}{2mm}
                        \addtolength{\belowdisplayskip}{2mm}
                        \addtolength{\abovedisplayshortskip}{2mm}
                        \addtolength{\belowdisplayshortskip}{2mm}}
\newcommand{\beq}{\begin{equation}\extraspace}
\newcommand{\eeq}{\end{equation}}

\newcommand{\bea}{\begin{eqnarray}\extraspace}
\newcommand{\eea}{\end{eqnarray}}
\newcommand{\beastar}{\begin{eqnarray*}\extraspace}
\newcommand{\eeastar}{\end{eqnarray*}}

\newsavebox{\uuunit}
\sbox{\uuunit}
    {\setlength{\unitlength}{0.825em}
     \begin{picture}(0.6,0.7)
        \thinlines
        \put(0,0){\line(1,0){0.5}}
        \put(0.15,0){\line(0,1){0.7}}
        \put(0.35,0){\line(0,1){0.8}}
       \multiput(0.3,0.8)(-0.04,-0.02){12}{\rule{0.5pt}{0.5pt}}
     \end {picture}}

\newfont{\twolineletters}{msbm10}

\def\lefthook{{\vrule height5pt width0.4pt depth0pt}}
\def\righthook{{\vrule height5pt width0.4pt depth0pt}}
\def\leftrighthookfill{$\mathsurround=0pt \mathord\lefthook
     \hrulefill\mathord\righthook$}
\def\underhook#1{\vtop{\ialign{##\crcr$\hfil\displaystyle{#1}\hfil$\crcr
      \noalign{\kern-1pt\nointerlineskip\vskip2pt}
      \leftrighthookfill\crcr}}}

\newcommand{\delr}{\raise.3ex\hbox{$\stackrel{\leftarrow}{\partial}$}
{}}
\newcommand{\dell}{\raise.3ex\hbox{$\stackrel{\rightarrow}{\partial}$}
{}}

\newcommand{\tr}{\mbox{tr}}

\newcommand{\NP}[1]{Nucl.\ Phys.\ {\bf #1}}
\newcommand{\CQG}[1]{Class.\ Quant.\ Grav.\ {\bf #1}}
\newcommand{\PL}[1]{Phys.\ Lett.\ {\bf #1}}

\newcommand{\PR}[1]{Phys.\ Rev.\ {\bf #1}}
\newcommand{\PRL}[1]{Phys.\ Rev.\ Lett.\ {\bf #1}}

\newcommand{\IJMP}[1]{Int.\ Jour.\ of\ Mod.\ Phys.\ {\bf #1}}
\newcommand{\JHEP}[1]{J.\ High\ Energy\ Phys.\ {\bf #1}}
\newcommand{\ATMP}[1]{Adv.\ Theor.\ Math.\ Phys.\ {\bf #1}}

\def\one{{\hbox{ 1\kern-.8mm l}}}
\def\gh{{\rm gh}}
\def\sgh{{\rm sgh}}
\def\NS{{\rm NS}}
\def\R{{\rm R}}
\def\ii{{\rm i}}

\def\tr{{\rm tr\,}}
\newcommand{\N}{{\cal N}}
\newlength{\bredde}
\def\slash#1{\settowidth{\bredde}{$#1$}\ifmmode\,\raisebox{.15ex}{/}
\hspace*{-\bredde} #1\else$\,\raisebox{.15ex}{/}\hspace*{-\bredde} #1$\fi}

\newcommand  {\Rbar} {{\mbox{\rm$\mbox{I}\!\mbox{R}$}}}

\newcommand {\Cbar}
    {\mathord{\setlength{\unitlength}{1em}
     \begin{picture}(0.6,0.7)(-0.1,0)
        \put(-0.1,0){\rm C}
        \thicklines
        \put(0.2,0.05){\line(0,1){0.55}}
     \end {picture}}}
\newsavebox{\zzzbar}
\sbox{\zzzbar}
  {\setlength{\unitlength}{0.9em}
  \begin{picture}(0.6,0.7)
  \thinlines
  \put(0,0){\line(1,0){0.6}}
  \put(0,0.75){\line(1,0){0.575}}
  \multiput(0,0)(0.0125,0.025){30}{\rule{0.3pt}{0.3pt}}
  \multiput(0.2,0)(0.0125,0.025){30}{\rule{0.3pt}{0.3pt}}
  \put(0,0.75){\line(0,-1){0.15}}
  \put(0.015,0.75){\line(0,-1){0.1}}
  \put(0.03,0.75){\line(0,-1){0.075}}
  \put(0.045,0.75){\line(0,-1){0.05}}
  \put(0.05,0.75){\line(0,-1){0.025}}
  \put(0.6,0){\line(0,1){0.15}}
  \put(0.585,0){\line(0,1){0.1}}
  \put(0.57,0){\line(0,1){0.075}}
  \put(0.555,0){\line(0,1){0.05}}
  \put(0.55,0){\line(0,1){0.025}}
  \end{picture}}
\newcommand{\Zbar}{\mathord{\!{\usebox{\zzzbar}}}}
\def\Im{{\rm Im ~}}
\def\Re{{\rm Re ~}}
\newcommand{\bra}[1]{\langle{#1}|}
\newcommand{\ket}[1]{|{#1}\rangle}
\newcommand{\braket}[2]{\langle{#1}|{#2}\rangle}
\newcommand{\ena}{\end{eqnarray}}
\newcommand{\beqa}{\begin{eqnarray}\extraspace}
\newcommand{\eeqa}{\end{eqnarray}}
\newcommand{\sect}[1]{Section~\ref{#1}}
\newcommand{\shalf}{\frac{1}{2}}
\newcommand{\eq}[1]{Eq.~(\ref{#1})}
\newcommand{\fig}[1]{Fig.~\ref{#1}}
\newcommand{\chap}[1]{Chapter~\ref{#1}}
\newcommand{\be}{\begin{equation}\extraspace}
\newcommand{\ee}{\end{equation}}
\newcommand{\si}{\sigma}
\def\sac{\, , \qquad}
\newcommand{\ba}{\begin{array}}
\newcommand{\ea}{\end{array}}
\newcommand{\nn}{\nonumber\\}

\def\narx {{\vec{\nabla} X}}
\def\piv {\mbox{$\vec{\Pi}$}}

\def\sac{\, , \qquad}

\def\fs{field strength }

\def\bg{background }
\def\su{supergravity }

\newcommand{\dsl}{\pa \kern-0.5em /}

\newcommand{\pa}{\partial}
\renewcommand{\t}{\theta}
\newcommand{\T}{\Theta}

\newcommand{\vp}{\varphi}

\begin{document}
\pagestyle{fancy}

% titlepage and so on:
% file: kaft.tex
% date: 15/2/99

\thispagestyle{empty}
\begin{figure}
  \epsfig{file=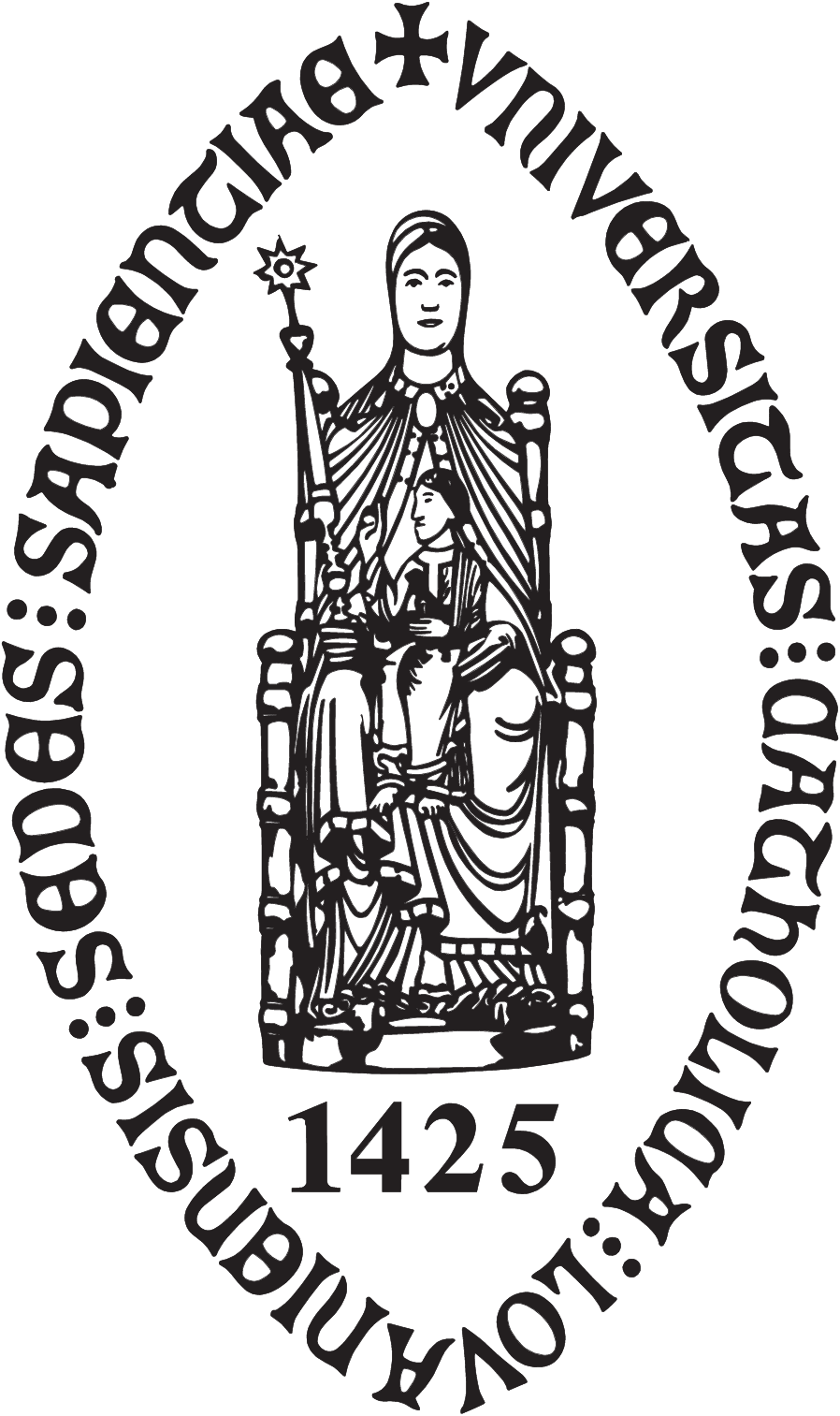,height=2cm,angle=0,trim=-4130 0 0 0}
\end{figure}
\begin{center}
\begin{minipage}{12.5cm}
\vspace{-3.8cm}
\begin{flushleft}
{\large Katholieke Universiteit Leuven\\
 Faculteit der Wetenschappen\\
 Instituut voor Theoretische Fysica\\}
\end{flushleft}
\end{minipage}
\vfill \vfill \vfill \vfill
\begin{minipage}{12.5cm}
\vspace{-1cm}
\begin{center}
{\huge\bf D-branes and boundary states}
\vskip 5mm
{\huge\bf in closed string theories}
\end{center}
\end{minipage}
\vfill \vfill \vfill
\begin{minipage}{12.5cm}
\begin{center}
{\large\bf Ben Craps \\ }
\end{center}
\end{minipage}
\vfill \vfill \vfill \vfill
\begin{minipage}{12.5cm}
\begin{minipage}[t]{6cm}
\begin{flushleft}
Promotor: Prof. Dr. W. Troost
\end{flushleft}
\end{minipage}
\hfill
\begin{minipage}[t]{6cm}
\begin{flushright}
       Proefschrift ingediend voor\\
       het behalen van de graad van\\
       Doctor in de Wetenschappen\\
\end{flushright}
\end{minipage}
\end{minipage}
\vfill
\begin{minipage}{12.5cm}
\begin{center}
{\large 2000}
\end{center}
\end{minipage}
\end{center}
\newpage

% file: voorw.tex
% date: 15/2/99
\selectlanguage{dutch}

\thispagestyle{empty}
\begin{center}
\vspace*{15,5cm}
\begin{minipage}{12,5cm}
\begin{flushleft}
De auteur is Aspirant van het
\\ Fonds voor Wetenschappelijk Onderzoek -- Vlaanderen.
\end{flushleft}
\end{minipage}
\end{center}
\newpage

\thispagestyle{empty}
\setlength{\parindent}{.7cm}
\parskip 5pt plus 1pt minus 1pt
\noindent {\centerline{\large\bf Voorwoord}}\\ \vspace{.5cm}

In de eerste plaats wil ik Walter Troost, mijn promotor, bedanken voor de 
manier waarop hij me de voorbije jaren begeleid heeft. Hij liet me vrij zelf 
mijn onderwerpen te kiezen, maar stond altijd klaar om te helpen als ik hem 
daarom vroeg. Zijn gave om een probleem onmiddellijk tot zijn essentie te 
herleiden, maakte onze talrijke discussies meestal zeer verhelderend. In het 
bijzonder ben ik hem dankbaar voor talloze nuttige raadgevingen bij het 
schrijven van deze tekst. Ook Toine Van Proeyen kan ik niet genoeg bedanken 
voor de kansen die hij me geboden heeft. Zijn voortdurende inspanningen om 
onze onderzoeksgroep in een Europees geheel te kaderen hebben me met veel 
buitenlandse collega's in contact gebracht.  Ook hij was altijd bereid gaten 
in mijn kennis op te vullen en raad te geven wanneer ik moeilijke beslissingen 
moest nemen. Walter en Toine, bedankt, ook voor alles wat jullie me op 
niet-wetenschappelijk vlak hebben bijgebracht!

Deze thesis zou er ongetwijfeld helemaal anders hebben uitgezien als ik niet was 
beginnen samenwerken met Fred Roose, toen 
we bijna vijf jaar geleden een trimester in Groningen studeerden. Ontelbaar 
zijn de uren die we sindsdien 
doorbrachten voor een volgeschreven, uitgeveegd en opnieuw volgeschreven bord. 
Het zal dan ook geen verbazing wekken dat Fred bijgedragen heeft tot bijna 
alle resultaten in deze thesis. Ik wil hem dan ook hartelijk bedanken voor de 
vruchtbare en zeer aangename samenwerking en voor al die zinnige en minder 
zinnige conversaties buiten de fysica. Marco Bill\'o bedank ik van harte voor 
de prettige samenwerking, voor de leuke tijd toen ik in Leuven zijn 
bureaugenoot was en voor de Italiaanse woorden die hij me toen leerde. I would 
also like to thank Quim Gomis and David Mateos for an enjoyable collaboration 
and for their hospitality in Barcelona.

Jan Troost wil ik bedanken voor de interessante gesprekken en discussies die 
we gehad hebben, onder andere over fysica; Frederik Denef voor zijn interesse 
in het reilen en zeilen in Leuven, voor het doorgeven van nuttige info over 
van alles en voor de bijhorende geestige e-mails; Alex Sevrin voor de pizza die 
ik hem verschuldigd ben; Ruud Siebelink en Stefan Vandoren voor verhalen uit de 
VS; Pieter-Jan De 
Smet voor vele vragen en antwoorden; Raymond Gastmans voor aanwijzingen ter 
verbetering van deze tekst; Kor Van Hoof, Joris Raeymaekers, Piet Claus, 
Martijn Derix, 
Jeanne De Jaegher, Sorin Cucu en Dirk Olivi\'e voor aangename conversaties. 
I am also grateful to Laura Andrianopoli, 
Alberto Santambrogio, Franco Cordaro, Andrea Refolli and Daniela Bigatti for 
contributing to the nice working atmosphere in our group.   

Andr\'e Verbeure bedank ik voor de soepele manier waarop hij het instituut de 
voorbije jaren geleid heeft; Mark Fannes voor de vlotte en aangename 
samenwerking bij de oefenzittingen; Christine Detroije en Anita Raets voor hun 
vriendelijke 
effici\"entie; Martijn Derix, Johan Andries en Pascal Spincemaille voor het 
beperkt houden van de gevolgen van mijn computeronkunde en voor hun geduld 
daarbij. Ik ben iedereen op `het zesde' dankbaar voor de vele gezellige 
momenten, gaande van geanimeerde koffiekamerdiscussies tot barbecues, 
wandelingen, basketbalmatchen, bridge-middagen en gewoon het eten in de Alma. 
Speciaal wil ik hierbij Dirk Olivi\'e en Mieke De Cock vermelden, die er 
meestal wel iets mee te maken hadden als er iets georganiseerd werd.     

Als mijn doctoraatsjaren ook buiten de fysica de moeite waard waren, dan is 
dat voor een groot deel te danken  aan de vrienden met wie ik in Leuven een 
leuke tijd beleefd heb. In de eerste plaats denk ik daarbij aan Gerrit Geens, 
Liesbeth Sabbe, Roby Roels, Roel Dausy, Caroline L'abb\'e, Gunther 
Fleerackers, An Van Houdt en Johan Cambr\'e. Bedankt voor de gezellige 
avonden, de sappige verhalen, de ernstiger gesprekken, de ongezouten 
meningen, de periodieke onderdompeling in moderne geluiden en de redding van 
de hongerdood door die vegetarische schotel die dinsdag. 

Mijn zusje Meena en mijn broers Stef en Jan wil ik bedanken voor hun steun, 
plagerijen, aanmoediging, bezorgdheid, taaladvies en bijwijlen zelfs wijze 
raad. Mijn ouders ben ik immens dankbaar voor alles wat ze voor mij gedaan 
hebben, doen en ongetwijfeld nog zullen doen. Het zou waanzin zijn aan een 
opsomming te beginnen, dus houd ik het maar bij een welgemeend: bedankt, pa 
en ma.

\newpage

\selectlanguage{english}

%\thispagestyle{empty}
%\begin{center}
%\begin{minipage}{10cm}
%\end{minipage}
%\end{center}
%\newpage
%
%\thispagestyle{empty}
%\begin{center}
%\begin{minipage}{10cm}
%\end{minipage}
%\end{center}
%\newpage

% initialisations for the table-of-contents:
\thispagestyle{empty}
\parskip 6pt plus 1pt minus 1pt
\lhead[\fancyplain{}{}]{\fancyplain{}{}}
\rhead[\fancyplain{}{}]{\fancyplain{}{}}
\chead{\fancyplain{}{\bf Contents}}
\cfoot{\fancyplain{}{}}
\tableofcontents
\newpage

% initialisations for the main text:
\parskip 5pt plus 1pt minus 1pt
\lhead[\fancyplain{}{\bf\thepage}]{\fancyplain{}{}}
\chead[\fancyplain{}{\bf\leftmark}]{\fancyplain{}{\bf\rightmark}}
\rhead[\fancyplain{}{}]{\fancyplain{}{\bf\thepage}}
\cfoot{\fancyplain{\bf\thepage}{}}

%Introduction
\chapter{Introduction}
\section{String theory}\label{intro:strings}

A good physical theory explains many experimental observations in
a consistent way from a small number of fundamental principles.

One example is the theory of general relativity. It implements the
principle that gravitation is due to the curvature of spacetime
and explains such apparently diverse phenomena as planetary
motion, the gravitational attraction of the Earth and the bending
of light by the Sun.
Another example is the standard model of elementary particle
physics. The fundamental principle is that the electromagnetic and
nuclear forces are due to gauge symmetries. Many predictions of
the standard model have been confirmed in accelerator experiments,
sometimes to an astonishing precision.

Thus, general relativity is a good theory of the gravitational
interaction, while the standard model is a good theory of the
strong, weak and electromagnetic interactions. Do we get a good
theory of all four fundamental interactions by just combining
general relativity and the standard model?
The problem is that general relativity is a classical theory,
whereas the standard model is a quantum theory. Put together, they
do not give a {\it consistent} description of nature. One could
try to quantize general relativity, but all straightforward
attempts to do so have failed.

Nevertheless, there does exist a quantum theory of gravity. It was
discovered by accident in the late sixties and early seventies, when people 
were looking for a theory of the strong interactions. In an attempt to
understand some experimentally observed properties of strongly interacting
particles, the idea was raised that the strong interactions
might be described by a theory of strings. This idea got less popular when
quantum chromodynamics (QCD) turned out to be a good theory of the
strong interactions. However, people had found out that string
theory includes a massless spin-2 particle, which is just what one
would expect for a particle mediating the gravitational
interaction in a theory of quantum gravity!

Motivated by the fact that it contained quantum gravity, string
theory%
\footnote{In \chap{superstrings}, we give an introduction to string theory,
focusing on the concepts we need in this thesis.}  
was proposed as a theory that could unify all interactions.
Indeed, it was discovered that string theory also
contains the gauge symmetries of the standard model. So, string
theory unifies gravity and the gauge symmetries of the standard
model in a consistent way!
However, consistency alone does not make string theory a good
physical theory. Such a theory should also explain many
experimental observations from a small number of fundamental
principles.

As for the experimental observations, the first requirement is
that string theory should reproduce the predictions of general
relativity and the standard model in circumstances where the
latter are known to give a good description of nature. Finding a
string background for which this is realized is the subject of
string phenomenology. Although to this date the standard model has
not been exactly reproduced, there exist `semi-realistic' models
that come close.

As for the fundamental principles of string theory, much remains to be
uncovered, although we have seen various hints about the structure of the
theory.

Firstly, one thing that is clear is that the structure of
perturbative string theory is much more constrained than the one
of field theory: whereas one can think of many different
interactions in field theory, in perturbative string theory only
the joining-splitting interaction of strings is allowed. However,
a full, non-perturbative description of string theory is still
lacking. What we do know is that string theory is not only a
theory of strings: it also contains other extended objects, called
`branes', a generalization of membranes. It is not clear whether
the full theory can be formulated in terms of strings.

Secondly, supersymmetry,%
\footnote{See \sect{central:susy} for an introduction to supersymmetry.} 
a symmetry between bosons and fermions,
has played a major role in string theory. Supersymmetry could be
one of the organizing principles underlying string theory.
Nevertheless, there have recently been attempts to make sense of
non-supersymmetric string theories.

Finally, dualities%
\footnote{An introduction to dualities can be found in \sect{central:duality}.} 
are omnipresent in the current developments of
string theory. They are our main handle on non-perturbative string
theory.

So far, we have discussed string theory as a candidate for a Theory
Of Everything. However, there is another way in which string
theory could revolutionize our understanding of nature: we could
use it as an auxiliary theory to learn something about quantum
field theory, in particular about the gauge theories of the
standard model. For instance, there is hope that string theory may
be an appropriate tool to deal with the notorious strong coupling
problems of QCD, as we shall now discuss.

Quantum chromodynamics (QCD) is a theory of the strong
interactions. It is part of the standard model. At high energies,
the theory is weakly coupled, so that perturbation theory allows one to
make predictions for high energy scattering experiments. These
predictions have been found to agree with the experimental
results. At low energies, the theory is strongly coupled, so
perturbation theory is useless. It is simply not known how to do
computations in this regime. Nevertheless, such computations are
necessary if we want to understand phenomena like quark
confinement.

Recently, it has been proposed that the strong coupling limit of
certain field theories can be described by a weakly coupled dual
string theory \cite{maldacena}. This conjecture is called the AdS/CFT
correspondence.%
\footnote{A very brief but slightly more technical introduction to the AdS/CFT 
correspondence can be found in \sect{AdSCFT}.}
In its present form, it does not really apply to
QCD, but the idea that some `cousins' of QCD can be described by
specific string theories comes tantalizingly close to the original
strong interaction motivation for string theory!
%%%%%%%%%%%%%%%%%%%%%%%%%%%%%%%%%%%%%%%%%%%%%%%%%%%%%%%%%%%%%%%%%%%%%%%%%%%%%%%%
\section{D-branes}\label{intro:D}
\subsection{Generalities}
In most of the recent developments in string theory, a crucial role is played by
{\it D-branes}.%
\footnote{D-branes will be introduced in \sect{superstrings:D-branes}.}
D-branes can be introduced as hyperplanes on which open strings can end.
However, it turns out that they are really dynamical objects: they are free to
move and to change their shapes. D-branes are a subclass of the extended objects
we mentioned in \sect{intro:strings} when we were saying that string theory is
not only a theory of strings. What distinguishes D-branes from the other
extended objects is that, although they are not perturbative string states
themselves, they allow a description in perturbative string theory: their
dynamics is governed by the open strings ending on them. This property makes
them much easier to study than other extended objects in string theory. D-branes
have been important in the following ways.

First, they appear in many models of string phenomenology. The main
reason is that
D-branes have gauge fields living on their world-volumes, in particular
non-abelian gauge fields if several D-branes coincide. These gauge fields  
can play the role of the gauge fields of the standard model.  
Particularly interesting from a phenomenological point of view is the
observation that 
D-branes can play a role in supersymmetry breaking in string theory. For
instance, systems with D-branes and anti-D-branes%
\footnote{See \sect{nonbps:mot:brane} for some elementary considerations about
brane--anti-brane systems.} 
have recently led to semi-realistic models \cite{ibanez}.%
\footnote{We refer to \cite{fredd} and references therein for a related
application of D-branes: they can be used to derive from string theory
non-perturbative results about supersymmetric field theory weakly coupled to
gravity.} 

Second, they are omnipresent in string dualities. Often,%
\footnote{See \sect{superstrings:dualities:S}, \sect{type0:dual} and
\sect{nonbps:mot:testing} for examples.} 
D-branes in one
perturbative description of string theory correspond to fundamental strings in a
dual description of string theory. It is in this sense that D-branes are no less
fundamental than strings.

Third, D$p$-branes, where $p$ is the number of spatial directions along which
the D-brane is extended, are related to {\it black $p$-branes}, a
$p$-dimensional generalization of black holes. In fact, D-branes and black
$p$-branes are complementary descriptions of the same object. Which description
is appropriate depends on the value of the string coupling constant. This
correspondence can be used to count the number of states of supersymmetric
(BPS)%
\footnote{See \sect{central:susy} for the meaning of BPS.}
black holes. For such black holes, the number of states does not change when we
change the string coupling constant, so that their states can equivalently be
counted in the regime where the D-brane description is valid. This procedure has
led to a microscopic explanation for the entropy of black holes, at least for
certain supersymmetric ones \cite{stromvaf}.

Fourth, D-branes are at the heart of the AdS/CFT correspondence, as will
hopefully be clear from \sect{AdSCFT} and \sect{type0:intro}.

%%%%%%%%%%%%%%%%%%%%%%%%%%%%%%%%%%%%%%%%%%%%%%%%%%%%%%%%%%%%%%%%%%%%%%%%%%%%%%%
\subsection{Topics studied in this thesis}\label{topics}
Having motivated why D-branes are important in string theory, we now turn to the
specific topics that are considered in this thesis.

Most of this thesis is devoted to the study of D-branes. In particular, we 
are interested in the D-brane effective action. This is a low-energy effective
action for the massless excitations of the D-brane, i.e., for the massless 
modes of the open strings governing the D-brane dynamics. One could imagine
having ``integrated out'' all the massive modes of the open strings. Needless to
say, such an effective, ``macroscopic'' description is much easier to use for
computations than the full, ``microscopic'' descriptions in terms of open strings.

Our main tool to obtain, or at least to test, 
certain terms of the D-brane effective action is the
computation, in the ``microscopic'' string theory, of string scattering amplitudes
in the presence of the D-brane.  The formalism we use to perform most of these
computations is the {\it boundary state} formalism. In this formalism, a D-brane
is represented as a source of closed string states.

The specific terms in the D-brane action we are mainly interested in are 
called {\it anomalous couplings}. They are part of the D-brane Wess-Zumino
action and are responsible for the cancellation, via anomaly inflow, of
gauge and gravitational anomalies in certain D-brane configurations. As such,
they
are intimately related to the celebrated consistency of string theory. Our main
contribution in this respect is to explicitly check the presence of these
terms in the D-brane action. As a by-product, we find additional terms in the
D-brane action, which are new. 

We extend the anomaly inflow argument and the scattering computations to
D-branes in the non-supersymmetric type 0 string theories. We derive a
Wess-Zumino action that is strikingly similar to the one of BPS D-branes in the
supersymmetric type II string theories. The type 0 D-branes have been used to
construct non-supersymmetric versions of the AdS/CFT correspondence. We continue
our investigations of type 0 string theories by studying the spectrum of other
extended objects, NS-fivebranes, in these theories. This allows us to comment on
a recent duality proposal.

Taking a non-supersymmetric string theory as a starting point is not the only
way to obtain non-supersymmetric D-brane field theories. One can also consider
non-supersymmetric (non-BPS) D-brane configurations in a supersymmetric string 
theory. Such configurations can be obtained by taking both BPS D-branes and
their antibranes. Alternatively, one can consider non-BPS D-branes, which are
unstable objects in type II string theories. We investigate these non-BPS
D-branes and write down a Wess-Zumino action for them. This action partly
explains that BPS D-branes can be considered as monopole-like configurations
in the world-volume theory of non-BPS D-branes.

Finally, we use the effective action of BPS D-branes in type II string theory to
study D-branes in the background of other D-branes. Such configurations have
many applications. We focus on an application in the AdS/CFT correspondence,
namely the analogon in string theory of an external baryon in field theory.

In this thesis, we have chosen to mainly restrict our attention to closed string
theories in ten flat, non-compact dimensions. This implies that we do not discuss
some closely related topics we have published about. 
In particular, we do not treat special K\"ahler geometry
\cite{special},
anomalous couplings of orientifold planes \cite{benfred,normal}, 
the superalgebra approach to D-branes in D-brane backgrounds \cite{CGMV}, 
non-supersymmetric gauge theories form type 0 orbifolds \cite{BCR}
and the boundary state formalism for orbifolds \cite{BCR}. We refer the
interested reader to our papers on these topics.

To summarize, the main tools in our studies of D-branes are string scattering
amplitudes and consistency arguments. We compute most of these string 
scattering amplitudes
in the boundary state formalism.  The consistency arguments we use
are related to the cancellation of gauge and gravitational anomalies via the
anomaly inflow mechanism. We are interested in non-supersymmetric D-brane
configurations. In particular, we try to extend parts of the beautiful structure 
of supersymmetric strings and D-branes to non-supersymmetric situations.  
%%%%%%%%%%%%%%%%%%%%%%%%%%%%%%%%%%%%%%%%%%%%%%%%%%%%%%%%%%%%%%%%%%%%%%%%%%%%%%%%
\section{Outline and summary of results}
In \sect{intro:D}, we have given a brief overview of the topics treated in this 
thesis, emphasizing our motivations and tools. In the present section, we outline
the structure of the thesis and give a somewhat more detailed summary of the
results that are presented.

Most of these results have been published before in the papers
\cite{benfred, normal, CGMV, BCR, ournonbps, 0dual}. Our aim here is to present 
them together in a coherent way and to provide enough background material to 
make (at least some of) them and the underlying ideas accessible to non-experts.
In particular, we have in mind people familiar with quantum field theory but 
without a very detailed knowledge of this branch of string theory.      

On the one hand, in \chap{central} we introduce some of the key concepts 
underlying our research
in different contexts than string theory, in particular in quantum field theory.
These concepts are duality, supersymmetry and anomaly inflow. 
On the other hand, in \chap{superstrings} and \sect{anomalous:boundary} we try
to provide enough technical information on strings, D-branes and boundary states
to enable the motivated reader to understand the computations in
\chap{anomalous}, in particular the ones given in quite some detail in
\sect{anomalous:BC}. \chap{superstrings} is also a necessary preparation for
\chap{baryon}, \chap{type0} and \chap{nonbps}. 
Let us now give a chapter-by-chapter outline of the thesis, at the same time
summarizing the main results.

In \chap{central}, we introduce three key concepts in a mostly field theoretic
setting. In \sect{central:duality}, we give various examples of dualities, with
an emphasis on electromagnetic duality. In particular, we briefly review
the duality conjecture by Montonen and Olive, and some of the puzzles it raised.
In \sect{central:susy}, we show how these puzzles naturally get a solution in
supersymmetric field theory.  We introduce the supersymmetry algebra and the
notion of BPS states. In \sect{central:anomalies}, we try to give a
self-contained presentation of some results from the anomaly and
anomaly inflow literature. 
We focus on one example, where one explicitly sees the inflow mechanism at work.
The concept of anomaly inflow is crucial to this thesis, so we strongly 
encourage the reader to go through this section if he is not familiar with it.

\chap{superstrings} deals with superstrings and D-branes. Of course, it is
impossible to give a complete and self-contained review of string theory in
fifty or sixty pages. This being said, we do make an effort to introduce 
in a coherent way the concepts relevant to this thesis, in particular
to the computations in \chap{anomalous}. \chap{superstrings} does not 
contain original results. 

\sect{superstrings:strings:worldsheet} introduces strings from a world-sheet
point of view. We encourage the reader to study the difference between closed
and open strings, the different sectors (e.g. NS versus R for open strings) and 
the string spectra, including the GSO projection. Readers who are not interested
in technical details may want to skip almost anything referring to ghosts and 
BRST operators. \sect{superstrings:strings:spacetime} deals with the low-energy
effective description of string theory as a supergravity theory in a
ten-dimensional spacetime. The interplay between the world-sheet and spacetime 
descriptions of string theory will be one the main themes in this thesis.
\sect{superstrings:background} introduces superstrings in background fields. 

In \sect{superstrings:scattering}, some tools to compute string scattering
amplitudes are collected.  The important concept of vertex operator is
introduced. Quite some effort goes into some subtleties related to ghosts,
superghosts, pictures and BRST invariance. From a technical point of view, it is
important to properly deal with these subtleties. However, readers who are not
interested in technicalities can safely skip them: these details will be
well-hidden in the explicit computations after \sect{anomalous:overview}. 
Roughly speaking, the details are only necessary to justify our choice of 
vertex operators. These vertex operators are given in
\sect{superstrings:scattering:closed}. 

In \sect{superstrings:D-branes}, D-branes and their effective actions are 
introduced. In particular, some of the terms in the D-brane Wess-Zumino action
are displayed in detail.
This section is essential to understand anything of what follows in the thesis. 
\sect{superstrings:inflow} deals with the anomaly inflow argument in the context
of D-branes and NS fivebranes. We do not give all the technical details, but the
ideas play a central role in the thesis. For one thing, these anomaly inflow
arguments will be applied to type 0 string theory in \chap{type0}.
In \sect{superstrings:dualities}, we discuss three dualities in string theory.
The most important one for our purposes may be type IIB S-duality, of which we
shall critically examine a generalization in \chap{type0}.

In \chap{anomalous}, we check the presence of some of the anomalous D-brane 
couplings \eq{WZ} by explicit string computations. These computions are done in
the boundary state formalism.
In \sect{anomalous:boundary}, we introduce the boundary state and give a 
precise set of rules to use it in sphere scattering amplitudes.%
\footnote{We do not try to derive these rules from first principles, though.} 
This section
does not contain much original material, though to our knowledge our particular
set of rules has not appeared before in the literature.  

In \sect{anomalous:overview}, we give an overview of the checks of the 
Wess-Zumino action \eq{WZ}. It turns out that there is indirect evidence for the
presence of all the terms in \eq{WZ}, but that for the terms involving NS-NS
fields direct evidence was lacking before our paper \cite{benfred} appeared. The
indirect arguments explicitly use the consistency of string theory, 
in particular the absence of gauge and gravitional anomalies.

The direct checks of some of the anomalous D-brane couplings are performed in
\sect{anomalous:BC}, \sect{anomalous:RRC} and \sect{anomalous:D:non}. 
In \sect{anomalous:BC}, we check the presence of the term%
\footnote{See \sect{superstrings:D-branes} for our notation.}
\beq
\frac{T_p}{\kappa}\int_{p+1}\hat C_{p-1}\wedge \hat{B}
\eeq
in the D-brane Wess-Zumino action. Thanks to our preparatory work in
\chap{superstrings} and especially \sect{anomalous:boundary}, we can give the
computation in quite some detail. \sect{anomalous:BC} is an expanded version of
a section of our paper \cite{benfred}.

In \sect{anomalous:RRC}, we check the presence of the terms
\beq
\frac{T_p}{\kappa}\int_{p+1}\hat C_{p-3}\wedge \frac{(4\pi^2\a ')^2}{384\pi^2}
(\tr R_T^2-\tr R_N^2)~,
\eeq 
which involve a four-form constructed from curvature two-forms. These computations were
first done in our papers \cite{benfred, normal}.

In \sect{anomalous:D:non}, which is based on our paper \cite{normal}, 
we consider the 
terms 
\beq
\frac{T_p}{\kappa}\int_{p+1}\hat C_{p-7}\wedge 
\left(\frac{(4\pi^2\a ')^4}{294912\pi^4}(\tr R_T^2-\tr R_N^2)^2+
\frac{(4\pi^2\a ')^4}{184320\pi^4} (\tr R_T^4-\tr R_N^4)\right)~,
\eeq 
involving an eight-form built from curvature two-forms. Using results from 
\cite{stefanski}, we check the presence of all these terms and,
as a nice by-product, we find new, non-anomalous D-brane couplings.

In \chap{baryon}, we discuss an application of the D-brane world-volume action: 
we show how it can be used to study strings ending on D-branes in a D-brane
background. We study one particular configuration, which is largely motivated by
studies of the baryon vertex in the AdS/CFT correspondence, a recently proposed 
duality between conformal field theories and string theories in certain 
backgrounds. 

\sect{baryon:intro} contains a lightning review of the AdS/CFT correspondence
and the baryon vertex, and a brief introduction to the method we use to study
the baryon vertex: the BPS method. The rest of \chap{baryon} closely
follows our paper \cite{CGMV}. In \sect{baryon:review}, the BPS method is
explained in a somewhat simpler setting than the one we shall consider. In
\sect{baryon:ham}, we study the baryon vertex configuration and a related
configuration in our formalism. We reproduce the BPS equations obtained and
conjectured in \cite{Imamura,Callan}. A nice aspect of our approach is that we 
can derive results
closely related to supersymmetry without explicitly considering fermions.%
\footnote{This is also the case for the BPS bound for magnetic monopoles, see
\sect{central:duality} and \sect{central:susy}.}
Another advantage of our approach is that it makes it possible to reinterpret 
the BPS equations in terms of superalgebras \cite{CGMV}. However, the latter
reinterpretation will not be discussed in this thesis.   

\chap{type0} is devoted to type 0 string theory. Compared to type II strings,
type 0 strings have a tachyon, twice as many R-R fields and
no spacetime fermions 
in their perturbative spectra. Because of the problems related to the
presence of a tachyon, type 0 string theory became popular only one and a half 
year ago. Then, it was noticed that the tachyon may not be a big problem when one
is interested in D-brane world-volume field theories, as in the AdS/CFT
correspondence.

In \sect{type0:intro}, we give a brief review of this motivation to study type 0
string theory. Type 0 strings are introduced in \sect{type0:strings}. In
\sect{type0:D}, we introduce type 0 D-branes. As could be anticipated from the
doubling of the R-R spectrum compared to type II string theory, the number of
different D-branes is also doubled. Following our paper
\cite{BCR}, we derive a Wess-Zumino
action for type 0 D-branes. We use an anomaly inflow argument very similar to 
the one used for type II D-branes. Again, the different terms in the Wess-Zumino
action can be checked via boundary state computations.

In \sect{type0:NS5}, which is based on our paper \cite{0dual}, we study NS fivebranes in
type 0 string theory. We derive their massless spectra and find that they are
non-chiral and purely bosonic for both type 0A and type 0B. Type IIA NS
fivebranes have a chiral, anomalous spectrum. The anomaly is cancelled by
anomaly inflow from the bulk of spacetime. We compute that for both type 0A and
type 0B there is no such inflow from the bulk. This is consistent with the
non-chiral and thus non-anomalous spectra of type 0 NS fivebranes. We propose a
speculative interpretation of the type 0 NS fivebrane spectra in terms of ``type
0 little strings''.
In \sect{type0:dual}, we combine our studies of type 0 D-branes and NS fivebranes
to comment on the recently proposed type 0B S-duality.

\chap{nonbps} deals with non-BPS D-branes. These are objects on which open
strings can end, but that preserve no supersymmetry. In some string theories,
such objects can be stable, but in ten-dimensional type II string theory they are
not. In \sect{nonbps:mot}, we review how
Sen's work motivated the study of non-BPS branes.  In \sect{nonbps:anomalous}, we
propose a Wess-Zumino action for the unstable non-BPS D-branes of type II string
theory \cite{ournonbps}. On the one hand, we argue that the action we propose
is consistent with the interpretation of BPS D-branes as topologically
non-trivial tachyon configurations on a non-BPS D-brane. On the other hand, we
check the presence of the terms we propose by explicit string computations.

%Central concepts
\chapter{Some key concepts}\label{central}

This thesis deals with string theory. However, an important role will be played
by three key concepts, whose importance is not limited to string theory. These
concepts are: duality, supersymmetry and anomaly inflow.  The aim of this 
chapter is to introduce them in a simpler context. 
The first section is based on \cite{harvey, olive} and the second on
\cite{olive}. Although the third section does not contain original material, we
have made an effort to give a consistent%
\footnote{
We shall use the following conventions. The four-dimensional spacetime metric 
is $\eta_{\m\n}={\rm diag}(-1,1,1,1)$. The gamma matrices satisfy the Clifford
algebra
\beq
\{\g^\m,\g^\n\}=2\eta^{\m\n}~.
\eeq
Further, we define $\g^5=\ii\g^0\g^1\g^2\g^3$, so that $(\g^5)^2=1$. The
conjugate $\bar\psi$ of a Dirac spinor $\psi$ is defined by
\beq
\bar\psi=\psi^\dagger\ii\g^0~.
\eeq 
}
presentation of some results from the anomaly and
anomaly inflow literature. The main references we have used are
\cite{zumino,alvaginsp,bardzum} for anomalies and \cite{CH,naculich} for anomaly
inflow. 
%%%%%%%%%%%%%%%%%%%%%%%%%%%%%%%%%%%%%%%%%%%%%%%%%%%%%%%%%%%%%%%%%%%%%%%%%%%%%%%
\section{Duality}\label{central:duality}
A physical system is said to exhibit ``duality'' if there are two complementary
formulations of the theory. Often, each description is useful in a particular
region of parameter space. Computations that are hard in one formulation of the
theory may be easy in a dual description.  

One example  of a system exhibiting duality is the Ising model of statistical
mechanics. Take a square two-dimensional lattice with spins $\s_i$ taking the
values $\pm 1$ and interacting ferromagnetically with their nearest neighbours
with strength $J$. The partition function at temperature $T$ is%
\footnote{We omit all subtleties related to boundary conditions for a finite
lattice, or to the infinite volume limit; our only aim is to give a heuristic
idea.}
\beq\label{ising}
Z(K) = \sum_{\sigma} \exp ( K \sum_{(ij)} \sigma_i \sigma_j )~,
\eeq
where the sum on $i,j$ runs over all nearest neighbors, the sum
on $\sigma$ over all spin configurations, and $K = J/k_B T$. Expanding the
exponential in \eq{ising} gives rise to an expansion in powers of $K$, which is
appropriate for weak coupling (or high temperature). This expansion can be
represented by certain ``bond graphs'' on the lattice, where a bond is a line
connecting two lattice sites. It turns out that there is an alternative way to
expand the partition function \eq{ising}, namely around a maximally ordered 
configuration. This expansion, which is a strong coupling (low temperature)
expansion, can be represented in terms of bond graphs on the dual lattice (the
square lattice whose vertices are the centers of the faces of the original
lattice). Using the graphical representation of both expansions, one can relate
the original partition function (with coupling $K$) to the partition function on
the dual lattice with coupling $K^*$, where $\sinh 2K^* = 1/(\sinh 2K)$:
\beq
Z(K)=Z_D(K^*)2^{1-N_D}(2\sinh 2K)^{Nq/4}~,
\eeq
where $Z_D$ denotes the partition function on the dual lattice, $N$ and $N_D$
are the number of lattice points of the original and dual lattice, respectively,
and $q$ is the number of nearest neighbours of any point of the original
lattice.
Note
that strong coupling on the original lattice is mapped to weak coupling on the
dual lattice, and vice versa.

So far, the discussion would not change much if we allowed more general lattices
than square ones. However, if the lattice is  square, then so is
the dual lattice, which implies that the strong and weak coupling regimes of
the {\it same} theory are related. 
A remarkable consequence is that, if the system is
to have a single phase transition, it must occur at the self-dual point with
$K=K^*$, or $\sinh(2J/k_B T_c) = 1$ \cite{kramerswannier}.  

Another non-trivial duality occurs in two-dimensional relativistic field theory.
The sine-Gordon model is defined by the action
\beq
S_{SG} = \int d^2 x \left(
-\shalf \partial_\mu \phi \partial^\mu \phi
+ {\alpha \over \beta^2} \left( \cos \beta \phi -1 \right) \right) ~.
\eeq
The theory contains the quantum excitations of the field $\phi$ as ``obvious''
(or ``fundamental'') particles, with mass $\sqrt{\a}$. In addition to these 
fundamental particles, there 
are the (less obvious) 
solitons interpolating between different minima of the potential, with mass
$8\sqrt\a/\b^2$. These solitons owe there stability to the topologically
non-trivial vacuum structure. 
By expanding the potential to
quartic order, we see that $\beta^2$ acts as the coupling
constant for this theory. Thus, the soliton mass is large (compared
to the one of the fundamental excitations) at weak coupling.

This theory is completely equivalent \cite{skyrme,coleman}
to an apparently unrelated theory of
interacting fermions, the massive Thirring model.
The action of the Thirring model is
\beq
S_T = -\int d^2 x \left( \bar \psi  \gamma^\mu \partial_\mu \psi
+ m \bar \psi \psi +
{g \over 2} (\bar \psi \gamma^\mu \psi)(\bar \psi \gamma_\mu \psi) \right) ~.
\eeq
The map  between the two theories relates the couplings through
\beq\label{coupling}
{\beta^2 \over 4 \pi} = {1 \over 1 + g/\pi } 
\eeq
and maps the soliton of the SG theory to the fundamental fermion
of the Thirring model and the fundamental particles of the sine-Gordon theory
to fermion anti-fermion bound states. 
We see from \eq{coupling} that strong coupling in one theory (i.e., large
$g$) is mapped to weak coupling (small $\beta $ ) in the other theory.
(Actually, the correspondence is only valid for $\b^2<8\pi$ since for larger
values the Hamiltonian of the sine-Gordon model turns out to be unbounded from
below.)
Thus, duality provides a means of performing strong coupling
calculations in one theory by mapping them to weak coupling
calculations in a dual theory. 

Another feature of this particular model is the interchange of fundamental 
quanta and solitons: sine-Gordon solitons can be viewed as being created by
massive Thirring fields in the dual formulation.
This feature is shared with many other dualities.

\paragraph{Electromagnetic duality}
Maxwell's classical equations for the free electromagnetic field,
\begin{eqnarray}
\nabla \cdot (\mathbf{E} + \ii \mathbf{B}) &=& 0 \ ;\\
\nabla \wedge (\mathbf{E} + \ii \mathbf{B}) - \ii\frac{\partial}{\partial t}
(\mathbf{E} + \ii \mathbf{B})&=&0\ ,
\end{eqnarray}
are invariant under electromagnetic duality rotations
\begin{equation}
\label{dirac:dualitytf}
(\mathbf{E} + \ii\mathbf{B}) \rightarrow e^{\ii\phi}(\mathbf{E} + \ii\mathbf{B})\ .
\end{equation}
This symmetry is broken if electric charges $q$ are introduced, but can
be restored by also introducing magnetic charges $g$. The duality rotation
\eq{dirac:dualitytf} is then supplemented by
\beq\label{dualrot}
q+\ii g\rightarrow  e^{\ii\phi} (q+\ii g)~.
\eeq

Quantum mechanically, when residing on different particles, these charges should 
satisfy Dirac's quantisation condition \cite{dirac}
\beq \label{dirac:qnts}
qg = 2\pi n\hbar \ ; \qquad n \in \Zbar \ .
\eeq 
Zwanziger and Schwinger \cite{schwingzwan}
extended this condition to dyons, which are particles
with both electric and magnetic charges:
\begin{equation}
q_{1}g_{2}-q_{2}g_{1} \in 2\pi\hbar\Zbar \ .
\end{equation}
This condition is invariant under duality rotations applied to dyons 1 and 2
simultaneously.

In unified gauge theories, magnetic charges show up in a natural way. Consider,
for instance, SU(2) Yang-Mills theory with scalars $\phi^a$ transforming in the
adjoint representation:
\begin{equation} \label{SO3lagr}
{\cal L} = -\frac{1}{4}F^{\mu\nu a} F^a_{\mu\nu} -
\frac{1}{2} D^\mu \phi^a D_{\mu} \phi^a -V(\phi)\ ,
\end{equation}
where 
\bea
F^a_{\mu\nu}&=&\partial_\mu A^a_\nu-\partial_\nu A^a_\mu+q\e^{abc}A^b_\mu
A^c_\nu~;\label{Fq}\\
D_\mu\phi^a&=&\partial_\mu\phi^a+q\e^{abc}A^b_\mu \phi^c~~,
\eea
and we take the $SU(2)$ invariant potential 
\begin{equation} \label{VSO3}
V(\phi) = \lambda^2(\phi^2-a^2)^2
\end{equation}
with $a \in \Rbar^+$. 
We have taken the coupling constant% 
\footnote{We put $\hbar=1$.}
inside the covariant derivative to be $q$.
In the vacuum, the Higgs \cite{higgs} field $\phi^a$ has an expectation value 
with magnitude
$a$. Expanding around this vacuum, one of the gauge fields remains massless: it
is identified with the U(1) gauge field of electromagnetism. The other two
gauge fields (the ``W-bosons'') carry electric charges $\pm q$ and have masses 
$aq$. Further, there is one scalar field, the physical Higgs field. 

In addition to
these excitations of the elementary fields, the theory contains
solitonic particles \cite{hooftpol}. These correspond to topologically 
non-trivial finite-energy
solutions of the classical equations of motion. It turns out that these solitons
carry magnetic charges $\pm g$ with
\beq\label{gq}
g=\frac{4\pi}{q}~, 
\eeq
hence they are called magnetic monopoles. In general,
there is an inequality \cite{bog} between their masses and charges, $M\ge ag$. 
This Bogomol'nyi bound can be saturated, $M=ag$, in the
limit $\lambda\rightarrow 0$ (where the physical Higgs particle becomes 
massless), which is called the Prasad-Sommerfield \cite{prasadsommerfield}
limit. In that limit, the conditions for the bound $M\ge ag$ to be saturated
constitute a set of first order differential equations, the Bogomol'nyi equations
\cite{bog}.

In fact, a universal
lower bound exists for the mass of a particle with electric charge $Q$ and
magnetic charge $G$:
\beq\label{bogbound}
M\ge a\sqrt{Q^2+G^2}~.
\eeq 
Again, this Bogomol'nyi bound can only be saturated in the Prasad-Sommerfield
limit. The field configuration
should satisfy a set of first order differential equations. These are often
easier to solve than the (second order) equations of motion, which is one of the
virtues of Bogomol'nyi's approach. In the supersymmetric models to be discussed
in \sect{central:susy}, the Bogomol'nyi bound and the conditions under which it
is saturated will get a nice interpretation.

In the limit of vanishing potential, all the particles mentioned above (photon,
Higgs, W-bosons $W^{\pm}$ and magnetic monopoles $M^{\pm}$) saturate the 
Bogomol'nyi bound, which is invariant under duality rotations \eq{dualrot}. 
All particles satisfy the universal mass formula
\beq\label{universalmass}
M= a\sqrt{Q^2+G^2}~.
\eeq
This is  remarkable, given the very different way in which the
various particles are described (fundamental excitations versus solitons). 

We have mentioned that sine-Gordon solitons can be viewed as being created by
massive Thirring fields in a reformulation of the theory. In view of this, one may
wonder whether a similar reformulation exists for the gauge theory \eq{SO3lagr}.
In such a dual formulation, the quantum fields should correspond to the monopoles
rather than the original gauge particles. This question was addressed in
a more general context, e.g., for a more general unbroken gauge group than U(1),
in \cite{GNO}. It was conjectured that the dual theory is
a gauge theory whose potential couples to the magnetic charge. Conjectures for
the structure of the magnetic, or dual, gauge group were made. A special
case, \eq{SO3lagr} with gauge group SU(2) and in the limit of vanishing 
potential, was considered by Montonen and Olive.
This is the case to which we now return.  

The duality invariance of the universal mass formula \eq{universalmass}
led Montonen and Olive \cite{montol} to the conjecture that the whole theory is 
{\it invariant} under interchanging the W-bosons with the monopoles and at the 
same time $q$ and $g$.
This means that there would be a dual, ``magnetic'' description of this
particular theory, 
which is {\it identical} to the original, ``electric'' description 
(up to the value of the coupling constant). It is an SU(2) 
gauge theory with the gauge group
spontaneously broken to U(1). The monopoles appear as quantum excitations of the
massive gauge fields, whereas the original W-bosons are solitons 
in the dual description. 

Note that this {\it self-duality} is special to the case Montonen and Olive were
considering. It is by no means a general feature that the dual theory is the same
as the original one.

This conjecture by Montonen and Olive is quite nontrivial and immediately raises 
a number of puzzles:
\begin{enumerate}
\item Will the duality symmetry of the mass spectrum survive quantum
corrections? 
\item The W-bosons have spin one, whereas the monopoles are rotationally
invariant. How do the monopoles acquire spin, so that they can be interpreted as
W-bosons in the dual description of the theory?
\item Since $q$ and $g$ are inversely proportional, the conjectured duality
relates two theories, one of which is necessarily at strong coupling. Is it
possible to test this conjecture at all? 
\end{enumerate}
The solution of these puzzles involves supersymmetry, as we shall now discuss.
%%%%%%%%%%%%%%%%%%%%%%%%%%%%%%%%%%%%%%%%%%%%%%%%%%%%%%%%%%%%%%%%%%%%%%%%%%%%%%%%
\section{Supersymmetry and BPS states}\label{central:susy}
A supersymmetry is a symmetry
with a fermionic generator, which means that bosons and fermions are transformed
into each other. Supersymmetry turns out to impose severe restrictions on
physical theories.%
\footnote{We refer the reader to our work \cite{special} (and references
therein), where some of the constraints imposed by supersymmetry are studied 
for a particular class of theories.}
%This tight structure results, for instance, in various non-renormalization
%theorems.
An obvious constraint is that the particles of a theory should fall into
supersymmetry multiplets. The supersymmetry algebra is such that the
supersymmetry generators commute with the momenta, which implies that each
particle is degenerate in energy with its superpartner(s). In what follows, the
supersymmetry algebra will be of great use in solving the puzzles raised at the
end of \sect{central:duality}.

Indeed, the theory discussed in the previous section can be extended in several 
ways before quantizing it.
It is possible to add fermionic and other fields to the original theory. These
play no role in the solutions discussed so far if putting them to zero is
consistent with the new equations of motion. It turns out that the theory can
be extended in such a way that it becomes supersymmetric \cite{DHD}. 

We introduce the supercharges $Q^i_\a$ carrying a Dirac spinor index $\a$
running from 1 to 4 and an internal symmetry index $i$ running from 1 ot ${\cal
N}$. We
impose a Majorana condition on these supercharges. Choosing a Majorana basis for
the gamma matrices (which means that the gamma matrices are real), this Majorana
condition means that the supercharges are hermitian. The simplest
anticommutation relation these supercharges can satisfy is
\beq\label{QQP}
\{Q^i_\a,Q^j_\b\}=2(\g^\m P_\m\g^0)_{\a\b}\d^{ij}~.
\eeq  
Taking the trace in the spinor indices $\a$ and $\b$ and choosing $i$ equal to
$j$ without summing, this implies that the Hamiltonian is
\beq
H=P^0=\frac 1 4 \sum_{\a=1}^4 (Q^i_\a)^2,\ \ \ \ i=1,2\ldots {\cal N}~.
\eeq
Since the supercharges are hermitian, the Hamiltonian is non-negative, and only
the vacuum (zero energy) state is annihilated by all supercharges. 

For a massive particle state, in a rest frame in which $P^\m=(M,0,0,0)$,  
\eq{QQP} becomes
\beq\label{QQPmassive}
\{Q^i_\a,Q^j_\b\}=2M\d_{\a\b}\d^{ij}~.
\eeq 
This is a Clifford algebra in $4{\cal N}$ Euclidean dimensions and has a unique
irreducible (complex) representation, which is of complex dimension 
$2^{2{\cal N}}$. For even ${\cal N}$, there exists a real irreducible
representation of real dimension  $2^{2{\cal N}}$. 

For a massless particle state, in a frame in which
$P^\m=(E,E,0,0)$, \eq{QQP} reads
\beq\label{QQPmassless}
\{Q^i_\a,Q^j_\b\}=2E(1+\g^1\g^0)_{\a\b}\d^{ij}~.
\eeq  
The matrix $(1+\g^1\g^0)$ has two eigenvalues equal to 0 and the other two equal to
2; in a certain Majorana basis it equals diag(0,2,0,2). Thus, half of the 
supercharges annihilate the massless state: it is supersymmetric. The other half
generate a Clifford algebra in $2{\cal N}$ dimensions. The unique irreducible
(complex) representation of this algebra has complex dimension $2^{\cal N}$, 
the square root of the dimension of a massive representation. A real
representation of real dimension $2^{\cal N}$ exists unless ${\cal N}=2\ {\rm
mod}\ 4$. 
The representation of dimension $2^{\cal N}$ is called ``short'', the one of 
dimension $2^{2{\cal N}}$ ``long''.

Furthermore, it can be shown that the range of helicities in a massless
representation is at least ${\cal N}/2$; in a massive representation this range 
is at least ${\cal N}$. 

Now, we return to the spontaneously broken gauge theory discussed in
\sect{central:duality}. For 
${\cal N}\ge 2$, the massless vector (with helicities $\pm 1$) and the massless
Higgs field (helicity 0) belong to a single gauge supermultiplet 
(for ${\cal N}= 4$, this multiplet is really irreducible, for
${\cal N}= 2$ it is the sum of two multiplets with helicity ranges from $-1$ 
to 0 and from 0 to 1, respectively). There are two
other such multiplets (containing the vectors that are going to become massive
through the Higgs mechanism and the scalar fields that are going to be ``eaten
up'' by the vectors in that process). The states in the latter multiplets acquire
a mass by the Higgs mechanism. This leads to a paradox: the Higgs mechanism
should not change the number of helicity states, whereas we have just discussed 
that the dimensionalities of the massless and massive representations differ
drastically. The resolution of this paradox is that the supercharges satisfy 
anticommutation relations that are not quite \eq{QQP}: extra terms occur on 
the right hand side. For instance, if ${\cal N}= 2$, we have
\beq\label{QQPcentral}
\{Q^i_\a,Q^j_\b\}=2(\g^\m P_\m\g^0)_{\a\b}\d^{ij}+2\ii
\e^{ij}\left((Z_1-\ii\g_5Z_2)\g^0\right)_{\a\b}~,
\eeq     
where $\e^{ij}$ is the anti-symmetric symbol with $\e^{01}=1$. The charges $Z_1$
and $Z_2$ are called ``central'' because they commute with all the generators of
the algebra. Their inclusion radically changes the 
representations, as we shall now show. Consider the algebra acting on a massive
particle at rest carrying charges $Z_1$ and $Z_2$:
\beq\label{QQPcentralmassive}
\{Q^i_\a,Q^j_\b\}=2M\d_{\a\b}\d^{ij}+2
\e^{ij}\left((Z_1-\ii\g_5Z_2)\g^0\right)_{\a\b}~.
\eeq 
As $\left((Z_1-\ii\g_5Z_2)\g^0\right)^2=-(Z_1^2+Z_2^2)$, the right hand side of 
\eq{QQPcentralmassive} has eigenvalues $2(M\pm\sqrt{Z_1^2+Z_2^2})$, each with
fourfold multiplicity. As the left hand side of \eq{QQPcentralmassive} is a
non-negative matrix (because of the hermiticity of the supercharges), the mass
satisfies the bound
\beq\label{susybound}
M\ge\sqrt{Z_1^2+Z_2^2}~.
\eeq
When this bound is saturated, half of the eigenvalues of the right hand side of 
\eq{QQPcentralmassive} vanish, so that the irreducible representation is short,
like the massless irreducible representation of \eq{QQP}. Thus, for the centrally
extended algebra \eq{QQPcentral},
the Higgs mechanism is possible without changing the number of states, provided 
the bound \eq{susybound} remains saturated.
Since the central charges are conserved (because the supercharges are) and
\eq{susybound} is reminiscent of \eq{bogbound}, it is
plausible that the central charges are related to the electric and magnetic 
charges. Indeed, it has been explicitly checked \cite{wittenolive}
that $Z_1=aQ$ and $Z_2=aG$ (where $Q$ and $G$ are the electric
and magnetic charges, respectively, and $a$ is the magnitude of the expectation
value of the Higgs field). Thus, from this perspective, the Bogomol'nyi bound
\eq{bogbound} is just a consequence of the supersymmetry algebra! The fact that
the massive particles appearing in the previous section saturate this bound
means that they preserve part of the supersymmetry. Such objects are called BPS,
after Bogomol'nyi, Prasad and Sommerfield.

For  ${\cal N}= 4$ rather than ${\cal N}= 2$, there are more central charges, but
the conclusions are essentially unchanged.

We shall now see how the observations in this section shed light on the puzzles
raised at the end of \sect{central:duality}.
\begin{enumerate}
\item In ${\cal N}= 4$, quantum corrections are under control. This ensures that
BPS states constructed at weak coupling evolve smoothly to states at strong
coupling. 

For both ${\cal N}= 2$ and ${\cal N}= 4$, the universal mass formula 
$M= a\sqrt{Q^2+G^2}$  survives as the
coupling constant is increased, because otherwise a short representation should
suddenly turn into a long one.

\item The argument that the states of a given momentum must represent the
supersymmetry algebra applies in particular to magnetic monopoles. For  
${\cal N}= 4$ the smallest massive representation is the gauge multiplet with
helicities ranging from $-1$ to 1. Thus, we have found monopoles with spin 1
\cite{osborn}. 
The gauge multiplet is a short representation, which, as used in the previous
point, implies the saturation of Bogomol'nyi's bound \eq{bogbound}. For 
${\cal N}= 2$, the monopole states form a hypermultiplet, which has helicities
ranging from $-1/2$ to $1/2$.

\item Even though the duality we would like to test relates a strongly coupled
theory to a weakly coupled one, supersymmetry does make some tests possible. 
As mentioned above, for ${\cal N}= 4$  BPS states cannot disappear as 
the coupling constant is increased. Thus, if duality predicts the presence of a
certain BPS state in a certain strongly coupled theory, a corresponding state 
should exist for weak coupling. Such BPS states can be predicted by extending
the duality symmetry to SL(2,$\Zbar$). A concrete test of such a prediction has
been performed in \cite{sen9402}.
\end{enumerate}

\section{Anomalies and anomaly inflow}\label{central:anomalies}
It often happens in quantum field theories that a classical symmetry is broken
at the quantum level. The symmetry in question is then called ``anomalous''.
The presence of an anomaly in a certain symmetry means that the corresponding
current is not conserved, and thus that conservation of the associated charge
is violated. This charge could be, for instance, the electric charge in a U(1) 
gauge theory, or energy and momentum in the case of gravitational anomalies. 
Anomalies in global symmetries are nothing to worry about, but if a gauge
symmetry is anomalous the theory becomes inconsistent. Usually, anomalies in
gauge symmetries are cancelled directly by adding extra fermion species with
appropriate quantum numbers. A famous example is the standard model of
elementary particle physics: for each generation, the anomalies due to the 
different chiral particles cancel. 
In this section, we will describe an alternative way
in which an anomalous gauge theory may be made consistent. This mechanism,
called ``anomaly inflow'', was developed in \cite{CH} and \cite{naculich}.  It
applies to certain defects, like strings or domain walls. If charge is not
conserved on such a defect, this can be compensated by charge inflow from the
outside world. 

Since anomaly inflow, in a different context, is going to play an important role in
this thesis, we shall now supplement this scenario with formulas. In this
section, we will only discuss gauge anomalies caused by the presence of chiral
fermions; the formalism can be extended to gravitational anomalies, and
to theories with a more general chiral matter content.

The gauge field $A$ is a Lie-algebra valued one-form,
\beq
A=A_\mu dx^\mu=A^a_\mu\,t_a\,dx^\mu~,
\eeq
with $t_a$ the Hermitean generators of the gauge group in the representation 
$\r$. They satisfy the commutation relations
\beq
[t_a,t_b]=\ii {C^c}_{ab}t_c~,
\eeq
where ${C^c}_{ab}$ are the (real) structure constants of the gauge group, e.g., 
${C^c}_{ab}=\e_{cab}$ for SU(2).
The generators satisfy
\beq\label{tnormalization}
\tr(t_at_b)=c\,\d_{ab}~,
\eeq
where $c$ is the {\it index\/}%
\footnote{For instance, the spin $j$ representation of SU(2) has index
$j(j+1)(2j+1)/3$.} 
of the representation $\r$.  
The field strength 
\beq
F=\frac 12 F_{\mu\nu}dx^\mu dx^\nu=\frac{1}{2} F^a_{\mu\nu}\,
t_a\,dx^\mu dx^\nu
\eeq
is defined as
\beq
F=dA-\ii A\wedge A=dA-\frac \ii 2 [A,A]~,
\eeq
which reads in components 
\beq\label{convBen}
F^a_{\mu\nu}=\partial_\mu A^a_\nu-\partial_\nu A^a_\mu+{C^a}_{bc}A^b_\mu
A^c_\nu~.
\eeq
For SU(2), this agrees with \eq{Fq} for $q=1$; this means that we have absorbed
the charge $q$ in the gauge field.
The  infinitesimal gauge transformation of $A$ is 
\beq
\d_v A=dv-\ii[A,v]\equiv Dv~,
\eeq 
where $D$ is the gauge covariant derivative.

Consider a (complex) Weyl fermion in $D$ dimensions
coupled to the gauge field $A$ of $F$, and
let $\Gamma[A]$ be the effective action for the gauge field, obtained by 
integrating out the fermion (there is only a classical and a one-loop 
contribution):
\beq
e^{\ii\G[A]}=\int\,[d\,\bar\psi\;d\,\psi]\,e^{\ii S[A,\psi,\bar\psi]}~,
\eeq
where 
\beq
S=-\int\bar\psi\g^\mu(\partial_\mu-\ii A_\mu)\psi~.
\eeq
Then, it is known from topological arguments \cite{alvaginsp} that the gauge 
variation of $\G$ is given by
\beq\label{deltaGamma}
\d\,\G=2\pi\int\,[ch_\r(F)]^{(1)}_D~,
\eeq
where $ch_\r(F)$ is the Chern character in the representation $\r$ of the gauge
group in which the fermion transforms,
\beq
ch_\r(F)=\tr_\r\exp\left(\frac{F}{2\pi}\right)=\sum_{j\ge 0}\frac{1}{j!}\tr
\left(\frac{F}{2\pi}\right)^j~,
\eeq 
and the subscript $D$ denotes that the $D$-form part of the expression in square
brackets should be taken. The superscript (1) in \eq{deltaGamma} refers to the
descent procedure, which we now introduce.

Consider any closed, gauge invariant polynomial $I$ of the gauge field strength 
$F$, which is a two-form on space-time. Being closed, $I-I_0$, where $I_0$
denotes the zero-form term, is exact
(at least locally, but global issues will be neglected throughout this 
discussion):
\beq\label{descent1}
I-I_0=d\,I^{(0)}~.
\eeq   
Moreover, the gauge variation of $I^{(0)}$ is exact:
\beq\label{descent2}
\d\,I^{(0)}=d\,I^{(1)}~.
\eeq

As a simple example of the
descent procedure, it is not difficult to check that for a Weyl fermion in $D=2$
\bea
\tr F^2&=&d\tr(AF+\frac \ii 3 A^3)~;\nonumber\\
\d_v\tr(AF+\frac \ii 3 A^3)&=&d\,\tr (v\,dA)~,\label{descentD=2}
\eea
(the wedge product between forms is suppressed)
so that for a Weyl fermion in $D=2$
\beq\label{anomD=2}
\d_v\G=\frac{1}{4\pi}\int\,\tr (v\,dA)~.
\eeq

Gauge invariance of $\G$ is related to covariant conservation of the current
\beq\label{current}
j^{\mu a}=\frac 1c\frac{\d\G}{\d A^a_\mu}~,
\eeq
as can be seen as follows:
\beq
\d_v\G=\int\,D_\mu v^a\,\frac{\d\G}{\d A^a_\mu}=
-\int c\, v^a\,D_\mu j^{\mu a}~.
\eeq
Thus, using \eq{anomD=2}, the gauge current in $D=2$ satisfies
\beq
D_\mu j^{\mu a}=-\frac{1}{4\pi}\,\epsilon^{\mu\nu}\partial_\mu A^a_\nu~.
\eeq
This is the familiar expression for the anomaly.

Now that we have developed enough machinary, we go on to describe the anomaly
inflow in the model of \cite{CH, naculich}. The model describes a Dirac fermion
in $D=4$, coupled to a gauge field $A$ and to a complex scalar field
$\Phi(x^\mu)=\Phi_1+\ii \Phi_2=f(x^\mu)e^{\ii \t(x^\mu)}$ with vacuum
expectation value $|\Phi|=\mu>0$, where $\t$ is called the axion field. A string
configuration is described by $f(x^\mu)=f(r),\ \t(x^\mu)=\phi$, where $r$,
$\phi$ are polar coordinates in the $yz$-plane (or 23-plane). The 
function $f$ is such that $f(r)\rightarrow 0$ as $r\rightarrow 0$ and 
$f(r)\rightarrow \mu$ as $r\rightarrow \infty$. Such a string is called an
``axion string''. The Lagrangian describing the
fermions is
\beq\label{4dimtheory}
L=-\bar\psi\gamma^\mu(\partial_\mu-\ii A_\mu)\psi
-\bar\psi(\Phi_1+\ii \gamma^5\Phi_2)\psi~,
\eeq
where $\g^5=\ii \g^0\g^1\g^2\g^3$.
Consider the Dirac equation for zero background gauge field, in the axion string
background:
\beq
-\gamma^\mu\partial_\mu\psi-f(r)(\cos\phi+\ii\sin\phi\,\g^5)\psi=0~.
\eeq
It is easy to check that this equation is solved by
\beq\label{ansatz}
\psi=\eta(x^0,x^1)\,\exp\left(-\int_0^rf(s)ds\right)
\eeq
if
\bea
-\gamma^\a\partial_\a\eta&=&0~;\label{masslessdirac}\\
\g^0\g^1\eta&=&\eta~;\label{chiral}\\
\g^2\eta&=&\eta~,\label{half}
\eea
where $\a$ runs over the values 0,1. To interpret this solution, first note that
\eq{ansatz} implies that the wave function is localized near the string (at
$r=0$): since $f(r)\rightarrow \mu>0$ for $r\rightarrow \infty$
the wave function is exponentially suppressed far away from the string. It
follows from \eq{masslessdirac} that the solution satisfies the massless Dirac
equation in two dimensions, while \eq{chiral} states that it is chiral in two
dimensions. The other equation, \eq{half}, implies that we have precisely one
(complex) Weyl spinor in two dimensions. Since the net number of chiral fermions
(by which we mean the number of chiral minus the number of antichiral fermions)
is invariant under continuous deformations of the background (it is given by an
index theorem), this chiral fermion will not disappear when a (topologically
trivial) background gauge field is turned on. 

This raises a puzzle. The original four-dimensional theory \eq{4dimtheory} does 
not have a gauge
anomaly, because the gauge fields have no chiral couplings. However, the 
fermion zero modes on the string, 
described by a Weyl fermion in two dimensions, give rise to a gauge
anomaly, meaning that the gauge charge they carry is not conserved. For the 
current
\beq
j^{\a a}=\frac1c\frac{\d\G_2}{\d A_\a^a}
\eeq
derived from the two-dimensional action for the fermion zero modes, this means
\beq\label{2dimanom}
D_\a j^{\a a}=-\frac{1}{4\pi}\,\epsilon^{\a\b}\partial_\a A^a_\b~.
\eeq
As charge is
conserved in the full, four-dimensional theory, the resolution must be that
charge flows from the bulk of spacetime onto the string, thus compensating for
the non-conservation of charge on the string, as we shall now describe.

The anomaly in \eq{2dimanom} tells us that the part of the fermion determinant
$\G$ coming from the fermion zero modes (denoted $\G_2$ above) is not invariant
under gauge transformations. The rest of the fermion determinant (coming from
the massive degrees of freedom which live off the string) must cancel this
variation and restore gauge invariance of the full theory. Indeed, 
in the thin string approximation ($f(r)=\mu$ for $r>0$) these massive
modes can be shown to mediate the following effective interaction between the 
axion field and the background gauge field:
\bea
S_\t&=&-\frac{1}{8\pi^2}\int\,d\t\,\wedge\,\tr\left(AF+\frac \ii3 A^3\right)
\nonumber\\
&=&-\frac{1}{8\pi^2}\int\,d^4x\,\e^{\m\n\r\s}\partial_\m\t\;\tr\left(A_\n
\partial_\r A_\s-\frac {2\ii}3 A_\n A_\r A_\s\right)~.
\eea
In this approximation, the string is singular, but this can be accounted for by
considering $d^2\t$ as  $2\pi$ times a delta-function, or rather a two-form with
delta-function support, in the
two dimensions transverse to the string, e.g.,
\beq
\int_{D^2}\,d^2\t=\int_{S^1}\,d\t=2\pi~.
\eeq
Here, $D^2$ denotes a disc around the origin in the two dimensions transverse 
to the string and $S^1$ is its boundary. Using this and \eq{descentD=2}, one
sees that the variation of $S_\t$ precisely cancels the anomalous
variation \eq{2dimanom} of $\G_2$.

This resolves the puzzle mentioned above. It is instructive to go one step
further and actually compute the current $J_\t$ induced by $S_\t$. The
result is
\bea
{J_\t}^{\m a}&=&\frac1c\frac{\d S_\t}{\d A^a_\m}\nonumber\\
&=&\frac{1}{8\pi^2}[\e^{\m\n\r\s}\partial_\n\t\,F^a_{\r\s}-
\e^{\m\n\r\s}\partial_\s\partial_\nu\t\,A^a_\r]\nonumber\\
&=&\frac{1}{8\pi^2}[\e^{\m\n\r\s}\partial_\n\t\,F^a_{\r\s}-
2\pi\d_\perp\,\e^{\m\n}\,A^a_\n]\nonumber\\
&\equiv&J_\infty+\D J~,
\eea
where $\e^{\m\n}=\e^{\a\b}$ for $\m,\n=0,1$ and vanishes otherwise.
We would like to make a few comments about this result. First, as a cross-check,
it can be checked explicitly that $j+J_\t$ is indeed covariantly conserved.
Second, the current $J_\t$ consists of two pieces. The first piece, $J_\infty$,
is localized in the whole spacetime, is covariant, and describes the inflow of 
charge from infinity onto the string. The second piece, $\D J$, is localized on
the string; it is the contribution of the massive fermion modes to the current
flowing on the string, and should be added to $j$, the contribution of the zero
modes. The individual pieces $j$ and $\D J$ are non-covariant, but it is easy to
see that their sum has a covariant divergence (the current itself is also known
to be covariant). 

This has a beautiful interpretation in terms of ``consistent'' and ``covariant''
anomalies. A consistent anomaly is the gauge variation of an effective action
and as such satisfies certain consistency conditions. These conditions imply
that the gauge anomaly cannot have a covariant form
(except in the case of an abelian gauge symmetry, where it is accidentally
covariant) and thus that
the corresponding current as defined in \eq{current} cannot be covariant. 
On the other hand, it is possible to define a different current that
is covariant and whose covariant divergence gives a covariant expression for the
``anomaly''. The latter ``anomaly'', which cannot be the gauge variation of an
effective action, is called the ``covariant anomaly''. Its physical meaning was
less clear, until the work we have just described appeared  \cite{naculich}. The term
$\D J$, which appeared above in a natural way, turns out to be precisely the
term one has to add to the consistent current to transform it into the covariant
one. The total current flowing on the string is the current that gives rise to
the covariant anomaly, thus providing a physical realization of the latter. Note
that the covariant anomaly is not the variation of a two-dimensional action:
part of it is provided by the variation of a higher-dimensional action.

%Type II superstrings
\chapter{Strings and branes}\label{superstrings}
This chapter is not meant to give an extensive review of string theory: much
more space and time would be needed for that. Rather, we want to highlight those
features of string theory that will play a prominent role in the second half of
this thesis, where our own results are presented. Readers who are interested in
more thorough introductions to string theory are referred to the many good text
books on the subject, e.g., \cite{polchinski, gsw, LustTheisen, kiritsis}.

One concrete aim is to enable the reader who has some knowledge of quantum field
theory and is willing to work his way through the present chapter, to understand
the computations in \chap{anomalous}. Readers who are not interested in all the
technical details can safely skip anything referring to ghosts, superghosts and
pictures: although these concepts are essential to justify the concrete form of
the amplitudes we compute, they do not play a significant role in the actual 
computations in \chap{anomalous}.  

%Henceforth we shall be using slightly different conventions than before. We take
%the metric signature to be ($-+\ldots+$) rather than ($+-\ldots-$). We shall use
%a new gauge potential, which is related to the old one by
%\beq
%A_{\rm new}=\ii A_{\rm old}~.
%\eeq
%Its field strength and the gauge transformation now read
%\bea
%F&=&dA-\ii A\wedge A\\
%\d A&=&d\l-\ii[A,\l]~,
%\eea
%where the new parameter of the gauge transformation is related to the old one by
%\beq
%\l=\ii v~.
%\eeq
%FIXED in previous chapter
%%%%%%%%%%%%%%%%%%%%%%%%%%%%%%%%%%%%%%%%%%%%%%%%%%%%%%%%%%%%%%%%%%%%%%%%%%%%%%%%
\section{Strings}\label{superstrings:strings}
\subsection{World-sheet point of view}\label{superstrings:strings:worldsheet}
\subsubsection{The bosonic string}
\label{superstrings:strings:worldsheet:bosonic}
A bosonic string in a $D$-dimensional trivial
Minkowski background is described by the Polyakov action
\beq \label{polyakov+chi}
S=S_X+\lambda\chi
\eeq
with
\beq \label{polyakov}
S_X=\frac{1}{4\pi\a'}\int_M d^2\s \sqrt{g} g^{ab}\partial_a X^\mu\partial_b
X^\nu \eta_{\mu\nu}~.
\eeq
In the previous formulae, $\l$ is a constant, the meaning of which will become
clear soon, $\chi$ is the Euler number of the world-sheet $M$ and $\eta_{\m\n}$ is
the
spacetime Minkowski metric of signature $(-++\cdots +)$. Further, $\s\equiv
(\s^1, \s^2)$ denotes the world-sheet coordinates. The fields
$X^\mu(\s)$ describe the embedding of the world-sheet in spacetime, and
$g_{ab}(\s)$ is a Euclidean world-sheet metric of signature $(+,+)$. Note that
this metric is non-dynamical; eliminating it from the action by using its classical
equation of motion gives rise to the Nambu-Goto action, which is
proportional to the area of the world-sheet. The constant $\a'$ has the
dimension of spacetime-length-squared and is related to the string tension $T$
by $T=1/2\pi\a'$.

The action \eq{polyakov} is invariant under the following symmetries:
\begin{enumerate}
\item $D$-dimensional Poincar\'e transformations, acting on $X^\mu$.
\item Diffeomorphisms of the worldsheet (under which $X^\mu$ transforms as a
scalar and $g$ as a covariant tensor), in short ``diff invariance''.
\item Weyl transformations, i.e., $\s$-dependent rescalings  of $g_{ab}$. This
symmetry is typical of a two-dimensional world-sheet.
\end{enumerate}

The variation of the action with respect to the metric $g$ defines the
energy-momentum tensor:
\bea
T^{ab}(\s)&=&-\frac{4\pi}{g(\s)^{1/2}}\frac{\d}{\d g_{ab}(\s)}S\nonumber\\
\label{energymomentum}
&=&-\frac{1}{\a'}(\partial^aX^\mu\partial^bX_\mu-\frac 12
g^{ab}\partial_cX^\mu\partial^cX_\mu)~.
\eea
% - sign agrees with (2.3.15), not with (3.4.5)
Diff invariance implies that $T$ is conserved, while it is traceless as a
consequence of Weyl invariance.

\subsubsection{Path integral quantization}
\label{superstring:strings:worldsheet:path}
\begin{figure}
\begin{center}
\epsfig{file=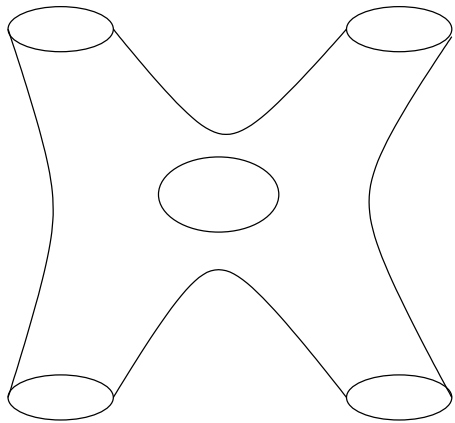}
\end{center}
\caption{A loop amplitude in string theory.}\label{fig:WS}
\end{figure}

A path integral prescription can be given to quantize the string. Define an
amplitude in string theory by summing over all world-sheets interpolating
between given initial and final curves (see \fig{fig:WS}). 
Each world-sheet $M$ is weighted by
$\exp(-S)$, leading to the naive path integral
\beq\label{naivepath}
\int[dX\,dg]e^{-S}~.
\eeq
Before refining this expression, we pause a moment to interpret the $\l\chi$
term in \eq{polyakov+chi}. As the Euler number $\chi$ is a topological
invariant, the $\exp(-\l\chi)$ factor in the path integral governs the relative
weighting of different topologies in the sum over world-sheets. For instance,
adding a handle to a given world-sheet reduces the Euler number by 2 and thus
adds a factor $\exp(2\l)$. Such a handle is the closed string analogue of a loop
in a Feynman diagram in field theory. Thus, it follows that the closed string
coupling constant is proportional to  $\exp(\l)$.

The reason why we called the path integral \eq{naivepath} naive is that it
includes an (infinite) overcounting: if $(X, g)$ and  $(X', g')$ are related by
a diff $\times$ Weyl transformation they represent the same physical
configuration. This overcounting can formally be accounted for by dividing out
the (infinite) volume of the group of diff $\times$ Weyl transformations:
\beq\label{path}
Z\equiv\int\frac{[dX\,dg]}{V_{{\rm diff}\times{\rm Weyl}}}e^{-S}~.
\eeq
Roughly speaking, the idea is now to use the diff $\times$ Weyl symmetry to fix
the metric to
\beq\label{unit}
g_{ab}(\s)=\d_{ab}
\eeq
and then to drop the integration over metrics. The gauge \eq{unit} is called
{\it unit gauge}. This gauge choice is possible at least locally on the
world-sheet (global issues will show up later).

For now, we will study the action we get from \eq{polyakov} by simply
replacing $g$ by $\d$ as in \eq{unit}. Some of the things that will have to be
supplemented later are the following.
\begin{enumerate}\label{remarks}
\item \label{remark1}
Varying the action \eq{polyakov} with respect to $g_{ab}$ gives rise to
the equation of motion $T^{ab}=0$. If we use the gauge-fixed action, this should
be imposed as a constraint.
\item \label{remark2}
The diff $\times$ Weyl leaves the Polyakov action invariant, but this does
not imply that it is really a symmetry of the quantum theory. Indeed, whereas it
is possible to define the quantum theory in a way that preserves diff invariance,
this will generically not be the case for the Weyl invariance. The requirement
that the quantum theory be gauge invariant imposes strong restrictions on the
theory (for instance, the Polyakov action \eq{polyakov} suffers from a Weyl
anomaly unless the spacetime dimension is 26).
\item \label{remark3}
Once we have fixed the metric to its reference value $\d$, for which we
have used the diff $\times$ Weyl symmetry of the action \eq{polyakov}, we might
hope to have eliminated the overcounting in the naive path integral. This is
not quite correct, as we shall see in the next paragraph.
\item
In fact, all of these issues can be dealt with by introducing 
{\it ghost fields}. This will be done later in this section. 
\end{enumerate}

\subsubsection{Conformal invariance}
Using a complex world-sheet coordinate $z$, which equals $\s^1+\ii\s^2$ or a
holomorphic function of it, and its complex conjugate $\bar z$, the action
\eq{polyakov} in unit gauge reads%
\footnote{
We are using the convention that
\beq
d^2z=dx\,dy
\eeq
for a complex variable  $z=x+\ii y$. This differs from the conventions in
\cite{polchinski} by a factor of 2.
}
\beq\label{polyakovzzbar}
S=\frac{1}{\pi\a'}\int d^2z\,\partial X^\mu\bar\partial X_\mu~.
\eeq

Before discussing the precise way in which to implement the constraints, we
study the gauge-fixed action \eq{polyakovzzbar} by itself. As long as we do not
worry about global issues, there is more gauge freedom than we needed to bring
the metric in unit gauge: there exist diffeomorphisms that leave the metric
invariant up to a Weyl transformation. These diffeomorphisms are called
conformal transformations. The corresponding diff $\times$ Weyl transformations
preserve the unit gauge. In terms of the complex coordinate $z$ introduced 
above,
the conformal transformations are the holomorphic coordinate transformations
\beq \label{conformalzzbar}
z'=f(z)~;~\bar z'=\bar f(\bar z)~.
\eeq
Actually, if we regard $z$ and $\bar z$ as independent coordinates, which
corresponds to letting $\s^1$ and $\s^2$ take complex values, the symmetry is
enlarged to independent transformations of $z$ and $\bar z$ (i.e., $f$ and $\bar
f$ need not be related by complex conjugation).

An alternative way of arriving at the conformal symmetry is by starting from the
energy-momentum tensor \eq{energymomentum}. In complex coordinates, its
tracelessness is expressed by
\beq
T_{z\bar z}=0~,
\eeq
while its conservation is then equivalent to
\beq
\bar\partial T_{zz}=\partial T_{\bar z\bar z}=0~,
\eeq
implying that
\beq
T(z)\equiv T_{zz}(z)\ {\rm and}\ \tilde T(\bar z)\equiv T_{\bar z\bar z}(\bar z)
\eeq
are respectively holomorphic and antiholomorphic. Now, it is obvious that,
because of the tracelessness of $T_{ab}$, the currents
\beq\label{confcurrents}
j(z)=\ii v(z)T(z)~,\ \ \tilde j(\bar z)=\ii \bar v(\bar z)\tilde T(\bar z)
\eeq
are conserved for arbitrary holomorphic $v(z)$. Again, the symmetry is enlarged
to independent holomorphic and antiholomorphic transformations if we extend the
range of $z$ and $\bar z$ beyond the ``real surface'' where they are each other's
conjugates. It follows from \eq{confcurrents} that the generators of the
conformal symmetry are the moments $L_m$ and $\tilde L_m$ of the energy-momentum
tensors:
\beq\label{Tinmodes}
T(z)=\sum_{m=-\infty}^\infty L_m z^{-m-2}~,\ \
\tilde T(\bar z)=\sum_{m=-\infty}^\infty \tilde L_m \bar z^{-m-2}~.
\eeq
It can be shown that, for any conformal field theory, the conserved charges
$L_m$ and $\tilde L_m$ satisfy the {\it Virasoro algebra}
\beq\label{Virasoro}
[L_m, L_n]=(m-n)L_{m+n}+\frac{c}{12}(m^3-m)\d_{m,-n}~,
\eeq
and analogously for the $\tilde L_m$ (with central charge or {\it Virasoro
anomaly} $\tilde c$).
One can prove
%gsw p. 146
that world-sheet energy-momentum conservation necessarily
breaks down (and thus the theory
is inconsistent) unless $c=\tilde c$. Moreover, if $c=\tilde c$ and
energy-momentum is conserved, the classical Weyl invariance is broken at the
quantum level unless $c=\tilde c=0$.  The theory as we have described it so far (by the action
\eq{polyakovzzbar}) is not Weyl invariant ($c=\bar c\neq 0$), but Weyl
invariance will be restored when the full theory (including ghosts) is
considered.

A special case of a conformal transformation is
\beq\label{scale}
z'=\zeta z
\eeq
for constant complex $\zeta$. These transformations are generated by $L_0$.
Usually, one chooses a basis of local operators that transform as follows under
\eq{scale}:
\beq\label{eigenop}
{\cal A}'(z',\bar z')=\zeta^{-h}\bar\zeta^{-\tilde h}{\cal A}(z,\bar z).
\eeq
The $(h,\tilde h)$ are known as the {\it conformal weights} of ${\cal A}$. An
important special class of operators are the {\it tensor operators} or {\it
primary fields} ${\cal O}$, on which a general conformal transformation acts as
\beq\label{tensor}
{\cal O}'(z',\bar z')=(\partial_zz')^{-h}(\partial_{\bar z}\bar z')^{-\tilde h}
{\cal O}(z,\bar z)~.
\eeq

\subsubsection{Superstrings}
So far, we have only discussed the bosonic string, which will not play a
significant role in this thesis. It is possible to generalize the above
discussion to the superstring. The corresponding action before gauge fixing
would have local worldsheet supersymmetry%
\footnote{ 
Note that this is not the same as spacetime supersymmetry, which is not
manifest in this formalism, which is called Ramond-Neveu-Schwarz formalism. 
There exists a formulation with manifest
spacetime supersymmetry, called the Green-Schwarz formulation. We shall refer to
the latter in \sect{type0:dual}.
}
invariance in addition to the diff
$\times$ Weyl invariance of the bosonic string.
The action in gauge-fixed form
is
\beq\label{gaugefixed}
S=\frac{1}{2\pi}\int d^2z\,\left(\frac{2}{\a'}\partial X^\mu\bar\partial X_\mu
+\psi^\mu\bar\partial\psi_\mu+\tilde\psi^\mu\partial\tilde\psi_\mu\right)~.
\eeq
The fields $\psi^\mu$ and $\tilde\psi^\mu$ are world-sheet Majorana-Weyl spinors
of opposite chirality, and spacetime vectors. As before, spacetime indices are
raised and lowered with the Minkowski metric $\eta_{\mu\nu}$.
Again, the gauge-fixed action should be accompanied by constraints. 
In addition to the constraint that the world-sheet energy-momentum tensor should
vanish, we now have to impose the vanishing of the {\it world-sheet
supercurrents}
\beq
T_{\rm F}(z)=\ii\sqrt{\frac{2}{\a'}}\psi^\mu(z)\partial X_\mu(z)~;\ \
\tilde T_{\rm F}(\bar z)=\ii\sqrt{\frac{2}{\a'}}\tilde\psi^\mu(\bar z)
\bar\partial X_\mu(\bar z)~.
\eeq
Weyl invariance now requires the
spacetime dimension to be 10, to be compared with the 26-dimensional spacetime
of the bosonic string.

\subsubsection{Closed and open strings}\label{closed-open}
\begin{figure}
\begin{center}
\epsfig{file=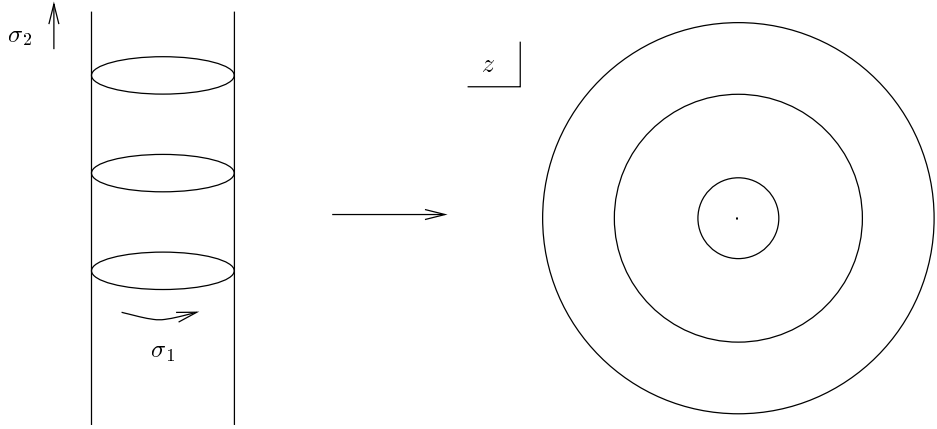}
\end{center}
\caption{Conformal map from the cylinder to the complex plane. Some equal time
contours are displayed.}\label{fig:closed}
\end{figure}

Until now, we have not
specified the topology of the world-sheet. For a {\it closed string},
identify the
world-sheet space coordinate $\s^1$ periodically: $\s^1\sim\s^1+2\pi$. With a
Euclidean time coordinate $-\infty<\s^2<\infty$, the world-sheet becomes an
infinite cylinder. This cylinder can be described with the complex coordinate
$w=\s^1+\ii\s^2$ satisfying $w\sim w+2\pi$, or by the complex coordinate
$z=e^{-\ii w}=e^{-\ii\s^1+\s^2}$, which is related to $w$ by a conformal
transformation. This conformal transformation maps the cylinder to the complex
plane, such that $\s^2=-\infty$ corresponds to the origin of the $z$-plane and
$\s^2=+\infty$ to the point at infinity of the compactified $z$-plane (see
\fig{fig:closed}).

We shall later apply canonical quantization to the action \eq{gaugefixed}, defined
on the world-sheet of the string. To do that we shall use the coordinate $\s_2$ as
the time coordinate. Thus time translations correspond on the complex plane to the
transformations \eq{scale} with real $\zeta$. As a consequence, time translations 
are generated by $L_0+\tilde L_0$. In this way,  $L_0+\tilde L_0$ gets an
interpretation as a Hamiltonian on the world-sheet. 

The action \eq{gaugefixed} leads to the equations of motion
\bea
\partial\bar\partial X^\mu(z,\bar z)&=&0~;\label{Xeom}\\
\bar \partial\psi(z,\bar z)&=&0~;\label{psieom}\\
\partial\tilde\psi(z,\bar z)&=&0~.\label{psibareom}
\eea
The equation of motion for $X^\mu$ implies that $\partial X^\mu$ is holomorphic
and $\bar \partial X^\mu$ antiholomorphic, so we have the following Laurent
expansions:
\bea
\partial X^\mu(z)&=&-\ii\sqrt{\frac{\a'}{2}}\sum_{m=-\infty}^\infty
\a^\mu_mz^{-m-1}~;\nonumber\\
\bar\partial X^\mu(\bar z)&=&-\ii\sqrt{\frac{\a'}{2}}\sum_{m=-\infty}^\infty
\tilde\a^\mu_m\bar z^{-m-1}~.\label{dXinmodes}
\eea
The 1 in the exponent of $z$ is the holomorphic conformal weight of $\partial X$
($\partial X$ is a primary field with $(h,\tilde h)=(1,0)$). Its appearance in
this expansion is due to the transformation properties of $\partial X$ under
the conformal transformation between the cylinder and
the plane (see \eq{tensor}). We always choose the modes such that they are the
usual Fourier modes on the cylinder. For instance, on the cylinder the mode
corresponding to $\a_0$ is $\s^1$-independent. Note that this also explains the
2's in \eq{Tinmodes}: up to the Virasoro anomaly $T(z)$ is a $(2,0)$ tensor
operator.

The equations \eq{dXinmodes} can be integrated to give
\bea
X^\mu(z,\bar z)&=&x^\mu-\ii \frac{\a'}{2}p^\mu\ln|z|^2
+\ii\sqrt{\frac{\a'}{2}}\sum_{\stackrel{m=-\infty}{m\neq 0}}^\infty
\frac 1m(\a_m^\mu z^{-m}+\tilde\a_m^\mu\bar z^{-m})\nonumber\\
&\equiv&X^\mu(z)+\tilde X^\mu(\bar z)~,\label{XplusXbar}
\eea
where $p^\mu\equiv\sqrt{2/\a'}\a_0\equiv\sqrt{2/\a'}\tilde\a_0$, and $x^\mu$ is
equally distributed between $X^\mu(z)$ and $\tilde X^\mu(\bar z)$. The overall
motion of the string is described by $x^\mu$ and $p^\mu$, while the ``oscillator
modes'' $\a_m^\mu$ and $\tilde\a_m^\mu$ describe the oscillations of the string.

For the world-sheet fermions, we allow more general periodicity conditions. We
start with the coordinate $w=\s^1+\ii\s^2$ on the cylinder. The fermion action
\beq
\frac{1}{2\pi}\int d^2w(\psi^\m\partial_{\bar w}\psi_\mu+
\tilde\psi^\m\partial_{w}\tilde\psi_\mu)
\eeq
must be invariant under the periodic identification of the cylinder, $w\sim
w+2\pi$. This condition plus ten-dimensional Lorentz invariance leaves two 
possible periodicity
conditions for $\psi^\mu(w)$:
\bea
{\rm Ramond \ (R)}&:&\psi^\mu(w+2\pi)=+\psi^\mu(w)~;\nonumber\\
{\rm Neveu-Schwarz \ (NS)}&:&\psi^\mu(w+2\pi)=-\psi^\mu(w)~,
\eea
with the same sign for all $\mu$. Similarly, there are two possible periodicities
for $\tilde\psi^\mu$. (For $X^\mu$ only the periodic periodicity condition is
possible in a theory with maximal Poincar\'e invariance.)
All in all, there are four different sectors, denoted NS-NS, NS-R, R-NS and R-R.

Taking into account the fact that $\psi^\mu$ and $\tilde \psi^\mu$ are
primary fields of weights
$(1/2,0)$ and $(0,1/2)$, respectively, the conformal transformation from $w$
to $z=e^{-\ii w}$ acts as
follows:
\bea
\psi^\mu(z)&=&(\partial_zw)^{1/2}\psi^\m(w)=\ii^{1/2}z^{-1/2}\psi^\m(w)~;
\nonumber\\
\tilde\psi^\mu(\bar z)&=&(\partial_{\bar z}\bar w)^{1/2}\tilde\psi^\m(w)=
\ii^{-1/2}\bar z^{-1/2}\tilde\psi^\m(\bar w)~.
\eea
These fields have ``Laurent'' expansions
\beq\label{psiinmodes}
\psi^\mu(z)=\sum_{r\in\Zbar+\nu}\psi^\m_rz^{-r-1/2}~;\ \
\tilde\psi^\mu(\bar z)=\sum_{r\in\Zbar+\tilde\nu}\tilde\psi^\m_r
\bar z^{-r-1/2}~,
\eeq
where $\nu=0$ in the Ramond sector and $\nu=1/2$ in the Neveu-Schwarz sector.

Canonical quantization gives rise to the (anti-)commutation relations
\bea
[\a_m^\m,\a^\n_n]&=&[\tilde\a_m^\m,\tilde\a^\n_n]=m\eta^{\m\n}\d_{m+n}~;\ \
[x^\m,p^\n]=\ii\eta^{\m\n}~;\label{bosonalgebra}\\
\{\psi^\m_r,\psi^\n_s\}&=&\{\tilde\psi^\m_r,\tilde\psi^\n_s\}=\eta^{\m\n}
\d_{r+s}~.\label{fermionalgebra}
\eea

The supercurrents have the Laurent expansions
\beq\label{TFinmodes}
T_{\rm F}(z)=\sum_{r\in\Zbar+\nu}G_rz^{-r-3/2}~,\ \
\tilde T_{\rm F}(\bar z)=\sum_{r\in\Zbar+\tilde\nu}\tilde G_r\bar z^{-r-3/2}~.
\eeq

\begin{figure}
\begin{center}
\epsfig{file=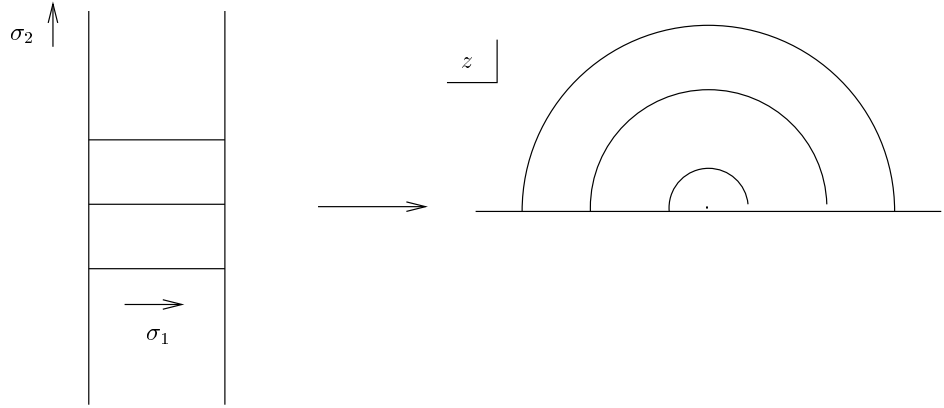,width=12.5cm}
\end{center}
\caption{Conformal map from the strip to the upper half-plane. Some lines of
equal time are displayed.}\label{fig:open}
\end{figure}

For {\it open strings}, let $0\leq\Re w\leq\pi$, which corresponds to $\Im z\geq
0$ if we now define $z=-e^{-\ii w}$. This conformal transformation maps the
strip to the upper half-plane (see \fig{fig:open}). 
The upper half-plane can also be conformally
mapped to the unit disc.

For open strings, the requirement that the action \eq{gaugefixed} be stationary
still implies the equations of motion \eq{Xeom}, \eq{psieom} and \eq{psibareom},
but in addition the fields should satisfy appropriate boundary conditions, so
that the boundary terms in the variation of \eq{gaugefixed} vanish. We will
now discuss these boundary terms, first for the bosonic fields $X^\m$, and then
(page~\pageref{fermionicbc}) for the fermionic fields $\psi^\m$.

If we take the world-sheet to be the upper half-plane, the boundary terms
involving $X^\mu$ are proportional to
\beq
\int dx \d X^\mu(\partial X_\mu-\bar\partial X_\mu)~.
\eeq
One way to make this vanish is by choosing the Neumann boundary condition
$\partial X^\mu=\bar \partial X^\mu$, which preserves the Poincar\'e symmetry.
Another possible choice is the Dirichlet boundary condition $\d X^\m=0$, which
means that the endpoints of the open string are fixed in the $\m$-direction. 
The boundary conditions
can be chosen independently for each value of $\m$ and for both boundary
components (the negative and the positive real axes, corresponding to $\s^1=0$
and $\s^1=\pi$ in the original coordinates on the strip). For instance, we can
consider an open string whose first endpoint is constrained to a
($p+1$)-dimensional hyperplane $X^\m=x^\m,\;\m=p+1,\ldots,9$, and whose second
endpoint must lie on a
($p'+1$)-dimensional hyperplane $X^\n=y^\n,\;\n=p'+1,\ldots,9$.

When strings with Dirichlet boundary conditions are introduced, the Poincar\'e
symmetry is broken by these boundary conditions. The hyperplanes to which the
endpoints of certain open strings are confined are called {\it D-branes}. These
objects, which are going to play a prominent role in this thesis, will be
introduced in more detail in \sect{superstrings:D-branes}. 

Open string fields have only one set of oscillators. We explain this in detail
for a field $X^\m$ with Neumann boundary
conditions at both endpoints (NN). The boundary conditions require $\a^\m_m=
\tilde\a^\m_m$ in the expansion \eq{XplusXbar}. The momentum
is now%
\footnote{The difference with the closed string is that the world-sheet space 
coordinate $\s^1$ of the open string takes values in $[0,\pi]$ rather than 
$[0,2\pi]$. The momentum one computes using Noether's theorem gives $p^\m$ in
both cases.}
$p^\m=\a_0^\m/\sqrt{2\a'}$. The expansion is
\beq
X^\m(z,\bar
z)=x^\m-\ii\a'p^\m\ln|z|^2+\ii\sqrt{\frac{\a'}{2}}\sum_{\stackrel{m=-\infty}{m\neq
0}}^{\infty}\frac{\a^\m_m}{m}(z^{-m}+\bar z^{-m})~.
\eeq
The modes are normalized so as to satisfy the usual commutation relation, given
in \eq{bosonalgebra}.

These boundary conditions can equivalently be imposed by using a {\it doubling
trick}.%
\footnote{Such a trick will be used in \sect{nonbps:anomalous}.}
First, use the equation of motion \eq{Xeom} to write $X^\mu(z,\bar z)$
as the sum of a holomorphic and an antiholomorphic part. To avoid confusion, we
write this as follows in terms of real coordinates $(x,y)$ on the upper
half-plane $y\geq 0$ ($z=x+\ii y$):
\beq\label{xy}
X^\mu(x,y)=X^\mu_{\rm hol}(x,y)+X^\mu_{\rm antihol}(x,y)~.
\eeq
Because of the Neumann boundary conditions, $X^\mu_{\rm hol}$ and
$X^\mu_{\rm antihol}$ can be combined into a single holomorphic field
$X^\mu_{\rm hol}$ defined on the whole plane by extending it as follows:
\beq
X^\mu_{\rm hol}(x,y)\equiv X^\mu_{\rm antihol}(x,-y)\ \ {\rm for}\ y<0~.
\eeq
Going back to the complex coordinate $z$, this extended holomorphic field has
mode expansion
\beq
X^\m(z)=\frac{x^\m}{2}-\ii\a'p^\m\ln z+\ii\sqrt{\frac{\a'}{2}}
\sum_{\stackrel{m=-\infty}{m\neq
0}}^{\infty}\frac{\a^\m_m}{m}z^{-m}~.
\eeq
This is very reminiscent of the holomorphic closed string field $X^\mu(z)$ in
\eq{XplusXbar}. 
For many purposes open strings are almost equivalent to
one sector (say the holomorphic one) of closed strings.

If the field $X^\m$ satisfies different boundary conditions, it is still
described by a single set of oscillators, but their moding varies.
For DD boundary conditions, $X^\m$ remains integer moded but the zero mode
fluctuations $x^\m$ and $p^\m$ are absent. For ND or DN boundary conditions, the
mode numbers are half-integer (and thus there are no zero modes).

\label{fermionicbc}
We now turn to the boundary conditions on the fermionic fields.
In $w$ coordinates, the condition that the boundary terms 
of the variation of the fermion action vanish takes the form
\beq
(\psi^\m\d\psi_\m+\tilde \psi^\m\d\tilde\psi_\m)|_{\rm boundary}=0~,
\eeq
which is solved in a Lorentz invariant way by imposing
\beq\label{fermionBC}
\psi^\m(0,\s^2)=\ii e^{2\pi\ii\n}\tilde\psi^\m(0,\s^2)~;\ \
\psi^\m(\pi,\s^2)=\ii e^{2\pi\ii\n'}\tilde\psi^\m(\pi,\s^2)
\eeq
at the boundary. Here, $\n$ and $\n'$ can take the values 0 and $1/2$. This
choice can be made independently for both boundary components, but by redefining
$\tilde\psi^\m\rightarrow e^{-2\pi\ii\n'}\tilde\psi^\m$ we can set $\n'=0$.
Therefore, there are two possibilities, named R sector ($\n=0$) and NS sector 
($\n=1/2$). It is
convenient to combine $\psi^\m$ and $\tilde\psi^\m$ in a single holomorphic
field with extended range $0\leq\s^1\leq 2\pi$ by defining
\beq
\psi^\m(\s^1,\s^2)=\ii\tilde\psi^\m(2\pi-\s^1,\s^2)
\eeq
for $\pi<\s^1\leq 2\pi$. This field is periodic in the R sector and antiperiodic
in the NS sector. Therefore, we have a single set of R or NS oscillators and the
corresponding algebra \eq{fermionalgebra}, as opposed to the two sets we had for
closed strings (see \eq{psiinmodes}). This is analogous to the situation for the
bosonic fields.

For open strings, the energy-momentum tensor satisfies the boundary condition
\beq
T_{ab}n^at^b=0~,
\eeq
where $n^a$ and $t^b$ are normal and tangent vectors to the boundary of the
world-sheet. This expresses that (world-sheet) translation invariance is
unbroken in the tangential direction. In terms of the $(z,\bar z)$ coordinates,
this means
\beq
T_{zz}=T_{\bar z\bar z}\ \ {\rm for}\ \Im z=0~.
\eeq
The fields $T_{zz}(z)$ and $T_{\bar z\bar z}(\bar z)$ can be combined into a
single holomorphic field $T_{zz}$ on the whole $z$-plane:
\beq\label{extended}
T_{zz}(z)\equiv T_{\bar z\bar z}(\bar z')\ \ {\rm for}\ \Im z<0~.
\eeq
Here, the prime denotes complex conjugation ($z'=\bar z$).%
\footnote{
Readers who find the notation confusing, may want to write everything in terms 
of real coordinates, as in \eq{xy}.
} 
There is a single set
of Virasoro generators
\bea
L_m&=&\frac{1}{2\pi\ii}\int_C(dz\,z^{m+1}T_{zz}-d\bar z\,\bar z^{m+1}T_{\bar
z\bar z})\nonumber\\
&=&\frac{1}{2\pi\ii}\oint dz\,z^{m+1}T_{zz}(z)~.
\eea
In the first line, the contour $C$ is a semi-circle centered on the origin; in
the second line this has been rewritten in terms of a closed contour and the
extended holomorphic field \eq{extended}. 

Summarizing the oscillator mode analysis, we may say that the open string modes
are analogous to the holomorphic sector modes of the closed string. The latter
contains in addition an independent set of antiholomorphic sector modes.

\subsubsection{Ghosts}
Now, we are ready to discuss the gauge fixing procedure, in particular how the
constraint that the energy-momentum tensor should vanish is imposed. The
procedure we follow is BRST quantization, which is a standard tool in
quantizing gauge theories. The reader is referred to 
\cite{FMS, polchinski, gsw} for the
details of the construction and for a systematic introduction to it; we just
mention some of the key points.

First, one adds Faddeev-Popov ghost fields to the theory: reparametrization
ghosts ($c$ and $\tilde c$) and antighosts ($b$ and $\tilde b$), and
superconformal ghosts ($\gamma$ and $\tilde \gamma$) and antighosts
($\b$ and $\tilde \b$). The first four are often called ``the ghosts'', the last
four ``the superghosts''. Sometimes, all eight are collectively called ``ghosts''.
The gauge-fixed action \eq{gaugefixed} is supplemented with the ghost action
\beq\label{Sgh}
S_{\rm gh}=\frac1\pi\int d^2z\,(b\bar\partial c+\tilde b\partial\tilde
c+\b\bar\partial\gamma+\tilde\b\partial\tilde\gamma)~.
\eeq
The fields $c$, $b$, $\gamma$ and $\b$ are primary fields of conformal
dimensions $(-1,0)$, $(2,0)$, $(-1/2,0)$ and $(3/2,0)$, respectively. The
ghosts $c$ and $b$ are anticommuting scalar fields; the superghosts $\gamma$
and $\b$ are commuting spinors. Analogous remarks apply to the tilded fields.

Although in this gauge the ghost and matter actions are decoupled, the ghosts
and superghosts do couple to the world-sheet metric of a non-flat world-sheet.
Thus, they contribute to the energy-momentum tensor, even if the world-sheet
metric is trivial! It turns out that if the spacetime dimension is 10 (for the
bosonic string this would be 26) the modes of the total energy-momentum tensor
satisfy the Virasoro algebra {\it without central charge}. As a consequence, in
this dimension the Weyl symmetry of the action before gauge-fixing is not
anomalous (see the discussion after \eq{Virasoro}), as anticipated in remark~%
\ref{remark2} on page \pageref{remarks}. 
The ghosts and superghosts also
contribute to the world-sheet supercurrents.

The equations of motion derived from \eq{Sgh} state that the untilded ghosts
and superghosts are holomorphic and the tilded ones antiholomorphic. For closed
strings, the holomorphic and antiholomorphic fields are independent; for open
strings, they can be combined into a single holomorphic field on the complex
plane. Hence, we restrict our attention to holomorphic ghost and superghost
fields on the complex plane.

In $z$ coordinates, the superghosts are periodic in the NS sector and
antiperiodic in the R sector (the
ultimate justification will be that the BRST current is then periodic, so that
it can be integrated to give the BRST charge). The mode expansions are
\bea
b(z)&=&\sum_{m=-\infty}^\infty b_mz^{-m-2}~;\label{binmodes}\\
c(z)&=&\sum_{m=-\infty}^\infty c_mz^{-m+1}~;\label{cinmodes}\\
\b(z)&=&\sum_{r\in\Zbar+\nu}\b_rz^{-r-3/2}~;\label{betainmodes}\\
\gamma(z)&=&\sum_{r\in\Zbar+\nu}\gamma_rz^{-r+1/2}\label{gammainmodes}
\eea
and similarly for the antiholomorphic fields. The (anti)commutation relations
are
\bea
\{b_m,c_n\}&=&\d_{n,-m}\nonumber\\
{}[\gamma_r,\b_s ]
&=&\d_{r,-s}~.\label{ghostalgebra}
\eea

The total (matter + ghost) holomorphic superconformal generators are
\bea
\label{Lm}
L_m&=&\frac12\sum_{n\in\Zbar}:\a^\mu_{m-n}\a_{\mu\,n}:+\frac14\sum_{r\in\Zbar+\n}
(2r-m):\psi^\mu_{m-r}\psi_{\mu\,r}:\\
&&+\sum_{n\in\Zbar}(m+n):b_{m-n}c_n:+\sum_{r\in\Zbar+\n}\frac12(m+2r):\b_{m-r}
\gamma_r:+a\d_{m,0}~;\nonumber\\
G_r&=&\sum_{n\in\Zbar}\a^\mu_n\psi_{\mu\,r-n}-\sum_{n\in\Zbar}[\frac12(2r+n)
\b_{r-n}c_n+2b_n\gamma_{r-n}]~.\label{Gr}
\eea

Here, $:~:$ denotes {\it creation-annihilation normal ordering}, placing all
lowering operators to the right of all raising operators, with a minus sign
whenever anticommuting operators are switched. The lowering operators are the
ones with positive subscript (which lower the $L_0$ eigenvalue) and, for this
purpose, $\a_0$, $b_0$, $\b_0$ and half of the $\psi_0^\mu$ ($\b_0$ and 
$\psi_0^\mu$ only
occur in the Ramond sector). The normal ordering constant $a$ is $-1/2$ in the
NS sector and 0 in the R sector (if spacetime is ten-dimensional).

\subsubsection{Fock space}
As a first step in determining the physical Hilbert space of the (free) string,
we represent the matter and ghost algebras \eq{bosonalgebra},
\eq{fermionalgebra} and \eq{ghostalgebra}. 
We start with eigenstates $|0,k\rangle$ of $p^\m$ (with 
momentum $k^\mu$) that are annihilated by all the oscillator
modes with a positive mode
number. We first discuss the representation of the algebra of the holomorphic
modes; for the closed string, we shall have to reintroduce the antiholomorphic
sector later.

In the NS sector, the ground state with given momentum is uniquely specified if we
further impose
\beq\label{Siegel}
b_0|0,k\rangle_{\rm NS}=0~.
\eeq
The state $|0,k\rangle_{\rm NS}$ is degenerate%
\footnote{The term ``degenerate'' will get its usual meaning once we implement the
constraint $L_0=0$.}
with $c_0|0,k\rangle _{\rm NS}$. (The latter state will never be present in physical
Hilbert spaces.) The full Fock space is built by applying raising
operators (modes with negative subscript) to these states.

In the R sector, there are more zero modes. The ambiguity in the definition of the
ground state due to the superghost zero modes can be removed by demanding that
the ground state be annihilated by $\b_0$. The degeneracy due to the $\psi^\m_0$
modes has more profound consequences. It is clear from \eq{fermionalgebra} that
the $\psi^\m_0$ satisfy the SO(1,9) Clifford algebra.
States span a representation space for this algebra. The irreducible real
representation is 32-dimensional. The operators $\psi^\m_0$ transform
ground states in ground states, so there must be (at least) 32 ground states
transforming as a non-chiral Majorana spinor of SO(1,9). 
Half
of this degeneracy will persist in the physical Hilbert space. This is how
spacetime fermionic states appear in this formalism. As in the NS sector, the
full Fock space is obtained by acting with raising modes on the ground states
(and the states with which they are degenerate).

Thus, for the open string, we have the NS sector with a ground state that 
transforms as a
scalar  under SO(1,9), and the R sector with an SO(1,9) spinor ground state. For
the closed string, we have to reintroduce the antiholomorphic fields, thus finding
four sectors. The ground state is a scalar in the NS-NS sector, a spinor in the
NS-R and R-NS sectors and a bispinor in the R-R sector.

\subsubsection{Physical Hilbert space}
We are now ready to build the physical Hilbert space out of these Fock spaces.
The crucial observation is that the gauge-fixed matter plus ghost action is
invariant under the {\it BRST-transformation}, which is generated by
\beq
Q_{\rm B}=\frac{1}{2\pi\ii}\oint dz\,j_{\rm B}
\eeq
for the open string, and by
\beq\label{QB}
Q_{\rm B}=\frac{1}{2\pi\ii}\oint (dz\,j_{\rm B}-d\bar z\,\tilde j_{\rm B})
\eeq
for the closed string. Here, $j_{\rm B}$ is defined by
\beq
j_{\rm B}=cT^{\rm m}+\gamma T_{\rm F}^{\rm m}+\frac12(cT^{\rm g}+\gamma
T_{\rm F}^{\rm g})~,
\eeq
where the superscripts denote the matter and ghost contributions to the
respective quantities, and $\tilde j_{\rm B}$ is defined analogously. The symbol
$\oint$ denotes a contour integral along a circle around the origin of the
complex plane.

In the critical spacetime dimension ($D=10$ for the superstring), the BRST charge
$Q_{\rm B}$ is nilpotent:
\beq
\{Q_{\rm B},Q_{\rm B}\}=0~.
\eeq
We now impose that {\it physical states} should be annihilated by $Q_{\rm B}$, in
other words they should be {\it BRST-closed}:
\beq\label{BRSTclosed}
Q_{\rm B}|{\rm phys}\rangle =0~.
\eeq
We anticipate that the inner product on the Fock space will be defined in such a
way that
\beq
Q_{\rm B}^\dagger=Q_{\rm B}~.
\eeq
Then, it follows that {\it BRST-exact} states (obtained by acting with $Q_{\rm B}$
on another state) are orthogonal to all physical states. Hence, two physical
states that differ by an exact state are physically equivalent. The {\it physical
Hilbert space} is defined as a set of equivalence classes of BRST-closed modulo
BRST-exact states. This is called the cohomology of $Q_{\rm B}$. It turns out
that this procedure is not sufficient as it stands: we have to impose the
additional conditions
\beq\label{b0}
b_0|{\rm phys}\rangle =0
\eeq
and, in the R sector,
\beq\label{beta0}
\b_0|{\rm phys}\rangle =0~,
\eeq
plus the analogous conditions for the tilded modes in the case of closed strings.
These conditions can be imposed on the Fock space states before applying the BRST
procedure. Since one can check that $\{Q_{\rm B},b_n\}=L_n$ and $[Q_{\rm B},
\b_r]=G_r$, the conditions \eq{BRSTclosed}, \eq{b0} and \eq{beta0} imply that
\beq\label{L0}
L_0|{\rm phys}\rangle =0
\eeq
and
\beq\label{G0}
G_0|{\rm phys}\rangle =0~.
\eeq

One can define a positive inner product on the space of physical states (as was
anticipated in our calling this space a Hilbert space). There are some subtleties
related to ghost and superghost zero modes, which we do not treat here. These 
subtleties will be dealt with in \sect{superstrings:scattering:ghosts}.

\subsubsection{String spectra}
Before we discuss the spectra of the free open and closed strings, we introduce
the operator
\beq\label{fermionnumber}
\exp(\pi\ii F)
\eeq
(or $(-1)^F$), where $F$ is the {\it world-sheet fermion number}. The operator
\eq{fermionnumber} is defined to anticommute with $\psi^\mu$ and the
(holomorphic) superghosts,
so that it would be more appropriate to call $F$ {\it world-sheet spinor
number}. It gives $-1$ on the NS ground state and acts as $\Gamma_{11}$ on the R
ground states (so it anticommutes with the $\psi^\mu_0$ modes, which act as
gamma-matrices). The operator \eq{fermionnumber} will allow us to perform
projections on the spectrum. Such projections are necessary for consistency of
the theory, and also to obtain  spacetime
supersymmetric theories. Finally, for the closed string there is also an operator
$\exp(\pi\ii\tilde F)$, the analogue of $\exp(\pi\ii F)$ in the antiholomorphic
sector.

\paragraph{Open strings}
The quantization prescription given above can be shown to lead to the following
spectra (we discuss only the lightest modes). For the {\it open string}, \eq{Lm}
for $m=0$ reads
\beq
L_0=\a'p^\mu p_\mu+N+a~,
\eeq
where $a=-1/2$ in the NS sector and $a=0$ in the R sector. $N$ is the excitation
number of the non-zero modes. The constraint \eq{L0} determines the mass of
a physical state in terms of its excitation number. 

In the  NS sector, the lowest state 
is $|0;k\rangle _{\rm NS}$. Its mass is given by
\beq\label{msquaredtachyon}
m^2=-k^2=-\frac{1}{2\a'}~,
\eeq
so the ground state is tachyonic.
As stated before, it has $\exp(\pi\ii F)=-1$. On the first excited level, we find
the physical states of a massless vector. This vector is obtained by letting 
$e_\mu\psi^\mu_{-1/2}$ act on a momentum state $|0;k\rangle _{\rm NS}$. 
Here, $e_\mu$ is the polarization vector. The unphysical
polarizations are removed from the physical Hilbert space by BRST invariance.
The mass shell condition \eq{L0} now implies $k^2=0$. These massless states have
$\exp(\pi\ii F)=+1$ and transform in the ${\bf 8}_v$ vector representation of the
SO(8) little group. 

In the R sector, the lowest states are
$u_A|A;k\rangle _{\rm R}$, where $A=1,\ldots,32$ is a spinor index, and $u_A$ is the
polarization (there is an implicit sum over $A$). The constraint \eq{G0} implies
that this state satisfies the massless Dirac equation, and thus is massless (as
can also be seen from \eq{L0}). The positive chirality states have
$\exp(\pi\ii F)=+1$ and transform in the ${\bf 8}_s$ spinor
representation of the SO(8)
little group, those with negative chirality  have $\exp(\pi\ii F)=-1$ and transform
in the ${\bf 8}_c$ conjugate spinor representation.

\paragraph{Closed strings}
For the {\it closed string}, we have
\bea
L_0&=&\frac{\a'}{4}p^\mu p_\mu+N+a~;\label{L0closed}\\
\tilde L_0&=&\frac{\a'}{4}p^\mu p_\mu+\tilde N+\tilde a~.\label{L0tildeclosed}
\eea
Again, $a$ and $\tilde a$ take values 0 or $-1/2$ depending on the sector.
The mass-shell condition can be written as
\beq\label{massshell}
\frac{\a'}{4}m^2=N+a=\tilde N+\tilde a~.
\eeq
The second equality is often called the {\it level-matching} condition. 

In the
NS-NS sector, we find a closed string tachyon with $m^2=-2/\a'$. The tachyon
has $(\exp(\pi\ii F),\exp(\pi\ii \tilde F))=(-1,-1)$. We include this information in
the notation by stating that the tachyon sits in the (NS$-$,NS$-$) sector. The
(NS+,NS$-$) and (NS$-$,NS+) sectors are empty because of the
level-matching condition. In the
(NS+,NS+) sector, we find 64 states in the ${\bf 8}_v\times{\bf 8}_v$
representation of the SO(8) little group: the graviton (${\bf 35}$),
antisymmetric tensor (${\bf 28}$) and dilaton (${\bf 1}$).

The R-R sectors contain massless antisymmetric tensor potentials,
transforming in the  ${\bf 8}_s\times{\bf 8}_s={\bf 1}+{\bf 28}+{\bf 35}_+$
for (R+,R+), in the ${\bf 8}_s\times{\bf 8}_c={\bf 8}_v+{\bf 56}_t$ for (R+,R$-$)
or (R$-$,R+), or in the ${\bf 8}_c\times{\bf 8}_c={\bf 1}+{\bf 28}+{\bf 35}_-$
for (R$-$,R$-$). For instance, the ${\bf 35}_+$ denotes a 4-form potential
$C_{\m\n\rho\sigma}$ with
selfdual field strength and the ${\bf 56}_t$ a 3-form potential $C_{\m\n\rho}$.

In the NS-R sectors, we find spinors and gravitini (vector-spinors), e.g.,
${\bf 8}_v\times{\bf 8}_s={\bf 8}_c+{\bf 56}$ for (NS+,R+) and
${\bf 8}_v\times{\bf 8}_c={\bf 8}_s+{\bf 56}'$ for (NS+,R$-$).

\subsubsection{Consistent string theories}
First, we focus on closed oriented strings. Not all of the states we have just found can
be present together in a (perturbatively) consistent string theory. We have not 
introduced the
necessary concepts to formulate the consistency conditions precisely
(we refer to section 10.6 of \cite{polchinski} for details), but we shall try to
indicate what kind of conditions are involved. First, scattering amplitudes, to
be introduced in \sect{superstrings:scattering}, have to be well-defined (``all
pairs of vertex operators have to be mutually local''). Second, poles in
scattering amplitudes of physical particles must correspond to physical 
particles (``the OPE must close''). Third, the theory must be ``modular
invariant'', which implies that UV-divergences can be removed without spoiling
space-time gauge invariances.

One could impose as an extra consistency condition the requirement that there 
should not be any excitations with negative mass squared (``tachyons'').
However, as in \cite{polchinski}, we shall not include this requirement in the
consistency conditions. The presence of a tachyon means that the starting point
of the perturbation series is not the real vacuum of the theory, so that the
perturbation theory does not make sense. Nevertheless, one may hope that the
theory does have a vacuum, so that it is really a consistent theory (although
the perturbation series around the original would-be vacuum did not make sense).
We refer to \sect{type0:intro} for a situation in which such a scenario has been
proposed.   

To obtain a consistent theory, one has to perform
projections on the spectrum obtained in the previous paragraph (``String
spectra''). 
These projections involve the world-sheet fermion
number operators and are called {\it GSO-projections}. Two inequivalent
projections
give rise to consistent theories without tachyons:%
\footnote{We shall encounter different GSO-projections in \chap{type0}.}
\beqa
{\rm IIB:}&&(NS+,NS+)\ ,~~(R+,R+)\ ,~~(R+,NS+)\ ,~~(NS+,R+)~;\nonumber\\
{\rm IIA:}&&(NS+,NS+)\ ,~~(R+,R-)\ ,~~(R+,NS+)\ ,~~(NS+,R-)~,\nonumber
\eeqa
where for instance R+ and R- are the Ramond sectors projected with 
$P_{\rm GSO}=(1 + (-)^F)/2$ and $P_{\rm GSO}=(1 - (-)^F)/2$, respectively,
as we have explained above. The presence of gravitini
indicates that these theories have local {\it spacetime supersymmetry}. In 
type
IIB, the gravitini have the same chirality, so the supersymmetry is chiral: there
are two supercharges transforming in say the ${\bf 16}$ of SO(1,9). On the other
hand, the IIA theory has non-chiral supersymmetry, with one supercharge in the
${\bf 16}$ and the other in the ${\bf 16}'$.

Before we proceed to study the type II theories from a spacetime point of view,
we introduce another supersymmetric string theory, which will show up in
\chap{nonbps}. The type IIB superstring, with the same chiralities on the holomorphic
and the antiholomorphic sides, has a world-sheet parity symmetry $\Omega$. This
symmetry can be gauged to obtain an unoriented closed string theory. This
results in the following changes for the spectrum.
In the NS-NS
sector, the antisymmetric tensor is eliminated, leaving the graviton and the
dilaton. Only the linear combination (NS-R)+(R-NS) of the two fermionic sectors
of type IIB survive the gauging, giving rise to massless states in the
${\bf 8}_c+{\bf 56}$. In particular, only one gravitino, and as a result only one
supercharge, is left. In the R-R sector, only the antisymmetric two-form potential
remains. It turns out that this closed string theory, called {\it type I closed
unoriented string theory} is inconsistent (for instance, there are spacetime
gravitational anomalies) unless one adds open strings to the
theory in a very precise way.

In order to add those open strings, we have to generalize the open strings we have
discussed so far. The two endpoints of an {\it oriented open string} are special
points, and distinct from each other. Therefore, it is possible to assume that the
open string carries ``charges'' at its end points. If we take the charge at the
first endpoint to transform in the ${\bf n}$ of a U($n$) symmetry group and the
one at the second endpoint in the $\bar{\bf n}$, then the open string as a whole
transforms in the  ${\bf n}\times\bar{\bf n}$, which is the adjoint of U($n$).
As a consequence, its U($n$) quantum numbers can be specified by giving a U($n$)
generator $\l$. In string scattering amplitudes, to be introduced 
in \sect{superstrings:scattering}, the
effect of the U($n$) charges will be to produce a {\it Chan-Paton}
group-theory factor $\tr \l_1\l_2\ldots\l_M$ whenever $M$ external strings are
attached to a disc in the cyclic order $(1\ 2\ 3\ldots M$). Although the charges
have trivial world-sheet dynamics, they have a crucial influence on spacetime
physics. For instance, one finds a  U($n$) gauge symmetry rather than the U(1)
gauge symmetry associated to the massless vector in the spectrum of the 
``ordinary''
open string. In fact, this one gauge field is replaced by a hermitean matrix of
gauge fields. Without Chan-Paton charges, the gauge field can be shown to be odd
under world-sheet parity, so it is not present on {\it unoriented open strings},
which are obtained from oriented ones by gauging world-sheet parity. However, if
an unoriented open string carries Chan-Paton charges, these also transform under
world-sheet parity. One possible action is $\l\rightarrow\l^T$ (the transposition
reflects the fact that world-sheet parity interchanges the endpoints). All in
all, the gauge fields with antisymmetric $\l$ survive. As a result, the U($n$)
gauge symmetry is reduced to O($n$).

Going back to the problem that type I closed unoriented string theory by
itself is inconsistent, it is known that this inconsistency can be removed by
including unoriented open strings with SO(32) gauge symmetry. The open string
GSO projection keeps the states with even world-sheet fermion number, thus
eliminating the tachyon.
The resulting
unoriented {\it type I open plus closed string theory} contains the massless
closed string states mentioned above, plus a gauge field and a spinor in the
adjoint of SO(32):
\beq
({\bf 8}_v+{\bf 8}_s)_{\rm SO(32)}~.
\eeq

We have discussed three tachyon-free and nonanomalous string theories: type IIA,
type IIB and type I SO(32). In addition to those, there exist two heterotic
string theories, which consist, roughly speaking, of the holomorphic side of a
bosonic string and the antiholomorphic side of a superstring. Heterotic strings
will only play a marginal role in this thesis (see \sect{nonbps:mot:testing}), 
so we do not introduce them in any detail.

%%%%%%%%%%%%%%%%%%%%%%%%%%%%%%%%%%%%%%%%%%%%%%%%%%%%%%%%
\subsection{Spacetime point of view}\label{superstrings:strings:spacetime}
\subsubsection{Low energy supergravity}
\label{superstrings:strings:spacetime:sugra}
The particle spectra of the superstring theories consist of a finite number of
massless particles and an infinite tower of massive ones. These massive string
states have masses of the order of $m_{\rm s}=\a'^{-1/2}$. Usually, the string
scale $m_{\rm s}$
is assumed to be of the order of $10^{18}$ GeV, though recently it has been
pointed out that string scales as low as 1 TeV cannot be excluded experimentally
(this bound
being set by the non-observation of string physics at present-day particle
accelerators). In any case, one is mainly interested in describing the light
(massless) degrees of freedom. If we had a second quantized description of
strings, based on an action which is a functional of the fields corresponding to
the excitations of a string (like the graviton, dilaton, etc.), we could imagine
integrating out the massive fields from this action (or, at the quantum level,
the quantum effective action). The resulting effective action for the massless
fields would be horrendously complicated and nonlocal, but useful formulas could
be obtained by restricting to the first few terms in a systematic expansion in
the number of derivatives (and fermions). This truncation would be called the
{\it low-energy effective action}. As we do not have such a second quantized
description of string theory, we cannot implement the programme we have just
outlined. What we can do is compute scattering amplitudes in perturbative
string theory (this will be explained in \sect{superstrings:scattering}). Starting 
from this on-shell
information, we can then construct a classical action for the massless fields
that reproduces these amplitudes.

For the superstring theories discussed above, the high degree of supersymmetry
completely determines the form of the low-energy effective action up to two
derivatives. We shall focus on the type II string theories, for which the low-%
energy effective actions are those of type IIA and type IIB supergravity.
However, the first action we shall write down is that of eleven-dimensional
supergravity, both for its relation to type IIA supergravity and for its
relation to M-theory, to be discussed in \sect{superstrings:dualities:M}.
We shall always restrict to the bosonic parts of the actions.

\subsubsection{Eleven-dimensional supergravity}
The eleven-dimensional supergravity theory contains two bosonic fields: the
metric $G_{MN}$ and a 3-form%
\footnote{\label{diffform}
We shall use differential forms to simplify the notation \cite{polchinski}. 
A $p$-form $A_p$
is a completely antisymmetric $p$-index tensor $A_{\m_1\ldots\m_p}$ with the
indices omitted (the $p$ subscript on  $A_p$ denotes the rank; we hope there
will be no confusion with the $p$'th component of a one-form). The wedge product
of a $p$-form $A_p$ and a $q$-form $B_q$ is defined by
\beq
(A_p\wedge B_q)_{\m_1\ldots\m_{p+q}}=\frac{(p+q)!}{p!q!}A_{[\m_1\ldots\m_p}
B_{\m_{p+1}\ldots\m_{p+q}]}~.
\eeq
(The wedge $\wedge$ is sometimes suppressed.)
Here, $[\;]$ denotes antisymmetrization with weight one. The exterior derivative
takes a $p$-form into a ($p+1$)-form:
\beq
(dA_p)_{\m_1\ldots\m_{p+1}}=(p+1)\partial_{[\m_1}A_{\m_2\ldots\m_{p+1}]}~.
\eeq

The integral of a $p$-form over a $p$-dimensional manifold,
\beq
\int A_p\equiv\int d^px A_{12\ldots p}~,
\eeq
is coordinate invariant. An important result is Stokes' theorem,
\beq
\int_{\cal M}dA_{p-1}=\int_{\partial{\cal M}}A_{p-1}~,
\eeq
where ${\cal M}$ is a $p$-dimensional manifold and $\partial{\cal M}$ its
boundary.

Using a metric $G_{\m\n}$ on a $D$-dimensional manifold, one defines the {\it Hodge
star} (or {\it dual}) of a $p$-form $A_p$ by
\beq
(*A)_{\m_1\ldots\m_{D-p}}=\frac{\sqrt{|G|}}{p!}\e_{\m_1\ldots\m_{D-p}}{}^{\n_1\ldots\n_p} 
A_{\n_1\ldots\n_p}~,
\eeq
where $G$ is the determinant of the metric, $\e_{\m_1\ldots\m_D}$ is the totally
antisymmetric symbol with $\e_{01\ldots D}=1$, and indices are raised with the
inverse of the metric $G_{\m\n}$.

It is convenient to represent differential forms by introducing an algebra of
anticommuting differentials $dx^\m$, writing
\beq
A_p=\frac1{p!}A_{\m_1\ldots\m_p}dx^{\m_1}\ldots dx^{\m_p}~,
\eeq
so that the wedge product corresponds to multiplying these differentials and the
exterior derivative is $d=dx^\n\partial_\n$.
}
potential $A_3$ with field strength
$F_4=dA_3$. In terms of the SO(9)
little group of the massless states, the metric gives a traceless symmetric
tensor (44 states) and the 3-form a rank 3 antisymmetric tensor (84 states).
Together with the 128 states of the SO(9) vector-spinor gravitino, these states
form a short multiplet of the supersymmetry algebra with 32 supercharges.

The bosonic part of the action is given by
\beq\label{11Dsugra}
2\k^2_{11}S_{11}=\int d^{11}x(-G)^{1/2}\left(R-\frac12|F_4|^2\right)-\frac16\int
A_3\wedge F_4\wedge F_4~,
\eeq
where in general
\beq
\int d^dx(-G)^{1/2}|F_p|^2=\int d^dx\frac{(-G)^{1/2}}{p!}G^{M_1N_1}\ldots
G^{M_pN_p}F_{M_1\ldots M_p}F_{N_1\ldots N_p}~.
\eeq

\subsubsection{Type IIA superstring}
To obtain the non-chiral, type IIA, supergravity action in ten dimensions,
compactify the 10-direction on a circle of radius $R$ and restrict to
configurations independent of this direction. The reduction of the metric is
given by
\bea\label{11IIA}
ds^2&=&G^{11}_{MN}(x^\mu)dx^Mdx^N\\
&=&\exp(-2\Phi(x^\mu)/3)G^{10}_{\m\n}(x^\mu)dx^\mu
dx^\nu+\exp(4\Phi(x^\mu)/3)[dx^{10}+C_\nu(x^\mu)dx^\nu]^2.\nonumber
\eea
Here, $M,N$ run from 0 to 10 and $\m,\n$ from 0 to 9. The superscript on the
metrics distinguishes between the 11-dimensional metric appearing in the
previous paragraph and the ten-dimensional metric, which will appear henceforth.
The superscript 10 will be omitted. The eleven-dimensional metric has reduced to
a ten-dimensional metric, a gauge field $C_1$ (note that in \eq{11IIA} $C_\nu$
denotes the $\n$ component of this one-form)
and a scalar $\Phi$. The potential
$(A_3)_{MNP}$ reduces to a three-form potential $C_3$ and a two-form potential
$B_2$:
\bea
C_{\m\n\rho}&=&A_{\m\n\rho}\\
B_{\m\n}&=&A_{10\m\n}~.
\eea

The action \eq{11Dsugra} gives rise to the type IIA supergravity
action \bea S_{\rm IIA}&=&S_{\rm NS}+S_{\rm R}+S_{\rm
CS}~,\label{IIAsugra}\\ S_{\rm NS}&=&\frac{1}{2\k^2_{10}}\int
d^{10}x(-G)^{1/2}e^{-2\Phi}\left(R+4\partial_\mu\Phi\partial^\mu\Phi-\frac12
|H_3|^2\right)~,\label{IIANS}\\ S_{\rm
R}&=&-\frac{1}{4\k^2_{10}}\int
d^{10}x(-G)^{1/2}\left(|F_2|^2+|\tilde
F_4|^2\right)~,\label{IIAR}\\ S_{\rm
CS}&=&-\frac{1}{4\k^2_{10}}\int B_2\wedge F_4\wedge F_4~,
\label{IIACS} 
\eea 
where the notation is as follows.
We have grouped the terms according to whether
the fields are in the NS-NS or R-R sector of type IIA string
theory; the Chern-Simons action $S_{\rm CS}$ contains both. The
constant $\k_{10}$ is related to $\k_{11}$ and the radius of
compactification by \beq \k_{10}^2=\frac{\k_{11}^2}{2\pi R}~. \eeq
The field $\Phi$ is the dilaton. The field $B_2$ is the NS-NS two-form
potential and $H_3$ its field strength: \beq H_3=dB_2~. \eeq We
have denoted the R-R potentials and field strengths by $C_p$ and
$F_{p+1}$: \beq F_{p+1}=dC_p~. \eeq We have defined
\beq\label{tildeF4} \tilde F_4=dC_3+H_3\wedge C_1~. \eeq

There are several terms in the action \eq{IIAsugra} that contain
potentials rather than their exterior derivatives. The term
\eq{IIACS} is gauge invariant because of the Bianchi identities
for the field strengths. The reason why the term involving $\tilde
F_4$ is gauge invariant is that the gauge variation of the second
term in \eq{tildeF4},
\beq
H_3\wedge d\l_0=-d(H_3\wedge \l_0)~,
\eeq
is cancelled by the transformation \beq\label{lambda0}
\d'C_3=H_3\wedge \l_0~, \eeq which supplements the usual $\d
C_3=d\l_2$. The $\l_0$ gauge transformation has its origin in
reparametrizations of $x^{10}$: \eq{lambda0} is simply part of the
eleven-dimensional tensor transformation. The gauge invariant
combination $\tilde F_4$ is to be regarded as the physical field
strength.

All the terms in \eq{IIAsugra} can be interpreted as tree level effects in type
IIA string theory, which means that they arise from sphere amplitudes. As we shall
argue later, one expects sphere amplitudes to contain a factor $e^{-2\Phi}$ as in
\eq{IIANS}. To obtain this dilaton dependence for \eq{IIAR} and \eq{IIACS},
and thus to make the relation to string theory more explicit, one
could redefine
\beq\label{RRredef}
C_p=e^{-\Phi}C'_p~,
\eeq
at the cost of complicating the Bianchi identity and gauge transformations.
Because of these drawbacks, the action is usually written as in \eq{IIAsugra}.

\subsubsection{Type IIB superstring}
Type IIB supergravity contains a self-dual field strength $\tilde F_5=*
\tilde F_5$. There
is no simple covariant action for such a field. However, we can use the
following action if we impose the self-duality of $\tilde F_5$ as an added constraint
on the {\it solutions}:
\bea
S_{\rm IIB}&=&S_{\rm NS}+S_{\rm R}+S_{\rm CS}~,\label{IIBsugra}\\
S_{\rm NS}&=&\frac{1}{2\k^2_{10}}\int
d^{10}x(-G)^{1/2}e^{-2\Phi}\left(R+4\partial_\mu\Phi\partial^\mu\Phi-\frac12
|H_3|^2\right)~,\label{IIBNS}\\
S_{\rm R}&=&-\frac{1}{4\k^2_{10}}\int d^{10}x(-G)^{1/2}\left(|F_1|^2+|\tilde
F_3|^2+\frac12|\tilde F_5|^2\right)~,\label{IIBR}\\
S_{\rm CS}&=&\frac{1}{4\k^2_{10}}\int (C_4+\frac12 B_2\wedge C_2)
\wedge H_3\wedge F_3~,
\label{IIBCS}
\eea
where
\bea
\tilde F_3&=&F_3+ H_3\wedge C_0~;\label{tildeF3}\\
\tilde F_5&=&F_5+H_3\wedge C_2~.\label{tildeF5}
\eea
As in type IIA  (see the discussion around \eq{lambda0}),
the action \eq{IIBsugra} is invariant under the modified R-R gauge
transformations
\beq\label{RRgaugetransf}
\d C=d\l+H_3\wedge\l~,
\eeq
where now we have combined the R-R potentials and the gauge
parameters in formal sums of forms of different degree:
\bea
C&=&C_0+C_2+C_4+C_6+C_8+C_{10}\label{formalsumC}\\
\l&=&\l_1+\l_3+\l_5+\l_7+\l_9~.\label{formalsumlambda}
\eea
Here, only $C_2$, $C_4$, $\l_1$ and $\l_3$ are relevant; we
have included the other forms for later use.

The higher forms do not contain independent degrees
of freedom. The relation to the lower forms appears as follows, as clearly
explained in \cite{bachas}. 
The R-R sector ground state
before the GSO projection is a $32 \times 32$ component bispinor of 
SO(1,9): one spinor index comes from
the holomorphic sector, the other from the antiholomorphic sector.
This bispinor can be expanded in products of gamma matrices:
\beq\label{FAB}
F_{AB}=\sum_{n=0}^{10}\frac1{n!}F_{\m_1\ldots\m_n}(C\G^{\m_1\ldots\m_n})_{AB}~,
\eeq
where $C$ is the charge conjugation matrix (to be introduced around
\eq{Cproperties}) and $\G^{\m_1\ldots\m_n}=\G^{[\m_1}\ldots\G^{\m_n]}$.
The coefficients $F_{\m_1\ldots\m_n}$ in
this expansion are the R-R field strengths. The holomorphic and
antiholomorphic GSO projections imply on the one hand that only
even/odd form field strengths occur in type IIA/B, and on the
other hand that those field strengths satisfy a Hodge duality
relation $F_{10-p}=\pm* F_{p}$ (the sign depending on $p$). 
This is why the five-form field
strength of type IIB is self-dual, for instance.

As an example of \eq{RRgaugetransf}, its four-form part reads
\beq
\d C_4=d\l_3+H_3\wedge\l_1~.
\eeq
To see that \eq{IIBCS} is invariant under \eq{RRgaugetransf}, one
has to make use of the fact that $H_3\wedge H_3=0$.
Note that \eq{RRgaugetransf} is also valid for type IIA, with the obvious
changes to \eq{formalsumC} and \eq{formalsumlambda}. The form in which we have
written the action
\eq{IIBsugra} differs slightly from the one in \cite{polchinski}. The
reason why we have chosen this form of the action is that it is
written in terms of the R-R potentials $C$ which will appear in
the D-brane Wess-Zumino action \eq{WZ}. (See the discussion on
gauge invariance in point \ref{WZgaugeinv} in
\sect{superstrings:D-branes}.)

An important property of the action \eq{IIBsugra} is that it has an
SL(2,$\Rbar$) symmetry. To see this, define
\bea
G_{{\rm E}\m\n}&=&e^{-\Phi/2}G_{\m\n}~,\ \ \tau=C_0+\ii e^{-\Phi}~;\nonumber\\
{\cal M}_{ij}&=&\frac{1}{\Im \tau}\left[\begin{array}{cc}|\tau|^2&\Re\tau\\
\Re\tau&1\end{array}\right]~,\ \ F_3^i=\left[\begin{array}{c}H_3\\F_3\end{array}
\right]~;\nonumber\\
{\cal C}_4&=&C_4+\frac12B_2\wedge C_2~.
\eea
Then,
\beq
\tilde F_5=d{\cal C}_4-\frac12C_2\wedge H_3+\frac12B_2\wedge F_3
\eeq
and
\bea
S_{\rm IIB}&=&\frac{1}{2\k^2_{10}}\int d^{10}x(-G_{\rm E})^{1/2}\left(R_{\rm E}
-\frac{\partial_\m\bar\tau\partial^\m\tau}{2(\Im\t)^2}-\frac{{\cal
M}_{ij}}{2}F^i_3\cdot F^j_3-\frac14|\tilde F_5|^2\right)\nonumber\\
&&-\frac{\e_{ij}}{8\k_{10}^2}\int {\cal C}_4\wedge F_3^i\wedge F^j_3~,
\eea
where the {\it Einstein metric} $G_{\rm E}$ is used everywhere. This action is
invariant under the following SL(2,$\Rbar$) symmetry:
\bea
\tau'&=&\frac{a\tau+b}{c\tau+d}~;\\
{F_3^i}'&=&\Lambda^i{}_jF_3^j~,\ \ \Lambda^i{}_j=\left[\begin{array}{cc}d&-b\\-c&a
\end{array}\right]~;\\
{\cal C}_4'&=&{\cal C}_4~,\ \
\tilde F_5'=\tilde F_5~;\nonumber\\ G'_{{\rm E}\m\n}&=&G_{{\rm E}\m\n}~,
\eea
with $a$, $b$, $c$ and $d$ real numbers satisfying $ad-bc=1$.
The SL(2,$\Rbar$) invariance of the $\tau$ kinetic term is
familiar, and that of the $F_3$ kinetic term follows from
\beq
{\cal M}'=(\L^{-1})^T{\cal M}\L^{-1}~.
\eeq
In type IIB string theory, the SL(2,$\Rbar$) invariance is broken to a discrete
SL(2,$\Zbar$) subgroup; see \sect{superstrings:dualities:S}.
%%%%%%%%%%%%%%%%%%%%%%%%%%%%%%%%%%%%%%%%%%%%%%%%%%%%%%%%%%%%%%%%%%%%%%%%%%%%%%%
\section{Strings in background fields}
\label{superstrings:background}
In \sect{superstrings:strings:worldsheet}, we have studied strings moving in a 
flat, Minkowski spacetime. Now, we
would like to generalize the discussion to curved spacetimes. For simplicity, we
restrict our attention to the bosonic sector of the superstring. 
It is natural to replace the action
\eq{polyakov} by
\beq\label{polyakovG}
S_X(G)=\frac{1}{4\pi\a'}\int_M d^2\s \sqrt{g} g^{ab}\partial_a X^\mu\partial_b
X^\nu G_{\mu\nu}(X)~.
\eeq
Unlike \eq{polyakov}, this does not reduce to a free field theory in the unit
gauge \eq{unit}. From the two-dimensional point of view, $G_{\mu\nu}(X)$ is a
``coupling functional'' (which can be viewed as an infinite number of coupling
constants by expanding $G_{\mu\nu}(X)$ around a fixed spacetime point
$X^\mu(\s)=x^\m_0$).

To see the relation between this gravitational background and the graviton
excitation of the closed string,
%\footnote{We have not discussed the spectrum of the bosonic string. Let us just
%note that the light states are very similar to those of the NS-NS and NS sectors
%of the superstring before
%GSO projection: for closed bosonic strings, one finds a tachyonic ground state and
%a massless graviton, dilaton and antisymmetric tensor; the open string has a
%tachyon and a massless vector. The critical spacetime dimension is 26 rather
%than 10.}
consider a spacetime that is close to flat
\beq
G_{\mu\nu}(X)=\eta_{\m\n}-2\k h_{\m\n}(X)
\eeq
with $2\k h_{\m\n}$ small ($\k$ will be defined in \eq{kappa}). Then, the factor
$\exp(-S_X(G))$ in the world-sheet path integral can be expanded as
\beq\label{expansionG}
\exp(-S_X(G))=\exp(-S_X(\eta))\left(1+\frac{\k}{2\pi\a'}\int_Md^2\s
\sqrt{g} g^{ab}h_{\m\n}(X)\partial_a X^\mu\partial_b X^\nu+\ldots\right).
\eeq
As we shall see in \sect{superstrings:scattering:closed}, 
%(at least for superstrings)
\beq\label{gravvertbos}
\frac{\k}{2\pi\a'}\int_Md^2\s
\sqrt{g} g^{ab}h_{\m\n}(X)\partial_a X^\mu\partial_b X^\nu
\eeq
is the bosonic part of the vertex operator for the graviton state of the 
string. Thus, \eq{expansionG}
means that turning on a background metric as in \eq{polyakovG} corresponds to
inserting a coherent state of gravitons.

From this perspective, it is quite plausible that backgrounds of the other
massless string states can be included as well. Restricting to the closed 
string, this means that the action \eq{polyakovG} should be extended to include%
\footnote{We only consider backgrounds of NS-NS fields; it is much harder to
turn on a R-R background.}
the antisymmetric tensor $B_{\m\n}$ and the dilaton $\Phi$. The result is
\beq\label{background}
S=\frac{1}{4\pi\a'}\int_M d^2\s \sqrt{g}\left[\left(g^{ab}G_{\m\n}(X)+\ii\e^{ab}
B_{\m\n}\right)\partial_a X^\mu\partial_b X^\nu+\a'R\Phi(X)\right]~,
\eeq
where $R$ is the world-sheet Ricci scalar.
We have called this action $S$ rather than $S_X$ because, as will become clear
soon, it extends
\eq{polyakov+chi} rather than \eq{polyakov}: it already contains the string
coupling constant.

The action \eq{background} is invariant under field redefinitions $X'^\mu(X)$
with $G_{\m\n}$ and $B_{\m\n}$ transforming as tensors. The Lagrangian density
changes by a total derivative under the spacetime gauge transformation
\beq\label{Bgaugetransf}
\d B_{\m\n}(X)=\partial_\m\zeta_\n(X)-\partial_\n\zeta_\m(X)~,
\eeq
so that \eq{Bgaugetransf} leaves the action invariant if the world-sheet does not
have a boundary. For world-sheets with boundary, gauge invariance is restored by
adding a boundary term to \eq{background}, as we shall discuss in
\sect{superstrings:D-branes}.
The $B_{\m\n}$ term in \eq{background} describes
the minimal coupling of the string world-sheet to the gauge potential $B_{\m\n}$,
much like the world-line of an electron couples to a photon. The analogue of the
electric charge is the string winding number.
%\footnote{For instance, if one of the directions of spacetime is periodically
%identified, an oriented closed string can have an arbitrary integer winding
%number along  the corresponding circle, the sign corresponding to the
%orientation. Non-compact space can be considered as a decompactification
%limit of a torus. The analogue of the winding number is the number of infinitely 
%extended strings with a certain orientation minus the number of infinitely
%extended strings with the opposite orientation.}
Invariance under \eq{Bgaugetransf} corresponds to conservation of string winding
number. In this respect, it is clear that once open strings (world-sheets with
boundary) are added to the theory and closed strings are allowed to be cut into
open strings, the gauge invariance \eq{Bgaugetransf} will have to be
reconsidered.

The dilaton term in \eq{background} has an important consequence. For constant
dilaton field, $\Phi(X)=\Phi_0$, it equals $\Phi_0\chi$, where
\beq
\chi=\frac{1}{4\pi}\int_M d^2\s\sqrt{g}R
\eeq
is the Euler number of the world-sheet $M$. This means that the constant $\l$ in
\eq{polyakov+chi} can be shifted by choosing a different vacuum expectation value
for the dilaton. As we have seen in \sect{superstring:strings:worldsheet:path},
$\l$ determines the string coupling constant. Here, it is the constant mode of
one of the fields.
What we have learned is that the
string coupling constant is not a free parameter of string theory. Rather,
different values of the coupling constant correspond to different vacua of the
same theory.

Some of the results of this section 
have an interpretation in terms of the spacetime
effective actions of \sect{superstrings:strings:spacetime:sugra}. The type II
string theories are theories of closed strings only;%
\footnote{This statement and the following conclusion will be altered in
\sect{superstrings:D-branes}.}
the tree level actions \eq{IIAsugra} and \eq{IIBsugra} are designed to reproduce
scattering amplitudes due to world-sheets with sphere topology. These actions are
manifestly invariant under the $B$-gauge transformations \eq{Bgaugetransf},
consistent with the fact that the sphere has no boundary. Further, we have
argued
that the actions \eq{IIAsugra} and \eq{IIBsugra} can be brought to a form where
the dilaton appears as a global factor $e^{-2\Phi}$ (in fact, this is the only
way in which the dilaton appears without derivatives). This means that the
coupling constant $\k_{10}$ in front of the action has no invariant meaning: it
can be shifted by redefining the dilaton field (so that it gets a different
vacuum expectation value). We shall only consider backgrounds with a constant
expectation value for the dilaton. In that case, we can redefine the dilaton 
field by a constant and make its expectation value vanish. 
Then, $\k_{10}$ (which we will henceforth
denote $\k$) is related to the ten-dimensional Newton's constant
(see \eq{kappa} in \sect{superstrings:D-branes}).

To conclude this section, we just mention the important fact that the condition
that the action \eq{background} be Weyl invariant reproduces the classical
equations of motion of the background fields, including $\a'$ corrected Einstein
equations. The reader is referred to \cite{gsw, polchinski} for details about this
essential consistency condition.

%%%%%%%%%%%%%%%%%%%%%%%%%%%%%%%%%%%%%%%%%%%%%%%%%%%%%%%%%%%%%%%%%%%%%%%%%%%%%%%
\section{String scattering amplitudes}
\label{superstrings:scattering}

\subsection{Vertex operators}\label{superstrings:scattering:onshell}

In \sect{superstrings:strings:worldsheet}, we defined amplitudes in
perturbative string theory by summing over world-sheets
interpolating between given initial and final curves. In practice,
this idea is very difficult to implement for generic initial and final curves.
It turns out that the situation simplifies when on-shell amplitudes (i.e.
scattering amplitudes, or S-matrix elements) are considered. 

A heuristic argument for this simplification goes as follows \cite{polchinski}.
First, one argues \cite{polchinski}
that scattering amplitudes of on-shell states are due to
world-sheets with incoming and outgoing legs that are semi-infinite cylinders.
Each of those legs can be described with a complex coordinate $w$,
\beq\label{cylinder}
-2\pi t\leq\Im w\leq 0~,\ \ w\approx w+2\pi~,
\eeq
in the limit $t\rightarrow\infty$. The end $\Im w=0$ fits onto the rest of the
world-sheet. The external state is specified by a weight factor in the path
integral depending on the configuration at $\Im w=-2\pi t$.

\begin{figure}
\begin{center}
\epsfig{file=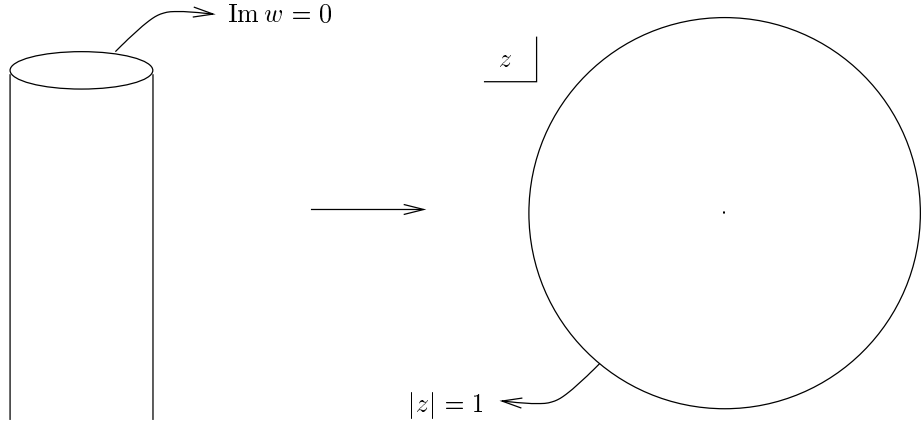}
\end{center}
\caption{Conformal map of a semi-infinite cylinder to the unit disc. The
``boundary at infinity'' is mapped to a single point.}\label{fig:vertex}
\end{figure}

Then, one notes that the 
cylinder \eq{cylinder} can be conformally mapped into the annular region
described with a coordinate $z$,
\beq
z=\exp(-\ii w)~,\ \ \exp(-2\pi t)\leq|z|\leq1~.
\eeq
In the $t\rightarrow\infty$ limit, this region becomes the unit disc (see
\fig{fig:vertex}). The
external state is now specified by the insertion in the path integral of a 
{\it local operator} at $z=0$.

Finally, one can always find a conformal transformation that maps each of the 
initial and final curves to finite points. In that way, the original world-sheet
is conformally mapped into a compact one. The on-shell external states are
represented by the insertion of local operators in the path integral.
These local operators are called {\it vertex operators}.

Readers who do not feel comfortable with this heuristic argument can just skip
it and {\it define} string scattering amplitudes as expectation values of
local vertex operators. They will find out that, for on-shell external states,
such a prescription is compatible with the local world-sheet symmetries, such as
conformal symmetry. Vertex operators for off-shell external states would break
some local world-sheet symmetries. Since these symmetries are necessary to
eliminate negative norm states from the physical string spectrum, at this point
the readers will conclude that they can only treat on-shell amplitudes. 

Thus, in practice, we can only compute the S-matrix in perturbative string 
theory. 
Off-shell questions, like the potential of massive or tachyonic particles,
are extremely difficult to answer in this framework. There have been attempts to
understand off-shell string theory, but these have only been
partially successful. We shall not deal with them in this thesis.
%%%%%%%%%%%%%%%%%%%%%%%%%%%%%%%%%%%%%%%%%%%%%%%%%%%%%%%%%%%%%%%%%%%%%%%%%%%%%%%
\subsection{Hilbert space interpretation}
\label{superstrings:scattering:Hilbert}

In \sect{superstrings:scattering:onshell}, we found a path integral prescription 
to compute string scattering amplitudes. The prescription instructs us to sum
over all compact world-sheets, with vertex operators inserted in the path
integral. To be more precise, the vertex ``operators'' are really local
functionals of the fields appearing in the world-sheet action.

If the world-sheet is a sphere (as it will be in most of this thesis),%
\footnote{
The other world-sheets we shall consider are the disc and the torus. In both
cases, a similar Hilbert space interpretation is possible.
}
the amplitudes can be given a Hilbert space interpretation. First, we map the
sphere to the (compactified) complex plane, described with a complex coordinate 
\beq
z=\exp(\tau+\ii\s)~.
\eeq 
We choose $\tau$ as our world-sheet time coordinate.%
\footnote{In fact, this is the time
coordinate that appeared naturally in \sect{superstrings:strings:worldsheet},
where we obtained the complex plane via a conformal transformation on the
cylinder.} 
Then, we can apply canonical
quantization, which, in fact, is what we did in 
\sect{superstrings:strings:worldsheet}.
The resulting algebra of the modes of the various fields in the world-sheet
action can then be represented on a Hilbert space, after dealing with the
subtleties we shall consider in \sect{superstrings:scattering:ghosts}.%
\footnote{In fact, in this thesis we shall not need to explicitly construct this 
Hilbert space with its positive scalar product; see
\sect{superstrings:scattering:inner}.
}

The local insertions in the path integral now get an interpretation as {\it
operators} on a Hilbert space. 

The conclusion is that the states of the free string Hilbert space correspond to
local operators. This reflects an important feature of conformal field
theory (CFT) on a cylinder: 
there is a complete correspondence between operators and
states. This correspondence is called the {\it state--operator mapping}. We
refer to \cite{polchinski} for more details on this construction. 

To illustrate the state--operator mapping, consider the conformal field theory 
describing
the NS sector of the superstring. Let us ask which state $|1\rangle $
corresponds to inserting the unit operator at the origin, i.e., to doing the path
integral without an insertion at the origin. To
answer this question, we find out by which modes the state is annihilated.
The modes $\a^\mu_m$ are defined by (see \eq{dXinmodes}):
\beq\label{alpha}
\a^\m_m=\left(\frac{2}{\a'}\right)^{1/2}\oint\frac{dz}{2\pi}z^m\partial
X^\m(z)~.
\eeq
When they act on the state created by an operator inserted at the origin, the
contour integral is along a small circle around the origin.
With no operator inserted at the origin (except for the unit
operator), the integrand of \eq{alpha} is perfectly holomorphic
inside the contour for $m\geq 0$. Thus, as far as the CFT describing the fields 
$X$ is
concerned, the state $|1\rangle $ is just the vacuum $|0;0\rangle $ with zero
momentum. Analogously, the state $|1\rangle $ is annihilated by all
$\psi^\m_r$ with $r\geq 1/2$ (see \eq{psiinmodes}).
Further, it is clear from \eq{Tinmodes} that $|1\rangle $ is annihilated
by the Virasoro generators $L_m$ and $\tilde L_m$ with $m\geq -1$,
in particular by the SL(2,$\Cbar$) subgroup generated by
$L_{0,\pm 1},\tilde L_{0,\pm 1}$. Therefore, it is called the {\it
SL(2,$\Cbar$) vacuum}. However, the SL(2,$\Cbar$) vacuum is not
the vacuum (or ground state, the state with zero occupation number in the Fock
spaces) we defined around \eq{Siegel}: it
follows from
\eq{cinmodes} and \eq{gammainmodes} that $|1\rangle $ is not annihilated
by the lowering operators $c_1$ and $\gamma_{1/2}$. As far as the
ghosts (as opposed to the superghosts) are concerned, there is an
easy relation between the SL(2,$\Cbar$) vacuum and the ground
state defined before: the ground state is obtained from the
SL(2,$\Cbar$) vacuum by acting with $c_1$.
%Incidentally, this is
%what causes the ground state of the bosonic string to be
%tachyonic.
For the superghosts, the relation between both vacua is far less
obvious. To discuss it, one would need the material introduced 
in \sect{superstrings:scattering:ghosts}.

As another example of the state--operator mapping, what is the
operator corresponding to the state $c_1|1\rangle $? Acting with $c_1$ on 
$|1\rangle$ amounts to inserting
\beq\label{c1action}
\frac{1}{2\pi\ii}\oint dz c(z)z^{-1}
\eeq
in the path integral, where the contour integral is along a small circle around
the origin. The integrand is holomorphic except at the origin. By Cauchy's
theorem, inserting \eq{c1action} is equivalent with inserting $c(0)$ at the
origin. Thus, the state $c_1|1\rangle $ is created by the operator $c(0)$.

Applying the mode $x_0^\m$ corresponds to inserting the operator
$X^\m(0,0)$. One can check that the state with several
creation modes excited corresponds to the {\it normal-ordered
product} of the operators associated to the modes. Here, by
normal-ordered we mean {\it conformally normal-ordered}, which is
generically different from the creation-annihilation normal ordering we
encountered before. The normal-ordered product of two operators
differs from the ordinary product in that it satisfies the naive
equations of motion. It is obtained from the ordinary product by
subtracting terms that become singular as the operators come close
to one another. As an example, in the CFT of the $X$ fields the state
$|0;k\rangle =e^{\ii k.x_0}|0;0\rangle $ corresponds to
operator $:e^{\ii k.X(0,0)}:$, where we denote conformal normal
ordering by $:~:$. Henceforth, we shall usually drop the normal
ordering symbols.

\begin{figure}
\begin{center}
\epsfig{file=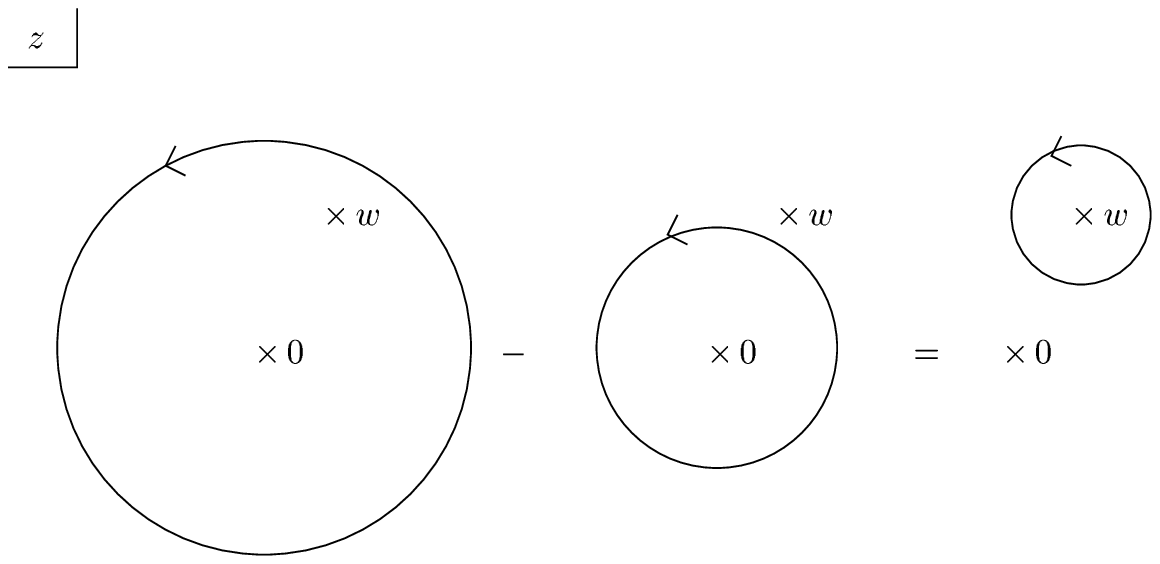,width=12.5cm}
\end{center}
\caption{Transition between \eq{defcomm} and \eq{contourdef}.}\label{fig:comm}
\end{figure}

As a useful tool, we define the commutator of a charge
\beq
Q=\frac 1 {2\pi\ii}\oint dz j(z)
\eeq
associated to a holomorphic current $j(z)$, and an operator $V(w)$
inserted at a point $w$. We define the commutator $[Q,V(w)]$ by
\bea
[Q,V(w)]&=&\left[\oint_{C_{+}}\frac{dz}{2\pi\ii}-\oint_{C_{-}}
\frac{dz} {2\pi\ii}\right]j(z)V(w)\label{defcomm}\\
&=&\oint_{C_{w}}\frac{dz}{2\pi\ii}j(z)V(w)~.\label{contourdef}
\eea
Here, $C_{+}$ is a circle centered at the origin with radius
slightly greater than $|w|$ and $C_{-}$ is one with slightly smaller
radius than $|w|$. The small circle $C_{w}$ is centered at $w$.
In \eq{contourdef} we have deformed the contour using holomorphicity
(see \fig{fig:comm}).
We have assumed that no other operators are inserted in the annulus between
$C_-$ and $C_+$.
All the holomorphicity arguments we are using
are only valid inside a path integral, which is where we
shall make use of them. From that point of view, the definition
\eq{defcomm} is very natural. In a Hilbert space interpretation of the
path integral, the operators come out time ordered. In the way we
quantized the sigma model on the sphere, time ordering is the same
as radial ordering, so the ordering of the operators on the right
hand side of \eq{defcomm} is as suggested by the left hand side.

We pauze to mention an important consequence of
the state--operator mapping, the {\it operator product expansion
(OPE)}. This states that inserting in a path integral two local
operators close to one another (compared to the other insertions)
is equivalent to inserting a sum of local operators at one of
both points. For operators ${\cal A}_i,{\cal
A}_j$ satisfying transforming as \eq{eigenop}, the OPE reads
\beq
{\cal A}_i(z,\bar z)=\sum_k(z-w)^{h_k-h_i-h_j}
(\bar z-\bar w)^{\tilde h_k-\tilde h_i-\tilde h_j}{c^k}_{ij}
{\cal A}_k(w, \bar w)~,
\eeq
where $k$ labels a complete set of operators transforming as \eq{eigenop},
and the equality holds inside correlation functions. OPEs are very
handy to compute commutators with charges associated to holomorphic
currents: as we have just seen, these involve circle integrals,
which can often be computed using Cauchy's theorem. As another
application, consistency with OPEs puts strong restrictions on the
form of correlation functions. For instance, OPEs are of great
help in evaluating correlation functions like the ones appearing
in \sect{nonbps:anomalous}.

We end this digression by giving two examples of useful OPEs. The
transformation property \eq{tensor} of primary fields is encoded
in the OPE
\beq\label{TO}
T(z){\cal O}(w,\bar w)=\frac h {(z-w)^2} {\cal O}(w,\bar w)+
\frac1{z-w} \partial {\cal O}(w,\bar w)+\ldots~,
\eeq
where $\ldots$ denotes non-singular terms, which
for many applications are not important. From this, one can compute the
commutator of $L_m$ (defined in \eq{Tinmodes}) with
${\cal O}(w,\bar w)$. The OPE of the energy-momentum tensor with
itself reads
\beq
T(z)T(w)\sim \frac c{2(z-w)^4}+\frac
2{(z-w)^2}T(w)+\frac1{z-w}\partial T(w)~,
\eeq
which states that the energy-momentum tensor is not a primary
field in a theory with non-vanishing central charge. The ``$\sim$''
instead of ``$=$'' denotes the omission of non-singular terms.
This OPE encodes the Virasoro algebra \eq{Virasoro}.

%%%%%%%%%%%%%%%%%%%%%%%%%%%%%%%%%%%%%%%%%%%%%%%%%%%%%%%%%%%%%%%%%%%%%%%%%%%%%%%

\subsection{Inner product}\label{superstrings:scattering:inner}
If we want to give the correlation functions computed from the
path integral an interpretation as expectation value in a Hilbert
space, we have to define an inner product on the space of states.
To be more precise, we should take the space of physical states
defined in \sect{superstrings:strings:worldsheet} and endow it
with a positive sesquilinear form. We shall not go
all that way, though. What we shall do is define a sesquilinear
form (``inner product'') on the Fock space of
the string oscillators, and introduce
a bracket notation for it. This will be sufficient for our purpose
of rewriting scattering amplitudes in a ``Hilbert space''
formalism. The step of explicitly extracting a positive inner
product on the physical Hilbert space will be omitted.

It is consistent with the commutation relations to ask that the
inner product be such that the modes satisfy the Hermiticity
properties
\bea
(\a^\m_m)^\dagger&=&\a^\m_{-m}~;\label{hermprop}\\
(b^\m_m)^\dagger&=&b^\m_{-m}~;\nonumber\\
(c^\m_m)^\dagger&=&c^\m_{-m}~;\nonumber\\
(\psi^\m_r)^\dagger&=&\psi^\m_{-r}~;\nonumber\\
(\gamma^\m_r)^\dagger&=&\gamma^\m_{-r}~;\nonumber\\
(\beta^\m_r)^\dagger&=&-\beta^\m_{-r}~,\nonumber
\eea
and analogously for the antiholomorphic modes. 
The first line in \eq{hermprop} expresses, for instance, 
the fact that the operators $\a^\m_m$ correspond to the modes of the
real classical field $X^\mu$.

If, for the time being, we denote the inner product of two states
$|\varphi\rangle $ and $|\chi\rangle $ by $(|\varphi\rangle ,|\chi\rangle )$, the Hermitian
conjugate of an operator $A$ is defined by
\beq\label{Adagger}
(A^\dagger|\varphi\rangle ,|\chi\rangle )=(|\varphi\rangle ,A|\chi\rangle )~.
\eeq

To define the sesquilinear form $(.,.)$, one could proceed as follows. First,
define $(\ket{1},\ket{\chi})$ to be the path integral on the sphere with the
vertex operator corresponding to the state  $\ket{\chi}$ inserted at the origin
and no other insertions. Then, arbitrary brackets $(\ket{\varphi},\ket{\chi})$
can be defined using sesquilinearity and the hermiticity properties
\eq{hermprop}.

Now, we introduce the bracket notation%
\footnote{ \label{inout}
The ``bra'' $\bra{\varphi}$ is often called the {\it conjugate state} of 
the ``ket'' $\ket{\varphi}$. Also, $\ket{\varphi}$ is called an {\it in-state} and
$\bra{\varphi}$ an {\it out-state}. The latter terminology can be motivated as
follows. We are studying a Euclidean field theory on the sphere, which we map to
the cylinder with coordinates $(\tau,\sigma)$, where $\tau$ is the
Euclidean time coordinate. Consider an operator 
that would be Hermitean in a Minkowski field theory obtained
from our Euclidean theory via Wick rotation. The Hermitean conjugate of such a
operator inserted at a point with coordinates $(\tau,\sigma)$, 
corresponds to the same operator inserted at a point
with coordinates $(-\tau,\sigma)$. We have seen in
\sect{superstrings:scattering:onshell} that an in-state $\ket{\varphi}$
is created by an operator inserted at $\tau=-\infty$:
\beq
\ket{\varphi}=V_\varphi\ket{1}~,
\eeq
where $V_\varphi$ is composed of Hermitean operators at $\tau=-\infty$. According
to \eq{Adagger}, we may write
\beq
\bra{\varphi}=\bra{1}V^\dagger_\varphi~,
\eeq
where $V^\dagger_\varphi$ is composed of Hermitean operators at $\tau=+\infty$.
}
by defining
\beq
\langle\varphi|\chi\rangle =(|\varphi\rangle ,|\chi\rangle )~.
\eeq

%%%%%%%%%%%%%%%%%%%%%%%%%%%%%%%%%%%%%%%%%%%%%%%%%%%%%%%%%%%%%%%%%%%%%%%%%%%%%%%

\subsection{Ghosts}\label{superstrings:scattering:ghosts}
The ghost and superghost systems have some peculiarities with
important consequences. We first discuss the ghost system.
One can define \cite{FMS} a ghost number operator $j_0^{\rm gh}$
that counts $+1$ for $c$ and $-1$ for $b$. 
(It is the charge associated to the
current $j^{\rm gh}=-bc$.)
It turns out \cite{FMS} that
\beq\label{ghostnumberdagger}
(j_0^{\rm gh})^\dagger=-j_0^{\rm gh}+3~,
\eeq
where the 3 is related to the anomaly in the ghost number current.
Suppose we want to compute the expectation value in the SL(2,$\Cbar$)
vacuum of an operator ${\cal O}$ with ghost number $q^{\rm gh}$,
\beq\label{ghostnumbercharge}
[j_0^{\rm gh},{\cal O}]=q^{\rm gh}{\cal O}~.
\eeq
Since the SL(2,$\Cbar$) vacuum $|1\rangle $ does not carry ghost charge, we can
use \eq{ghostnumberdagger} and \eq{ghostnumbercharge} to derive
\beq
\langle 1|j_0^{\rm gh}{\cal O}|1\rangle =3\langle 1|{\cal O}|1\rangle =q^{\rm gh}\langle 1|{\cal
O}|1\rangle ~,
\eeq
so that only operators with ghost number
\beq\label{qgh3}
q^{\rm gh}=3
\eeq
can have a
non-vanishing expectation value!

For the superghost system, one can analogously
define a superghost number operator $j_0^{\rm sgh}$
that counts $\gamma=+1,\ \b=-1$ charge. (It is the charge associated to the
current $j^{\rm sgh}=-\b\gamma$.) Now, \cite{FMS}
\beq\label{superghostnumberdagger}
(j_0^{\rm sgh})^\dagger=-j_0^{\rm sgh}-2
\eeq
(where $-2$ is related to the anomaly in the superghost number current),
so that only operators with superghost number
\beq\label{qsgh-2}
q^{\rm sgh}=-2
\eeq
can have a non-vanishing expectation value.

Of course, the same reasoning goes through for the antiholomorphic
ghosts and superghosts, so that in addition to \eq{qgh3} and
\eq{qsgh-2} non-vanishing amplitudes have
\beq\label{qantihol}
\tilde q^{\rm gh}=3~,\ \ \tilde q^{\rm sgh}=-2~.
\eeq

It is convenient to write the superghosts $\gamma$ and $\b$ in
terms of new variables as \cite{FMS}
\bea
\gamma(z)&=&e^{\phi(z)}\eta(z)~;\nonumber\\
\b(z)&=&e^{-\phi(z)}\partial\xi(z)~,\label{bosonizesgh}
\eea
where 
\beq
-\partial \phi(z)=j^{\rm sgh}(z)
\eeq
($j^{\rm sgh}$ has been defined above
\eq{superghostnumberdagger}), and $\eta$ and $\xi$ are
holomorphic fermions with dimensions 1 and 0, respectively.
The OPEs are
\beq
\phi(z)\phi(w)\sim -\ln(z-w)~;\ \ \eta(z)\xi(w)\sim\frac 1 {(z-w)}~.
\eeq

We can associate the following charge to the $\eta,\xi$ system:
\beq
j_0^{\eta,\xi}=\oint\frac{dz}{2\pi\ii}j^{\eta,\xi}=-\oint\frac{dz}{2\pi\ii}
\eta\xi~,
\eeq
which satisfies \cite{FMS}
\beq\label{etaxinumberdagger}
(j_0^{\eta,\xi})^\dagger=-j_0^{\eta,\xi}+1~.
\eeq
A non-vanishing amplitude can be obtained if one $\xi(z)$ is inserted in the path
integral over $\phi,\eta$ and $\xi$. This can also be seen as follows. The $(0,0)$
field $\xi$ has a single  zero mode $\xi_0$ on the sphere, while the $(1,0)$ field 
$\eta$ has none. The Grassmann integral over  $\xi_0$ vanishes unless a factor of
$\xi$ is inserted. Note, in passing, that this also makes it clear that the 
insertion point does not matter. 

However, it is a peculiar feature of \eq{bosonizesgh} that only $\partial\xi$ occurs,
not $\xi$ itself. For computations involving $\b$ and $\gamma$, we can 
therefore restrict to the reduced algebra with $\xi_0$
omitted (not integrated over in the path integral). If this is done, no $\xi$ 
insertion is needed. Working in the ``large'' algebra is useful when discussing {\it
pictures}. The picture operator $P$ is defined as
\beq\label{defP}
P=j_0^{\rm sgh}+j_0^{\eta,\xi}=\oint\frac{dz}{2\pi\ii}(-\partial\phi-\eta\xi)~.
\eeq
For instance, inserting $\xi,\eta,\gamma$ or $\b$ inside the contour contributes 
$1,-1,0$ or $0$ to $P$, respectively. It is clear from \eq{qsgh-2}
and \eq{etaxinumberdagger} that, in the large algebra, only operators with
total superghost number $q^{\rm sgh}=-2$ and $\eta,\xi$ charge $q^{\eta,\xi}=1$ can
have a nonzero expectation value. In the reduced algebra, this means that the total
superghost number should be $-2$ and the total picture should be 
\beq\label{Ptot-2}
P^{\rm tot}=-2~.
\eeq

Of course, the same can be done for the antiholomorphic
superghosts. In the $\phi,\eta,\xi$ variables, the superghost number corresponds
to the power of $e^\phi$. The SL(2,$\Cbar$) vacuum $|1\rangle $ has
superghost number zero. Vacua with other superghost numbers can be
defined by
\beq\label{sghvacua}
|q^{\rm sgh},\tilde q^{\rm sgh}\rangle \equiv e^{q^{\rm sgh}\phi(0)}
e^{\tilde q^{\rm sgh}\tilde \phi(0)}|1\rangle ~.
\eeq
%%%%%%%%%%%%%%%%%%%%%%%%%%%%%%%%%%%%%%%%%%%%%%%%%%%%%%%%%%%%%%%%%%%%%%%%%%%%%%%

\subsection{BRST invariance}\label{superstrings:scattering:BRST}
On vertex operators, the equivalent of the physical state condition
\eq{BRSTclosed} reads
\beq\label{BRSTclosedop}
[Q_{\rm B}, V_{\rm phys}]=0~.
\eeq
%It follows from \eq{BRSTclosedop} and the BRST invariance of the
%SL(2,$\Cbar$) vacuum that
%\beq
%Q_{\rm B} V_{\rm phys}(0)|1\rangle =\lim_{w\rightarrow 0}Q_{\rm B}
%V_{\rm phys}(w)|1\rangle =0~.
%\eeq
BRST exact states are created by
\beq\label{BRSTexactop}
V_{\rm exact}=[Q_{\rm B},{\cal O}]~,
\eeq
with ${\cal O}$ any operator.
From all this, it is clear that BRST exact states decouple from
physical sphere amplitudes: write the BRST exact vertex operator
as in \eq{BRSTexactop} and use \eq{BRSTclosedop} to commute
$Q_{\rm B}$ to the left or the right until it annihilates $\langle 1|$
or $|1\rangle $.

Using \eq{Lm} and \eq{Gr}, the BRST charge \eq{QB} can be
decomposed as follows into pieces with definite ghost number:
\beq
Q_{\rm B}=Q_0+Q_1+Q_2\ \ +{\rm antiholomorphic}~,
\eeq
where
\bea
Q_0&=&\sum_{m\in\Zbar} c_{-m}(L_m^{{\rm m}+{\rm sgh}}
+\frac12L_m^{\rm gh})~;\label{Q0}\\
Q_1&=&\frac12\sum_{r\in\Zbar+\n}\gamma_{-r}G_r^{\rm m}~;\\
Q_2&=&-\sum_{q,r\in\Zbar+\n}\gamma_{-q}\gamma_{-r}b_{q+r}~.
\eea
Note that due to a contribution of the supercurrent piece of
$Q_{\rm B}$ there is no $1/2$ in front of $L_m^{\rm sgh}$ in
\eq{Q0}.

All vertex operators we shall use have definite ghost number. Such
vertex operators must commute with the three pieces of
$Q_{\rm B}$ independently. In the following, we shall focus on the
condition
\beq
[Q_0,V_{\rm phys}]=0~.
\eeq
Take an $(h,\tilde h)=(1,1)$ primary field  ${\cal V}(z,\bar z)$ that
does not contain ghost fields (superghost fields are allowed).
Then, it follows from \eq{TO} (and the fact
that the OPE between $c$ and non-ghost fields is non-singular)
that
\beq
[Q_0,{\cal V}(z,\bar z)]=\partial(c(z){\cal V}(z,\bar z))~,
\eeq
and analogously
\beq
[\tilde Q_0,{\cal V}(z,\bar z)]=\bar\partial(\tilde c(\bar z)
{\cal V}(z,\bar z))~,
\eeq
so that, in real coordinates $\sigma^a$,
\beq
[Q_0+\tilde Q_0,{\cal V}]=\partial_a(c^a{\cal V})~.
\eeq
This vanishes if we integrate ${\cal V}$ over the sphere.%
\footnote{On world-sheets with boundary, like the disc, the
boundary terms vanish due to the boundary condition on the ghost
field, which we have not discussed explicitly.} Analogously, one
can use \eq{Q0} and the fact that the $c(z)c(w)$ OPE is
non-singular to show that $c\tilde c{\cal V}$ is invariant under
$Q_0$ and $\tilde Q_0$.

Open string vertex operators, which we shall use
in \sect{nonbps:anomalous}, are inserted on a boundary of the
world-sheet. To construct a BRST invariant vertex operator, one
takes a dimension 1 primary field and either integrates it over
the boundary or adds a ghost field. In most of this thesis, we
shall focus on closed string vertex operators, to which we now
return.

We have found two kinds of candidate vertex operators: fixed
vertex operators with ghost number one, and integrated vertex
operators with ghost number zero,
\bea
V_{\rm fixed}(z_0,\bar z_0)&=&c(z_0)\tilde c(\bar z_0)
{\cal V}(z_0,\bar z_0)~;\label{Vfixed}\\
V_{\rm integrated}&=&\int d^2z\,{\cal V}(z,\bar
z)~.\label{Vintegrated}
\eea
In both cases, ${\cal V}(z,\bar z)$ is an $(h,\tilde h)=(1,1)$
primary field of the Virasoro algebra. Note that the structure of
\eq{Vfixed} is consistent with the observations around
\eq{c1action} and the fact that physical states
are built by applying matter and superghost oscillators on
the vacuum $|0;k\rangle $.

Now, we combine our new knowledge with the observations summarized
in \eq{qgh3}, to conclude that, in a scattering amplitude on the
sphere, three vertex operators should be in the fixed form
\eq{Vfixed} and the others should come in the integrated form
\eq{Vintegrated}. We could also have obtained this result by carefully
gauge fixing a path integral like \eq{path}. The three fixed
insertion points serve to gauge-fix the six real parameter family
of diff $\times$ Weyl transformations that were not fixed by the
gauge choice \eq{unit}. Indeed, the sphere has six conformal
Killing vectors!

Of course, in order for these candidate vertex operators to really
be physical vertex operators, they should also commute with $Q_1$,
$Q_2$ and their antiholomorphic counterparts. The ideas involved
are similar to the ones developed above, but technically more
complicated. We shall not go through the details.

Starting from an $(h,\tilde h)=(1,1)$ tensor operator ${\cal V}$ (satisfying
certain conditions), we have found two different BRST invariant
vertex operators, one with ghost numbers one, the other with ghost
number zero. In computing scattering amplitudes, it does not matter
which vertex operators we fix and where we fix them: all
possibilities correspond to equivalent ways of fixing the gauge.
Independently of the gauge fixing, we determined the number of
ghost number one vertex operators in a non-vanishing amplitude
from the ghost number properties of the vacuum, \eq{qgh3}.
Analogously, one can show that each string state corresponds
to more than one operator ${\cal V}$. In fact, there is an infinite
number of them for each state. They are said to be in different 
{\it pictures}, i.e., they carry different values of $P$ (see \eq{defP}) or
$\tilde P$. Independently,
each of these operators ${\cal V}$ can be fixed or integrated, subject to
the constraint that precisely three of them should be fixed.
There is another constraint coming from \eq{Ptot-2} and its antiholomorphic
counterpart: the holomorphic and antiholomorphic pictures of the vertex 
operators in an amplitude
should add up to $(-2,-2)$. By introducing {\it picture changing
operators} and using contour deformation arguments, one can show
that this is the only constraint. As long as the pictures add up
to $(-2,-2)$, it does not matter which vertex operator is in which
picture.

Of these products of vertex operators, only terms consistent with \eq{qsgh-2} and its
antiholomorphic counterpart contribute to the amplitude.
%%%%%%%%%%%%%%%%%%%%%%%%%%%%%%%%%%%%%%%%%%%%%%%%%%%%%%%%%%%%%%%%%%%%%%%%%%%%%%%

\subsection{Closed string vertex operators}
\label{superstrings:scattering:closed}
This is a good point to write down the explicit form of the closed string vertex
operators we shall use in this thesis. We shall always compute scattering amplitudes
in a trivial Minkowski background
\beq
G_{\m\n}=\eta_{\m\n}~,\ \phi=0~,\ B_{\m\n}=0~,\ C=0~,
\eeq
where the formal sum $C$ was defined in \eq{formalsumC}. The closed string vertex
operators describe fluctuations of the fields around their background values:
\bea
G_{\m\n}(X)&=&\eta_{\m\n}-2\k h_{\m\n}(X)=
\eta_{\m\n}-2\k \zeta_{\m\n}^Ge^{\ii k\cdot X}~;\\
B_{\m\n}(X)&=&-\sqrt 2 \k\zeta_{\m\n}^Be^{\ii k\cdot X}~;\\
C_{\m_1\ldots\m_{m+1}}(X)&=&\sqrt 2 \k c_{\m_1\ldots\m_{m+1}}e^{\ii k\cdot X}~.
\eea
The graviton vertex operator in the $(0,0)$ picture is given by
\beq\label{GVO}
{\cal V}_g^{(0,0)}=\frac{2\kappa}{\pi\a'}\,\zeta_{\mu\nu}^G\,(\partial X^{\mu}
-\ii\frac{\a'}2 \,k\cdot\psi\,\psi^{\mu}) (\bar\partial X^{\nu}-\ii\frac{\a'}2 
k\cdot\tilde\psi\,\tilde\psi^{\nu})\,e^{\ii k\cdot X}~.
\eeq
Note that \eq{gravvertbos} is reproduced when the fermions are put to zero. For the
NS two-form, the $(0,0)$ vertex operator reads
\beq\label{BVO}
{\cal V}_B^{(0,0)}=\frac{\sqrt 2\kappa}{\pi\a'}\,\zeta_{\mu\nu}^B\,(\partial X^{\mu}
-\ii\frac{\a'}2 \,k\cdot\psi\,\psi^{\mu}) (\bar\partial X^{\nu}-\ii\frac{\a'}2 
k\cdot\tilde\psi\,\tilde\psi^{\nu})\,e^{\ii k\cdot X}~.
\eeq
The vertex operator for a R-R potential in the $(-3/2,-1/2)$ picture can be found in
\cite{billo9802}. Its expression is slightly complicated. However, in this thesis
we shall always be able to insert the R-R vertex operator at infinity on the
complex plane. 
%We remind the reader that an operator inserted at the origin of
%the complex plane creates an initial (``in'') state, denoted by a ket. Analogously, 
%a vertex operator inserted at infinity creates a final (``out'') state, which we 
%shall denote by a bra.%
%\footnote{Because the world-sheet theory is free, ``in'' and ``out'' states are the
%same, so our associating a bra with an `out' state is consistent with how the
%bra was introduced in \sect{superstrings:scattering:inner}.}
It turns out \cite{billo9802} that the ``out'' state%
\footnote{See footnote \ref{inout}.}
obtained by inserting a R-R
vertex operator at infinity has a relatively simple expression:
\beq\label{RRstate}
\bra{{\cal C}_{m+1};k}=\shalf (\bra{{\cal C}_{m+1},+;k}+\bra{{\cal C}_{m+1},-;k})
\eeq
with
\beq\label{Czm1}
\bra{{\cal C}_{m+1},\eta';k}=e^{\ii\eta'\b_0\tilde\gamma_0}
\N^{(\eta')}_{AB}\;{}_{-3/2}\bra{A;k}\; {}_{-1/2}\bra{\tilde B;k}~,
\eeq
where we define%
\footnote{Note that the matrix $C$ in the following equation is the charge
conjugation matrix, as we shall explain shorthly. It has nothing to do with a
formal sum of R-R potentials, for which unfortunately we have
used the same symbol.}
\beq\label{Czm2}
\N^{(\eta')}_{AB}=\frac{1}{2\sqrt{2}\,(m+1)!}\,
\left[C\Gamma^{\mu_1\cdots\mu_{m+1}}
\left(\frac{1-\ii\eta'\Gamma_{11}}{1+\ii\eta'}\right)\right]_{AB}
c_{\mu_1\cdots\mu_{m+1}}~.
\eeq
Note that $\eta'$ just denotes $+$ or $-$ and should not be confused with the field in
\eq{bosonizesgh}. Further, ${}_{-3/2}\bra{A;k}$ denotes the holomorphic R vacuum with 
spinor index $A$ (introduced after \eq{msquaredtachyon}), momentum $k$ and superghost
charge $q^{\rm sgh}=-3/2$ (see \eq{sghvacua}). 
The matrix $C$ is the charge conjugation matrix. Conventions on the RR zero
modes and gamma matrices can be found in the Appendix of \cite{dv9707}. Here, we 
only note that 
\bea
\left(\Gamma^\mu\right)^T &=& - C\,\Gamma^\mu\,C^{-1}~;\nonumber \\
C^T&=&-C~;\nonumber\\
\bra{A}B\rangle&=&(C^{-1})^{AB}~;\nonumber\\
\bra{\tilde A}\tilde B\rangle&=&(C^{-1})^{AB}\label{Cproperties}
\eea
and
\bea\label{convgamma}
\psi_0^\mu\, |A\rangle |{\widetilde B}\rangle
&=& \frac{1}{\sqrt{2}} \left(\Gamma^\mu\right)^A_{~C}
\,\left(\!\one\, \right)^B_{~D}|C\rangle\, |{\widetilde D}\rangle~;
\nonumber\\
{\tilde \psi}_0^\mu \,|A\rangle |{\widetilde B}\rangle
&=& \frac{1}{\sqrt{2}} \left(\Gamma_{11}\right)^A_{~C}
\,\left(\Gamma^\mu\right)^B_{~D}\,
|C\rangle |{\widetilde D}\rangle~~.
\ena

In \chap{nonbps}, we shall also use open string vertex operators (to be introduced
there) and $(-1/2,-1/2)$ R-R vertex operators. The latter involve spin fields, a
concept we shall not introduce in any detail. The reader is referred to the
references for more information.

%%%%%%%%%%%%%%%%%%%%%%%%%%%%%%%%%%%%%%%%%%%%%%%%%%%%%%%%%%%%%%%%%%%%%%%%%%%%%%%%%%%%%
\section{D-branes}\label{superstrings:D-branes}
\begin{figure}
\begin{center}
\epsfig{file=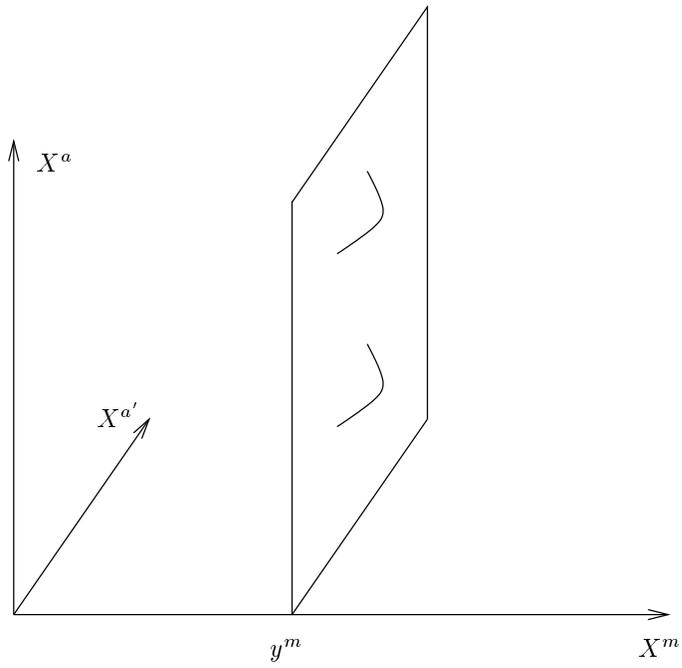}
\end{center}
\caption{A D-brane and two open strings ending on it.}\label{fig:D-brane}
\end{figure}

In \sect{closed-open}, we noted that Neumann boundary conditions are not the only
possible boundary conditions for open strings. Consider, first in a trivial
Minkowski background, open strings satisfying
Neumann boundary%
\footnote{As in \sect{superstrings:strings:worldsheet} we have chosen the real
axis to coincide with the boundary of the open string world-sheet.} 
conditions in the directions 0 to $p$
\beq\label{NeumannX}
\partial X^a-\bar\partial X^a=0\ \ {\rm for}\ \ \Im z=0,\ \ a=0,\ldots p~,
\eeq
and Dirichlet boundary conditions in the other directions
\beq\label{DirichtletX}
X^m=y^m\ {\rm for}\ \ \Im z=0,\ \ m=p+1,\ldots 9~.
\eeq
Here, $y^m$ are constants. These boundary conditions describe open strings whose
end-points are confined to the ($p+1$)-dimensional plane $X^m=y^m$ (see
\fig{fig:D-brane}). Such a plane,
on which open strings can end, is called a D-brane, or D$p$-brane. 

The boundary conditions \eq{NeumannX} and \eq{DirichtletX} have to be
supplemented with appropriate boundary conditions on the fermions:
\bea
\psi^a-\eta\tilde\psi^a=0&\ \ {\rm for}\ \ &\Im z=0,\ \ a=0,\ldots p
\label{BCfermionN}\\
\psi^m+\eta\tilde\psi^m=0&\ \ {\rm for}\ \ &\Im z=0,\ \ m=p+1,\ldots 9~.
\label{BCfermionD}
\eea
Here, $\eta$ equals $1$ or $-1$. In fact, $\eta$ can be chosen independently for
each string endpoint, giving rise to NS and R sectors as in
\sect{superstrings:strings:worldsheet}.
The absence of factors of $\ii$ (present in
\eq{fermionBC}) is due to our working in $z$ rather than $w$ coordinates.  The
different sign in \eq{BCfermionD} compared to \eq{BCfermionN} can be argued for
using T-duality and its interpretation as a one-sided parity transformation (see
\sect{superstrings:dualities:T}).

The
quantization of the open strings goes as described above, the only difference
being that the momenta only have components along the brane (there are no $x^m$
and $p^m$ modes for $m$ a direction in which Dirichlet boundary conditions are
imposed). As a result, the states of the open string will correspond to fields
depending only on the coordinates along the brane, unlike for instance the
metric and gauge fields discussed in the previous section, which depend on all
spacetime coordinates. Before the GSO projection,  the open strings have a
tachyonic ground state, and at the massless level a vector $A_a$, $9-p$
scalars $\phi^m$ and some fermions (the vector and scalars are the reduction of
a ten-dimensional vector to $p+1$ dimensions). As before, the GSO projection
eliminates the tachyon and keeps the massless bosonic modes.%
\footnote{The other GSO projection, keeping the tachyon and eliminating the
massless bosonic modes, also plays a role in string theory, as we shall see in
\chap{nonbps}. In that chapter, we shall also encounter D-branes described by
open strings {\it without} GSO projection.}

The scalars $\phi^m$ can be interpreted as describing local fluctuations in the 
position of the D-brane in transverse space. This is a crucial axiom in the study
of D-branes: it implies that D-branes are dynamical objects. 
Even if we start from a brane world-volume
(which is the generalization of the world-line of a particle and the world-sheet
of a string) which is just a plane, we find excitations (described by open
strings ending on the brane) that describe fluctuations in the shape of the
brane.

\subsection{Action for a single D-brane}
At leading order, the low-energy effective action describing the massless degrees
of freedom of a D$p$-brane is the dimensional reduction of the ten-dimensional
U(1) super-Yang-Mills action to $p+1$ dimensions. As usual, there are higher
order corrections in $\a'$ (since $\a'$ has dimensions of length squared, these are
suppressed at low energies). To leading order in derivatives of the gauge field
strength $F_{ab}=\partial_aA_b-\partial_bA_a$, but to all orders in the field
strength itself, the effective action is given by the {\it Dirac-Born-Infeld
action}
\beq\label{BIflat}
S_{\rm BI}=-\frac{T_p}{\k}\int d^{p+1}x \sqrt{-\det(\eta_{ab}+\partial_aX^m
\partial_bX^m+2\pi\a'F_{ab})}~,
\eeq
where $X^m=2\pi\a'\phi^m$ and $T_p/\k$ is the tension of a D$p$-brane:
\bea
T_p&=&\sqrt{\pi} (2\pi\sqrt{\alpha'})^{3 - p}~;\label{Tp}\\
\k&=&8\pi^{7/2}\a '^2g_s=\sqrt{8\pi G_N}~.\label{kappa}
\eea
Here, $g_s$ is the string coupling constant and $G_N$ Newton's constant.

In the presence of NS-NS background fields, the action \eq{BIflat} is modified to
\beq\label{BIbackground}
S_{\rm BI}=-\frac{T_p}{\kappa}\int d^{p+1}\xi\,
e^{-\Phi}\sqrt{-\det [\hat{G}_{ab}+\hat{B}_{ab}
+2\pi\a '\,F_{ab}]}~,
\eeq
where $\xi^a$ are arbitrary coordinates on the D-brane world-volume, and
$\hat G_{ab}$ and $\hat B_{ab}$ denote the pullbacks to the D-brane worldvolume
of  the bulk fields $G_{\m\n}$ and $B_{\m\n}$. This action receives perturbative
and non-perturbative curvature corrections, (some of) which were computed in
\cite{BBG}.
The action \eq{BIflat} is written in static gauge: $\xi^a=X^a,\ a=0,\ldots p$.

If the R-R background is nontrivial, the Dirac-Born-Infeld action
\eq{BIbackground} should be supplemented with the {Wess-Zumino action}
\beq\label{WZ}
S_{\rm WZ}=\frac{T_p}{\kappa}\int_{p+1}\hat C\wedge e^{2\pi\a '\,F+\hat{B}}
\wedge\sqrt{\frac{\hat{A}(R_T)}{\hat{A}(R_N)}}~.
\eeq
Before we comment on the structure of this action, we explain the notation.
First, $R_T$ and $R_N$ are the curvatures of the tangent and normal bundles of
the D-brane world-volume. Then, $\hat{A}$ denotes the A-roof genus:
\beqa
\label{Aroof}
\sqrt{\frac{\hat{A}(R_T)}{\hat{A}(R_N)}}&=&1+\frac{(4\pi^2\a ')^2}{384\pi^2}
(\tr R_T^2-\tr R_N^2)
+\frac{(4\pi^2\a ')^4}{294912\pi^4}(\tr R_T^2-\tr R_N^2)^2\nonumber\\&&+
\frac{(4\pi^2\a ')^4}{184320\pi^4} (\tr R_T^4-\tr R_N^4)+\ldots~,
\eeqa
where the second term is a four-form, the third and the fourth terms are 
eight-forms and
higher terms are irrelevant because they have at least twelve antisymmetrized
indices, whereas the dimension of any D-brane world-volume is at most 10, the
dimension of spacetime.

By $\hat C$, we mean a formal sum of pullbacks of all R-R potentials (i.e., all
$C_q$ with $q$ odd/even in type IIA/IIB), see \eq{formalsumC}. The two-form
$\hat{B}$ is the pullback of the the NS-NS two-form $B_2$ (we shall often
omit the subscript 2 on the latter).
The integrand in \eq{WZ} is a formal
sum of forms of different degree. The $\int_{p+1}$ sign instructs us to pick
only the ($p+1$)-form part and to integrate it over the ($p+1$)-dimensional 
world-volume.

The Wess-Zumino action \eq{WZ} plays a central role in this thesis. Therefore, we
shall now display some of its terms more explicitly.

Taking the 1 from both the exponential in \eq{WZ} and the expansion
\eq{Aroof}, we find the well-known coupling \cite{Pol} between a D$p$-brane and 
a ($p+1$)-form R-R potential:
\beq
\frac{T_p}{\kappa}\int_{p+1}\hat C_{p+1}~.
\eeq
This coupling just means that D$p$-branes are charged under the ($p+1$)-form R-R
potential.

Taking the linear part of the exponential in \eq{WZ} and the 1 from the expansion
\eq{Aroof}, we find the following coupling between a D$p$-brane and a
($p-1$)-form R-R potential:
\beq\label{FB}
\frac{T_p}{\kappa}\int_{p+1}\hat C_{p-1}\wedge(2\pi\a '\,F+\hat{B})~.
\eeq
The term involving $F$ implies, for instance, that a D$p$-brane with a magnetic
flux carries D($p-2$)-brane charge. The other term says that the same is true 
for a D$p$-brane with $B$-flux.
\sect{anomalous:BC} will be devoted to the part of \eq{FB} that 
involves the $B$-field.

Taking the 1 from the exponential and the four-form part of \eq{Aroof}, we find
the coupling 
\beq\label{4form}
\frac{T_p}{\kappa}\int_{p+1}\hat C_{p-3}\wedge \frac{(4\pi^2\a ')^2}{384\pi^2}
(\tr R_T^2-\tr R_N^2)~.
\eeq 
The first term implies, for instance, that a D$p$-brane wrapped around certain
topologically nontrivial manifolds carries D($p-4$)-brane charge.
The terms \eq{4form} will be derived in \sect{anomalous:RRC}.

Taking the 1 from the exponential and the eight-form part of \eq{Aroof}, we find
the terms 
\beq
\frac{T_p}{\kappa}\int_{p+1}\hat C_{p-7}\wedge 
\left(\frac{(4\pi^2\a ')^4}{294912\pi^4}(\tr R_T^2-\tr R_N^2)^2+
\frac{(4\pi^2\a ')^4}{184320\pi^4} (\tr R_T^4-\tr R_N^4)\right)~,
\eeq 
which will be the subject of \sect{anomalous:D:non}.

There are perturbative \cite{normal} and non-perturbative \cite{BBG} curvature 
corrections to the Wess-Zumino action \eq{WZ}. The perturbative corrections will 
be discussed in \sect{anomalous:D:non}.

Let us pause to comment on the structure of the actions \eq{BIbackground} and
\eq{WZ}.
\begin{enumerate}
\item
If we just consider the spacetime metric and the map embedding the brane in the
spacetime, 
the action \eq{BIbackground} is proportional to the induced volume of the
($p+1$)-dimensional brane world-volume. This is similar to the Nambu-Goto 
action for the string (see \sect{superstrings:strings:worldsheet:bosonic}).
\item
The dilaton dependence of \eq{BIbackground} is as expected for an open string
tree level action (see \sect{superstrings:background}; the Euler number of a disc
is one). From \eq{WZ}, we would also get this dilaton dependence if we redefined
the R-R potentials $C$ as in \eq{RRredef}.
\item
The Born-Infeld field strength $F_{ab}$ and the
pullback $\hat B_{ab}$ of the bulk field $B_{\m\n}$ always appear in the 
combination
\beq\label{BplusF}
\hat B_{ab}+2\pi\a'F_{ab}~.
\eeq
This is related to the comments on $B$-gauge invariance after \eq{Bgaugetransf},
and can be understood as follows. The closed string field $B_{\m\n}$ and the open
string field $A_a$ appear in the world-sheet action of an open string with
end-points on the D-brane as
\beq\label{combination}
\frac{\ii}{2\pi\a'}\int_M B+\ii\int_{\partial M}A~,
\eeq
where by $B$ and $A$ we mean the pullbacks of the respective fields to the
string world-sheet ${\cal M}$ or its boundary $\partial{\cal M}$
(we are using differential form notation, see footnote \ref{diffform} in
\sect{superstrings:strings:spacetime}).
The first term was introduced in
\eq{background}. The second term means that the end-points of the string are
charged under the gauge field on the brane (in fact, both end-points carry
opposite charges). This term is necessary for gauge invariance, as
we now show. Under a gauge transformation \eq{Bgaugetransf}, the first term
changes by
\beq
\frac{\ii}{2\pi\a'}\int_M d\zeta=\frac{\ii}{2\pi\a'}\int_{\partial M}\zeta~.
\eeq
This non-invariance can be compensated if the transformation \eq{Bgaugetransf} 
is accompanied by
\beq\label{BgaugetransfA}
\d A_a=-\frac{\zeta_a}{2\pi\a'}~,
\eeq
where $\zeta_a$ is the pullback of $\zeta_\m$ to the D-brane world-volume. We
conclude that in \eq{combination} only the combination  
is invariant under the gauge transformation
\eq{Bgaugetransf}, \eq{BgaugetransfA}.
\item\label{WZgaugeinv}
To see that the Wess-Zumino action \eq{WZ} is invariant under the
R-R gauge transformations \eq{RRgaugetransf}, note that, as $H_3=dB$,
\beq\label{RRgaugeinv}
\d C\wedge e^B=(d\l+H_3\wedge\l)\wedge e^B=d(\l\wedge e^B)~.
\eeq
Since all other forms appearing in \eq{WZ} are closed, the
variation of the integrand is a total derivative.
%\item
%The first term in the expansion of \eq{WZ} describes the minimal
%coupling of the D$p$-brane to the ($p+1$)-form R-R potential
%$C_{p+1}$. The fact that D-branes carry R-R charge \cite{Pol}
%has been one of
%the crucial insights in the recent developments in string theory.
%The term proportional to $C_{p-1}\wedge F$ implies that a magnetic
%flux induces D($p-2$)-brane charge on a D$p$-brane. One of the
%consequences of the curvature terms is that lower brane charges
%can be induced when branes are wrapped around a manifold with
%non-trivial topology. For instance, a D4-brane wrapped around a K3
%manifold is charged under both $C_5$ and $C_1$.
\end{enumerate}

%\subsection{Supergravity description}
%D$p$-branes are objects in string theory, which at low energies is
%well approximated by supergravity. Therefore, it is no big surprise
%that D-branes have also been found in supergravity: they
%correspond to classical solutions of the supergravity equations of
%motion. These solutions are independent of $p+1$ directions. They
%describe how spacetime is curved due to the presence of D-branes,
%and how the R-R potentials are excited.
%We shall not give the explicit form of the solutions. We would
%like to comment, without going into any technical details,
%on the interpretation of the massless excitations
%of the branes from a supergravity perspective. We believe that this
%gives a nice intuitive understanding of ``why'' these massless
%excitations are present.

%%%%%%%%%%%%%%%%%%%%%%%%%%%%%%%%%%%%%%%%%%%%%%%%%%%%%%%%%%%%%%%%%%%%%%%%%%%%%%%
\subsection{Multiple D-branes}\label{multiple}
\begin{figure}
\begin{center}
\epsfig{file=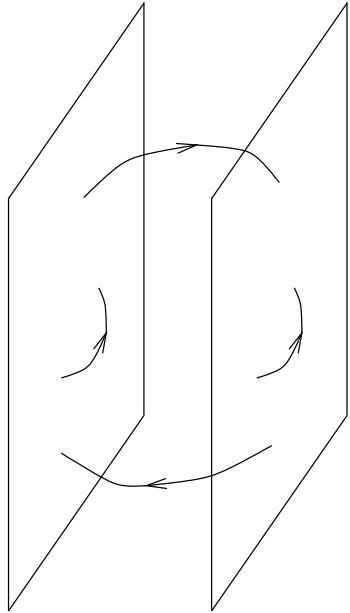}
\end{center}
\caption{Two D-branes. The four open strings are distinguished by the different
boundary conditions they satisfy.}\label{fig:multiple}
\end{figure}

When we consider more than one D-brane, we find more degrees of
freedom than the ones of the individual D-branes. The additional
degrees of freedom are provided by open strings stretching between
different D-branes (see \fig{fig:multiple}). 
For simplicity, consider $N$ coincident
D$p$-branes. Then, the massless degrees of freedom constitute an ${\cal N}=1$, 
$D=10$ U($N$) vector multiplet dimensionally reduced to $p+1$ dimensions. Now, 
$A_a$ is a nonabelian gauge
field and the scalars $\Phi^m$ transform in the adjoint
representation of U($N$). If the D-branes are moved away from each
other, the gauge symmetry is broken to the U(1)$^{N}$ of the
individual branes. The ``off-diagonal'' degrees of freedom get
masses proportional to the distance between the branes to which
their end points are confined.

At the end of \sect{superstrings:strings:worldsheet}, we introduced
Chan-Paton factors, leading to nonabelian gauge groups. Here, we
see that considering multiple D-branes provides an alternative
way to get nonabelian gauge symmetry from string theory. However, it can be
argued via T-duality%
\footnote{T-duality will be introduced in \sect{superstrings:dualities:T}.}
that these two mechanisms are closely related. In fact, we shall refer to
indices for the gauge group on multiple D-branes as Chan-Paton indices.

The leading terms in the low energy action of this system of
D$p$-branes should correspond to the reduction to $p+1$ dimensions
of the $D=10$ U($N$) super-Yang-Mills theory. However, it is
a nontrivial (and, in fact, not completely solved) problem%
\footnote{See, for instance, \cite{DST} for a recent discussion of the
nonabelian Born-Infeld action and for more references.}
to find
the nonabelian generalization of \eq{BIbackground} and \eq{WZ}.
For \eq{WZ} one can, as a ``first approximation'', include a trace
over the gauge (Chan-Paton) indices of the nonabelian field strength which now
appears in the exponential factor. However, it was shown in
\cite{myers} that the full story is more involved and interesting.
For instance, there are additional commutator terms that couple a
D$p$-brane to R-R potentials of degree higher than $p+1$
(remember that the action \eq{WZ} couples this brane to the
potentials up to degree $p+1$). Nevertheless, in this thesis we
shall stick to the ``first approximation'', keeping in mind that there
will be corrections to our results. We shall not discuss the
nonabelian generalization of \eq{BIbackground} in this thesis.

%%%%%%%%%%%%%%%%%%%%%%%%%%%%%%%%%%%%%%%%%%%%%%%%%%%%%%%%%%%%%%%%%%%%%%%%%%%%%%%
\section{Branes and anomaly inflow}\label{superstrings:inflow}
\subsection{D-branes}\label{superstrings:inflow:D}
The D-brane Wess-Zumino action \eq{WZ} is essential for the consistency of
D-branes, in particular for the consistency of intersections of D-branes 
(also called {\it I-branes}). In fact, when intersecting branes (or branes in 
certain nontrivial
backgrounds) are involved, the WZ action may not be invariant under gauge or
coordinate transformations. It turns out \cite{GHM,CY} that the anomalous
variation of this action precisely cancels anomalies of chiral degrees of freedom
living on the intersection (or on the brane itself). This is an example of the
anomaly inflow mechanism, which we introduced in \sect{central:anomalies} in a
field theory context. In this subsection, we shall give a hint of how the inflow
mechanism on I-branes works. We refer the reader to \cite{GHM,CY} for more
rigorous treatments.

A D$p$-brane couples electrically to $C_{p+1}$ and magnetically to the Hodge dual
$*C_{p+1}$, which equals $C_{7-p}$, possibly up to a sign. In configurations
where both electric and magnetic charges are present, it is more
appropriate to work with the field strengths rather than the potentials.
Therefore, we shall first rewrite the WZ action in terms of the gauge invariant
field strengths $\tilde F$, by which we mean a formal sum of R-R field strengths
(cf. \eq{formalsumC}):
\beq\label{deftildeF}
\tilde F=dC+H_3\wedge C
\eeq
(see \eq{tildeF4}, \eq{tildeF3} and \eq{tildeF5}).

The structure of \eq{WZ} and its naive nonabelian generalization discussed
briefly in \sect{multiple} is the following:
\beq\label{WZschematic}
\int_{p+1} C\wedge e^B\wedge I~,
\eeq
where $I$ is a closed, gauge invariant polynomial constructed out of the gauge
field strength and curvature two-forms (to be precise, $I$ is invariant under
gauge transformations of the vector field on the brane and under local Lorentz 
rotations but not under
$B$ gauge transformations).%
\footnote{Note that we have dropped the hats on $C$ and $B$. It
should be clear from the context when a pullback is meant.} 
Writing, as in 
\sect{central:anomalies},
\bea
I-I_0&=&d\,I^{(0)}~;\nonumber\\
\d\,I^{(0)}&=&d\,I^{(1)}~,\label{I0I1}
\eea
where $I_0$ is the leading zero-form term of $I$,
we can integrate \eq{WZschematic} by parts to obtain
\beq\label{WZfieldstr}
\int_{p+1} (C\wedge e^B I_0+I^{(0)}\wedge\tilde F\wedge e^B)~.
\eeq
Note that the $\d$ in \eq{I0I1} can now be a gauge transformation of the vector
field living on the brane, or a local Lorentz rotation in the tangent or normal
bundle to the brane. The first is really a gauge symmetry on the brane. The local
Lorentz rotations are gauged in spacetime, so that there must not be any
anomalies in them either. Incidentally, it is a matter of choice whether one
discusses anomalies in local Lorentz rotations or in general coordinate
transformations (one can shift them from one to the other). Therefore, we refer 
to the anomalies in local Lorentz transformations as gravitational anomalies.

It follows from \eq{deftildeF} that
\beq\label{bianchi}
d(\tilde F\wedge e^B)=0
\eeq
(this can be considered as the Bianchi identity for $\tilde F$). Combining this
with \eq{I0I1}, it is clear that \eq{WZfieldstr} is invariant under gauge and
local Lorentz transformations (as it should be, being equal to
\eq{WZschematic},
which is manifestly invariant). Analogously, \eq{WZfieldstr} is still invariant
under $B$ gauge transformations $\d_BB=d\Lambda$ if this is combined with
$\d_B I^{(0)}=-\Lambda\wedge I$:
\beq
\d_B(e^B\wedge I^{(0)})=-e^B\wedge\Lambda I_0+e^Bd(\Lambda\wedge I^{(0)})~.
\eeq
The first term in the variation of the second term in \eq{WZfieldstr} will 
cancel the variation of the first term in \eq{WZfieldstr}; 
the second term in the variation of the second term in \eq{WZfieldstr} 
is zero upon using \eq{bianchi}.   
 
Just as in electrodynamics, the Bianchi identity is modified in the presence of
magnetic charges, \eq{bianchi} will undergo changes in the presence of D-branes.
This will give rise to noninvariance of \eq{WZfieldstr} under gauge and local
Lorentz transformations in certain backgrounds. As in \sect{central:anomalies}, 
its gauge variation will be cancelled by an anomaly due to localized chiral
fermions \cite{GHM,CY}.

As is usual at present, we shall put $B=0$, though it would be interesting to see
how a nonzero $B$ field should be incorporated. Also, in \cite{myers} the
question was raised what happens to $B$ gauge invariance in the presence of
intersecting branes. We shall not deal with these issues here.

The argument that anomalies due to chiral zero modes on intersections of D-branes
(and on D-branes in certain nontrivial gravitational backgrounds) are precisely
cancelled by anomaly inflow can be found in \cite{GHM,CY}. We shall give the
reader a rough idea on two points: which chiral zero modes live on D-brane 
intersections and
how is the variation of \eq{WZfieldstr} localized on brane intersections? 
We shall do this for the special case of two D5-branes intersecting on a string. 

\begin{figure}
\begin{center}
\epsfig{file=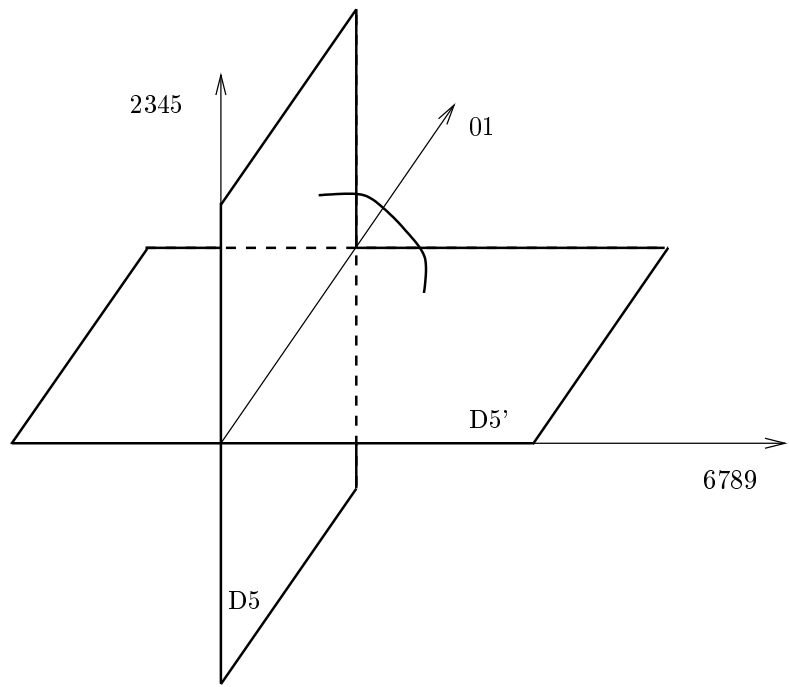}
\end{center}
\caption{Two D5-branes intersecting on a string. The strings stretching from one
fivebrane to the other give rise to chiral fermions on the intersection.}
\label{fig:inflow}
\end{figure}

Consider a D5-brane (denoted D5) stretched along the directions 012345 and another
D5-brane (D5') along the directions 016789. They intersect on a string (I-brane)
in the 01 directions (see \fig{fig:inflow}). 
The usual degrees of freedom of both D5-branes are obtained
as in \sect{superstrings:D-branes} by quantizing open strings with their 
end-points confined to the D-brane in question. For D5-branes in flat space (as we
are considering), these degrees of freedom are nonchiral. However, two intersecting
branes have more degrees of freedom than those associated to the individual
branes: there are open strings with one end-point on D5 and the other on D5'. Upon
quantization, these strings give rise to extra degrees of freedom localized on the
intersection. It turns out that, in the situation we are considering here, these 
degrees of freedom are chiral fermions, charged under the gauge symmetry on both 
branes. These chiral fermions give rise to gauge and gravitational anomalies.

The reader is invited to compare this situation with the one in   
\sect{central:anomalies}. In both cases, we have chiral fermions 
localized on a submanifold of spacetime, although the full bulk theory is (assumed
to be) consistent, i.e., free of anomalies in gauge symmetries. In both cases, the
anomalies due to the localized chiral fermions will have to be cancelled by
inflow. From a ``practical'' point of view, there is one great difference between
both cases. In  \sect{central:anomalies}, we knew the degrees of freedom of the 
theory we started from, and the cancellation of anomalies was the result of a
somewhat artificial splitting of the effective action in two pieces. 
The chiral modes living on the axion string were just modes of a field in the
action.
In the present case, we start from a theory of which we do not know the fundamental
degrees of freedom. We do not know what ``fundamental field'' the chiral fermions
are modes of, we just found them using the fact that open strings can end on
D-branes. The anomaly inflow argument can be used to extract information about the
theory away from the intersection.

We still have to indicate how the WZ action \eq{WZfieldstr} can cancel anomalies
localized on D-brane intersections. Going back to our example, consider the
\beq\label{D5D5}
\int_{\rm D5} I^{(0)}_3\wedge\tilde F_3
\eeq 
term of \eq{WZfieldstr}, which would be equal to
\beq
\int_{\rm D5}C_2 \wedge I_4
\eeq
in the absence of other branes. Note that we are considering the situation with
$B=0$, so that $\tilde F_3$ is just $dC_2$. In the presence of D5', \eq{bianchi}
changes into
\beq
d\tilde F_3=\d_{\rm D5'}
\eeq  
(up to a constant),
where in general $\d_{\rm brane}$ is the {\it current} associated to the 
world-volume of the D-brane in question.%
\footnote{%  
For instance, for a D$q$-brane stretched along the first
$q+1$ coordinate directions the associated current reads
\beq
\d_{{\rm D}q}=\d(x^{q+1})dx^{q+1}\wedge\ldots\wedge\d(x^9)dx^9~.
\eeq
In general, the current is defined such that for any $q$-form $\zeta$
\beq
\int_{\rm brane}\zeta=\int_{\rm spacetime}\d_{\rm brane}\wedge\zeta~.
\eeq}
Using $\d\,I^{(0)}_3=d\,I^{(1)}_2$ (see \eq{I0I1}) we thus find
\beq
\d\int_{\rm D5} I^{(0)}_3\wedge\tilde F_3=-\int_{\rm D5}I^{(1)}_2\wedge 
\d_{\rm D5'}~,
\eeq
which shows that the variation of \eq{D5D5} is indeed localized on the
intersection with D5'. Analogously, a similar term on D5' will also have a 
variation localized on the intersection. 

In \eq{WZfieldstr}, the first term is expressed in terms of the R-R
potentials rather than the field strengths. In the case we are considering
(only D5-branes and $B=0$), this
term describes the coupling of the D5-branes to $C_6$.% 
\footnote{This term cannot be expressed in terms of a field strength.} 
It turns out \cite{GHM,CY} that, in the presence of branes, the R-R potentials
have an anomalous gauge variation. The variation of the terms
\beq
\int_{\rm D5}C_6\ \ {\rm and}\ \ \int_{\rm D5'}C_6
\eeq 
is also localized on the intersection of both branes.

All in all \cite{GHM,CY}, the total variation of the Wess-Zumino actions for D5
and D5' precisely cancels the anomaly due to the chiral fermions on the
intersection. In fact, most of the curvature terms in \eq{WZ} were discovered 
using anomaly inflow arguments. 

The terms displayed in \eq{WZ} have been called {\it
anomalous D-brane couplings} \cite{GHM}. 

This anomaly inflow argument and, in particular, the curvature terms in \eq{WZ}
fixed by it, occupy a central position in this thesis. 
In \chap{anomalous}, we shall find more direct evidence for the presence of
these curvature terms. In \chap{type0}, we repeat the anomaly inflow argument for
D-branes in non-supersymmetric type 0 string theories. In this way, we fix the
whole Wess-Zumino action for those D-branes. The Wess-Zumino action can then
be checked directly repeating the computations of \chap{anomalous}.
%%%%%%%%%%%%%%%%%%%%%%%%%%%%%%%%%%%%%%%%%%%%%%%%%%%%%%%%%%%%%%%%%%%%%%%%%%%%%%%%
\subsection{NS fivebranes}\label{superstrings:inflow:NS5}
We have seen that type II string theories contain fundamental strings, which are 
electrically charged under the $B$ field (see \eq{background}), and D-branes, which
are electrically and magnetically charged under certain R-R potentials. String
perturbation theory gives an explicit descriptions of the dynamics of these objects
(at weak string coupling). However, these are not the only extended objects in the
type II theories. This subsection deals with the NS fivebrane, which is a solitonic
object (a localized classical solution to the field equations). It is magnetically
charged under the $B$ field. 

To describe the dynamics of this object, one can look at
fluctuations of the fields around the classical solution \cite{CHS}.%
\footnote{This is how we found the chiral fermion living on the axion string in
\sect{central:anomalies}.}
In both type IIA and type IIB, there are normalizable massless modes corresponding to
translations in the four transverse directions. In addition, there is a vector on the
the IIB NS5-brane and a self\-dual
antisymmetric tensor and an additional scalar on the IIA NS5-brane. These massless 
degrees of freedom are accompanied by fermionic superpartners, which are chiral for 
IIA and nonchiral for IIB. 

Incidentally, we shall derive these results in \sect{type0:NS5:spectra} by
first T-dualizing to Kaluza-Klein (KK) monopoles, which are other extended objects 
in type
II string theories. Of course, in that derivation we use T-duality properties of
NS5-branes. Although T-duality will be introduced in \sect{superstrings:dualities:T},
we shall not derive the fact that NS5-branes and KK monopoles are T-dual. Note that
dualities are often helpful in determining the excitations of extended objects.%
\footnote{This often works the other way around: knowing the field
content living on certain extended objects adds evidence to duality conjectures.}
For instance, the modes living on the IIB NS5-brane are predicted by (or add evidence
to) type IIB S-duality, to be introduced in \sect{superstrings:dualities:S}.

As the reader may have guessed, in this subsection we shall be interested in the
chiral degrees of freedom living on the IIA NS5-brane: a selfdual antisymmetric
tensor and a number of chiral fermions. It is well-known that this spectrum gives
rise to a gravitational anomaly in six dimensions: the 
energy-momentum tensor of the chiral modes living on the NS5-brane is
not conserved at the quantum level.%
\footnote{We are implicitly making the choice that the anomaly is in the general
coordinate transformations rather than the local Lorentz transformations.} 
The anomaly, i.e., the divergence of the energy-momentum tensor,
is localized on the NS5-brane world-volume. Nevertheless, the full theory
we started with (type IIA string theory) is non-chiral and thus free of 
gravitational anomalies. The reader who has read \sect{central:anomalies} and the 
previous subsection will not be surprised by the resolution of this puzzle
\cite{duff}. The anomaly should be cancelled by a bulk term of the form
\beq\label{Homega}
-\int_{10}H_3\wedge \omega_7~,
\eeq 
where
\bea
X_8&=&d\omega_7~,\nonumber\\
\d\omega_7&=&dJ_6\label{X8}
\eea
for a specific closed, invariant polynomial $X_8$ of the curvature twoform. 
Here, $\d$ is a general coordinate transformation.
Needless to say, \eq{X8} is just an example of the descent relations  \eq{descent1}
and \eq{descent2}. \eq{Homega} is often written as $\int B_2\wedge X_8$, although
$B_2$ is not well-defined in the presence of an NS5-brane. Using the fact that
NS5-branes are magnetically charged under $B_2$,
\beq
dH_3=\d_{\rm NS5}
\eeq
(up to a factor),
it is clear from \eq{X8} that the variation of \eq{Homega} is localized on the
NS5-brane:
\beq
-\d\int_{10}H_3\wedge \omega_7=-\int_{10}dH_3\wedge J_6=-\int_{\rm NS5}J_6~.
\eeq
It turns out \cite{duff} that this variation of \eq{Homega} indeed cancels the 
anomalous variation due to the chiral zero modes on the type IIA NS5-brane.

This picture is quite satisfactory, but it would be nice to have more direct evidence
for the presence of the term \eq{Homega} in the type IIA supergravity action (note
that this term is higher order in momenta than the terms shown in
\sect{superstrings:strings:spacetime}). In fact, it was explicitly computed to be
present in \cite{vafawitten}, before this anomaly inflow argument was made. Some
details of the computation of \cite{vafawitten} will be given in
\sect{type0:NS5:anomalies}.

In \sect{type0:NS5:anomalies}, we shall also repeat the computation of 
\cite{vafawitten} for type 0 string theories. We shall find that in those
theories there is no bulk term like \eq{Homega}. From that observation, we are
then able to
conclude that the NS5-branes in those theories must have a non-chiral spectrum.
This will be confirmed by a more direct argument.
%%%%%%%%%%%%%%%%%%%%%%%%%%%%%%%%%%%%%%%%%%%%%%%%%%%%%%%%%%%%%%%%%%%%%%%%%%%%%%%
\section{Dualities}\label{superstrings:dualities}
\subsection{T-duality}\label{superstrings:dualities:T}
We start from type IIA or type IIB string theory in a trivial
Minkowski background and compactify one of the directions (say the
9-direction) on a circle of radius $R$. Just as in field theory,
the first effect is that the momentum is quantized:
\beq
k=\frac{n}{R}~,\ \ n\in\Zbar~.
\eeq
The second effect has no counterpart in field theory: strings can
wind around the compact direction:
\beq
X^9(\s^1+2\pi,\s^2)=X^9(\s^1,\s^2)+2\pi Rw~,\ \ w\in\Zbar ~.
\eeq
The integer $w$ is the {\it winding number}. The mode expansion
\eq{XplusXbar} for $\m=9$ is replaced by
\beq
X^9(z,\bar z)=X^9_L(z)+X^9_R(\bar z)~,
\eeq
with
\bea
X^9_L(z)&=&x^9_L-\ii\frac{\a'}{2}p^9_L\ln z+\ii\sqrt{\frac{\a'}{2}}
\sum_{\stackrel{m=-\infty}{m\neq 0}}^\infty
\frac 1m\a_m^9 z^{-m}~;\nonumber\\
X^9_R(\bar z)&=&x^9_R-\ii\frac{\a'}{2}p^9_R\ln\bar z+\ii\sqrt{\frac{\a'}{2}}
\sum_{\stackrel{m=-\infty}{m\neq 0}}^\infty
\frac 1m\tilde\a_m^9\bar z^{-m}~.
\eea
Here,
\bea
p^9_L&=&\frac{n}{R}+\frac{wR}{\a'}~;\nonumber\\
p^9_R&=&\frac{n}{R}-\frac{wR}{\a'}~.
\eea
Instead of \eq{massshell}, we now have
\beq
m^2=-\sum_{\m=0}^8k^\m k_\mu=(k_L^9)^2+\frac4\a'(N+a)=
(k_R^9)^2+\frac4\a'(\tilde N+\tilde a)~,
\eeq
or
\bea
m^2&=&\frac{n^2}{R^2}+\frac{w^2R^2}{\a'^2}+\frac{2}{\a'}(N+\tilde
N+a+\tilde a)~;\\
0&=&nw+N-\tilde N+a-\tilde a~.
\eea
This mass formula is invariant under
\beq
R\rightarrow R'=\frac{\a'}{R}~,\ \ n\leftrightarrow w~.
\eeq
%In fact, this equivalence extends to the interactions as well.
Interchanging $n$ and $w$ can be achieved by replacing $X^9(z,\bar
z)=X^9_L(z)+X^9_R(\bar z)$ by
\beq
X'^9(z,\bar z)=X^9_L(z)-X^9_R(\bar z)~.
\eeq
By superconformal invariance, this ``one-sided parity
transformation'' includes a reflection of $\tilde\psi^9(\bar
z)$:
\beq
\tilde\psi'^9(\bar z)=-\tilde\psi^9(\bar z)~.
\eeq
This has an interesting consequence: as the zero mode
$\tilde\psi^9_0$, which acts as a gamma matrix in the
antiholomorphic R sector, changes sign, the chirality of the
antiholomorphic R sector ground state is reversed. Therefore,
interchanging winding and momentum takes IIA to IIB and vice
versa. What we have learned is that, for instance, the type IIA
theory compactified on a circle of small radius has the same
spectrum as the type IIB theory on a circle of large radius.
In fact, this equivalence extends to the interactions as well.
It is called {\it T-duality}.

Thus, we see that the uncompactified type IIA and IIB theories are
just different limits of a single space of compactified theories.
A remarkable consequence of T-duality is that strings cannot tell
the difference between large and small compactification radii: it
looks like the string length $\sqrt{\a'}$ constitutes a minimal
length scale%
\footnote{This discussion requires some modification when D-brane
probes are considered: they can probe smaller distances than
strings.}. This dramatically shows how our intuitive notions of
spacetime break down at the string scale.

As T-duality interchanges IIA and IIB, it should turn a R-R
potential with an odd number of indices into one with an even
number of them, and vice versa. Up to signs, the result of
T-duality in the 9-direction is to remove the index 9 if it is
present and to add it if it is not. For instance,
\bea
C_9&\rightarrow& C~;\nonumber\\
C_{\m}&\rightarrow&C_{\m 9}~.\label{RRTduality}
\eea

So far, we have only considered closed strings. Let us now
introduce D-branes and open strings and see how T-duality acts on
them. It is easy to see that if the open string coordinate
$X^9(z,\bar z)$ satisfies Neumann boundary conditions, then the
``one-sided parity transformed'' coordinate $X^9(z,\bar z)$
satisfies Dirichlet boundary conditions, and vice versa. Thus, a
D-brane extended in the 9-direction is transformed into one which
is localized in that direction, and vice versa. Note that this is
consistent with the D-branes being charged under the R-R
potentials and the transformation \eq{RRTduality} of the latter.

%%%%%%%%%%%%%%%%%%%%%%%%%%%%%%%%%%%%%%%%%%%%%%%%%%%%%%%%%%%%%%%%%%%%%%%%%%%%%%%
\subsection{Type IIB S-duality}\label{superstrings:dualities:S}
String theory is not as well-understood as field theory: we are lacking a fundamental
description of the theory. One tool we have is string perturbation theory, which only
makes sense when the string coupling constant is small. Even in principle, it is not
known how to do computations at generic values of the string coupling constant.
However, in the past few years some powerful tools have been developed that allow us
to understand the physics at least at some regions of parameter space. These tools
are called {\it dualities}. In \sect{central:duality}, we have introduced this concept
mainly in the context of field theory. We have seen examples where a strongly coupled
theory could equivalently (but with much more computational power!) be described by a
weakly coupled dual theory. The picture of string theory that has emerged in the last
five or six years is that there is one ``big'' theory (called M-theory, see also the
next subsection) of which all five known consistent string theories are perturbative
descriptions, which are each useful in a certain corner of parameter space. One
example was given in the previous subsection: weakly coupled type IIA and type IIB 
theory in ten noncompact dimensions are continuously related to one another. The
parameter interpolating between the two theories is the size of a compactified
direction. 

We now give a different, more involved example.
We start from weakly coupled type IIB string theory and ask what happens when
the coupling constant grows strong.

The conjecture is that type IIB string theory is {\it selfdual}, i.e., that the theory
at strong coupling looks precisely the same as at weak coupling! In 
\chap{central}, we
have seen an example of a selfdual field theory, and in fact the situation for type IIB is
quite similar to that one. 

There are two strings in the theory: the fundamental string and the D-string
(D1-brane). The ratio of their tensions is $\tau_{\rm F1}/\tau_{\rm D1}=g_s$, so that
at weak coupling the fundamental string is much lighter than the D-string. At strong
coupling, this is the other way around. The duality conjecture is that at strong
coupling the theory is equivalent to the weakly coupled type IIB theory, with the
D-string now playing the role of the fundamental string. Type IIB S-duality,
as it 
is called, reverses the string coupling and exchanges the two strings. 

The evidence for S-duality is largely based on supersymmetry. For instance,
supersymmetry allows to follow the ground states of, say, the D-string to strong
(fundamental string) coupling and compare them to those of a weakly coupled
fundamental string. Another piece of evidence is that the low energy effective action
is fixed by its large amount of supersymmetry.

S-duality leaves the potential $C_4$ invariant, so it should take the D3-brane to
itself. As it exchanges the fundamental string and the D-string, it should also
exchange the potentials $B_2$ and $C_2$ they couple to, and the corresponding
magnetically charged objects, the NS5-brane and the D5-brane.

In fact, the S-duality group is bigger than just the $\Zbar_2$ transformation we
have considered so far. It is enlarged to a discrete SL(2,$\Zbar$) subgroup of the 
SL(2,$\Rbar$) symmetry group of type IIB supergravity (see
\sect{superstrings:strings:spacetime}).

In \sect{type0:dual}, we shall critically examine a similar duality conjecture for 
the nonsupersymmetric type 0B string theory.
%%%%%%%%%%%%%%%%%%%%%%%%%%%%%%%%%%%%%%%%%%%%%%%%%%%%%%%%%%%%%%%%%%%%%%%%%%%%%%%%%%%
\subsection{M-theory}\label{superstrings:dualities:M}
M-theory in the strict sense is the strong coupling limit of type IIA string theory.
It is to a large extent unknown what this theory looks like, though it is known that
it is eleven-dimensional and that its low energy limit is eleven-dimensional
supergravity. Thus, the eleven-dimensional supergravity theory we described in 
\sect{superstrings:strings:spacetime} appears in string theory in a remarkable way. 

There are many more dualities between the different consistent string theories. For
instance, in \sect{nonbps:mot:testing} we briefly introduce heterotic/type I S-duality.
The ``big picture'' that emerged in the last half of the nineties is that all
consistent supersymmetric string theories correspond to different corners of moduli
space (parameter space) of a single underlying theory, which is also called M-theory.
Uncovering the fundamental degrees of freedom of M-theory (and thus of string theory)
is one of the most important goals in this branch of physics. Nevertheless, M-theory
will not appear explicitly in the rest of this thesis. 
%%%%%%%%%%%%%%%%%%%%%%%%%%%%%%%%%%%%%%%%%%%%%%%%%%%%%%%%%%%%%%%%%%%%%%%%%%%%%%%%%%%%
%\subsubsection{M-theory origin of anomalous couplings}
%\label{superstrings:dualities:M:origin}
%{\bf XXX} Misschien zet ik hier Mukhi's oplossing voor een probleem dat wij
%opwierpen. Misschien komt hier gewoon niets...

%Anomalous couplings from string computations
\chapter{Anomalous couplings from string computations}\label{anomalous}
In this chapter, we check the presence of some of the anomalous D-brane 
couplings \eq{WZ} by explicit string computations. These computations are done in
the boundary state formalism. In \sect{anomalous:boundary}, we introduce the
boundary state and give an explicit set of rules to use it in sphere  
scattering amplitudes. In \sect{anomalous:overview}, we give an overview of the
checks of the Wess-Zumino action \eq{WZ}. The actual checks are performed in
\sect{anomalous:BC}, \sect{anomalous:RRC} and \sect{anomalous:D:non}. As a nice
by-product, we find new, non-anomalous D-brane couplings.

\section{Boundary states}\label{anomalous:boundary}
\subsection{Introduction}

In \sect{superstrings:D-branes}, we have introduced D-branes as
spacetime defects on which open strings can end. Adding%
\footnote{Actually, it is likely \cite{polchinski} that D-branes would be present from
the start in a full, nonperturbative definition of type II string theory. In
that sense we are not really adding anything.} 
a D-brane to a theory of
closed strings means adding open strings with their end-points confined to the
D-brane. These open string degrees of freedom interact with the closed strings
in the bulk. For instance, one can imagine two open strings joining and forming
a closed string, which can escape from the D-brane. Another example is the end-%
points of a single open string coming together, thus transforming the open
string in a closed string, which is no longer confined to the brane. One is
often interested in the resulting interaction of the D-brane with the gravitons,
gauge potentials and other closed string states in the bulk. These interactions
determine, amongst other things, the spacetime quantum numbers of the D-brane,
such as its mass, charges etc. From the closed string point of view, D-branes
are sources for closed string states: they are objects emitting (off-shell)
gravitons, gauge potentials etc. The boundary state formalism, which goes back
to \cite{callan87}%
\footnote{The careful reader may note that D-branes were not known at the time
\cite{callan87} was written. The boundaries in that reference describe the
interaction with open strings of the type I theory of open and closed strings.
The open strings, which live in the bulk of spacetime, can be reinterpreted in the D-brane
language as living on a number of spacetime-filling D9-branes.}
and even further to \cite{ademollo},
implements this point of view. 
%A boundary ``state'' is not a state in the
%physical Hilbert space of the string, because it is not normalizable. One way to
%think about it is as a linear operator on the Fock space of the string,
%associating to each state its expectation value on the disc. However, in
%practice it is convenient to use a notation in which the boundary ``state'' 
%appears to be a state. The properties of this ``state'' just reflect properties
%of the operator associated to it. In fact, these considerations need not concern
%us too much, as we are going to use the boundary state as a practical devise: it
%gives a prescription for computing the interaction of closed strings with 
%D-branes. This
%prescription is equivalent with prescriptions which are more straightforward
%from the open string point of view. For instance, a sphere diagram with a
%boundary state inserted is the same as a disk diagram with the appropriate
%boundary conditions on the worldsheet fields.

A boundary state is, roughly speaking, a coherent closed string state. Inserting
it in an amplitude has the same effect as cutting a disc out of the world-sheet
and imposing appropriate boundary conditions on the world-sheet fields. For
instance, a sphere diagram with a boundary state inserted is the same as a disk 
diagram with the appropriate boundary conditions on the world-sheet fields.

\begin{figure}
\begin{center}
\epsfig{file=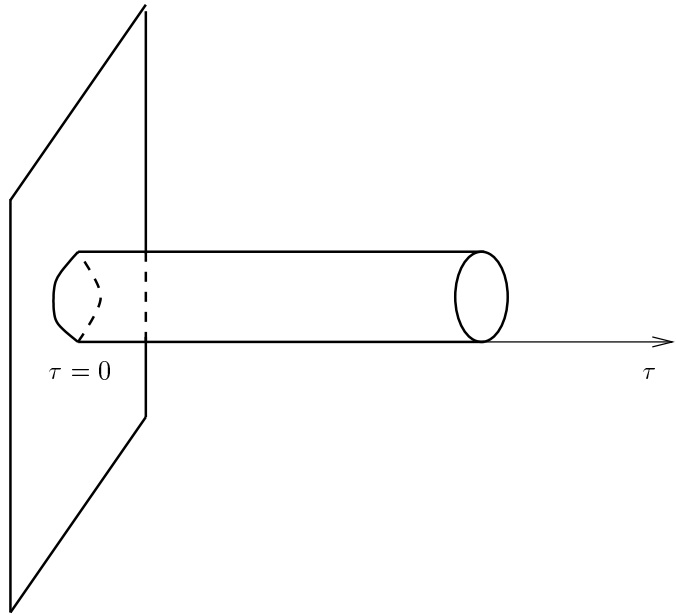}
\end{center}
\caption{The boundary state describes closed strings emitted by a D-brane.}
\label{fig:BS}
\end{figure}

We shall choose world-sheet coordinates such that the boundary is at world-sheet
time $\tau=0$ and that the part of the world-sheet that has not been cut out is
described by $\tau\geq0$ (see \fig{fig:BS}).%
\footnote{The part of the world-sheet that has been cut out is thus described by
$\tau<0$. In the coordinate $z=\exp(\tau+\ii\s)$, this corresponds to the unit
disc.} 
The boundary state then describes the (off-shell) closed string
states emitted by the D-brane. One can think of a coherent state, created at
$\tau=-\infty$, which has the property that it is annihilated by the quantized
boundary conditions at $\tau=0$.

The boundary conditions that are imposed on the bosonic world-sheet fields
are the following: 
the boundary is tied to the brane (Dirichlet boundary
conditions in the transversal directions) but free to move along the brane
(Neumann boundary conditions in the longitudinal directions). We shall now give
the explicit form of the boundary state.
%%%%%%%%%%%%%%%%%%%%%%%%%%%%%%%%%%%%%%%%%%%%%%%%%%%%%%%%%%%%%%%%%%%%%%%%%%%%%%%

\subsection{Explicit form}\label{precise}

We construct the boundary state as in \cite{billo9802}. The boundary state
$|B\rangle$
is a BRST invariant state
%\footnote{Strictly speaking, one should say that the corresponding operator
%associates zero expectation value to BRST exact closed string states, see the
%discussion above.}
that enforces the boundary conditions imposed by a D-brane. It exists
in both the NS-NS sector and the R-R sector of the closed superstring. 
In either sector, it can be written as
\beq\label{BS}
|B\rangle=|B_{\rm m}\rangle|B_{\rm g}\rangle~,
\eeq
where $|B_{\rm m}\rangle$ and $|B_{\rm g}\rangle$ are the matter and ghost parts of the
boundary state, which in turn factorize into $X$ and $\psi$ field (resp. ghost
and superghost) dependent factors:
\bea
|B_{\rm m}\rangle&=&|B_{X}\rangle|B_{\psi}\rangle~;\\
|B_{\rm g}\rangle&=&|B_{\rm gh}\rangle|B_{\rm sgh}\rangle~.
\eea
We shall first construct a boundary state that describes a boundary at $\tau=0$.
The condition that the string world-sheet start on the brane (positioned at
$y^i$, $i=p+1,\ldots 9$) is implemented by specifying that the boundary state is
an eigenstate of the string position operators
\beq\label{defBSXd}
(X^i-y^i)|_{\tau=0}|B_{X}\rangle =0~.
\eeq
Similarly, the condition that this initial position be freely movable along the
brane is translated as
\beq\label{defBSXn}
\partial_\tau X^\a|_{\tau=0}|B_{X}\rangle=0~,\ \ \a=0,\ldots p~.
\eeq
We also impose that the boundary state be annihilated by the boundary conditions
\eq{BCfermionN} and \eq{BCfermionD} on the fermionic fields:%
\footnote{Note that the $\psi$'s are expressed in $w=\s+\ii\tau$ coordinates, such
that the boundary is located at $\Im w=0$. Therefore, the form of the boundary
conditions are the same as in \eq{BCfermionN} and \eq{BCfermionD}, which,
however, were expressed in terms of $z=\exp(\tau+\ii\s)$ coordinates.}  
\beq\label{defBSfermion}
(\psi^\a-\eta\tilde\psi^\a)|_{\tau=0}|B_{\psi},\eta\rangle =0\ ,\ \ 
(\psi^i+\eta\tilde\psi^i)|_{\tau=0}|B_{\psi},\eta\rangle =0~.
\eeq 
%\label{defBS}
%\footnote{`Operators that vanish due to the boundary conditions should have
%vanishing expectation value on the disc.'}

In \eq{defBSfermion}, 
$\eta$ can take the values $\pm 1$, but it turns out that the
GSO projection on the boundary state
%\footnote{`Only GSO projected states can have a nonzero expectation value on the
%disc'; in physical terms: a D-brane in type IIA or IIB only emits states with the
%appropriate GSO projection.}  
selects a specific combination of $|B,+\rangle $ and $|B,-\rangle$.

Expanded in modes, \eq{defBSXd}, \eq{defBSXn} and \eq{defBSfermion} read
\bea
(\a_n+S\cdot\tilde\a_{-n})|B_{X}\rangle &=&0 \ \ (n\neq 0)~;\nonumber\\
p^\a|B_{X}\rangle &=&0 ~;\nonumber\\
(x^i-y^i)|B_{X}\rangle &=&0 ~;\nonumber\\
(\psi_m-\ii\eta S\cdot\tilde\psi_{-m})|B_\psi,\eta\rangle &=&0~,\label{defBSinmodes}
\eea
where $m$ is integer in the R-R sector and half-odd-integer in the NS-NS sector,
and we define
\begin{equation}
S_{\mu \nu} = ( \eta_{\alpha \beta} , - \delta_{ij} )~.   
\label{smunu}
\end{equation}

Now, we include the ghosts and the superghosts.
The relations \eq{defBSinmodes} imply that $|B_{\rm m},\eta\rangle $ is annihilated by 
the following combinations of the matter super-Virasoro generators:
\beq\label{superVirasoroBS}
(L^{\rm m}_n-\tilde L^{\rm m}_{-n})|B_{\rm m},\eta\rangle =0~,\ \ (G^{\rm m}_m
+\ii\eta\tilde G^{\rm m}_{-m})|B_{\rm m},\eta\rangle =0~.
\eeq
Because of BRST invariance, i.e.,
\beq
Q_{\rm B}|B,\eta\rangle =0~,
\eeq
the boundary state must be annihilated by the following combinations of ghost
fields:
\bea
(c_n+\tilde c_{-n})|B_{\rm gh}\rangle =0&,& (b_n-\tilde b_{-n})|B_{\rm gh}\rangle =0~;
\nonumber\\
(\gamma_m+\ii\eta\tilde\gamma_{-m})|B_{\rm sgh}\rangle =0&,& (\b_m+\ii\eta\tilde\b_{-m})
|B_{\rm sgh}\rangle =0~.\label{defghostBS}
\eea

The solution of \eq{defBSinmodes} and \eq{defghostBS} can be written as follows
\cite{billo9802}:
\begin{equation}
\label{bs3}
\ket{B,\eta}_{\rm R,NS} = {T_p\over 2}
\ket{B_X}\, \ket{B_\gh}\,{\ket{B_\psi,\eta}}_{\rm R,NS}
  \,{\ket{B_\sgh,\eta}}_{\rm R,NS}~~,
\end{equation}
where
\begin{equation}
\label{bs5}
\ket{B_X} = \delta^{(d_\bot)}(x - y)
\exp\biggl[-\sum_{n=1}^\infty \frac{1}{n}\,
\a_{-n}\cdot S\cdot
\tilde \a_{-n}\biggr]\,
\ket{1}_X
\end{equation}
and
\begin{equation}
\label{bs6}
\ket{B_\gh} = \exp\biggl[\sum_{n=1}^\infty
(c_{-n}\tilde b_{-n}
 - b_{-n} \tilde c_{-n})\biggr]\,{c_0 +\tilde c_0\over 2}\,c_1\tilde c_1|1\rangle _{\rm
 gh}~~. 
\end{equation}
In the NS sector in the $(-1,-1)$ picture,
\begin{equation}
\label{bs7}
\ket{B_\psi,\eta}_{\rm NS} = \exp\biggl[\ii\eta\sum_{m=1/2}^\infty
\psi_{-m}\cdot S \cdot \tilde \psi_{-m}\biggr]
\,\ket{1}_\psi
\end{equation}
and
\begin{equation}
\label{bs8}
\ket{B_\sgh,\eta}_{\rm NS} =
\exp\biggl[\ii\eta\sum_{m=1/2}^\infty(\gamma_{-m}
\tilde\beta_{-m} - \beta_{-m}
  \tilde\gamma_{-m})\biggr]\,
  \ket{P=-1}\,\ket{\tilde P=-1}~.
\end{equation}
In the R sector in the $(-1/2,-3/2)$ picture,
\begin{equation}
\label{bs9}
\ket{B_\psi,\eta}_\R = \exp\biggl[\ii\eta\sum_{m=1}^\infty
\psi_{-m}\cdot S \cdot \tilde \psi_{-m}\biggr]
\,\ket{B_\psi,\eta}_\R^{(0)}
\end{equation}
and
\begin{equation}
\label{bs10}
\ket{B_\sgh,\eta}_\R =
\exp\biggl[ \ii\eta\sum_{m=1}^\infty(\gamma_{-m}
\tilde\beta_{-m} - \beta_{-m}
\tilde\gamma_{-m})\biggr]\,
 \ket{B_\sgh,\eta}_\R^{(0)}~.
\end{equation}
If we define
\begin{equation}
\label{bs14}
{\cal M}^{(\eta)} = C\Gamma^0\Gamma^{l_1}\ldots
\Gamma^{l_p} \,\left(
\frac{1+\ii\eta\Gamma_{11}}{1+\ii\eta}\right)~~,
\end{equation}
the zero mode parts of the boundary state are
\bea
\label{bsr0}
\ket{B_\psi,\eta}_\R^{(0)} &=&
{\cal M}_{AB}^{(\eta)}\,\ket{A} \ket{\tilde B}~~,\\
\label{bsrsg0}
\ket{B_\sgh,\eta}_\R^{(0)} &=&
\exp\left[\ii\eta\gamma_0\tilde\beta_0\right]\,
  \ket{P=-{1/ 2}}\,\ket{\tilde P=-{3/ 2}}~~.
\eea

The matrix $S_{\mu \nu}$ was defined in \eq{smunu}
and the normalization factor $T_p$ in \eq{Tp}.

One can show that the type IIA or IIB (depending on whether $p$ is even or odd)
GSO projection selects the following combinations:
\begin{equation}
\label{bs22ab}
\ket{B}_\NS
= {1\over 2} \Big( \ket{B,+}_\NS - \ket{B,-}_\NS \Big)~~,
\end{equation}
\beq
\label{bs22bb}
\ket{B}_\R  =
    {1\over 2} \Big( \ket{B,+}_\R + \ket{B,-}_\R\Big)~~.
\end{equation}
%%%%%%%%%%%%%%%%%%%%%%%%%%%%%%%%%%%%%%%%%%%%%%%%%%%%%%%%%%%%%%%%%%%%%%%%%%%%%%%%
\subsection{Comments}

We pause to comment on some features of the boundary states we have just
constructed.
\begin{enumerate}

\item
In this section, we have made explicit the tensor product structure of the Fock 
space of the string. For instance, the SL(2,$\Cbar$) vacuum $|1\rangle $ is now
decomposed as
\beq
|1\rangle =|1\rangle _{X}|1\rangle _{\psi}|1\rangle _{\rm gh}|1\rangle _{\rm sgh}~,
\eeq
where for instance $|1\rangle _{\psi}$ could be further decomposed into a holomorphic
and an antiholomorphic part. The subscripts are often omitted,
since it is usually clear from the context which $|1\rangle $ is meant.

\item
The delta function in the $d_\bot=9-p$
transverse directions (see \eq{bs5}) localizes the initial string position to 
be on the D-brane.  
(In that equation, $x^i$ are operators, the constant modes 
of $X^i$, whereas $y^i$ are the coordinates of the brane in its transverse
directions.) 
The algebra $[x^i,p^i]=\ii$ (see \eq{bosonalgebra}) is
represented on the space of functions of $x^i$ and the ``wavefunction'' is
$\d(x^i-y^i)$. The state $|1\rangle _{X}$ is the ground state with zero 
momentum.

Because we can write
\bea\label{BSmomentum}
\delta^{(d_\bot)}(x - y)\ket{1}_X &=& \frac 1 {(2\pi)^{d_\bot}} \int d^{d_\bot} 
k^\bot e^{\ii k^\bot\cdot x}\ket{1}_X\nonumber\\
&=&\frac 1 {(2\pi)^{d_\bot}} \int d^{d_\bot} k|0;k^\bot\rangle_X ~,
\eea
the boundary state is a superposition of states with zero longitudinal momentum
and arbitrary transverse momentum. What this really means is that insertions need
only obey momentum conservation in the longitudinal directions, 
as we shall now show in more detail.

\item
We have normalized the vacua with momenta $k',k$ and ``complementary'' pictures
$P,-2-P$ and $\tilde P, -2-\tilde P$ (see \eq{qsgh-2}) such that
\beq\label{normalizationvacua}
\langle k';P,\tilde P|c_{-1}\tilde c_{-1}c_0\tilde c_0 c_1\tilde c_1|k;-2-P,-2-\tilde P\rangle 
=(2\pi)^{10}\d^{(10)}(k-k')~.
\eeq  
Consider the contribution of all zero modes except those from the $\psi$ fields
to an amplitude with vertex operators with momenta $k_j$ inserted ($j$ labels the
vertex operators):
\bea\label{delta}
&&\langle k';P,\tilde P|c_{-1}\tilde c_{-1}c_0\tilde c_0 c_1\tilde c_1 \prod_je^{\ii
k_j\cdot x}|k;-2-P,-2-\tilde P\rangle \nonumber\\
&&=(2\pi)^{10}\d^{(10)}(k+\sum_jk_j-k')~.
\eea  
This is the usual momentum conservation delta function, which is well known from
field theory.
Using \eq{BSmomentum}, we find the following momentum factor for amplitudes 
involving one boundary state:
\beq\label{deltamomentum}
\frac 1 {(2\pi)^{d_\bot}} \int d^{d_\bot} k (2\pi)^{10}\d^{(10)}(k+\sum_jk_j-k')=
(2\pi)^{p+1}\d^{(p+1)}(\sum_jk^\Vert_j-k'^\Vert)~.
\eeq
This shows that, in the presence of a D-brane, only longitudinal momentum is
conserved. Of course, this is precisely what one would expect, since the D-brane
explicitly breaks translation invariance in the transverse directions.
\end{enumerate}

%%%%%%%%%%%%%%%%%%%%%%%%%%%%%%%%%%%%%%%%%%%%%%%%%%%%%%%%%%%%%%%%%%%%%%%%%%%%%%%%
\subsection{Scattering amplitudes}\label{anomalous:scattering}

In the previous section, we have, somewhat arbitrarily, defined a boundary state
that enforces boundary conditions at $\tau=0$. We could as well have chosen any
other value of $\tau$. The effect would be to change \eq{bs5} into
\begin{equation}
\label{bs5tau}
\ket{B_X,\tau} = \delta^{(d_\bot)}(x - y)
\exp\biggl[-\sum_{n=1}^\infty \frac{1}{n}\,e^{2n\tau}\,
\a_{-n}\cdot S\cdot
\tilde \a_{-n}\biggr]\,
\ket{1}_X~~,
\end{equation}
and analogously for the other components of the boundary state.
This can be written as follows:
\beq\label{BStau}
\ket{B,\tau}=e^{\tau(L_0+\tilde L_0)}\ket{B}~.
\eeq 
(Note that, as we observed
before \eq{Xeom}, $L_0+\tilde L_0$ is the world-sheet Hamiltonian.)
All these boundary states play a role when we use the boundary state formalism to
compute string scattering amplitudes, as we shall explain shortly. Before going
into that, we have to say a couple of words about the string propagator.

We introduce the closed string propagator as in \cite{gsw}. The propagator of an
ordinary bosonic field obeying the Klein-Gordon equation $(-\partial_\m
\partial^\m+m^2)\phi=0$ has
propagator $(-\partial_\m\partial^\m+m^2)^{-1}$. The closed string analogue of 
the Klein-Gordon
equation is the mass-shell condition $(L_0+\tilde L_0)\ket{\phi}=0$. 
Taking into account \eq{L0closed} and \eq{L0tildeclosed},  
the propagator is
\beq
\Delta=\frac{\a'}2(L_0+\tilde L_0)^{-1}=\frac{\a'}2\int_1^\infty d\r \r^{-L_0
-\tilde L_0-1}~.
\eeq
Physical closed string states satisfy the level matching condition 
$(L_0-\tilde L_0)\ket{\phi}=0$. We can modify the propagator to one that only
propagates states satisfying this constraint:
\beq
\Delta=\frac{\a'}{4\pi}\int_1^\infty d\r \int_0^{2\pi}d\phi\,\r^{-L_0 
-\tilde L_0-1}e^{-\ii\phi(L_0-\tilde L_0)}~.
\eeq
If we define $w=\r e^{\ii\phi}$, this becomes
\beq\label{propagator}
\Delta=\frac{\a'}{4\pi}\int_{|w|\geq 1}\frac{d^2w}{|w|^2}w^{-L_0}\bar w^{-\tilde
L_0}~.
\eeq
From the OPE \eq{TO}, one derives that $[L_0,z{\cal V}(z,\bar z)
]=z\frac\partial{\partial z}(z{\cal V}(z,\bar z))$ for an $(h,\tilde h)=(1,1)$
tensor ${\cal V}(z,\bar z)$. From this equality and a similar one for $\tilde
L_0$, one can derive
\beq\label{commuteL0}
w^{L_0}\bar w^{\tilde L_0}{\cal V}(z,\bar z)w^{-L_0}\bar w^{-\tilde L_0}
=|w|^2{\cal V}(wz,\bar w\bar z)~.
\eeq

It has been argued (see, e.g., \cite{callan87,frau}) that the boundary state
is connected to the rest of a string diagram by a closed string propagator.
Though we shall not try to give an analysis from first principles (like
factorizing disc diagrams in the closed string channel \cite{frau}), we shall now
find out what precisely this statement means, i.e., we shall give a precise
set of rules for how to use the boundary state in a scattering amplitude. 
The result will indeed be to insert an expression like \eq{propagator}, but some
care is needed in determining the integration domain. 

Our set of rules for computing a scattering amplitude of closed string states
on a world-sheet with sphere topology in the presence of a boundary is as 
follows.
\begin{enumerate}
\item
Fix one vertex operator at $\infty$: 
\beq
c(\infty)\tilde c(\infty){\cal V}(\infty,\infty)~.
\eeq
\item
Fix another vertex operator at an arbitrary point of the sphere: 
\beq
c(z) \tilde c(\bar z){\cal V'}(z,\bar z)~.
\eeq
\item
Insert the boundary state \eq{BS}. We can consider it as an insertion 
at the origin. Intuitively,
this is because it cuts out a disc centered at the origin. In formulas, the
$c_1\tilde c_1$ zero mode insertion in \eq{bs6} is the same one would have for 
a vertex operator fixed at the origin.% 
\footnote{\label{footnoteb0}
Of course, there is an extra zero mode insertion $(c_0+\tilde c_0)/2$ 
in \eq{bs6}. We shall deal with this in rule \ref{ruleb0}.} 
Thus, by now we have fixed enough vertex operators, so that all other insertions
should be integrated over the whole plane.
\item
%As we remarked at the beginning of this subsection, there is no reason to only
%consider boundary states inserting a boundary at $\tau=0$. In fact, we should
Integrate \eq{BStau} over all ``allowed'' values of $\tau$. What are the allowed
values of $\tau$? It is a very plausible guess that those values of $\tau$ are
allowed  for which no insertions lie on the disc that is cut out, i.e., we
integrate over $\tau$ as far as we can without ``hitting'' the first insertion
point. Thus, we insert
\beq\label{propagatorBS}
\frac{\a'}2\int_{e^\tau<|z_{\rm min}|}d\tau e^{\tau(L_0+\tilde L_0)}\ket{B}~, 
\eeq
where $z_{\rm min}$ is the insertion point closest to the origin, and we have
chosen the normalization by hand.%
\footnote{In principle, it should be fixed by factorization arguments along the
lines of \cite{frau}.} 
Performing a change of integration variable $\r=e^{-\tau}$, this becomes
\beq
\frac{\a'}2\int_{\r>1/|z_{\rm min}|}d\r \r^{-L_0-\tilde L_0-1}\ket{B}~,
\eeq
which, as $\ket{B}$ satisfies the level matching condition, can also be written
as
\beq
\frac{\a'}{4\pi}\int_{|w|>1/|z_{\rm min}|}\frac{d^2w}{|w|^2}w^{-L_0}
\bar w^{-\tilde L_0}\ket{B}~.
\eeq
As promised, this is the same as inserting the propagator \eq{propagator}, except
for the different integration region.

\item\label{ruleb0}
Insert a $b_0+\tilde b_0$ factor together with the
propagator connecting the boundary state with the rest of the diagram. 
This is to get rid of the extra ghost zero mode insertion alluded to in 
footnote \ref{footnoteb0}.
Though $b_0$
insertions are certainly not unfamiliar for people who know about higher
genus amplitudes in string theory, we shall not try to justify it any further
here. However, we do want to make one comment about it.

The reason why we put the $(c_0+\tilde c_0)/2$ factor in \eq{bs6} is that
otherwise the boundary state would not be BRST invariant. The reason why we want
the boundary state to be BRST invariant is that we want unphysical states to
decouple from disc amplitudes. Now, if we effectively remove the 
$(c_0+\tilde c_0)/2$ by inserting $b_0+\tilde b_0$, do we lose BRST invariance
and thus the decoupling of unphysical states? Happily, the answer is no. Using
$\{Q_{\rm B}, b_0+\tilde b_0\}=L_0+\tilde L_0$ (as before \eq{L0}) and the form of
\eq{propagatorBS}, it is clear that the BRST variation of \eq{propagatorBS} with
the $b_0+\tilde b_0$ insertion is the $\tau$ integral of a derivative with
respect to $\tau$. This is a well-known phenomenon in string theory. 
The resulting boundary terms usually vanish due to
analytic continuation arguments, or can equivalently be cancelled by adding
contact terms \cite{greenseiberg, gutperle}. It should be possible to make this
argument more precise for this case too.

\item\label{rulenorm}
Multiply by an overall normalization factor. The rules to
compute this factor are the following.%
\footnote{We do not motivate these rules here. The way to derive them is via
factorization arguments.} 
\begin{enumerate}
\item\label{C0}
There is a factor $4\pi^3/\a'\k^2$, which is common to all sphere amplitudes (not
only those with boundary states).
\item\label{normfactBS}
There is a factor $\k/\pi$ associated to the insertion of the boundary state.
\item
Analogous factors have already been absorbed in the vertex operators \eq{GVO} and
\eq{BVO}~.
\item\label{normRR}
An additional factor of $\k/\pi$ has to be included if we use the R-R state
\eq{RRstate}.
\end{enumerate}

\end{enumerate}

Now that we have provided all the components of the construction, let us write
down the result. Up to the normalization factor of rule \ref{rulenorm}, 
the tree level scattering amplitude of $n+2$ closed string
vertex operators $V, V', V_1,\ldots V_n$ in the presence of a D-brane described
by a boundary state $\ket{B}$ is given by
\bea
&&\bra{V}c_{-1}\tilde c_{-1} \int d^2z_1{\cal V}_1(z_1,\bar z_1)\ldots
\int d^2z_n{\cal V}_n(z_n,\bar z_n)\,c(z)\tilde c(\bar z){\cal V'}(z,\bar z)\,
\nonumber\\&&
\frac{\a'}{4\pi}(b_0+\tilde b_0)
\int_{|w|>{\rm max}(\frac1{|z|},\frac1{|z_i|})}\frac{d^2w}{|w|^2}w^{-L_0}
\bar w^{-\tilde L_0}\ket{B}~.\label{BSamplitudepropagator}
\eea

We now simplify this formula for practical use. First, we deal
with the ghost part. As mentioned in rule \ref{rulenorm}, 
the effect of the $b_0+\tilde b_0$
insertion is to cancel the $(c_0+\tilde c_0)/2$ hidden in $\ket{B}$. Then, as the
$c_{-1},\tilde c_{-1},c_1$ and $\tilde c_1$ ghost zero modes are provided by the
vertex operator fixed at $\infty$ (denoted as an out-state in 
\eq{BSamplitudepropagator}) and by the boundary state, the vertex
operator fixed at $z$
will have to provide the $c_0$ and $\tilde c_0$ modes. Therefore, we can
replace $c(z)\tilde c(\bar z)$ by $c_0\tilde c_0|z|^2$. All in all, we end up
with the combination of ghost zero modes present in \eq{normalizationvacua}, so
we need not care about the ghost sector any more. 

Making repeated use of \eq{commuteL0}, we can commute the propagator to the left
until $L_0+\tilde L_0$ annihilates the out-state (see \eq{L0}). The final result
is that \eq{BSamplitudepropagator} becomes
\beq\label{BSamplitude}
\frac{\a'}{4\pi}\bra{V}\int_{|y|>1}d^2y\int_{|y_1|>1}d^2y_1\ldots
\int_{|y_n|>1}d^2y_n\,{\cal V}_1(y_1,\bar y_1)\ldots {\cal V}_n(y_n,\bar y_n)
{\cal V'}(y,\bar y)\ket{B}~,
\eeq
where by $\ket{B}$ we mean, in fact, the boundary state with the ghost part
omitted. Note that now all vertex operators except the one at infinity are
integrated, and that the integration region is the complement of the unit disc.

Before going into specific examples, we shall learn how to compute correlators of 
the form
\beq
\langle :e^{\ii k_1\cdot X(z_1,\bar z_1)}:\ldots:e^{\ii k_n\cdot X(z_n,\bar z_n)}:
\rangle 
\eeq
in the presence of a boundary state.
Here, the dots denote conformal normal ordering (see
\sect{superstrings:scattering:Hilbert}). For the $X$ conformal field theory on the
$z$-plane, this is the same as the creation-annihilation normal ordering
introduced after \eq{Gr} in
\sect{superstrings:strings:worldsheet} (see \cite{polchinski} Vol. 1, p. 60 for a proof
of this fact). We remind the reader that for the  creation-annihilation normal 
ordering $p^\m$ are considered to be lowering operators.

Consider the case 
\beq\label{corrX2}
\langle :e^{\ii k_1\cdot X(z_1,\bar z_1)}:\;:e^{\ii k_2\cdot X(z_2,\bar z_2)}:
\rangle~, 
\eeq
first without a boundary state.
The correlator \eq{corrX2} factorizes into contributions from the zero modes
$x^\mu,p^\mu$ and the holomorphic and antiholomorphic nonzero modes $\a_n,
\tilde \a_n (n\neq 0)$. 
\begin{itemize}
\item
To compute the contribution of the nonzero modes (NZM), we use
the fact that for operators $A,B$ linear in harmonic oscillator creation and
annihilation operators
\beq\label{coherent}
\langle :e^A:\;:e^B:\rangle=e^{\langle AB\rangle}~.
\eeq
(See, for instance, \cite{gsw} Vol. 1, Appendix 7.A).
We compute
\bea
\langle X^\mu(z_1)X^\nu(z_2)\rangle_{\rm NZM}&=&-\frac{\a'}2\langle1|
\left(\sum_{n=1}^\infty \frac{\a^\mu_n}nz_1^{-n}\right)
\left(\sum_{n=1}^\infty \frac{\a^\mu_{-n}}{-n}z_2^{n}\right)|1\rangle\nonumber\\
&=&\frac{\a'}2\eta^{\m\n}\sum_{n=1}^\infty\frac1n\left(\frac{z_2}{z_1}\right)^n
=-\frac{\a'}2\eta^{\m\n}\ln(1-\frac{z_2}{z_1})~.\nonumber
\eea
Note that this derivation is only valid for $|z_2|<|z_1|$. However, when
a path integral expression is transformed into an operator language (as in the
first line of the previous equation) the insertions automatically come out time 
(i.e., radial) ordered. 

Using \eq{coherent} and doing the same for the antiholomorphic
modes, we find the following nonzero mode contribution to \eq{corrX2}:
\beq
\langle :e^{\ii k_1\cdot X(z_1,\bar z_1)}:\;:e^{\ii k_2\cdot X(z_2,\bar z_2)}:
\rangle_{\rm NZM}=\left|1-\frac{z_2}{z_1}\right|^{\a'k_1\cdot k_2}~.
\eeq

\item
To compute the zero mode contribution, one has to remember that the normal
ordering prescription is such that $p^\mu$ are considered to be lowering
operators. Then, one computes
\bea
\langle :e^{\ii k_1\cdot X(z_1,\bar z_1)}:\;:e^{\ii k_2\cdot X(z_2,\bar z_2)}:
\rangle_{\rm ZM}&=&\bra{1}e^{\ii k_1\cdot x}|z_1|^{\a'k_1\cdot p}
e^{\ii k_2\cdot x}|z_2|^{\a'k_2\cdot p}\ket{1}\nonumber\\
&=&|z_1|^{\a'k_1\cdot k_2}
\eea
(times a momentum-conservation delta function, which we have suppressed; see
\eq{delta}).
\end{itemize} 
Combining both contributions, we find
\beq\label{resultXX}
\langle :e^{\ii k_1\cdot X(z_1,\bar z_1)}:\;:e^{\ii k_2\cdot X(z_2,\bar z_2)}:
\rangle=|z_1-z_2|^{\a'k_1\cdot k_2}~.
\eeq

The generalization is
\beq
\langle :e^{\ii k_1\cdot X(z_1,\bar z_1)}:\ldots:e^{\ii k_n\cdot X(z_n,\bar z_n)}:
\rangle =\prod_{i<j}|z_i-z_j|^{\a'k_i\cdot k_j}~.
\eeq

If a boundary state is added to the correlator \eq{corrX2}, there are additional
contributions of two kinds. 
\begin{itemize}
\item
The first kind are zero mode contributions due to the 
momentum of the boundary state itself,
see \eq{BSmomentum}. These are easy to incorporate: one acts as if an extra
$e^{\ii k^\perp\cdot X}$ were inserted at the origin. We shall not display the
extra contribution explicitly here. 
 
\item
The other contributions due to the boundary state come from the nonzero modes in
\eq{bs5}. In practice, the boundary state allows to also ``contract'' holomorphic
with antiholomorphic fields, e.g.,
\bea
&&\langle X^\mu(z_1)\tilde X^\nu(\bar z_2)\ket{B_X}_{\rm NZM}\nonumber\\
&&=-\frac{\a'}2\langle1|
\left(\sum_{n=1}^\infty \frac{\a^\mu_n}nz_1^{-n}\right)
\left(\sum_{n=1}^\infty \frac{\tilde\a^\mu_{-n}}{-n}\bar z_2^{n}\right)
\exp\biggl[-\sum_{n=1}^\infty \frac{1}{n}\,
\a_{-n}\cdot S\cdot
\tilde \a_{-n}\biggr]
|1\rangle\nonumber\\
&&=
\frac{\a'}2\langle1|
\left(\sum_{n=1}^\infty \frac{\a^\mu_n}nz_1^{-n}\right)
\left(\sum_{n=1}^\infty \frac{\tilde\a^\mu_{-n}}{-n}\bar z_2^{n}\right)
\sum_{n=1}^\infty \frac{1}{n}\,
\a_{-n}\cdot S\cdot
\tilde \a_{-n}
|1\rangle\nonumber\\
&&=-\frac{\a'}2S^{\m\n}\ln(1-\frac1{z_1\bar z_2})~.
\eea
A reasoning analogous to the one leading to \eq{coherent} then implies that in 
the presence of a boundary state \eq{resultXX} should be multiplied by
\beq
\left(1-\frac1{|z_1|^2}\right)^{\frac{\a'}2k_1\cdot S\cdot k_1}
\left(1-\frac1{|z_2|^2}\right)^{\frac{\a'}2k_2\cdot S\cdot k_2}
\left|1-\frac1{z_1\bar z_2}\right|^{\a'k_1\cdot S\cdot k_2}
\eeq
(times the additional zero mode contribution mentioned in the previous item).
\end{itemize}
%%%%%%%%%%%%%%%%%%%%%%%%%%%%%%%%%%%%%%%%%%%%%%%%%%%%%%%%%%%%%%%%%%%%%%%%%%%%%%%%
\section{Checks of Wess-Zumino action: an overview}\label{anomalous:overview}
Sections \ref{anomalous:BC}, \ref{anomalous:RRC} and \ref{anomalous:D:non} are 
devoted to some explicit checks of the D-brane Wess-Zumino action \eq{WZ}. 
Those sections are based on our papers \cite{benfred, normal}. In the present
section, we give the reader an overview of how the different terms in \eq{WZ} were
found and of which tests have been performed on them.  

First, we summarize the situation as it was before \cite{benfred} appeared.
\begin{itemize}
\item
The term
\beq
\frac{T_p}{\kappa}\int_{p+1}\hat C_{p+1}
\eeq
was found in \cite{Pol} and recovered in the boundary state formalism in
\cite{Li,dv9707}. The boundary state computation amounts to computing the bracket
of a R-R potential with the boundary state, i.e., a one point function of a
closed string state in the presence of a D-brane.

\item
The couplings involving only a R-R potential and the gauge field,
\beq\label{eF}
\frac{T_p}{\kappa}\int_{p+1}\hat C\wedge e^{2\pi\a '\,F}~,
\eeq 
were derived in the boundary state formalism in \cite{Li,dv9707} and 
interpreted in terms of branes inside branes in \cite{douglas}. The boundary
state computation amounts to computing the bracket
of a R-R potential with a boundary state that is modified to describe a brane
with a constant magnetic field turned on. The presence of the terms \eq{eF} was
confirmed in \cite{GHM} using an anomaly inflow argument.

\item 
Given the terms \eq{eF}, the gauge invariance argument given after
\eq{combination} implies the presence of  
\beq\label{eB}
\frac{T_p}{\kappa}\int_{p+1}\hat C\wedge e^{\hat{B}}~.
\eeq 
The presence of this term had not been checked directly.

\item
One can use invariance under R-R gauge transformations \eq{RRgaugetransf} to
argue for the presence of \eq{eB} (see \eq{RRgaugeinv}). Reversing the gauge
invariance argument used in the previous item, this implies also the presence of
the coupling \eq{eF}.

\item
The term
\beq
\frac{T_p}{\kappa}\int_{p+1}\hat C_{p-3}\wedge\frac{(4\pi^2\a ')^2}{384\pi^2}
\tr R_T^2
\eeq
was found in \cite{bersadvaf} using a duality argument (see \sect{type0:dual}).
Its presence was confirmed in \cite{GHM} using an anomaly inflow argument. Direct
tests had not been performed on this term.

\item
The other curvature terms were found using anomaly inflow arguments
\cite{GHM,CY}. No direct tests had been performed on them.

\end{itemize}

It is clear from all these studies that the different terms in \eq{WZ} all have
to be present. They are needed for the consistency of string theory. However,
before \cite{benfred} appeared, only the couplings to $F$ had been explicitly
computed from string theory. Strong as it was, all the evidence 
for the couplings to the NS-NS fields ($B$ and the spacetime metric) was 
indirect. 

In \cite{benfred},
we started the project of directly computing these couplings from string theory.
This work was carried out further in \cite{MSS,stefanski,normal}.

In \sect{anomalous:BC}, we explicitly compute a string scattering amplitude to 
check the presence of the term
\beq
\frac{T_p}{\kappa}\int_{p+1}\hat C_{p-1}\wedge \hat{B}~.
\eeq
We have chosen to give these computations, the result of which appeared in our 
paper \cite{benfred}, in detail in this thesis. The reason is that they
are simple enough to be done explicitly in a reasonable amount of time and space,
and just complicated enough to illustrate the main features of scattering
amplitudes with boundary states.%
\footnote{
They have the disadvantage that they are less
``clean'' than the computations of the curvature terms. First, the limitation that 
only
on-shell questions can be addressed by string scattering amplitudes will make it 
hard to relate our computations directly to ``measurable'' quantities like cross
sections. A way out will be proposed in footnote \ref{Euclidean}. Second, 
the field theory amplitude
with which we shall compare the result of our string computation gets two
different contributions: one from the source term \eq{WZ} we are interested in,
but also one in which the bulk interactions \eq{IIAR} or   \eq{IIBR} play a
role. 

We invite the reader not to be distracted too much by these two subtleties, as 
they are not present for the computations of the curvature couplings. On the other
hand, this kind of problems occur rather often in string computations (see
\sect{nonbps:anomalous} for another example). Especially the fact that string
perturbation theory is blind for off-shell questions is one of its major
drawbacks. 
}

In \sect{anomalous:RRC}, we check the presence of the terms
\beq
\frac{T_p}{\kappa}\int_{p+1}\hat C_{p-3}\wedge \frac{(4\pi^2\a ')^2}{384\pi^2}
(\tr R_T^2-\tr R_N^2)~,
\eeq 
which involve a four-form constructed from the curvature two-forms. These computations were
done in \cite{benfred, normal}.

In \sect{anomalous:D:non}, we consider the terms 
\beq
\frac{T_p}{\kappa}\int_{p+1}\hat C_{p-7}\wedge 
\left(\frac{(4\pi^2\a ')^4}{294912\pi^4}(\tr R_T^2-\tr R_N^2)^2+
\frac{(4\pi^2\a ')^4}{184320\pi^4} (\tr R_T^4-\tr R_N^4)\right)~,
\eeq 
involving an eight-form constructed from the curvature two-forms. The computations for the
terms involving only the
curvature two-form $R_T$ of the tangent bundle were done in \cite{stefanski}.
They were generalized to include the normal bundle curvature $R_N$ in
\cite{normal}. In the latter paper, we also found additional couplings of a
D$p$-brane to a ($p-7$)-form R-R potential. These couplings are not anomalous,
which is why they were not found in \cite{CY}. They are to be added to \eq{WZ}.

The computations we discuss in this thesis follow
\cite{benfred,stefanski,normal}. However, the various terms in the Wess-Zumino
action \eq{WZ} were also checked in an alternative formalism in \cite{MSS},
which, in fact, appeared between \cite{benfred} and \cite{stefanski, normal}. 

%%%%%%%%%%%%%%%%%%%%%%%%%%%%%%%%%%%%%%%%%%%%%%%%%%%%%%%%%%%%%%%%%%%%%%%%%%%%%%%%
\section{The $\protect\hat B \wedge \protect\hat C_{p-1}$ interaction}
\label{anomalous:BC}  
This section is an expanded version of the second section of \cite{benfred}.
To probe the presence of the term
\beq \label{BCsource}
\frac{T_p}\k \,\int_{p+1}\hat C_{p-1}\wedge \hat{B}
\eeq
in \eq{WZ}, considered as a source term in \eq{IIAsugra} or \eq{IIBsugra}, we
shall compute the tree level string amplitude for a D$p$-brane to absorb a
$B$-field and emit a ($p-1$)-form R-R potential.%
\footnote{or to absorb a ($p-1$)-form R-R potential and emit a $B$-field.}
The low-energy limit of this string amplitude will be compared with the 
corresponding on-shell amplitude in supergravity.

We shall work in static gauge (see \sect{superstrings:D-branes}) and
take the D$p$-brane to be extended along the directions $0,\ldots, p$. We choose
the nonzero polarizations $c_{\a_1\ldots\a_{p-1}}$ and $\zeta_{\b_1\b_2}$ to be 
along the brane directions: $\a_i,\b_j\in\{0,\ldots, p\}$. 

\begin{figure}
\begin{center}
\epsfig{file=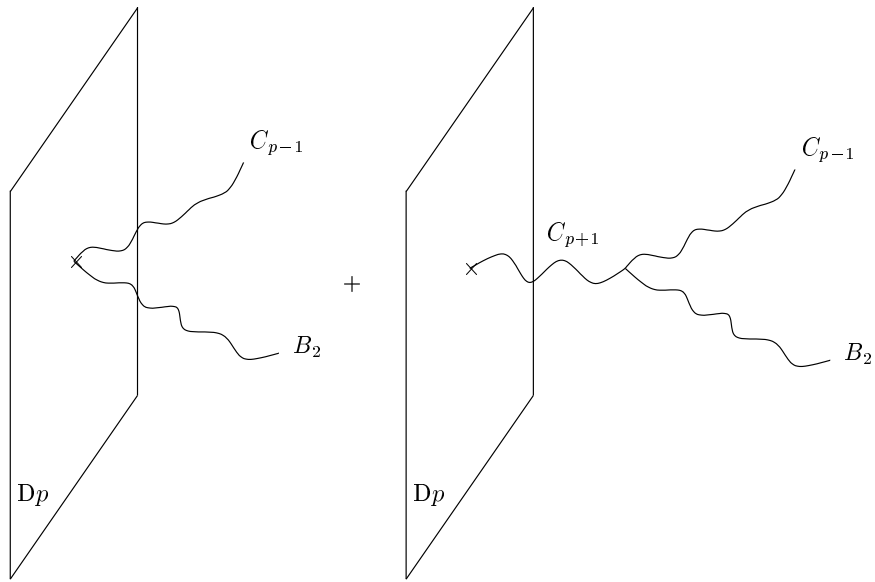,width=12.5cm}
\end{center}
\caption{In supergravity, there are two contributions to the amplitude
for a D$p$-brane to absorb a
$B$-field and emit a ($p-1$)-form R-R potential.}\label{fig:sugraA}
\end{figure}

First, we write down the supergravity amplitude ${\cal A}_{\rm sugra}$. It
receives two contributions (see \fig{fig:sugraA}). 
The first contribution comes from the term
\eq{BCsource} we are interested in. It is equal to
\beq\label{Asugra1}
{\cal A}_{\rm sugra}^{(1)}=
\frac{\kappa T_p}{(p-1)!} \epsilon^{\a_1\ldots\a_{p-1}
\b_1\b_2} c_{\a_1\ldots\a_{p-1}}\zeta_{\b_1\b_2}~,
\eeq
where here and below we omit the delta function corresponding to momentum
conservation in the directions along the D-brane.
The second contribution comes from the bulk supergravity action.
The field strength $\tilde F_{p+2}$ of $C_{p+1}$ is not just $dC_{p+1}$ but 
has the form \eq{deftildeF}. As a consequence, the corresponding kinetic term 
induces a three point bulk coupling between $C_{p+1}$, $C_{p-1}$ and $B$. Thus, 
the scattering amplitude we are considering also gets a contribution from the 
D$p$-brane emitting a $C_{p+1}$ potential which combines with the incoming
$C_{p-1}$ potential to give an outgoing $B$-field.
This contribution equals%
\footnote{In the computation that leads to this result, the mass-shell conditions
for the external fields $C_{p-1}$ and $B$ are used.}
\beq\label{Asugra2}
{\cal A}_{\rm sugra}^{(2)}=-\frac12
\frac{\kappa T_p}{(p-1)!} \epsilon^{\a_1\ldots\a_{p-1}
\b_1\b_2} c_{\a_1\ldots\a_{p-1}}\zeta_{\b_1\b_2}~.
\eeq
Summing \eq{Asugra1} and \eq{Asugra2}, we find the total amplitude in
supergravity:
\beq\label{Asugra}
{\cal A}_{\rm sugra}=
\frac{\kappa T_p}{2(p-1)!} \epsilon^{\a_1\ldots\a_{p-1}
\b_1\b_2} c_{\a_1\ldots\a_{p-1}}\zeta_{\b_1\b_2}~.
\eeq
The amplitude \eq{Asugra} will be compared to the low-energy limit of
the amplitude we shall now compute in string theory.

\begin{figure}
\begin{center}
\epsfig{file=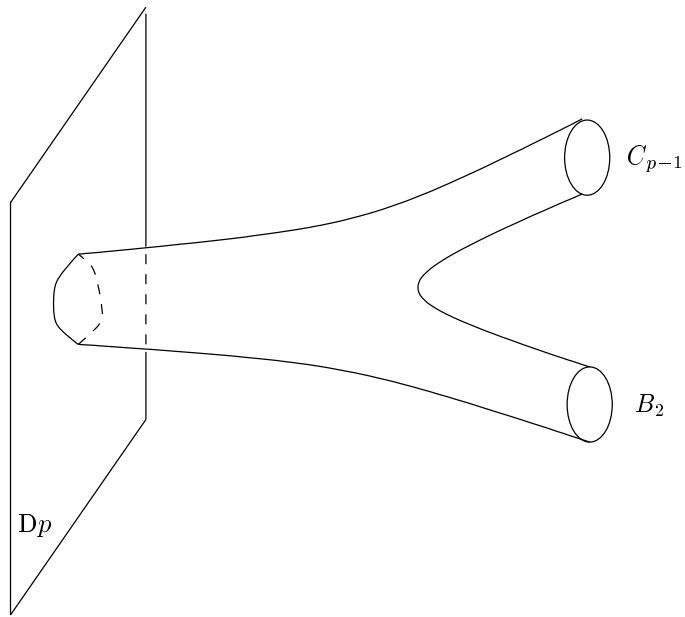}
\end{center}
\caption{String theory amplitude for a D$p$-brane to absorb a
$B$-field and emit a ($p-1$)-form R-R potential.}\label{fig:stringA}
\end{figure}

In string theory, the tree level amplitude for a D$p$-brane to absorb a
$B$-field and emit a ($p-1$)-form R-R potential is given by
\beq\label{BC:ampl}
{\cal A} = \bra{{\cal C}_{p-1};k_2} V_B^{(0,0)}(\zeta,k) \ket{B}_\R 
\eeq
(see \fig{fig:stringA}).
In \eq{BC:ampl}, the ``bra'' is the R-R state \eq{RRstate} and the ``ket'' is the 
boundary state in the R-R sector.
Because the pictures of the R-R state \eq{RRstate} and the boundary state \eq{bs22bb}
add up to ($-2$,$-2$), we have taken the (0,0) vertex operator \eq{BVO} for the
$B$-field (see \sect{superstrings:scattering} for an introduction to pictures):
\beq
V_B^{(0,0)}(\zeta,k)=\int_{|z|>1}d^2z\,{\cal V}_B^{(0,0)}(\zeta,k)~,
\eeq
where
\beq\label{VB}
{\cal V}_B^{(0,0)}(\zeta,k)=\frac{\sqrt 2\kappa}{\pi\a'}\,\zeta_{\mu\nu}\,
(\partial X^{\mu}
-\ii\frac{\a'}2 \,k\cdot\psi\,\psi^{\mu}) (\bar\partial X^{\nu}-\ii\frac{\a'}2 
k\cdot\tilde\psi\,\tilde\psi^{\nu})\,e^{\ii k\cdot X}
\eeq
and the integration region is as in \eq{BSamplitude}.

The $C_{p-1}$ potential has momentum $k_2$, the momentum of the $B$-field is $k$
and the momentum of the boundary state will now be denoted by $k_1$. As usual in
perturbative string theory (see \sect{superstrings:scattering:onshell}), 
the external states described by $C_{p-1}$ and $B$ are on-shell, which means
\beq\label{onshell}
k_2^2=k^2=0~.
\eeq
Momentum conservation (see \eq{delta} and \eq{deltamomentum}) says that 
\beq\label{momcon}
k_1=-k-k_2
\eeq 
and that the components of $k_1$  in the directions along the brane are zero.

Let us first comment on the fermion zero modes and the gamma matrices.
From \eq{Czm1}, \eq{Czm2}, \eq{convgamma}, \eq{bs14} and \eq{bsr0}, one can see 
that the result will involve a trace of a product of gamma
matrices. The boundary state provides $p+1$ gamma matrices (and possibly a
$\G_{11}$), whereas the R-R state
gives $p-1$ of these gamma matrices (and possibly a $\G_{11}$). 

To get a non-vanishing trace, we need two gamma matrices
from fermion zero modes. In $V_B^{(0,0)}(\zeta,k)$, each fermion carries a vector
index that is contracted with either $k$ or $\zeta$. The result of performing the
trace will be that the two indices contracted with the fermion zero modes are
antisymmetrized with the $p-1$ indices of $c_{\a_1\ldots\a_{p-1}}$. It is not
difficult to convince oneself that only the four fermion part of 
$V_B^{(0,0)}(\zeta,k)$ can contribute.%
\footnote{If one tried a two fermion piece, say the
two holomorphic fermions, the $\bar\partial X^{\b_2}$ would have nothing to
contract its index with (remember that $k^\m\zeta_{\m\n}=0$ from BRST invariance, 
that the boundary state carries only transverse momentum, and that contracting  
$\bar\partial X^{\b_2}$ with the momentum factor of the R-R potential inserted at
infinity would give an infinite denominator).}
The two fermions that are not needed in the trace will have to be contracted with
each other, either via the nonzero modes of the boundary state, or by symmetrizing
the gamma matrices coming from their zero modes. As we cannot contract $k$ with an
index on $\zeta$,%
\footnote{Again, this is because  $k^\m\zeta_{\m\n}=0$; as $\zeta$ has all its
indices along the brane, contracting indices via the matrix $S$ (see \eq{smunu})
does not make a difference.}
and as $k$ cannot be antisymmetrized with itself, the fermions contracted with the
indices of $\zeta$ will be used in the trace and $k$ will be contracted with
itself (via the matrix $S$). 

To be a little bit more explicit about the contribution from the fermion zero
modes in general, we write it down schematically:
\beq \label{factor}
\left(\frac{1}{\sqrt{2}}\right)^n \tr \left( C^{-1}{\cal M} R\,C^{-1} {\cal N}^T 
L \right)~~,
\eeq
where $L$ is a product of $\gamma$-matrices corresponding to the left-over 
left-moving fermi\-ons, $R$ a product of $\gamma$'s for the right-moving fermions 
and $n$ is the total 
number of $\gamma$'s in $LR$. The matrices ${\cal M} $
and ${\cal N}$ show up in the boundary state and the RR state respectively. 
Note that \eq{factor} is
schematic: in fact, keeping track of all factors of $\gamma_{11}$ and 
minus-signs is crucial to obtain correct results. We refer the reader to
\eq{zeromodes} for a more explicit expression, at least for the case we are
considering in this section. This ends the comment on the fermion zero modes and
the gamma matrices.

In general, to contract holomorphic with antiholomorphic fields, one proceeds 
as we did at the end of \sect{anomalous:scattering}: one pulls
the non-zero mode operators 
\beq
\exp\biggl[-\sum_{n=1}^\infty \frac{1}{n}\,
\a_{-n}\cdot S\cdot
\tilde \a_{-n}\biggr]\ ; 
\eeq
\beq \label{expferm}
\exp\biggl[\ii\eta\sum_{m=1}^\infty
\psi_{-m}\cdot S \cdot \tilde \psi_{-m}\biggr]
\eeq
from the boundary state to the left where they annihilate the out vacuum.%
\footnote{
In practice, one just supplements the usual two point functions
\bea
\langle X^\mu(z_1)X^\nu(z_2)\rangle_{\rm NZM}&=&
-\frac{\a'}2\eta^{\m\n}\ln(1-\frac{z_2}{z_1})~;\\
\langle\psi^\m(z_1)\psi^\n(z_2)\rangle_{\rm
R}&=&-\frac{\eta^{\m\n}}2\frac1{z_1-z_2}\frac{z_1+z_2}{\sqrt{z_1z_2}}
\eea
with the holomorphic-antiholomorphic contractions
\begin{eqnarray}
\langle X^{\mu}(z_1)\,\bar X^{\nu}(\bar z_2)\ket{B_X}&=&-\frac{\a'}2S^{\mu\nu}\ln
(1-\frac{1}{z_1\bar z_2})~,\label{XXbar}\\ \label{psipsibar}
\langle\psi^{\mu}(z_1)\,\bar\psi^{\nu}(\bar z_2)\ket{B_\psi,\eta}_{\rm R}&=&
\frac{S^{\mu\nu}}{2}
\frac{\ii\eta}{z_1\bar z_2-1}\frac{1+z_1\bar z_2}{\sqrt{z_1\bar z_2}}
\end{eqnarray}
and uses Wick's theorem. Note that $z_1$ and $\bar z_2$ take values on the
complement of the unit disc, so that the holomorphic-antiholomorphic 
contractions do not have singularities unless both fields approach the boundary.
}

As a last general remark before going into the concrete computations of 
\eq{BC:ampl}, we note that the ghost sector has been dealt with in 
\sect{anomalous:scattering}, whereas the superghost sector contributes a factor
1/2 \cite{billo9802} (this will also be the case for the amplitudes in
\sect{anomalous:RRC} and \sect{anomalous:D:non}).

We shall first compute the contribution from the fermion zero modes and the gamma
matrices, then the integral that remains after the contractions 
and finally the overall normalization factor.

We compute the trace of the gamma matrices for the amplitude \eq{BC:ampl}. 
In \eq{expferm} and \eq{psipsibar}, we have taken a boundary
state with a specific $\eta$, and the result of the contractions is proportional to
$\eta$. This $\eta$ also appears in the zero mode part \eq{bsr0} of the boundary state.
The GSO projection \eq{bs22bb} instructs us to average the final result over
$\eta=\pm1$. Analogously, the R-R state \eq{RRstate} is the average over $\eta'=\pm1$ of
a state depending on $\eta'$. We shall first compute the trace for specific values of
$\eta$ and $\eta'$ and in the end take the averages. 

Using \eq{Cproperties} and \eq{convgamma}, we find%
\footnote{
We shall not keep track of the overall sign of the amplitude \eq{BC:ampl}. 
In the following computations, ``$\doteq$'' means equality possibly up to a sign, 
and up to the normalization factor in \eq{Czm2}, which we restore later.
}
\bea
&&\eta\,\tr\left({\cal M}^{(\eta)}\G^{\b_2}C^{-1}({\cal N}^{(\eta)})^TC^{-1}(\G^{\b_1})^T
\G_{11}\right)\label{zeromodes}\\
&&\doteq\eta\,\tr\left(\G^{\b_1}\G_{11}C^{-1}{\cal M}^{(\eta)}\G^{\b_2}C^{-1} 
({\cal N}^{(\eta)})^T\right)\nonumber\\
&&\doteq\eta\,\tr\left(\G^{\b_1}\G_{11}\Gamma^0\Gamma^{1}\ldots
\Gamma^{p} \,\left(\frac{1+\ii\eta\Gamma_{11}}{1+\ii\eta}\right)
\G^{\b_2}\left(\frac{1+\ii\eta'\Gamma_{11}}{1+\ii\eta'}\right)\G^{\a_{p-1}}\ldots
\G^{\a_1}\right)\nonumber\\
&&\doteq\tr\left(\G^{\b_1}\Gamma^0\Gamma^{1}\ldots\Gamma^{p}\left[\eta\G_{11}
\left(\frac{1+\ii\eta\Gamma_{11}}{1+\ii\eta}\right)
\left(\frac{1-\ii\eta'\Gamma_{11}}{1+\ii\eta'}\right)\right]\G^{\b_2}
\G^{\a_{p-1}}\ldots\G^{\a_1}\right)~.\nonumber
\eea
Using
\bea
\frac{1+\ii\eta\Gamma_{11}}{1+\ii\eta}&=&\frac{1+\G_{11}}2-\ii\eta\frac{1-\G_{11}}2~;
\nonumber\\
\frac{1-\ii\eta'\Gamma_{11}}{1+\ii\eta'}&=&
\frac{1-\G_{11}}2-\ii\eta'\frac{1+\G_{11}}2~,
\eea
it is clear that the average over $\eta$ and $\eta'$ of the previous expression is
\beq\label{16}
\tr\left(\Gamma^0\Gamma^{1}\ldots\Gamma^{p}\G^{\a_1}\ldots\G^{\a_{p-1}}\G^{\b_1}
\G^{\b_2}\frac{1-\G_{11}}2\right)=16\epsilon^{\a_1\ldots\a_{p-1}\b_1\b_2}~.
\eeq
This is the contribution from the fermion zero modes and gamma matrices.

Now, we turn to the contractions.
It follows from the considerations at the end of \sect{anomalous:scattering} that
the contribution to \eq{BC:ampl} from the $X$ sector is proportional to
\beq
\left(1-\frac1{|z|^2}\right)^{\frac{\a'}2k\cdot S\cdot k}|z|^{-\a'k\cdot k_2}~,
\eeq
where the last factor comes from contracting $e^{\ii k\cdot X(z,\bar z)}$ with the
momentum factor $e^{\ii k_1\cdot x}$ of the boundary state, and using
$k_1=-k-k_2$ and $k^2=0$. 

As stated above, only the zero mode of the fermions contracted with $\zeta$
contribute: they give a factor $1/|z|$ (see \eq{psiinmodes}). The remaining two
fermions can be contracted either via the zero mode part or the nonzero mode part
of the boundary state $|B_\psi,\eta\rangle_{\rm R}$ (see \eq{bs9}). We shall be
interested in the nonzero mode part.%
\footnote{\label{zeromodecon}
The zero-mode contribution is proportional to 
$\frac{k\cdot S\cdot k}{k\cdot k_2}$ for small on-shell 
momenta. Further investigations suggest that this vanishes whenever the 
$B \wedge {\cal C}_{p-1}$ interaction could contribute to cross-sections, as we 
shall argue in footnote \ref{Euclidean}.}
This is proportional to
\bea
&&\ii\eta\langle k\cdot \psi \,k\cdot \tilde \psi \left(\sum_{m=1}^\infty
\psi_{-m}\cdot S\cdot\tilde\psi_{-m}\right)
\rangle\nonumber\\
&&=\ii\eta\, k_\m k_\n\langle \left(\sum_{m=1}^\infty\psi_{m}^\m z^{-m-1/2}\right)
\left(\sum_{m=1}^\infty\tilde\psi_{m}^\n \bar z^{-m-1/2}\right)
\left(\sum_{m=1}^\infty\psi_{-m}\cdot S\cdot\tilde\psi_{-m}\right)\rangle\nonumber\\
&&=\ii\eta (k\cdot S\cdot k)|z|^{-1}\sum_{m=1}^\infty |z|^{-2m}\nonumber\\
&&=\ii\eta (k\cdot S\cdot k)\frac1{|z|}\frac1{|z|^2-1}~.
\eea
Thus, the integral we have to do is
\beq\label{integraltodo}
\int_{|z|>1}d^2z\,
\left(|z|^2\right)^{-\frac{\a'}2k\cdot S \cdot k-\frac{\a'}2k\cdot k_2}
\left(|z|^2-1\right)^{\frac{\a'}2k\cdot S \cdot k}
\frac{\a'}2\frac{(k\cdot S\cdot k)}{|z|^2 \left(|z|^2-1\right)}~,
\eeq
which gives 
\beq\label{pifactor}
\pi\frac{\a'}2(k\cdot S \cdot k)B(1+\frac{\a'}2k\cdot k_2,\frac{\a'}2k\cdot S \cdot k)
\approx\pi~,
\eeq
where the approximation is for small momenta ($\a'\rightarrow 0$), which is the 
limit in which string theory and supergravity should agree. Thus, the integral
resulting from the contractions contributes a factor $\pi$.
 
Now, we compute the global normalization factor.
\begin{itemize}
\item
There is a factor $\a'/4\pi$ from \eq{BSamplitude}, a factor  $4\pi^3/\a'\k^2$
from item (\ref{C0}) on page~\pageref{C0}
and a factor $(\k/\pi)^2$ from items (\ref{normfactBS}) 
and (\ref{normRR}).
\item
The boundary state normalization factor is $T_p/2$ (see \eq{bs3}).
\item
We have to restore the $1/2\sqrt2(p-1)!$ factor from \eq{Czm2}.
\item
From \eq{VB}, we get a factor $\sqrt2\k/\pi\a'$ and two $\a'/2$ factors, one of which
has in fact been absorbed in the integral \eq{integraltodo}.
%\item
%The integral above gave a factor $\pi$ (\eq{pifactor}).
\item
There is a factor 16 from \eq{16}, 
a factor $1/2$ from the two fermion zero modes and
\eq{convgamma}, and another factor $1/2$ from the superghost sector
\cite{billo9802}.
\end{itemize}

The total string amplitude is thus given by
\bea
{\cal A}&=&\frac{\k T_p}{2\pi(p-1)!}\epsilon^{\a_1\ldots\a_{p-1}\b_1\b_2}
c_{\a_1\ldots\a_{p-1}}\zeta_{\b_1\b_2}\label{totalamplitudeBC}\\
&&\times\int_{|z|>1}d^2z\,
\left(|z|^2\right)^{-\frac{\a'}2k\cdot S \cdot k-\frac{\a'}2k\cdot k_2}
\left(|z|^2-1\right)^{\frac{\a'}2k\cdot S \cdot k}
\frac{\a'}2\frac{(k\cdot S\cdot k)}{|z|^2
\left(|z|^2-1\right)}\nonumber\\
&=&\frac{\k T_p}{2(p-1)!}\epsilon^{\a_1\ldots\a_{p-1}\b_1\b_2}
c_{\a_1\ldots\a_{p-1}}\zeta_{\b_1\b_2}~,\label{Astring}
\eea
where the last equality is valid for small momenta.%
\footnote{\label{Euclidean}
We would like to give a comment which is related to footnote \ref{zeromodecon}. 
Because of the decoupling of longitudinal polarizations, one can check that
the amplitude vanishes 
if $k$ (or equivalently $k'$) has a non-zero component along the brane 
directions. In computing 
cross-sections, one has to average over neighbouring momenta. So in order to 
have a non-vanishing cross-section, one needs the possible momentum components 
along the brane to be discrete, i.e., all worldvolume directions
should be compactified. This suggests considering Euclidean branes. In that 
case, one can have
on-shell momenta without a component along the brane world-volume. For such 
momenta, the numerator
of the expression in footnote \ref{zeromodecon} vanishes.
}

We see that, for small momenta, the string amplitude \eq{Astring} precisely 
matches the supergravity amplitude \eq{Asugra}. The computations we have done
thus provide direct evidence for the presence of the term \eq{BCsource} in the
Wess-Zumino action of the D-brane.

%%%%%%%%%%%%%%%%%%%%%%%%%%%%%%%%%%%%%%%%%%%%%%%%%%%%%%%%%%%%%%%%%%%%%%%%%%%%%%%%
\section{The $\tr R^2\wedge \protect\hat C_{p-3}$ interactions}
\label{anomalous:RRC}
In this section, which is based on results reported in our paper \cite{benfred}, 
we check the presence of the terms
\beq\label{checkRRC}
\frac{T_p}{\kappa}\int_{p+1}\hat C_{p-3}\wedge \frac{(4\pi^2\a ')^2}{384\pi^2}
(\tr R_T^2-\tr R_N^2)
\eeq 
in the Wess-Zumino action of a D$p$-brane (with $p\geq3$). We compute the tree
level string amplitude for a D$p$-brane to absorb two gravitons and emit a
($p-3$)-form R-R potential. The low energy limit of this string amplitude is 
compared to the corresponding amplitude in supergravity.

As in the previous section, we work in static gauge. The polarization of the R-R
potential is chosen to be nonzero only for all indices along the brane 
directions: $c_{\a_1\ldots\a_{p-3}}$ is only nonzero if $\a_i\in\{0,\ldots p\}$.

Although the terms \eq{checkRRC} involve curvature two-forms, we consider a flat,
infinitely extended D-brane in a flat, Minkowski background. We probe the
presence of the terms involving the curvature by scattering gravitons. Gravitons
are small fluctuations of the spacetime metric around its flat background value,
see the discussion around \eq{expansionG}. A graviton with polarization tensor
$\zeta_{\m\n}$ and momentum $k$ corresponds to the following fluctuation of the 
spacetime metric:
\beq
G_{\m\n}(X)=\eta_{\m\n}-2\k\zeta_{\m\n}e^{\ii k\cdot X}~.
\eeq 

\begin{figure}
\begin{center}
\epsfig{file=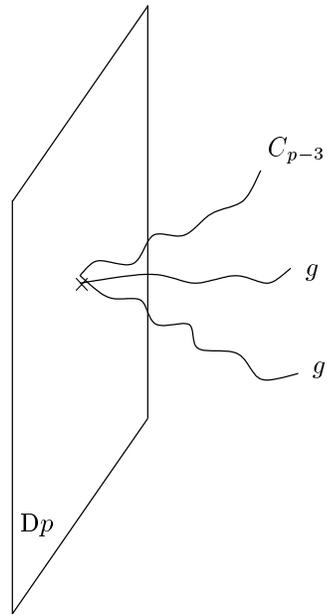}
\end{center}
\caption{The only diagram contributing to the supergravity amplitude for a 
D$p$-brane to absorb two gravitons and emit a $C_{p-3}$ potential.}
\label{fig:RRCsugra}
\end{figure}

In supergravity, the fact that we are considering small fluctuations around a
flat background allows us to use the linearized expressions for the curvature
two-forms. The supergravity amplitude for a D$p$-brane to absorb two gravitons
(with polarization tensors $\zeta_3$ and  $\zeta_4$, and momenta $k_3$ and $k_4$)
and emit a $C_{p-3}$ potential (with polarization tensor $c_{\a_1\ldots\a_{p-3}}$
and momentum $k_2$) only gets contributions from the term \eq{checkRRC} we are
interested in (see \fig{fig:RRCsugra}). 
The amplitude reads
\begin{eqnarray}\label{AsugraRRC}
{\cal A}_{\rm sugra}&=&\frac{\sqrt{2}\,\pi^2\,\kappa^2\,T_p\a '^2}{6\,(p-3)!}\,
\epsilon^{\a_1\cdots\a_{p-3}\b_1\cdots\b_4}\,c_{\a_1\cdots\a_{p-3}}\,k_{3\b_1}\,k_{4\b_3} \\ \nonumber
&&\times[(k^{\Vert}_4\cdot\zeta^{\Vert}_{3\b_2})(k^{\Vert}_3\cdot \zeta^{\Vert}_{4\b_4}) -
 (k^{\Vert}_4\cdot k^{\Vert}_{3})(\zeta^{\Vert}_{3\b_2}\cdot \zeta^{\Vert}_{4\b_4})\\ \nonumber 
&&-(k^{\perp}_4\cdot\zeta^{\perp}_{3\b_2})(k^{\perp}_3\cdot \zeta^{\perp}_{4\b_4})
+(k^{\perp}_4\cdot k^{\perp}_{3})(\zeta^{\perp}_{3\b_2}\cdot
\zeta^{\perp}_{4\b_4})]~~.
\end{eqnarray}
We explain the notation. A dot means contraction of one
Lorentz index, so when a graviton polarization tensor is dotted with another
tensor one free Lorentz index remains. The superscripts $\Vert$ or $\perp$ on two
vectors that are contracted with one another have the following meaning. A  
$\Vert$ means that the Lorentz index that is being summed over only runs over the
directions along the brane. A $\perp$ means that the Lorentz index that is 
being summed over only runs over the directions transverse to the brane. For
instance, for vectors $a_\m$ and $b_\m$ one has
\beq\label{dotvertperp}
a\cdot b=a^\Vert\cdot b^\Vert+a^\perp\cdot b^\perp~.
\eeq
In \eq{AsugraRRC}, the first two terms in square brackets come from the curvature
two-form $R_T$ of the tangent bundle to the D-brane world-volume. 
The last two terms come from the curvature two-form $R_N$ of the normal bundle.

\begin{figure}
\begin{center}
\epsfig{file=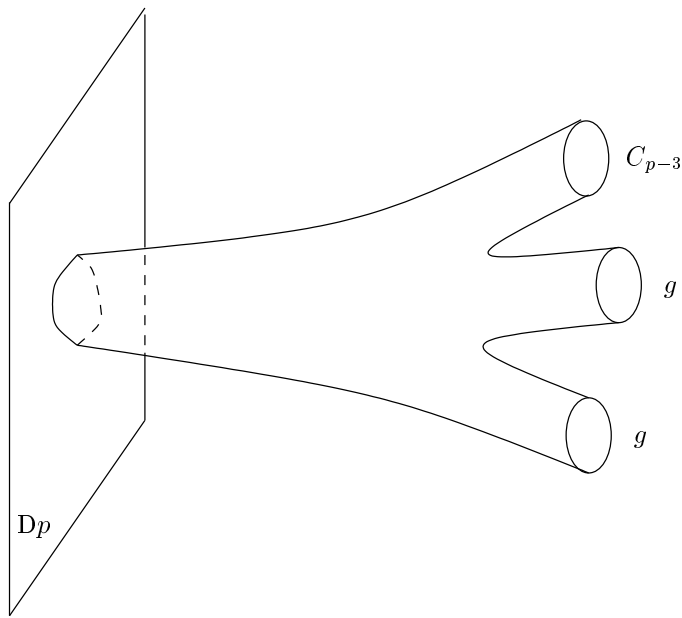}
\end{center}
\caption{String theory amplitude for a D$p$-brane to absorb two gravitons and 
emit a ($p-3$)-form R-R potential.}\label{fig:RRCstring}
\end{figure}

In string theory, the tree
level amplitude for a D$p$-brane to absorb two gravitons and emit a
($p-3$)-form R-R potential is given by 
\beq\label{RRCamp}
{\cal A}=\bra{{\cal C}_{p-3};k_2}V_g^{(0,0)}(\zeta_3,k_3)\,V_g^{(0,0)}
(\zeta_4,k_4)\ket{B}_{\R}
\eeq
(see \fig{fig:RRCstring}).
Here,
\beq
V_g^{(0,0)}(\zeta,k)=\int_{|z|>1}d^2z\,{\cal V}_B^{(0,0)}(\zeta,k)~,
\eeq
with, as in \eq{GVO},
\beq\label{Vg}
{\cal V}_g^{(0,0)}(\zeta,k)=\frac{2\kappa}{\pi\a'}\,\zeta_{\mu\nu}\,(\partial X^{\mu}
-\ii\frac{\a'}2 \,k\cdot\psi\,\psi^{\mu}) (\bar\partial X^{\nu}-\ii\frac{\a'}2 
k\cdot\tilde\psi\,\tilde\psi^{\nu})\,e^{\ii k\cdot X}~.
\eeq

As the computation is analogous to the one presented in \sect{anomalous:BC} 
but considerably longer, we shall not give all manipulations in detail. 
Instead, we shall first write down the result, and give some comments for readers
who would like to reproduce it. The result, in a form analogous to
\eq{totalamplitudeBC}, reads
\bea
{\cal A}&=&
\frac{\kappa^2\,T_p\a '^2}{4\sqrt{2}\,(p-3)!\,\pi^2}\,
\epsilon^{\a_1\cdots\a_{p-3}\b_1\cdots\b_4}\,c_{\a_1\cdots\a_{p-3}}\,
k_{3\b_1}\,k_{4\b_3}\label{AmpRRC}\\ 
&&\times[(k_4\cdot S\cdot\zeta_{3\b_2})(k_3\cdot\zeta_{4\b_4})-(k_3\cdot S\cdot k_4)(\zeta_{3\b_2}
\cdot\zeta_{4\b_4})\nonumber \\ 
&&+(k_4\cdot\zeta_{3\b_2})(k_3\cdot S\cdot\zeta_{4\b_4})
-(k_3\cdot k_4)(\zeta_{3\b_2}\cdot S\cdot\zeta_{4\b_4})]\nonumber\\
&&\times\int_{|z_3|,|z_4|>1}d^2z_3\,d^2z_4
(|z_3|^2-1)^{k_3\cdot S\cdot k_3}(|z_4|^2-1)^{k_4\cdot S\cdot k_4}
\nonumber\\&&\times
|z_3|^{-2\,k_3\cdot S\cdot k_3-2k_3\cdot S\cdot k_4-2} 
|z_4|^{-2\,k_4\cdot S\cdot k_4-2\,k_3\cdot S\cdot k_4-2}
\nonumber \\ &&\times 
|z_3-z_4|^{2\,k_3\cdot k_4-2}\,|z_3\bar z_4-1|^{2\,k_3\cdot S\cdot k_4-2}\,
(z_3\bar z_4-\bar z_3z_4)^2.\nonumber
\eea
The matrix $S$ was defined in \eq{smunu}. In terms of the $\Vert$ and $\perp$
superscripts, we have, for vectors $a_\m$ and $b_\m$, 
\beq\label{Svertperp}
a\cdot S\cdot b=a^\Vert\cdot b^\Vert-a^\perp\cdot b^\perp~.
\eeq

We now give some observations on how to obtain \eq{AmpRRC}.

The fermion zero modes have to provide the four gamma-matrices with Lorentz 
indices in the worldvolume directions complementary to the ones of
the RR potential that is being considered. The graviton vertex operator \eq{Vg}
naturally splits into four terms, according to whether one takes the bosonic or
the fermionic pieces of the holomorphic and the antiholomorphic sectors,
respectively. 
As a consequence, the amplitude \eq{RRCamp} naturally
splits into sixteen pieces. These pieces can be grouped according to the number
of fermions they contain. The piece with the maximal number of fermions (eight)
will be called the ``eight fermion part'', etc. 
  
The factor in square brackets in \eq{AmpRRC} involves four momenta and two 
graviton polarization tensors, together carrying eight Lorentz indices. Four of 
these are contracted with the $\epsilon$-symbol; the others are contracted with 
one another. Note that in each term one contraction involves the matrix $S$ and 
the other does not. This feature arises in the computations in a nontrivial way,
which we now explain.  

As a preliminary remark, the only contributions come from the four and eight
fermion parts.

To look for terms with precisely one $S$, one can do one left-right contraction
and one left-left or right-right contraction. For the terms where the graviton
polarizations are contracted with momenta, this is the whole story: the
contributions of the eight and four fermion parts combine into the corresponding
terms in \eq{AmpRRC}. 

For the terms involving graviton polarizations contracted with each other (and
consequently also momenta with each other), one does not find the complete 
result in 
this way. There are extra terms coming from doing two left--left or right--right
contractions in the eight fermion part, which thus, at first sight, do not 
involve the matrix $S$. (There are also terms involving two
mixed contractions, and thus twice the matrix $S$; that story is completely 
analogous, so we shall not discuss it in detail here.) 

The point is that the extra terms, which did not seem to involve the matrix $S$,
combine with extra terms from the four-fermion part. The combination allows 
an integration by parts that brings it to a form that fits in \eq{AmpRRC} after 
all.

To be explicit, we display the extra terms, which are proportional to
$\zeta_3\cdot\zeta_4$ (this factor will not be displayed). 
We have fixed the angular coordinate of $z_4$, and denote
the angle of $z_3$ by $\theta$. We shall only write down the $\theta$-dependent
factors of the integrand:
\beq
|z_3-z_4|^{2\,k_3\cdot k_4}|z_3\bar z_4-1|^{2(k_3\cdot S\cdot k_4)}
\left\{(k_3\cdot k_4)\frac{(z_3\bar z_4-\bar z_3z_4)^2}{|z_3-z_4|^4}
       -\frac{z_3z_4}{(z_3-z_4)^2}-\frac{\bar z_3\bar z_4}
       {(\bar z_3-\bar z_4)^2}\right\}.
\eeq 
The first term in braces comes from the eight fermion part, the other two from
the four fermion part. This expression can be written as
%\[|z_3-z_4|^{2(k_3\cdot k_4-2)}|z_3\bar z_4-1|^{2(k_3\cdot S\cdot k_4)}
%\left\{(k_3\cdot k_4-1)(z_3\bar z_4-\bar z_3z_4)^2 
%       -(z_3\bar z_4+\bar z_3z_4)|z_3-z_4|^2 \right\}
%\]
\beq 
|z_3\bar z_4-1|^{2(k_3\cdot S\cdot k_4)}\frac{\partial}{\partial\theta}
\left\{i(z_3\bar z_4-\bar z_3z_4)|z_3-z_4|^{2(k_3\cdot k_4-1)} \right\}~.
\eeq  
Integrating by parts, this leads to
\bea
&&- i(z_3\bar z_4-\bar z_3z_4) |z_3-z_4|^{2(k_3\cdot k_4-1)}
\frac{\partial}{\partial\theta}|z_3\bar z_4-1|^{2(k_3\cdot S\cdot k_4)}
\\
&&=-(k_3\cdot S\cdot k_4)(z_3\bar z_4-\bar z_3z_4)^2 |z_3-z_4|^{2(k_3\cdot k_4-1)}
|z_3\bar z_4-1|^{2(k_3\cdot S\cdot k_4-1)}~,\nonumber
\eea 
which is of the right form to appear in \eq{AmpRRC}. This ends the observations
on how to obtain \eq{AmpRRC}.

Since we want to compare the amplitude \eq{AmpRRC} to supergravity, we are 
interested in its low-energy behaviour.
Restricting to small momenta, the momenta in the exponents can be put to
zero, such that the integral simplifies to
\beq \label{Int}
{\cal I}=\int_{|z_3|,|z_4|>1}d^2z_3\,d^2z_4\,|z_3z_4|^{-2}\,
|z_3-z_4|^{-2}\,|z_3\bar z_4-1|^{-2}\,(z_3\bar z_4-\bar z_3z_4)^2~.
\eeq
As has been shown in \cite{stefanski}, the integral  then evaluates to 
$\frac{2 \pi^4}{3}$. 

Using \eq{dotvertperp} and \eq{Svertperp}, we can then, for small momenta, 
rewrite \eq{AmpRRC} as 
\begin{eqnarray} \label{stramp2}
{\cal A} &=&
\frac{\sqrt{2}\,\pi^2\,\kappa^2\,T_p\a '^2}{6\,(p-3)!}\,
\epsilon^{\a_1\cdots\a_{p-3}\b_1\cdots\b_4}\,c_{\a_1\cdots\a_{p-3}}\,k_{3\b_1}\,k_{4\b_3} \\ \nonumber
&&\times[(k^{\Vert}_4\cdot\zeta^{\Vert}_{3\b_2})(k^{\Vert}_3\cdot \zeta^{\Vert}_{4\b_4}) -
 (k^{\Vert}_4\cdot k^{\Vert}_{3})(\zeta^{\Vert}_{3\b_2}\cdot \zeta^{\Vert}_{4\b_4})\\ \nonumber 
&&-(k^{\perp}_4\cdot\zeta^{\perp}_{3\b_2})(k^{\perp}_3\cdot \zeta^{\perp}_{4\b_4})
+(k^{\perp}_4\cdot k^{\perp}_{3})(\zeta^{\perp}_{3\b_2}\cdot 
\zeta^{\perp}_{4\b_4})]~~.
\end{eqnarray}

The string amplitude \eq{stramp2} precisely agrees with the supergravity
amplitude \eq{AsugraRRC}! It is remarkable that, for small momenta, the string 
amplitude nicely splits into tangent and normal bundle contributions.

In this section, we have directly checked the presence of the terms \eq{checkRRC}  
in the D-brane Wess-Zumino action.

%%%%%%%%%%%%%%%%%%%%%%%%%%%%%%%%%%%%%%%%%%%%%%%%%%%%%%%%%%%%%%%%%%%%%%%%%%%%%%%%
\section[Four graviton amplitude and non-anomalous couplings]
{Four graviton amplitude and non-anomalous couplings}

\label{anomalous:D:non}
This section is based on the results of \cite{normal}. 
We want to check the presence of the eight-form curvature terms in \eq{WZ}:
\beq\label{eightformcheck}
\frac{T_p}{\kappa}\int_{p+1}\hat C_{p-7}\wedge 
\left(\frac{(4\pi^2\a ')^4}{294912\pi^4}(\tr R_T^2-\tr R_N^2)^2+
\frac{(4\pi^2\a ')^4}{184320\pi^4} (\tr R_T^4-\tr R_N^4)\right)~.
\eeq 
The stategy is precisely the same as in \sect{anomalous:BC} and
\sect{anomalous:RRC}: compute the amplitude of a D$p$-brane with $p\geq7$ to
absorb four gravitons and emit a ($p-7$)-form R-R potential in supergravity, and
compare the result with the low-energy limit of the corresponding tree level
amplitude in string theory.

As the explicit formulas are analogous to but much longer than the ones in 
\sect{anomalous:RRC}, we shall not give all of them in detail. Rather, we shall
give most results in words, but focus on an unexpected feature of the string
amplitude \cite{normal}: it encodes more eight-form curvature couplings than 
the ones present in \eq{eightformcheck}! 

Before we discuss these new couplings, let us describe the string amplitude one
has to compute. It reads 
\beq\label{RRRRCamp}
{\cal A}=\bra{{\cal C}_{p-7};k_5}V_g^{(0,0)}(\zeta_1,k_1)\,
V_g^{(0,0)}(\zeta_2,k_2)\, V_g^{(0,0)}(\zeta_3,k_3)\,V_g^{(0,0)}
(\zeta_4,k_4)\ket{B}_{\R}~.
\eeq
For notational convenience, we shall 
consider polarization tensors of the special form
\beq\label{halfpol}
\zeta_{\m\n}=\zeta_\m\zeta_\n~.
\eeq
However, this is not essential: the general results can always 
be obtained by replacing $\zeta_\m\zeta_\n$ by $\zeta_{\m\n}$ in the formulas.

In \cite{stefanski}, the amplitude we have just described was computed for the
special case of all momenta and polarizations along the D-brane. In this
section, we extend the computation to general momenta and graviton polarizations.%
\footnote{The polarization of the R-R potential is still taken to be along the
brane.}

The structure of the amplitude \eq{RRRRCamp} is as follows. From each graviton,
one momentum index and one polarization index is contracted with the $\e$-symbol,
giving rise to a universal factor
\beq
\epsilon^{\a_1\ldots\a_{p-7}
\mu_1\ldots\mu_8}c_{\a_1\ldots\a_{p-7}}\z_{1\mu_1}k_{1\mu_2}\ldots\z_{4\mu_7}
k_{4\mu_8}
\eeq
in the string amplitude.%
\footnote{ 
The eight momentum and polarization indices will correspond to the eight form-% 
indices of the eight-form in the curvatures.
}
Each graviton contributes one more momentum and one more
``half'' polarization,%
\footnote{where ``half'' is in the sense of \eq{halfpol}.} which each carry one
Lorentz index. These eight indices have to be contracted with one another.  
There are two substantially different ways in which these contractions can be
done.

First, one has terms in which the gravitons are divided in pairs and the
contractions are done within each pair, e.g.,
\beq
(k_1\cdot\zeta_2)(k_2\cdot S\cdot\zeta_1)(k_3\cdot\zeta_4)
(k_4\cdot S\cdot\zeta_3)~.
\eeq
The other terms of this kind can be obtained by permuting the gravitons and antisymmetrizing 
in polarizations and momenta of each graviton. It turns out that the part of the
string amplitude coming from this class of terms precisely reproduces the part of
the supergravity ampitude coming from the 
$(\tr R_T^2-\tr R_N^2)^2$ term in \eq{eightformcheck}. To derive this, one
can use the computations of \sect{anomalous:RRC}: for small momenta the integrand
of the integral over $z_1$, $z_2$, $z_3$ and $z_4$ factorizes into, say, the
integrand of \eq{Int} and the same integrand with $z_3,z_4$ replaced by
$z_1,z_2$.
  
Second, there are the terms for which such a factorization does not occur. Their
contribution to the string amplitude \eq{RRRRCamp} turns out to be
\beqa \label{trR4}
&&\frac{\sqrt{2}T_p \pi^4\kappa^4\a'^4}{128 (p-7)!}\epsilon^{\a_1\ldots\a_{p-7}
\mu_1\ldots\mu_8}c_{\a_1\ldots\a_{p-7}}\z_{1\mu_1}k_{1\mu_2}\ldots\z_{4\mu_7}
k_{4\mu_8}\nonumber\\&& 
       \times{\bf\{}\frac{14\pi^8}{45}
       [(\z_1\cdot S\cdot k_2)(\z_2\cdot k_3)(\z_3\cdot k_4)(\z_4\cdot k_1)
       \nonumber\\ &&
       +(\z_1\cdot k_2)(\z_2\cdot S\cdot k_3)(\z_3\cdot k_4)(\z_4\cdot k_1)%\right.
       \nonumber\\ && %\left.
       +(\z_1\cdot k_2)(\z_2\cdot k_3)(\z_3\cdot S\cdot k_4)(\z_4\cdot k_1)
       \nonumber\\ &&
       +(\z_1\cdot k_2)(\z_2\cdot k_3)(\z_3\cdot k_4)(\z_4\cdot S\cdot k_1)]%\right.
       \nonumber\\ &&%\left.
       +\frac{2\pi^8}{45}
       [(\z_1\cdot S\cdot k_2)(\z_2\cdot S\cdot k_3)(\z_3\cdot S\cdot k_4)(\z_4\cdot k_1)
       %\right.
       \nonumber\\ &&%\left.
       +(\z_1\cdot S\cdot k_2)(\z_2\cdot S\cdot k_3)(\z_3\cdot k_4)(\z_4\cdot S\cdot k_1)
       %\right.
       \nonumber\\ &&%\left.
       +(\z_1\cdot S\cdot k_2)(\z_2\cdot k_3)(\z_3\cdot S\cdot k_4)(\z_4\cdot S\cdot k_1)
       \nonumber\\ &&
       +(\z_1\cdot k_2)(\z_2\cdot S\cdot k_3)(\z_3\cdot S\cdot k_4)(\z_4\cdot S\cdot k_1)]
       {\bf\}}
\eeqa
plus terms obtained from this by permuting the gravitons and antisymmetrizing 
in polarizations and momenta of each graviton.

Using \eq{dotvertperp} and \eq{Svertperp} to write
\beqa
\z_i\cdot k_j&=&\z_i^{\Vert}\cdot k_j^{\Vert}+
\z_i^{\perp}\cdot k_j^{\perp}~;\nonumber\\
\z_i\cdot S\cdot k_j&=&\z_i^{\Vert}
\cdot k_j^{\Vert}-\z_i^{\perp}\cdot k_j^{\perp}~,
\eeqa
it is clear that the expression in braces contains 
\beq\label{trRT4}
\frac{64\pi^8}{45} (\z_1^{\Vert}\cdot k_2^{\Vert})(\z_2^{\Vert}\cdot k_3^{\Vert})
                   (\z_3^{\Vert}\cdot k_4^{\Vert})(\z_4^{\Vert}\cdot k_1^{\Vert})~,
\eeq
the part derived in \cite{stefanski}, and
\beq\label{trRN4}
-\frac{64\pi^8}{45} (\z_1^{\perp} \cdot k_2^{\perp})(\z_2^{\perp}\cdot k_3^{\perp})
(\z_3^{\perp}\cdot k_4^{\perp})(\z_4^{\perp}\cdot k_1^{\perp})~.
\eeq
It turns out that \eq{trRT4} reproduces the  $\tr R_T^4$ term in
\eq{eightformcheck} and that \eq{trRN4} reproduces the  $\tr R_N^4$ term.

Thus, we can conclude that all the terms in \eq{eightformcheck}
are seen in an explicit string scattering computation. 

However, this is not the full story. \eq{trR4} also contains terms
which are not accounted for by the anomalous gravitational couplings in \eq{WZ},
such as
\beq
\frac{8\pi^8}{15} (\z_1^{\perp} \cdot k_2^{\perp})(\z_2^{\Vert}\cdot k_3^{\Vert})
                   (\z_3^{\Vert}\cdot k_4^{\Vert})(\z_4^{\Vert}\cdot k_1^{\Vert})~.
\eeq
The other terms can be obtained from this one by obvious permutations, and by taking three
``transversal'' factors rather than one (the latter terms have an extra minus sign).

Is the presence of extra terms involving the metric, apart from the ones in 
\eq{WZ}, a contradiction with the anomaly inflow arguments of \cite{GHM,CY}?
These anomaly inflow arguments state that, in certain D-brane configurations, 
the variation under local Lorentz transformations of the curvature terms in 
\eq{WZ}  precisely cancels the gravitational anomaly due to certain chiral modes.
If the extra terms we have just found also had a variation under local Lorentz
transformations, we would be in trouble: string theory in certain D-brane
backgrounds would not be invariant under local Lorentz transformations.

The question thus is: are the extra terms invariant under local Lorentz
transformations? In fact, they need only be invariant under the Lorentz
transformations that leave the D-brane world-volume invariant: the other Lorentz
transformations are explicitly broken by the D-brane. Lorentz transformations
that leave the D-brane world-volume invariant are the ones that do not mix
directions tangent and perpendicular to the brane. 

Now, we check whether the extra terms are invariant under local Lorentz
transformations that do not mix tangent and perpendicular directions. As we know
form \sect{superstrings:inflow:D}, we should first express the couplings in terms
of the R-R field strengths rather than the potentials. In terms of the linearized
spacetime curvature twoform $R$, the extra
terms read, schematically, 
\beq
C\wedge\tr(P_{\Vert}\,R\, P_{\perp}\,R\, P_{\perp}\,R\, P_{\perp}\,R)~,
\eeq
where $P_{\Vert}$ and $P_{\perp}$ are constant matrices projecting on indices along and
perpendicular to the brane, respectively.
Integrating by parts, this becomes, again schematically,
\beq \label{CS}
F\wedge \tr(P_{\Vert}\,\omega\, P_{\perp}\,R \,P_{\perp}\,R\, P_{\perp}\,R)~,
\eeq
where $\omega$ is the linearized spin connection and $F$ is the R-R field
strength. Under local Lorentz
transformations that do not mix directions tangent and perpendicular to the brane
world-volume, the variation of $\omega$ is a
block-diagonal matrix, such that the variation of \eq{CS} vanishes.

We conclude that the extra terms we have found are non-anomalous, so their
presence is consistent with the anomaly inflow arguments.

\section{Conclusions}
Anomaly inflow arguments fix the different terms in the D-brane 
Wess-Zumino action \eq{WZ}. In this chapter, we have checked the presence of 
three of these terms by explicit string theory computations. These computations
were performed in the boundary state formalism. As a bonus of the explicit
computations, we have found extra terms in the D-brane Wess-Zumino action. These
extra terms are non-anomalous and are thus consistent with the anomaly inflow
arguments.

%AdS/CFT
%\include{AdS}

%Baryon vertex
\chapter{Baryon vertex in AdS/CFT}\label{baryon}
In this chapter, we discuss an application of the D-brane world-volume action: we
show how it can be used to study strings ending on D-branes in a D-brane
background. We study one particular configuration, which is largely motivated by
studies of the baryon vertex in the AdS/CFT correspondence, a recently proposed 
duality between conformal field
theories and string theories in certain backgrounds. Therefore, we include 
very brief introductions to the AdS/CFT correspondence in general and  to the
baryon vertex in particular.
%%%%%%%%%%%%%%%%%%%%%%%%%%%%%%%%%%%%%%%%%%%%%%%%%%%%%%%%%%%%%%%%%%%%%%%%%%%%%%%
\section{Introduction}\label{baryon:intro}
\subsection{AdS/CFT correspondence}\label{AdSCFT}
In \cite{maldacena} (see \cite{review} for an extensive review),
Maldacena conjectured that {\it ${\cal N}=4$ SU($N$) super-Yang-Mills theory%
\footnote{There is a subtlety about whether the gauge group is SU($N$) or 
U($N$) \cite{wittentopological, review}. Apparently, one has a choice here,
related to whether one includes certain degrees of freedom on the AdS side. We
shall not go into this and use the SU($N$) version of the conjecture.}
in 3+1 dimensions is dual to type IIB
superstring theory on $AdS_5\times S^5$}.

The evidence for this conjecture comes from studying a system of $N$ coincident
D3-branes in type IIB string theory from two points of view. In the first
description, the D3-branes are described by a solution to the type IIB
supergravity equations of motion \cite{HorStrom}. This solution, whose explicit
form will be given in \sect{baryon:ham}, involves a non-trivial metric and 
five-form R-R field
strength. In the second description, one adds the effective D-brane action (see
\sect{superstrings:D-branes}) to the bulk spacetime action (which is the 
supergravity
action plus higher derivative corrections). In the latter description, the
D-branes are described by (the effective field theory of the massless modes of) 
the open strings ending on them. 

It turns out that in a low-energy limit both descriptions give rise to two
decoupled systems, one of which is supergravity in flat space. In the first
description, the other system is string theory on $AdS_5\times S^5$ with
five-form flux
\beq\label{fluxN}
\int_{S^5}F_5=16\pi^4g_s\a'^2N~.
\eeq
This can be derived from \eq{RR} and \eq{R4}, using the fact that the volume of
the unit five-sphere is $\pi^3$. The space $AdS_5\times S^5$ is the near 
horizon geometry of the D3-brane background, i.e., the ``limit'' of the
geometry of the D3-brane background as one approaches the horizon. 
In the other description, the second system is
${\cal N}=4$  SU($N$) super-Yang-Mills theory in 3+1 dimensions. Thus, it is natural
to identify string theory on $AdS_5\times S^5$ with ${\cal N}=4$  SU($N$) 
super-Yang-Mills theory in 3+1 dimensions. 

This identification is a strong/weak duality, which means that if one
description is weakly coupled, the other one is at strong coupling (see
\sect{central:duality} and \sect{central:susy} for examples of strong/weak
dualities). 
Like most
strong/weak dualities, it is extremely hard to prove or disprove, but very
powerful. On the one hand, the gauge theory description gives a non-perturbative 
definition of string theory on a certain background. On the other hand, one can
use string theory to learn something about strongly coupled gauge theory. For
instance, in the limit where string theory is well-approximated by classical
supergravity explicit computations are fairly easy \cite{WittenAdS, GKP}. 
In the gauge theory, this
limit corresponds to the large $N$ limit at strong 't~Hooft coupling.

When string theory was originally discovered, it was meant to be a theory of the
strong interactions, which are nowadays described by quantum chromodynamics 
(QCD). The AdS/CFT
correspondence realizes the old idea of describing gauge theory by strings.
Unfortunately, so far it is not known whether there is a string theory dual of 
QCD. 

Without going into the details, we mention that
the gauge theory can be thought of as living on
the boundary of the anti-de Sitter (AdS) space \cite{WittenAdS}. 
It can be argued that the gauge
theory gives a holographic description of the string theory.
%%%%%%%%%%%%%%%%%%%%%%%%%%%%%%%%%%%%%%%%%%%%%%%%%%%%%%%%%%%%%%%%%%%%%%%%%%%%%%%%
\subsection{Baryon vertex}
Since ${\cal N}=4$ SU($N$) super-Yang-Mills theory only contains matter fields in 
the adjoint representation of the gauge group, there are no dynamical quark
fields in the theory. However, one can add external quarks to the theory. These
can be regarded as endpoints of fundamental strings in the AdS space
\cite{maldacenawilson, reyyee}. Indeed, it
is clear from \eq{combination} that the endpoints of a fundamental string are
electrically charged (with charge $\pm 1$) under the gauge field living on the
D-brane the string ends on. Thus, an endpoint of a fundamental string on a
D3-brane ``at the boundary of AdS'' will
correspond to a charged object in the dual gauge theory. Consider a fundamental
string connecting two points at the boundary of AdS. As the two
endpoints of a string are oppositely charged under the gauge field, such a
string corresponds to an external quark--antiquark pair.

Witten \cite{baryons} succeeded in constructing a ``baryon'' out of these
external quarks, i.e., a gauge-invariant combination of $N$ external quarks. 
Such
a configuration should consist of $N$ strings, oriented in the same way, with 
each one endpoint on the boundary and the other endpoints joined by a ``baryon
vertex'' in the interior of AdS. Witten argued that the baryon vertex is simply
a D5-brane wrapped on the $S^5$ of $AdS_5\times S^5$, i.e., a D5-brane with
world-volume   $S^5\times \Rbar$ with $\Rbar$ a timelike curve in $AdS_5$.

The argument goes as follows. The D5-brane WZ action contains a term
\beq
\frac{T_5}\k \int_{S^5\times\Rbar} \,2\pi\a' A \wedge F_5~,
\eeq
which, because of \eq{fluxN}, \eq{Tp} and \eq{kappa}, contributes $N$ units of
$A$-charge. As $S^5$ is closed, this charge should be cancelled by $-N$ units of
charge from some other source. As we have seen above, $N$ fundamental strings
(with the right orientation) ending on the D5-brane do the trick.

\sect{ham.anal.} is devoted to the study of this baryon vertex and an analogous
configuration in the full (as opposed to near-horizon) D3-brane background.

%%%%%%%%%%%%%%%%%%%%%%%%%%%%%%%%%%%%%%%%%%%%%%%%%%%%%%%%%%%%%%%%%%%%%%%%%%%%%%%%%
\subsection{BPS method}
World-volume techniques have been a powerful tool in analyzing
configurations of strings and D-branes. 
In~\cite{cm, g}, a fundamental string ending on
a D-brane was analyzed from the point of view of the D-brane
world-volume theory. That theory was shown to admit solutions allowing
an interpretation as fundamental strings ending on the brane. 
In~\cite{jqp}, these solutions were interpreted along
the lines of Bogomol'nyi's analysis in the context of magnetic monopoles
\cite{bog} (see \sect{central:duality}). 
The square of the energy density is written as a sum
of squares, such that putting one of the squares equal to zero yields
a first order differential equation implying the equations of motion.

In the case of monopoles, Bogomol'nyi's bound was reinterpreted by Witten
and Olive in terms of central charges in a supersymmetry algebra
\cite{wittenolive}. A similar
interpretation is possible here \cite{BGT}.

The world-volume analysis of \cite{cm,g} was performed for branes in a flat
background. Our aim is to extend the method to
the case of a brane in the background produced by another brane. We
have chosen the example of a D5-brane in the background of $N$
D3-branes, since this configuration is physically particularly
interesting. In the near horizon limit of the D3-brane geometry, it is
related to the baryon vertex in ${\cal{N}}=4, D=4$ SYM~\cite{baryons,
Imamura}. In the non-near horizon case, it is relevant for the
Hanany-Witten effect \cite{hanany, Callan}. 

Section~\ref{review} contains a review of the world-volume analysis of
a D5-brane in a flat background. The Bogomol'nyi argument will be treated in
detail. For the interpretation in terms of supersymmetry algebras, the reader
is referred to the literature \cite{BGT,BPvdS,CGMV}. 

In \sect{ham.anal.}, we write the energy density as (the square root
of) a sum of squares  along the lines of~\cite{jqp}, derive a BPS bound and 
interpret its physical meaning. For the interpretation in terms of supersymmetry
algebras, we refer the reader to our paper \cite{CGMV}.

What follows is mainly based on \cite{CGMV}. 
%%%%%%%%%%%%%%%%%%%%%%%%%%%%%%%%%%%%%%%%%%%%%%%%%%%%%%%%%%%%%%%%%%%%%%%%%%%%%%%
\section{BPS method for D-branes in a flat background}

\label{review}\label{baryon:review}
In this section, we will review the main points of the BPS method for
D-branes in a flat Minkowski background~\cite{BGT, jqp}. We will work with the
particular case of a D5-brane, since in the next section we will be
interested in extending this analysis to a D5-brane in a non-trivial
background.

Let us thus consider a D5-brane in a flat background. The action is
given by
\be\label{baryonBIflat}
S = - \frac{T_5}\k \int d^{5+1} \sigma \, \sqrt{-\det (\hat G+F)} \,,
\ee
where $T_5/\k=1/g_s\,{(2\, \pi)}^5\,{\a'}^3$ is the D5-brane tension
(from now on we set $T_5/\k\equiv 1$), $\hat G$ is the metric on the D-brane
world-volume induced by the Minkowski spacetime metric
and $F=dA$ is the field strength of the Born-Infeld (BI) gauge vector
$A$, in which we have absorbed the factor $2\pi\a'$ present in \eq{BIflat}.  
Let us fix the static gauge
\be
X^0 = \si^0 \sac X^4=\si^1 , \ldots , X^8=\si^5~,
\ee
corresponding to a brane extending (asymptotically) along directions
45678.

Now, we look for ``spike-like'' world-volume solutions
describing a fundamental string attached to the D5-brane.  We consider
configurations with only one scalar excited and with purely electric
BI field:
\be
X^1 = X^2 = X^3 = 0 \,,\quad X^9=X(\si) \,,\quad A_t = A_t(\si) \,,\quad
 \vec{A} \equiv (A_4,\ldots ,A_8)=\vec{A}(\si_0) \,.
\label{truncation}
\ee
From the ten-dimensional spacetime point of view, one might represent the
configuration we are considering by the following array:
\be
\ba{ccccccccccl}
 &&&&&&&&&&\quad \mbox{flat background}      \nn
{\rm D5}: &\_&\_&\_&4&5&6&7&8&\_&\quad \mbox{world-volume}      \nn
{\rm F1}: &\_&\_&\_&\_&\_&\_&\_&\_&9 &\quad \mbox{BPS solution.}
\ea
\ee

Since we are interested in bounds on the energy for this
configuration, it is useful to pass to the Hamiltonian formalism. Let
us denote the canonically conjugate momenta associated to $X$ and
$\vec{A}$ by $P$ and $\piv$ respectively.  We define a Hamiltonian density
\be
{\cal{L}} = \dot{\vec{A}} \cdot \vec{\Pi} + \dot{X} \, P - L =
\vec{E} \cdot \vec{\Pi} + \dot{X} \, P - L - A_t \,
\vec{\nabla} \cdot \vec{\Pi}  \,,
\ee
where $L$ is the original Lagrangian density, $\vec E=\dot {\vec
A}-\vec\nabla A_t$, and in the last step we have integrated
$\vec{\nabla}A_t \cdot \vec{\Pi}$ by parts. In terms of fields and canonical
momenta,
$\cal{L}$ is given by
\be
{\cal{L}} =  \sqrt{\left( 1+ (\narx)^2 \right)
\left(1+ P^2 \right)+
 \piv^2 + \left( \piv \cdot \narx \right)^2}
-   A_t \, \vec{\nabla} \cdot \vec{\Pi}   \,.
\ee
The first term is the desired expression for the energy density
\be
{\cal{H}}= \sqrt{\left( 1+  (\narx)^2 \right)
\left(1+ P^2 \right)+
 \piv^2 +  \left( \piv \cdot \narx \right)^2}\,,
\label{hamiltonian}
\ee
whereas the second term yields the Gauss law (constraint)
\beq\label{gausslaw}
\vec{\nabla} \cdot \vec{\Pi}=0~.
\eeq

For static configurations, $P=0$, the Hamiltonian reduces
to~\cite{jqp}
\be
{\cal{H}} = \sqrt{\left( 1 \pm  \narx \cdot \piv \right)^2
+ \left(\narx \mp \piv \right)^2}  \,.
\ee
In order to get a bound for the energy from the previous expression, we
should consider the ``vacuum'' or ground state solution of our flat
D5-brane and its energy density.  The ground state solution
corresponds to $X(\sigma)$ not being excited and $F=0$. 
This configuration is a solution of the equations of motion
and its energy density is \mbox{${\cal{E}}_{\rm gs}=1$}. Hence, the energy
density in \eq{hamiltonian} can be split as
${\cal{H}}={\cal{E}}_{\rm gs}+{\cal{E}}_{\rm def}$, where ${\cal{E}}_{\rm def}$ is
the deformation energy density of the brane, i.e., the energy density
of the brane relative to its ground state.  After this splitting, we
obtain a bound for the energy density ${\cal{E}}_{\rm def}$ given by
\be
{\cal{E}}_{\rm def}\geq  \left\vert \narx \cdot \piv \right\vert \,,
\ee
with equality when
\be
\narx = \pm \piv  \,.
\label{bound}
\ee
The bound on the density implies the following bound on the
deformation energy:
\bea
E&=&\int_\Sigma \, {\cal{E}}_{\rm def}  \geq  |Z_{\rm el}|~; \nonumber\\
Z_{\rm el}&\equiv& \int_\Sigma \, \piv \cdot \narx \,,
\label{zel}
\eea
where $\Sigma$ is the world-space of the D5-brane.

\begin{figure}
\begin{center}
\epsfig{file=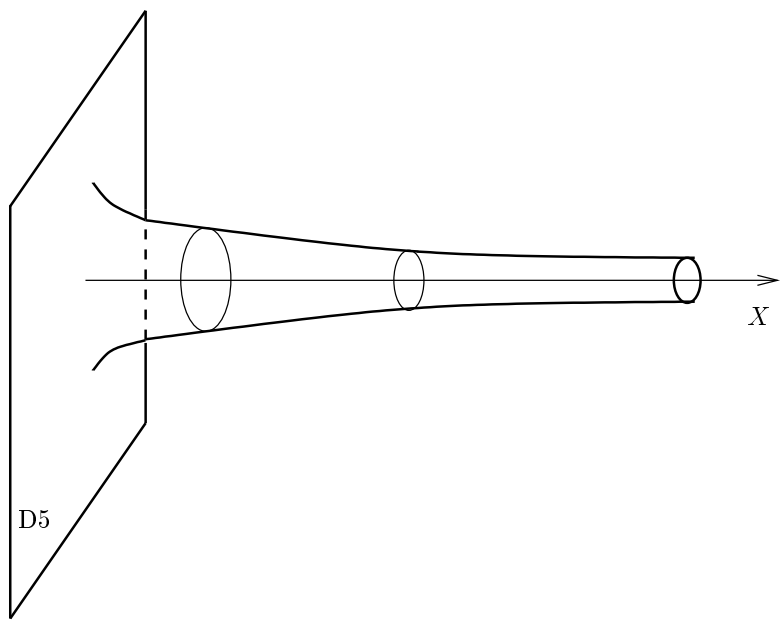}
\end{center}
\caption{A spike solution of the D5-brane world-volume theory is interpreted as
a fundamental string ending on the D5-brane.}\label{fig:Zel}
\end{figure}

Let us now figure out the physical meaning of $Z_{\rm el}$.  Because of the
Gauss law constraint, solutions of \eq{bound} correspond to solutions
of $\nabla^2 X=0$. Solutions with isolated singularities in $X$ are the BIons 
found in~\cite{cm, g}:
\be
X=\frac{q}{3V_{(4)} r^3}\,, \qquad r^2= \left(\si^1\right) ^2 +\cdots +
\left( \si^5\right) ^2 ,
\qquad V_{(4)}=\frac{8\pi^2}{3}\,,
\label{Bion}
\ee
corresponding to a charge $q$ at the origin. This charge at the origin
is the source of the BI vector.%
\footnote{\label{fn:Gauss}The action
\eq{baryonBIflat} has no source for the BI vector. The case of a
D5-brane in the background of D3-branes, studied in the next section, is
different since for a generic configuration there is a source for the
BI vector due to a term in the D-brane Wess-Zumino action.}  
For the configuration \eq{Bion}, $|Z_{\rm el}|$ and thus $E$ are infinite. However,
it is instructive to have a closer look at this divergence by considering the
energy in a region $r>\e$ for small $\e$:
\beq
|Z_{\rm el}(\e)|=\frac{q^2}{V_{(4)}}\int_\e^\infty r^4\,dr\,r^{-8}=\frac{q^2}{3\e^3
V_{(4)}}=qX(\e)~,
\eeq 
where $X(\e)$ represents the ``height'' of the spike at a distance $\e$
from the origin.  The charge $q$ should be quantized in the quantum
theory \cite{cm}.  Setting it equal to its minimal positive value
and restoring the factor $T_5/\k$, one finds \cite{cm}
\beq
|Z_{\rm el}(\e)|=T\,X(\e)~,
\eeq
where $T=1/2\pi\a'$ is the tension of a fundamental string.
This makes clear the interpretation of $|Z_{\rm el}|$ as the energy
of a fundamental string attached to the D5-brane (see \fig{fig:Zel}).

The analysis up to \eq{zel} is analogous to Bogomol'nyi's analysis \cite{bog}: 
the energy
is bounded from below by a ``topological'' charge (by which we mean a charge 
that is
invariant under small local variations of the fields; $Z_{\rm el}$ depends only on
the limiting behaviour of the solution, as is clear from \eq{zel} and
\eq{gausslaw}).
Configurations saturating the bound will automatically solve the equations of
motion. In order to find such configurations, a first order differential equation
(rather than a second order one) should be solved (see \eq{bound}). 

There also exists an analogue of the interpretation \cite{wittenolive}
of the Bogomol'nyi bound in terms of central charges of a supersymmetry algebra.
The charge $Z_{\rm el}$ can be identified with a charge appearing in the most 
general anticommutator of the supersymmetries left unbroken by the D5-brane
ground state. 
We refer the interested reader to the literature for more details
\cite{BGT,BPvdS,CGMV}.
%%%%%%%%%%%%%%%%%%%%%%%%%%%%%%%%%%%%%%%%%%%%%%%%%%%%%%%%%%%%%%%%%%%%%%%%%%%%%%%%
\section{D5-brane in D3-brane background}
\label{ham.anal.}\label{baryon:ham}

In this section, we derive a BPS bound on the energy of a D5-brane
in the \bg\ geometry of a stack of $N$ D3-branes, by showing that it
is bounded from below by a topological quantity.  Let us begin by
describing the D3-brane background.  The ten-dimensional metric is
\be
ds^2_{(10)} = H^{-1/2} d X_{\parallel}^2 + H^{1/2} \left( dr^2 + r^2
d \Omega_{(5)}^2 \right) . 
\ee 
Here, $X_{\parallel}=(X^0, X^1, X^2, X^3)$ are Cartesian coordinates on
$\Rbar^{(1,3)}$ and $d \Omega_{(5)}^2$ is the line element on a unit
five-sphere $S^5$, which we take to be parametrized by standard angular
coordinates $\T, \Phi^i, i=1, \ldots , 4$, where $\Phi^i$ are
angular coordinates on a four-sphere.  Thus, we have
\be
d \Omega_{(5)}^2 = d \T^2 + \sin^2 \T \, d \Omega_{(4)}^2 \,. 
\ee 
The
coordinate $r$ parametrizes the radial distance to the branes, so $r,
\T$ and $\Phi^i$ are spherical coordinates on the six-dimensional
space transverse to the branes. The function $H$ is a harmonic function 
given by
\begin{equation}
H= a +\frac{ R^4}{  r^4}  \, ,
\end{equation}
where two values for $a$ are of interest. 
The ``full'', asymptotically flat D3-brane background is described by $a=1$.
Its near-horizon ($r\rightarrow 0$) limit is obtained by putting $a=0$. 
The near-horizon geometry is $AdS_5 \times S^5$. 
The parameter $R$ is given by 
\beq\label{R4}
R^4=4\pi g_s\a'^2N~.
\eeq
In the case $a=0$, it coincides with the radius of both $AdS_5$ and $S^5$. 

Apart from the metric, the \su solution describing
the D3-branes involves a non-vanishing Ramond-Ramond (R-R)
five-form \fs
\bea F_5&=& - H^{-2} \, H' \, dX^0 \wedge dX^1 \wedge
dX^2 \wedge dX^3 \wedge dr + 4 \, R^4 \, \omega_{(5)} \label{RR}~; \\
\omega_{(5)} &=& \sin^4 \T \, d\T \wedge \omega_{(4)} \,, 
\eea 
where $\omega_{(n)}$ is the volume form on a unit $n$-sphere, $H'
\equiv \pa_r H$ and $F_5=dC_4$ (we are working in a background with $B_2=0$,
so $\tilde F_5=F_5$, see \eq{tildeF5}).

Now, let us consider a D5-brane coupled to the above
background. It is described by the action
\be
S = - \frac{T_5}\k \int_{\Sigma} d^6 \sigma \, \sqrt{-\det (\hat G+2\pi\a'F)} +
\frac{T_5}\k \int_{\Sigma} \,2\pi\a' A \wedge \hat F_5    \,.
\label{fullaction}
\ee
Note that the pullback $\hat F_5$ of the R-R field acts as source for the 
world-volume gauge field $A$
through its coupling in the Wess-Zumino term in \eq{fullaction}. For all the
configurations we shall consider, only the second term in \eq{RR} 
contributes to this coupling.  

We fix the static
gauge by choosing $\sigma=\{\t , \vp^i ; i=1, \ldots , 4\}$ as
coordinates on the D5-brane world-volume and by identifying
\be
X^0 = t \sac \T=\t \sac \Phi^i= \vp^i \,.
\ee
For simplicity, we restrict ourselves to the following type of configurations:
\be
X^1 = X^2 = X^3 = 0 \,,\quad r=r(t,\t) \,,\quad A_t = A_t(t,\t)
\,,\quad A_\t = A_\t(t,\t) \,,\quad A_{\vp^i} = 0 \,.
\label{truncation2}
\ee
Under these conditions, the D5-brane action reduces to
\be
S = \frac{T_5}\k \, V_{(4)} \, \int dt \, d\t \, \sin^4 \t \, \left[
- H \, r^4 \, \sqrt{r^2 + {r'}^2 - H \, {\dot{r}}^2 \,
r^2 - E^2} + 4 \, R^4 \, A_t \right] ,
\label{action}
\ee
where $\dot{r} \equiv \pa_t r \, , \, r' \equiv \pa_\t r \, , \, E
\equiv F_{0\t}$ and $V_{(4)}=8\pi^2/3$ is the volume of a unit
four-sphere. As in \sect{baryon:review}, we have absorbed a $2\pi\a'$ factor in
the gauge field.
From now on, we set $T_5  V_{(4)}/\k \equiv 1$.
  
We pass to the Hamiltonian formalism in order to
derive a bound on the energy.
The
canonical momenta $P$ and $\Pi$ conjugate to the fields $r$ and $A_\t$
in the action \eq{action} are
\bea
P &\equiv& \frac{\pa L}{\pa \dot{r}} \, = \,
\sin^4 \t \, \frac{H^2 \, r^6 \, \dot{r}}
{\sqrt{r^2 + {r'}^2 - H \, {\dot{r}}^2 \,
r^2 - E^2}}~; \nn
\Pi &\equiv& \frac{\pa L}{\pa \dot{A_\t}} \, = \,
\frac{\pa L}{\pa E} \, = \,
\sin^4 \t \, \frac{H \, r^4 \, E}
{\sqrt{r^2 + {r'}^2 - H \, {\dot{r}}^2 \,
r^2 - E^2}} \,.
\label{momenta}
\eea

We define a Hamiltonian density
\be
{\cal{L}} = \dot{A_\t} \, \Pi + \dot{r} \, P - L =
E\, \Pi + \dot{r} \, P - L - A_t \, \Pi'\,,
\ee
where in the last step we have integrated $A_t' \, \Pi$ by
parts. Inverting the relations~(\ref{momenta}), one can rewrite
$\cal{L}$ in terms of the fields and their canonical momenta:
\be
{\cal{L}}= \sqrt{\left( r^2 + {r'}^2 \right)
\left( \frac{P^2}{H \, r^2} + {\Pi}^2 + {\Delta}^2 \right)} +
A_t \, \left( - \Pi' - 4\, R^4 \, \sin^4\t \right) ,
\ee
where
\beq
\Delta \equiv H \, r^4 \, \sin^4\t\,.
\eeq
The first term is the desired energy density
\be
{\cal{H}}= \sqrt{\left( r^2 + {r'}^2 \right)
\left( \frac{P^2}{H \, r^2} + {\Pi}^2 + {\Delta}^2 \right)},
\ee
whereas the second term yields the ``modified'' Gauss law 
\beq\label{GLaw}
\pa_\t \Pi =
-4 \, R^4 \, \sin^4\t~. 
\eeq
Remark that, unlike in \sect{baryon:review}, where we
had the 
Gauss law $\vec{\nabla} \cdot \vec{\Pi}=0$, here we have a
source term in the Gauss law \eq{GLaw}, as a consequence of the 
Wess-Zumino coupling to
$F_5$ in the D5-brane action. This is the difference referred to
in footnote~\ref{fn:Gauss} on page \pageref{fn:Gauss}.

The Gauss law constraint is solved by~\cite{Callan}
\be
\Pi(\nu,\t)= \frac{1}{2} \, R^4 \, \left[ 3 \, (\nu \, \pi - \t) +
3 \, \sin\t \, \cos\t + 2\,\sin^3 \t \, \cos\t \right]\,,
\label{GaussLawSol}\ee
where $\nu$ is an integration parameter. Its meaning will become clear
below (see also~\cite{Callan}).  

Now, we shall look for static solutions $r(\theta)$. To find them, we shall use a
Bogomol'nyi trick. For static configurations,  
the energy density reduces to
\be\label{calH}
{\cal{H}} = \sqrt{\left( r^2 + {r'}^2 \right)
\left( \Pi^2 + \Delta^2 \right)}  .
\ee

The first key observation to prove the existence of a BPS bound on the
energy of the D5-brane is that \eq{calH} can be rewritten as
\be
{\cal{H}} = \sqrt{{\cal{Z}}_{\rm el}^2 + r^2 \, \left(\Delta \,
\cos\t - \Pi\sin\t \right)^2 \, \left(\frac{r'}{r} - f \right)^2}
\label{right}~, 
\ee 
where 
\bea {\cal{Z}}_{\rm el} &\equiv& r \,
\left(\Delta \, \cos\t - \Pi\sin\t \right) \, \left( 1 +
\frac{r'}{r}\, f \right), \label{calZ} \\ f(a,\nu;r,\t)&\equiv&
\frac{\Delta (a,r,\theta) \, \sin\t + \Pi(\nu,\t)\, \cos\t}
{\Delta(a,r,\theta) \,\cos\t - \Pi(\nu,\t)\, \sin\t}\,. 
\eea 
%Generically the function $r(\theta)$ has a limited range in $\theta$ {\bf XXX} 
%from $\theta_i$ to $\theta_f$. 

\begin{figure}
\begin{center}
\epsfig{file=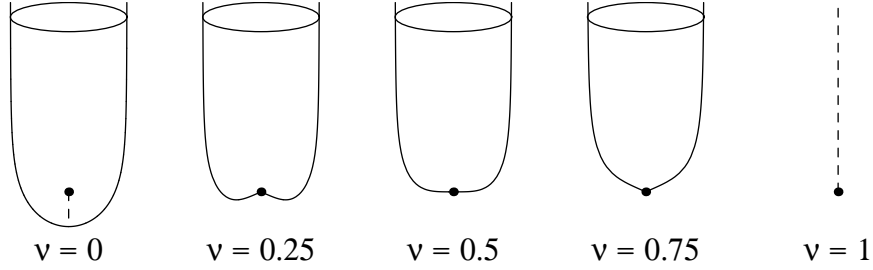}
\end{center}
\caption{Polar plots of $r(\t)$ for ``upper tube'' solutions to
the BPS equation \eq{BPScondition} for $a=0$ ($\t=\pi$ at the top of the plots). 
A tube will be interpreted as $(1-\n)N$ strings.}\label{fig:adstubes}
\end{figure}

\begin{figure}
\begin{center}
\epsfig{file=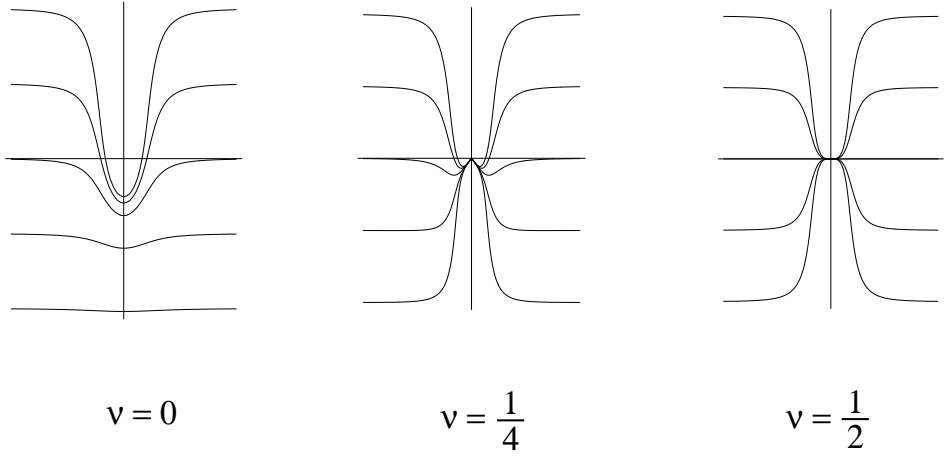,width=12.5cm}
\end{center}
\caption{Solutions to the BPS equation \eq{BPScondition} for $a=1$. 
The vertical axis displays $z\equiv-r\cos \t$. A D5-brane with asymptotic value
$z_\infty>0$ is connected to the $N$ D3-branes at the origin of the plot by 
$(1-\nu)N$ strings. A D5-brane with asymptotic value
$z_\infty<0$ is connected to the D3-branes by 
$-\nu N$ strings (the sign denotes the orientation of the strings). The fact
that the number of strings jumps by $N$ as a D5-brane is dragged upwards from
negative to positive $z_\infty$ is related
to the Hanany-Witten effect \cite{hanany}.}\label{fig:fig5}
\end{figure}

The second key observation is that ${\cal{Z}}_{\rm el}$ is a total derivative:%
\footnote{It is important to note that only the Gauss law 
solution~\eq{GaussLawSol} has to be used
to deduce~\eq{TotDeriv}. The field equation for
$r(\t)$ is not needed.}
\beq\label{TotDeriv}
{\cal{Z}}_{\rm el}=\frac{d}{d\t}\,\left\{\Pi\,r\cos\t+
\left(\frac{a}{5}+\frac{R^4}{r^4}\right)(r\sin\t)^5\right\}~.
\eeq
A solution $r(\t)$ to the equations of motion will be parametrized by an angle
$\t$ that takes values in a certain interval. If $(r,\t)$ are good coordinates
to describe the solution, $r(\t)$ is single-valued (see, for instance,
\fig{fig:adstubes}). In this case, the range of $\t$ is
\beq
\t_i\leq\t\leq\t_f~,
\eeq
where $\t_i$ and $\t_f$ are the ``initial'' and ``final'' angles.
It is also possible that $r(\t)$ is multiple-valued for a
certain solution, so that $\t$ may take values outside the interval
$[\t_i, \t_f]$ (see, for instance, \fig{fig:fig5}). In that case, it may be
better to describe the solution in different coordinates.

\eq{TotDeriv} implies that the integral
\be
Z_{\rm el} \equiv \int_{\t_i}^{\t_f} d\t \, {\cal{Z}}_{\rm el}
\label{Z}
\ee
only depends on the ``boundary values'' $r(\t_i)$ and $r(\t_f)$. This
means that, for fixed values of $r(\t_i)$ and $r(\t_f)$, $Z_{\rm el}$ is
invariant under local variations of the fields.  It is in this sense
that it is a ``topological'' quantity.  The importance of this will be that 
configurations saturating the
bound~\eq{BPSbound} we are about to derive, minimize the energy for fixed 
boundary conditions and therefore automatically solve the equations of
motion. 

From \eq{right}, we find the following BPS bound on the D5-brane energy:
\be\label{BPSbound}
E\equiv\int_{\t_i}^{\t_f} d\t \, {\cal{H}} \geq |Z_{\rm el}|\,,
\ee
through the following two inequalities
\be
E = \int_{\t_i}^{\t_f} d\t \, {\cal{H}} \geq
\int_{\t_i}^{\t_f} d\t \, |{\cal{Z}}_{\rm el}| \geq |Z_{\rm el}|\,.
\label{chain} 
\ee 

The BPS bound~\eq{BPSbound} is saturated if and only if both
inequalities in \eq{chain} are. We shall first discuss the equation implied by
the saturation of the first inequality, and its solutions. Then (on
p.~\pageref{pageSecond}), we shall examine
the condition that the second inequality be saturated, at least for the case
$a=0$. 

Saturation of the first inequality in \eq{chain}%
\footnote{If
$r\left(\Delta \, \cos\t - \Pi\sin\t \right)=0$, the introduction of
$f$ in \eq{right} is not allowed. In this case, saturation of the BPS
bound implies that the numerator of $f$ is zero too, leading to
$\Pi(\nu,\theta) = \Delta(a,r,\theta)=0$, which implies $\theta=0$ or
$\theta=\pi$. This case can thus be neglected.
}  
yields a first order
differential equation on the D5-brane embedding $r(\t)$, namely
\be
\frac{r'}{r} = f\,.
 \label{BPScondition}
\ee
This equation is valid for both the near-horizon case and the
asymptotically flat case (but the function $f$ is different for
the two situations). In the former case, it was derived in~\cite{Imamura}
by imposing the preservation of some fraction of global world-volume
supersymmetry. For the latter case, it was proposed in~\cite{Callan}
as a plausible generalization of the result in~\cite{Imamura} and
shown to imply the equations of motion of the D5-brane. In our approach,
both cases can be dealt with at once, and in both of them the BPS
equation emerges from a bound on the D5-brane energy.

For the near-horizon case ($a=0$), \eq{BPScondition} was solved analytically in
\cite{Callan}. There are two types of solutions.
Define
\be\label{etatheta}
\eta(\t) \equiv \t - \pi \, \nu - \sin\t\,\cos\t~,
\ee
which depends on the parameter $\n$ introduced in \eq{GaussLawSol}. Then,
the solutions of the first type, which will be called ``upper tubes'' for
reasons that will become clear soon, read
\beq\label{solutionr}
r(\t)=c\,\frac{\eta(\t)^{1/3}}{\sin \t}~,
\eeq
where $c$ is an arbitrary scale factor, reflecting the scale invariance of
the $AdS_5 \times S^5$ metric. Some of these solutions are plotted in
\fig{fig:adstubes} \cite{Callan}. 

In~\cite{Callan}, the
restriction was made to $\nu\in [0,1]$.  We make the same restriction
in the main text.  In this
case, there is a unique point $\t_0$ in the interval
$[0,\pi]$ for which $\eta(\t_0)=0$.
The solution \eq{solutionr} only makes sense for $\eta>0$,
i.e., for 
\beq
\t_0<\t<\pi~.
\eeq 

The solutions of the second type are ``lower tubes'', given by
\beq
r(\t)=c\,\frac{(-\eta(\t))^{1/3}}{\sin \t}~,
\eeq 
which makes sense for $0<\t<\t_0$. We shall restrict our attention to upper
tubes.\footnote{If one were to consider solutions with
$\nu$ outside the range $[0,1]$, $\eta$ would have a definite sign on the
interval $[0,\pi]$, so that,
for $a=0$, the solution $r(\t)$ can extend over the whole interval
$[0,\pi]$, developing spikes at both $\t=0$ and $\t=\pi$.}

For the ``full'', asymptotically flat D3-brane background ($a=1$),
\eq{BPScondition} was solved numerically in \cite{Callan} and analytically in
\cite{camino}. Some of these solutions are plotted in \fig{fig:fig5}
\cite{camino}. 
These solutions are labelled by $\n$ (introduced in \eq{GaussLawSol}) and by the
asymptotic value $z_\infty$ of $z\equiv-r\cos \t$ (see \fig{fig:fig5}).

Since the $a=0$ background can be considered as the $r\rightarrow 0$ limit of
the $a=1$ background, one expects the solutions for $a=0$ and $a=1$ to
approximately coincide for small $r$. Indeed, it was observed in \cite{camino}
that this is indeed the case if the parameters $c$ and $z_\infty$ are related 
by
\beq
c=\left(\frac{R^4}{2|z_\infty|}\right)^{1/3}~.
\eeq 

The condition that the second inequality in \eq{chain} be saturated states
that \label{pageSecond}
${\cal{Z}}_{\rm el}$ should not change sign in the integration region
in \eq{Z}.  
%This condition determines the possible values of
%$\t_i$ and $\t_f$ (for BPS solutions), where boundary conditions for
%$r$ should be supplemented.  
When \eq{BPScondition} holds, the sign
of ${\cal{Z}}_{\rm el}$ is determined by the sign of
\be
\Delta \cos\t-\Pi\sin\t = a\, r^4\, \sin^4\t\,\cos\t +
\frac{3}{2}\, R^4 \, \sin\t\, \eta(\t)\,,
\label{signfactor}
\ee
where the function $\eta$ is defined in \eq{etatheta}. 
%\footnote{Recall that
%$0\leq\t_i\leq\t_f\leq\pi$ since we identified $\t$ with $\T$, one of
%the angular coordinates on the~$S^5$.}  
For $a=0$, we precisely recover the condition that $\eta$ should have definite
sign. (We have not done the
analysis for $a=1$.)

%Therefore, in the case $a=0$, the second
%inequality in~\eq{chain} is saturated if
%\be
%\t_i=0\sac \t_f=\t_0
%\label{range0i}
%\ee
%or
%\be
%\t_i=\t_0\sac \t_f=\pi\,.
%\label{range0ii}
%\ee

The remainder of this chapter is devoted to the physical interpretation of 
$Z_{\rm el}$. It follows from \eq{TotDeriv} that
\be
Z_{\rm el} =\left[\Pi\,r\cos\t\right]_{\t_i}^{\t_f}
+\left[\left(\frac{a}{5}+\frac{R^4}{r^4}\right)
(r\sin\t)^5\right]_{\t_i}^{\t_f}~.
\label{Zintegrated}
\ee
We shall evaluate this expression for the solutions of \eq{BPScondition}, 
displayed in \fig{fig:adstubes} and \fig{fig:fig5}.

%We first consider the second term in \eq{Zintegrated}.

In the case $a=0$, the second term in~\eq{Zintegrated} vanishes for
all the solutions of~\eq{BPScondition}, as is suggested by \fig{fig:adstubes}.  
Using \eq{GaussLawSol} and restoring the factor $T_{5} \, V_{(4)}/\k$, the first
term gives the divergent result
\beq
Z_{\rm el}=(1-\n)NTL~,
\eeq
where $T$ is the tension of a fundamental string and $L$ is the (infinite)
coordinate distance between the D3-branes at the origin of \fig{fig:adstubes}
and the ``upper end'' of the D5-brane. However, the interpretation of this
divergence is clear: it is the tension of $(1-\n)N$ infinitely extended
fundamental strings.

For $a=1$, the second term in \eq{Zintegrated} is infinite. However, this infinity is to be expected
for an infinitely extended D-brane. In fact, it precisely equals the energy
of an unexcited D5-brane whose world-space is the $\t=\pi/2$ hyperplane in the 
transverse space to the D3-branes. Indeed, for  such a D5-brane one computes
\be
E\equiv E_0=\int_\Sigma d^5 \sigma\,\sqrt{-\det \hat G}=
\int_0^{r_f} d r \, r^4\,H(r)=
\left[\left(\frac{a}{5}+\frac{R^4}{r^4}\right)\, r^5\right]_
0^{r_f},
\ee 
which agrees with the second term in \eq{Zintegrated} evaluated for the
solutions in \fig{fig:fig5}.%
\footnote{
Incidentally, this unexcited D5-brane corresponds to one of the
$\n=1/2$ configurations drawn in \fig{fig:fig5}. 
It may sound counter-intuitive that an unexcited D5-brane is
connected to the D3-branes by $\pm N/2$ strings. However, this is also what 
has been found in other studies of this system \cite{DFKBG}. 
}
We shall just subtract the infinite second term in \eq{Zintegrated} from the
energy and study the deformation energy $E_{\rm def}$:
\beq
E=E_0+E_{\rm def}~.
\eeq

Thus, for $a=1$ we rewrite the bound \eq{BPSbound} as
\beq
E_{\rm def}\geq Z_9~,
\eeq 
where
\beq
Z_9\equiv\left[\Pi\,r\cos\t\right]_{\t_i}^{\t_f}=(\frac12-\n)NTz_\infty.
\eeq
Here, we have used \eq{GaussLawSol} and the solutions in \fig{fig:fig5}. To
interpret $Z_9$, note that for $z_\infty>0$, $Z_9$ corresponds to the tension 
of $(1/2-\n)N$ strings stretched between the D5-brane and the D3-branes.
At first sight, this might seem to be in contradiction with the fact that there
are really $(1-\n)N$ strings between the D5-brane and the D3-branes. However,
remember that an unexcited D5-brane is connected to the D3-branes via $N/2$
strings. Thus, one can imagine that $N/2$ of the $(1-\n)N$ strings
are needed to bring
the D5-brane in its unexcited state and that only the remaining $(1/2-\n)N$ 
strings exert a force on the D5-brane. By the way, note that this force is not
sufficient to make the infinitely heavy D5-brane move: we are really studying
static configurations.

To motivate the name $Z_9$, note that, from the ten-dimensional spacetime point
of view, we have been studying the triple intersection
\be
\ba{ccccccccccl}
{\rm D3}: &1&2&3&\_&\_&\_&\_&\_&\_ &\quad \mbox{background}    \nn
{\rm D5}: &\_&\_&\_&4&5&6&7&8&\_ &\quad \mbox{world-volume}     \nn
{\rm F1}: &\_&\_&\_&\_&\_&\_&\_&\_&9 &\quad \mbox{BPS solution.}
\ea
\label{triple}
\ee
In our analysis, the D3-branes are the background and the F1 plays the role of
a soliton of the D5-brane world-volume theory.\footnote{Triple intersections
from the point of view of world-volume theories have also been studied
in~\cite{BGT} and \cite{glw}.}
We have chosen the name $Z_9\!$ because of the direction in which the
string stretches. 

In our paper \cite{CGMV}, further evidence is provided in favour of the interpretation of 
$Z_{9}$ (or $Z_{\rm el}$ in the case $a=0$) as a charge
associated to fundamental strings ending on the D5-brane world-volume. The 
additional
evidence consists in the fact that $Z_{9}$ (or $Z_{\rm el}$) has the right 
quantum 
numbers to appear as a central charge in the D5-brane world-volume supersymmetry 
algebra. We refer the reader to \cite{CGMV} for details.
The analysis presented in this section has been generalized to other
configurations of intersecting branes in \cite{camino, GRST}.

%Type 0 string theory
\chapter{Type 0 string theory}\label{type0}
In this chapter, we study some aspects of type 0 string theory \cite{dixharv}. 
Compared to type II strings,
type 0 strings have a tachyon, twice as many R-R fields and
no spacetime fermions 
in their perturbative spectra. Because of the problems related to the
presence of a tachyon, type 0 string theory became popular only one year and a
half ago. Then, it was noticed \cite{polya,9811035} that the tachyon may not be 
a big problem when one is interested in D-brane world-volume field theories, as 
in the AdS/CFT correspondence. In \sect{type0:intro}, we give a brief review of 
this motivation to study type 0 string theory. Type 0 string theory has also been
studied by itself, which has led to some interesting speculations about type~0 
dualities \cite{berggab}.

In this chapter, we study the main building blocks in these 
developments: type 0 strings, D-branes and NS-fivebranes. In particular, we
derive the D-brane Wess-Zumino action \cite{BCR}, determine the massless 
spectrum of NS-fivebranes \cite{0dual} and combine these ingredients to comment 
\cite{0dual} on a type 0 duality \cite{berggab}.

In \sect{type0:strings}, we introduce type 0 strings and compare them to type II
strings \cite{dixharv}. In
\sect{type0:D}, we introduce type 0 D-branes \cite{bergmangaberdiel,9811035}. 
As could be anticipated from the
doubling of the R-R spectrum compared to type II string theory, the number of
different D-branes is also doubled. Following our paper
\cite{BCR}, we derive a Wess-Zumino action for type 0 D-branes. We use an anomaly 
inflow argument very similar to 
the one used for type II D-branes. Again, the different terms in the Wess-Zumino
action can be checked via boundary state computations.

In \sect{type0:NS5}, which is based on our paper \cite{0dual}, we study NS 
fivebranes in
type 0 string theory. We derive their massless spectra and find that they are
non-chiral and purely bosonic for both type 0A and type 0B. Type IIA NS
fivebranes have a chiral, anomalous spectrum. The anomaly is cancelled by
anomaly inflow from the bulk of spacetime. We compute that for both type 0A and
type 0B there is no such inflow from the bulk. This is consistent with the
non-chiral and thus non-anomalous spectra of type 0 NS fivebranes. We propose a
speculative interpretation of the type 0 NS fivebrane spectra in terms of ``type
0 little strings''.

In \sect{type0:dual}, we combine our studies of type 0 D-branes and NS fivebranes
to comment \cite{0dual} on the recently proposed type 0B S-duality 
\cite{berggab}.

One of the themes in our studies of type 0 string theories is that, although they
are not supersymmetric, they resemble the supersymmetric type II theories in many
respects. A striking example is the analysis of \sect{type0:D:inflow} \cite{BCR}.
However, we have to stress that, due to the presence of a tachyonic mode in the
bulk of spacetime, the results presented in this chapter are on a less firm
footing than the analogous ones in type II. 
%%%%%%%%%%%%%%%%%%%%%%%%%%%%%%%%%%%%%%%%%%%%%%%%%%%%%%%%%%%%%%%%%%%%%%%%%%%%%%%%
\section{Introduction}\label{type0:intro}
In \sect{AdSCFT}, we indicated how string theory can be used to study
the strong coupling limit of ${\cal N}=4$ SU($N$) super-Yang-Mills theory
in 3+1 dimensions.
Interesting as it is, studying the strong coupling limit of a supersymmetric
gauge theory is not the final aim of high energy physics. One of the real 
interests is non-supersymmetric gauge theory. There have been several approaches
towards this aim. In this section, we briefly introduce one that is
based on type 0 string theories.

At the end of \sect{superstrings:strings:worldsheet}, we restricted ourselves to
supersymmetric string theories. These theories satisfy consistency conditions
like modular invariance, and are free of tachyons, excitations with negative 
mass squared. However, there also exist modular invariant non-supersymmetric 
string theories.
Two of them are the type 0A and type 0B closed string theories, to be introduced
in \chap{type0}. They differ from the type IIA and type IIB string theories in
that they have different GSO projections. The main differences with the type II
theories are that the type 0 theories have a doubled R-R spectrum, a tachyon and 
no spacetime fermions. 

The presence of a tachyon means that one is doing perturbation theory around an
unstable ``vacuum''. This obviously makes it hard to do computations in these
theories. Nevertheless, the presence of a tachyon does not imply that the
theories are irrelevant to physics: it may well be that the tachyon potential
has a stable minimum in which the tachyon can condense. This, in turn, raises
another problem: determining the tachyon potential requires off-shell
information, which is not available in perturbative string theory.

Despite these problems, one can guess an off-shell extrapolation of on-shell
results and check the internal consistency of the resulting proposal. This
is what Klebanov and Tseytlin did in \cite{9811035}. They argued that in
certain backgrounds with a large five-form R-R flux the tachyonic instability 
may be cured.

Like the R-R spectrum, the number of different D-branes is doubled in type 0 with
respect to type II. For instance, D3-branes can be electrically or magnetically
charged under the unconstrained five-form R-R field strength. The configuration
studied in  \cite{9811035} involves a configuration of coincident electric
D3-branes. In the spirit of Maldacena's conjecture, it is proposed that the
field theory living on the D-branes has a string theory dual. A crucial observation is
that, although there is a closed string tachyon in the bulk of the
ten-dimensional spacetime, the field theory on the branes (describing the
low-energy excitations of the open strings ending on them) is tachyon-free.
Thus,
one may hope that D-branes will be stable objects once the closed string tachyon
has condensed.

The field theory one finds in this way is non-supersymmetric, non-conformal
(in fact, asymptotically free) and tachyon-free \cite{9811035,minahan,9812089}.

One can also study stacks of an equal number of electric and magnetic D3-branes
\cite{KT}, giving rise to field theories that are conformally invariant in the
limit of a large number of branes. We shall not go into these theories here,
though we refer the reader to our paper \cite{BCR}, where we study an orbifold
of these models.  

Type 0 string theories have also been studied independently of the AdS/CFT
correspondence. In particular, there have been some interesting type 0 duality
conjectures, one of which will be examined in \sect{type0:dual}.
It is unclear how to deal with the tachyonic instability of type
0 string theory in a flat background. Our point of view is that arguments based
on anomaly cancellation may be robust under continuous deformations of the 
theory. 
Therefore, one may hope that, even though the background one is expanding around
is unstable, one could be able to extract useful information from such anomaly
arguments. However, we repeat that the whole chapter is highly conjectural.

%%%%%%%%%%%%%%%%%%%%%%%%%%%%%%%%%%%%%%%%%%%%%%%%%%%%%%%%%%%%%%%%%%%%%%%%%%%%%%%%  
\section{Strings: from type II to type 0}\label{type0:strings}
In the Neveu-Schwarz-Ramond formulation, type II string theories are obtained 
by imposing 
%the same (or the opposite in the R sector) 
independent GSO projections on the left and right moving part. 
This amounts to keeping the following (left,right) sectors: 
\beqa\label{IIspectr}
{\rm IIB:}&&(NS+,NS+)\ ,~~(R+,R+)\ ,~~(R+,NS+)\ ,~~(NS+,R+)~;\nonumber\\
{\rm IIA:}&&(NS+,NS+)\ ,~~(R+,R-)\ ,~~(R+,NS+)\ ,~~(NS+,R-)~,\nonumber
\eeqa
where for instance R+ and R- are the Ramond sectors projected with 
$P_{\rm GSO}=(1 + (-)^F)/2$ and $P_{\rm GSO}=(1 - (-)^F)/2$, respectively,
$F$ being the world-sheet fermion number (see
\sect{superstrings:strings:worldsheet}).

There is another, equivalent choice for both theories 
(related to the first choice by a spacetime reflection):
\beqa
{\rm IIB':}&&(NS+,NS+)\ ,~~(R-,R-)\ ,~~(R-,NS+)\ ,~~(NS+,R-)~;\nonumber\\
{\rm IIA':}&&(NS+,NS+)\ ,~~(R-,R+)\ ,~~(R-,NS+)\ ,~~(NS+,R+)~.\nonumber
\eeqa
For the massless R-R sector, the difference between the primed and 
unprimed theories shows up in opposite chiralities of the bi-spinor 
containing the R-R field strengths. This implies a sign difference in the
Hodge duality relations among these field strengths, 
resulting in a selfdual five-form field strength 
in IIB and an antiselfdual one in IIB', for instance.

The type 0 string theories contain instead the following sectors:
\beqa\label{0spectr}
{\rm 0B:}&&(NS+,NS+)\ ,~~(NS-,NS-)\ ,~~(R+,R+)\ ,~~(R-,R-)~; \nonumber\\
{\rm 0A:}&&(NS+,NS+)\ ,~~(NS-,NS-)\ ,~~(R+,R-)\ ,~~(R-,R+)~.\nonumber
\eeqa

Here are some differences between type 0 and type II string theory. First,
type 0 theories do not contain bulk spacetime fermions, which would have to
come from ``mixed''
(R,NS) sectors.\footnote{However, fermions will occur when D-branes are
introduced \cite{bergmangaberdiel}. This will be crucial for our results, see
\sect{type0:D:inflow}.}    
Second, the inclusion of the NS-NS sectors with odd fermion numbers means that 
the closed string tachyon is not projected out. 
Third, the type 0 R-R spectrum is doubled compared to type II: the R-R 
potentials of the primed and unprimed type II theories 
are combined, resulting for instance in an unconstrained five-form
field strength in type 0B. 

We shall summarize the spectra of type 0 versus type II string theories in a
table at the end of \sect{type0:NS5:little}. 

%%%%%%%%%%%%%%%%%%%%%%%%%%%%%%%%%%%%%%%%%%%%%%%%%%%%%%%%%%%%%%%%%%%%%%%%%%%%%%%%
\section{D-branes}\label{type0:D}
D-branes in type 0 theories have been discussed in \cite{bergmangaberdiel} 
and  \cite{9811035}. First, we mainly review some of their results. Then, in 
\sect{type0:D:inflow}, we present a result from our paper \cite{BCR}.

Because of the doubling of the R-R spectrum, in type 0B there are two kinds
of D3-branes, named electric and magnetic in contrast 
to the selfdual D3-brane of type IIB. 

A configuration with equal numbers of coincident electric and magnetic branes 
is reminiscent of type II branes, which have been well studied. It will
turn out that many results can be transferred to type 0 almost without 
effort. For instance, we shall see in \sect{type0:D:inflow} that chiral 
fermions are present on certain 
intersections of electric and magnetic branes, leading to anomaly inflow on
the intersection via anomalous D-brane couplings. 
These, in turn, lead to the creation of a string when certain D-branes cross 
each other. 
 
As discussed in \sect{type0:strings}, the spectrum of type 0 theories contains two 
($p+1$)-form R-R potentials for each even (0A) or odd (0B) $p$. We will
denote these by $C_{p+1}$ and $C'_{p+1}$, referring to the unprimed and 
primed type II theories mentioned above. For our purposes, more convenient 
combinations are
\beq \label{C+-}
(C_{p+1})_\pm=\frac{1}{\sqrt{2}}(C_{p+1}\pm C'_{p+1})~.
\eeq 
For $p=3$, these are the electric ($+$) and magnetic ($-$) potentials 
\cite{9811035}. We will adopt this terminology also for other values of $p$. 

\begin{figure}
\begin{center}
\epsfig{file=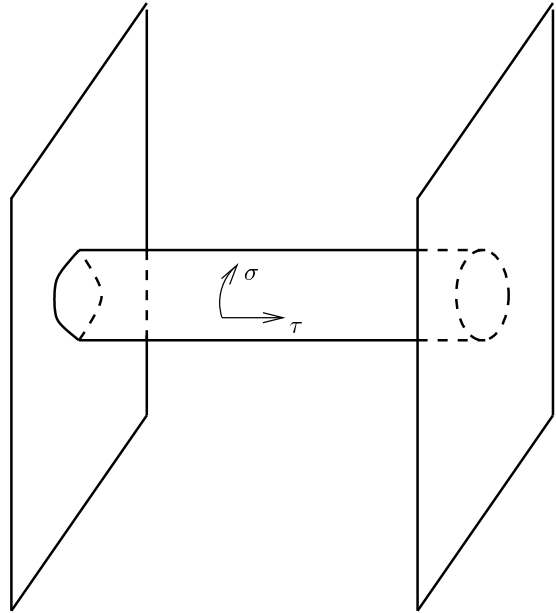}
\end{center}
\caption{The cylinder diagram between two D-branes can be interpreted in two
different ways. In the {\it open string channel}, i.e., if $\s$ is treated as
the world-sheet time coordinate, it represents a one-loop vacuum diagram. In the
{\it closed string channel}, i.e., if $\tau$ is treated as
the world-sheet time coordinate, it is a tree-level diagram: the two
D-branes exchange a closed string. One can derive the charges of D-branes by
computing cylinder diagrams in the open string channel and imposing that the
result should be reproduced in the closed string channel.}\label{fig:cylinder0}
\end{figure}

There are four types of ``elementary'' D-branes for each $p$: an electric and a 
magnetic one, i.e. charged under $(C_{p+1})_\pm$, and the
corresponding antibranes. 
In  \cite{9811035}, the interaction energy of two identical parallel 
($p+1$)-branes was derived by computing the relevant
%\footnote{%
%For instance, for like branes the contribution of the
%open string Ramond sector is absent: consistently with the absence of
%space-time fermions in the bulk, there are no worldvolume fermions.}
cylinder diagram (see \fig{fig:cylinder0})
in the open string channel, analogously to the Polchinski 
computation~\cite{Pol} in type II. 
Isolating, via modular transformation, the 
contributions due to the exchange of long-range fields in the closed string 
channel, it is found on the one hand that the tension of these branes is a 
factor $\sqrt 2$ smaller than for type II  branes. 
On the other hand, the R-R repulsive force between two like branes  
has twice the strength of the graviton-dilaton attraction \cite{9811035}; 
thus, the type 0 branes couple to the corresponding R-R potentials 
$(C_{p+1})_\pm$ with the same charge as the branes in type II couple to the  
potential $C_{p+1}$.

The cylinder diagram between two D-branes in type II can also be considered
as a tree-level diagram in which a closed string propagates
between two ``boundary states''. 
Recall from \sect{anomalous:boundary} that a boundary state is a particular 
BRST invariant closed string state
that describes the emission of  a closed string from a D-brane. It
satisfies conditions
that correspond to the boundary conditions for open strings ending on 
the D-brane. 
In particular, for the fermionic fields $\psi^\mu$ the boundary state 
$\ket{B,\eta}_{\rm NS,R}$, which depends on the sector, R or NS, 
and on an additional sign $\eta=\pm$, satisfies 
\beq
(\psi^\mu-\eta S^\mu_{~\nu}\tilde\psi^\nu)\ket{B,\eta}_{\rm NS,R}=0~,
\eeq
where $S^{\mu}_{~\nu}$ is diagonal, with entries $1$ in the worldvolume
and $-1$ in the transverse directions. 
In type II theories, the GSO projection requires a mixture of the two choices
$\eta=\pm 1$. 
Indeed, starting, for instance,  with $\eta=+1$, one finds that
the type II boundary state 
$\ket B$ $=P_{\rm GSO}\ket{B,+}_{\rm NS}$ 
$\oplus P_{\rm GSO}\ket{B,+}_{\rm R}$
is
\begin{equation}
 \label{am1}
 \ket B = 
 {1\over 2}\left((\ket{B,+}_{\rm NS} - \ket{B,-}_{\rm NS})\oplus 
 (\ket{B,+}_{\rm R} +\ket{B,-}_{\rm R})\right)~.
\end{equation} 

We remark that in the case of type 0 D-branes, 
the various cylinder amplitudes between an electric (or magnetic) D-brane 
and  an electric (or magnetic) D-brane can simply be reproduced using the 
following unprojected (and differently normalized) boundary states, whose sum is 
$\sqrt{2}$ times the type II boundary state: 
\beq \label{B+-}
\ket{B,\pm}=\frac{1}{\sqrt 2}\left(\pm\ket{B,\pm}_{\rm NS}\oplus 
\ket{B,\pm}_{\rm R}\right)~.
\eeq
Here, $\ket{B,+}$ represents an electric brane and $\ket{B,-}$ a magnetic one.

As a cross-check, with these boundary states one can compute the 
one-point function on the disc of a R-R potential (as in \cite{dv9707}).
Denoting by ${}_\pm\bra{C_{p+1}}$ the out-state corresponding to the 
$(C_{p+1})_\pm$ potentials, the one-point function describing its coupling
to a type 0 D$p$-brane will be
${}_\pm\braket{C_{p+1}}{B,\pm}$. 
As one can see from  \eq{C+-}, \eq{am1} and \eq{B+-},
this gives indeed  
the same charge as in type II, where one would compute
$\braket{C_{p+1}}{B}$.  
%%%%%%%%%%%%%%%%%%%%%%%%%%%%%%%%%%%%%%%%%%%%%%%%%%%%%%%%%%%%%%%%%%%%%%%%%%%%%%%%
\subsection{Anomaly inflow and Wess-Zumino action}\label{type0:D:inflow}

The D-branes we have just described show many similarities
to their type II cousins. In this section,
we will push the analogy further to include the whole Wess-Zumino action, 
i.e., all the anomalous D-brane
couplings\footnote{and the non-anomalous ones found in \cite{normal} (see
\sect{anomalous:D:non})}. 

The open strings stretching between two like branes are bosons, just like the 
bulk fields of type 0. However, fermions appear from strings between an 
electric and a magnetic brane \cite{bergmangaberdiel}. Thus, one could
wonder whether there are chiral fermions on the intersection of an electric 
and a magnetic brane. Consider such an orthogonal intersection with no overall 
transverse directions. If the dimension of the intersection is
two or six, a cylinder computation reveals that there are precisely enough 
fermionic degrees of freedom on the intersection 
to form one chiral fermion.

In type II string theory, the analogous computation shows that chiral fermions 
are present on two- or six-dimensional intersections of two orthogonal branes 
with no overall transverse directions. That observation has
far-reaching consequences. As we discussed in \sect{superstrings:inflow:D}, 
the presence of chiral fermions 
leads to gauge and gravitational anomalies on those intersections of D-branes 
\cite{GHM}. In a consistent theory, such anomalies should be cancelled by anomaly inflow \cite{CH}. In the
present case, the anomaly inflow is  provided by the anomalous D-brane couplings in the Wess-Zumino part of
the D-brane action \cite{GHM,CY}. These anomalous couplings have an anomalous variation localized on the
intersections with other branes.     

A careful analysis of all the anomalies \cite{CY} shows that the anomalous 
part of the D$p$-brane action is given, in terms of the formal sum $C$
of the various R-R forms, by
\beq\label{actionwz}
S_{\rm WZ}=\frac{T_p}{\kappa}\int_{p+1} C\wedge e^{2\pi\a '\,F+B}\wedge
\sqrt{\hat{A}(R_T)/\hat{A}(R_N)}~~.
\eeq
Here, $T_p/\kappa$ denotes the D$p$-brane tension, $F$ the gauge field on 
the brane and $B$ the NS-NS two-form.
Further, 
$R_T$ and $R_N$ are the curvatures of the tangent and normal 
bundles of the D-brane world-volume, and $\hat{A}$ denotes the A-roof genus.
We refer to \sect{superstrings:D-branes} for details on the action
\eq{actionwz}.

These anomalous D-brane couplings have various applications. To mention just 
one, using T-duality it has been 
argued \cite{BDG} that they imply the creation of a fundamental string whenever 
certain type II D-branes cross
each other. This string creation process is dual to the Hanany-Witten effect 
\cite{hanany}.

Let us now return to type 0 string theory. As stated above, here chiral fermions live on intersections of 
electric and magnetic type 0 D-branes.  
The associated gauge and gravitational anomalies on such intersections 
match the ones for type II D-branes. To cancel them, the minimal coupling of a
D$p$-brane to a ($p+1$)-form R-R potential should be extended to the following 
Wess-Zumino action \cite{BCR}:   
\beq
\label{WZaction}
S_{\rm WZ}=\frac{T_p}{\kappa}\int_{p+1}(C)_\pm\wedge 
{\rm e}^{2\pi\a '\,F+B}\wedge
\sqrt{\hat{A}(R_T)/\hat{A}(R_N)}~.
\eeq
The $\pm$ in \eq{WZaction} distinguishes between electric and magnetic 
branes. Note that $T_p/\kappa$ denotes the tension
of a type II D$p$-brane, which is $\sqrt 2$ times the type 0 D$p$-brane tension.

\begin{figure}
\begin{center}
\epsfig{file=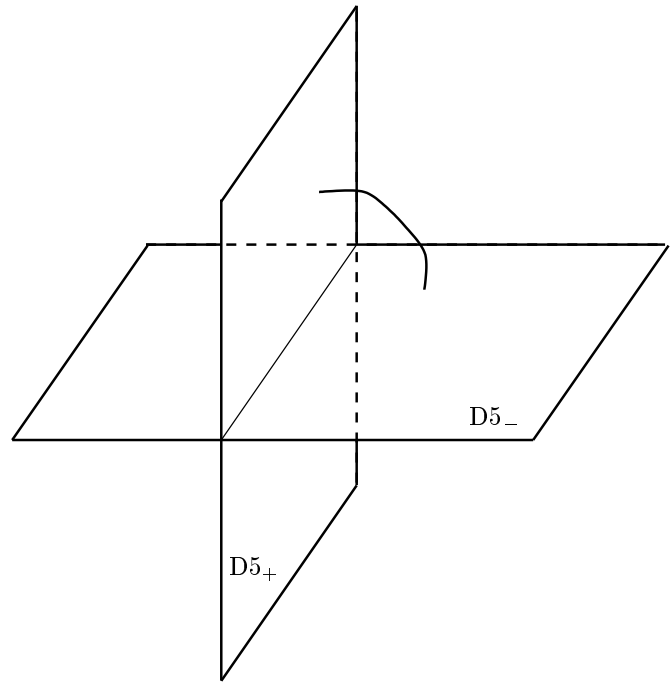}
\end{center}
\caption{Electric and magnetic D5-branes intersecting on a string. The open
strings stretching from one brane to the other give rise to chiral fermions
living on the intersection.}\label{fig:inflow0}
\end{figure}

The argument that the variation of this action\footnote{Again, to be precise, 
as in type II \cite{GHM, CY} one should use an action expressed in 
terms of the R-R field strengths instead of the potentials, 
which is different from \eq{WZaction}, see \sect{superstrings:inflow:D}.}
cancels the anomaly on the intersection is a copy of
the one described above in the type II case, apart from one slight
subtlety. For definiteness, consider the intersection of an electric 
D5-brane (denoted D5$_+$) and a 
magnetic D$5$-brane (D5$_-$) on a string (see \fig{fig:inflow0}).
Varying the electric D$5$-brane action (exhibiting the $(C_2)_+$ potential, 
or rather, its field strength $(\tilde F_3)_+$), one finds that the variation 
is localized on the intersection of the electric  D$5$-brane with
branes charged magnetically under the $(\tilde F_3)_+$ field strength. 
Using \eq{C+-}, the different behaviour under Hodge duality of the R-R field strengths 
of type II and II' shows that these are precisely the branes  
carrying (electric) $(\tilde F_7)_-$ charge, i.e., 
what we called the magnetic D$5$-branes. 
Schematically,
\bea
\delta \int_{{\rm D}5_+}\,I_3^{(0)}\wedge(\tilde F_3)_+
&=&-\int_{{\rm D}5_+}\,I_2^{(1)}\wedge d(\tilde F_3)_+\nonumber\\
&=&-\int_{{\rm D}5_+}\,I_2^{(1)}\wedge d*(\tilde F_7)_-\nonumber\\
&=&-\int_{{\rm D}5_+}\,I_2^{(1)}\wedge \delta_{{\rm D}5_-}~.
\eea
A completely analogous discussion goes through for the
variation of the magnetic D$5$-brane action. 

This anomaly inflow argument fixes (the anomalous part of) the Wess-Zumino 
action, displayed in \eq{WZaction}.
The presence of these terms (and of a similar non-anomalous one \cite{normal}) can be checked by a disc
computation, as in type II \cite{benfred, stefanski, normal}, see
\chap{anomalous}. 
In fact, the computation is practically the same, confirming the form of the
action \eq{WZaction}.

Assuming T-duality to hold between type 0A/B, the arguments of  \cite{BDG} 
indicate the creation of a fundamental
string when certain electric and magnetic branes cross each other. In type II, 
this is linked to the Hanany-Witten
effect \cite{hanany} by a chain of dualities. However, this chain involves 
type IIB S-duality. It is not clear whether S-duality exists in type 0B string
theory (see \sect{type0:dual}).

The conclusion of \sect{type0:D} is that type 0 D-branes are similar to type II
D-branes. In particular, their interaction with R-R potentials is
described by the Wess-Zumino action \eq{WZaction}.

%%%%%%%%%%%%%%%%%%%%%%%%%%%%%%%%%%%%%%%%%%%%%%%%%%%%%%%%%%%%%%%%%%%%%%%%%%%%%%%%
\section{NS-fivebranes}\label{type0:NS5}

This section is based on our paper \cite{0dual}.
First, we derive the spectra of type 0 NS5-branes using T-duality. These spectra
are non-chiral for both type 0A and type 0B. Then, we check that this
non-chirality is consistent with anomaly considerations. Finally, we propose a
speculative interpretation of these spectra in terms of ``type 0 little
strings''.   

\subsection{NS fivebrane spectra}\label{type0:NS5:spectra}

In type II string theories, the massless spectrum on NS5-branes can be derived via an analysis of the 
zero-modes of the
classical supergravity NS5-brane solutions \cite{CHS}. We will find it convenient to do the analysis for the
T-dual objects, since there the geometrical interpretation is manifest. 

Under T-duality, the type IIA/B NS5-branes are mapped to type IIB/A Kaluza-Klein 
(KK) monopoles (see, for instance, \cite{townsend}). A KK monopole
is described by a supergravity solution with six worldvolume directions and a Euclidean Taub-NUT (ETN) 
metric in 
the transverse space \cite{hawking}. 
The massless fields on the KK monopole correspond to the zero-modes of the bulk
supergravity fields on ETN. The normalizable harmonic forms on ETN consist of 
one selfdual two-form \cite{GauntlettLowe}. 
In the NS-NS sector, there are
three zero-modes from broken translation invariance and one scalar from the NS two-form. These four scalars
correspond to the translation zero-modes on the corresponding NS5-branes. The R-R fields give additional
bosonic zero-modes: on the IIB KK monopole, the R-R two-form gives rise to a scalar and the selfdual%
\footnote{
By abuse of language, we term the potential forms (anti-)selfdual, according to the \mbox{(anti-)}selfdual
 nature of their respective field strengths.} 
R-R
four-form potential leads to a selfdual two-form; the IIA KK monopole has a vector coming from the R-R
three-form. In addition, there are fermionic zero-modes from broken supersymmetry: two fermions with opposite
chirality for the IIA KK monopole and of the same chirality for IIB.

All in all, this leads to an ${\cal N}=(2,0)$ tensor multiplet on the IIA NS5 and an
${\cal N}= (1,1)$ vector multiplet on the IIB NS5.

Repeating this analysis for type 0 NS5-branes, it is clear that the fermionic 
zero-modes disappear (because there are
no fermions in the bulk). Nothing changes as far as the NS-NS zero-modes are concerned, 
but from the R-R fields we find extra zero
modes from the doubled R-R spectrum in the bulk. On the 0B NS5, there is an extra vector (compared to the
IIB NS5), whereas the 0A NS5  gets an extra scalar and an anti-selfdual tensor. 
In particular, the spectrum is
non-chiral on both branes. 

Note that, although type 0 NS5-branes are not supersymmetric, their massless
spectra consist of the bosonic part of ${\cal N}=2$ supermultiplets. 
At the end of \sect{type0:NS5:little}, we shall
summarize the massless spectra of type II and type 0 NS5-branes in a table.   
%%%%%%%%%%%%%%%%%%%%%%%%%%%%%%%%%%%%%%%%%%%%%%%%%%%%%%%%%%%%%%%%%%%%%%%%%%%%%%%%%%%%%%%%%%%%%%%%%%%%%%%%%%%%%%
\subsection{Absence of anomalies}\label{type0:NS5:anomalies}

We discussed in \sect{superstrings:inflow:NS5} that at one string loop the type 
IIA tree level supergravity action is supplemented with the
Wess-Zumino type term $\int B \wedge X_8(R)$ coupling the NS-NS two-form $B$ to four gravitons.
$X_8$ is a quartic polynomial in the spacetime curvature two-form.
The original derivation of this one-loop term \cite{vafawitten} provided a non-trivial consistency check
for six-dimensional heterotic-type II duality. Apart from that, this specific piece of the
action is also responsible for gravitational anomaly cancellation on IIA NS
fivebranes \cite{duff}. 

Let us first highlight some aspects of the 
direct calculation in type II, which will enable us to draw conclusions 
about the analogous 
type 0 amplitude almost without effort. First, as the interaction contains the ten-dimensional
$\epsilon$ symbol, only odd spin structures on the torus can contribute,
i.e., either the left-moving or
the right-moving fermions but not both, have to be in the odd spin structure. 
Furthermore, the even spin structures are summed over to achieve modular invariance. 
From a different perspective, this may also be seen as performing the GSO projection on the
closed string states in the loop \cite{SW}. In a Hamiltonian framework, the torus vacuum diagram is 
written as a
trace over the closed string Hilbert space, which decomposes in four different sectors as in
\eq{IIspectr}. Chiral traces over these sectors can be expressed in terms of traces over fermion sectors
with periodic (P) or antiperiodic (A) boundary conditions in the 
($\sigma$,$\tau$) world-sheet 
directions:
\begin{eqnarray*}
{\rm Tr}_{(NS,-)} &=& \frac 1 2 \left((A,A) - (A,P) \right) \ ; \\
{\rm Tr}_{(NS,+)} &=& \frac 1 2 \left((A,A) + (A,P) \right) \ ; \\
{\rm Tr}_{(R,-)} &=& \frac 1 2 \left((P,A) - (P,P) \right) \ ; \\
{\rm Tr}_{(R,+)} &=& \frac 1 2 \left((P,A) + (P,P) \right) \ . 
\end{eqnarray*}
Notice that the odd $(P,P)$ spin structure only occurs in the Ramond sector. 
From the
GSO-projected spectrum in \eq{IIspectr}, only (R,NS) and (R,R) sectors contain (odd,even) pieces,
yielding 
\begin{equation}\label{IIspinsum}
\frac 1 4 \left((P,P),(A,A) + (A,P) + (P,A) \right)\ .
\end{equation}
The right-moving combination is modular invariant, and in fact vanishes as a
consequence of Jacobi's ``abstruse identity''. If one calculates the amplitude for one
B-field and four gravitons, one inserts vertex operators in the appropriate pictures.
Still, the right-moving even structures are summed over as in \eq{IIspinsum}. 
In contrast with the
partition function, however, in this case a non-vanishing result is found 
\cite{vafawitten}. 
It is also argued there that in type IIB the
(odd,even) and (even,odd) contributions cancel out whereas they add in type IIA. 
So only in the latter case is the probed interaction present.

This difference between type IIA and IIB could have been inferred from their respective NS
fivebrane worldvolume spectra, a chiral one in type IIA and a non-chiral one in type IIB. The
chiral spectrum suffers from a gravitational anomaly in six dimensions, which is neatly cancelled
by the standard anomaly inflow \cite{CH} from the bulk through the above derived coupling \cite{duff}. 

Let us repeat the argument for the one-loop correlator in type 0 string
theories. Now, it is only (R,R) sectors that contribute to the (odd,even) piece in the one-loop
partition function, yielding
\begin{equation}\label{0spinsum}
\frac 1 4 \left( (P,P)-(P,P), (P,A) \right) 
\end{equation}
both for 0A and 0B. As in type II, one may now wish to insert vertex operators. However, it is
obvious that the obligatory sum over spin structures as in \eq{0spinsum} 
now yields a vanishing result for
the (odd,even) part by itself. Analogously, the (even,odd) part will vanish by itself. 
Whence the absence of the anomaly-cancelling term in both type 0A and 0B.

Reversing the anomaly inflow argument, in type 0A the  selfdual  two-form must be
supplemented with an anti-selfdual two-form, in order not to give rise to gravitational
anomalies. This nicely agrees with the unconstrained two-form on the NS
fivebrane, derived in \sect{type0:NS5:spectra}. 
From T-duality, an additional vector (compared to the IIB NS5-brane)
is then to be expected on the type 0B NS5-brane, which is also consistent with 
\sect{type0:NS5:spectra}.

%%%%%%%%%%%%%%%%%%%%%%%%%%%%%%%%%%%%%%%%%%%%%%%%%%%%%%%%%%%%%%%%%%%%%%%%%%%%%%%%
\subsection{A little string interpretation}\label{type0:NS5:little}
In type II string theories, the fivebrane worldvolume theories have been conjectured to be
properly described by ``little strings''\cite{little}. 
These non-critical closed string theories have 
hitherto remained mysterious, although some qualitative features may be understood. 
We briefly recall some of these, focusing on the type IIB case. However, we
prefer first to treat the more familiar example of a fundamental string, providing a guiding
reference when dealing with the little strings. 

A fundamental string solution in supergravity breaks half of the 32 bulk supersymmetries. These
broken symmetries give rise to fermion zero-modes on the string, of which 8 are left-moving 
and 8 right-moving. They are identified with the Green-Schwarz space-time fermion fields on
the world-volume. Upon quantization of these fermions, 128 bosons and 128 fermions are found,
together building the on-shell ${\cal N}$ = 2 supergravity multiplet. This 
multiplet is also
found in the RNS formalism used in this thesis, so the picture is consistent. 
In the latter formulation, it is
not hard to see that the NS-NS sector fields together with, say, the R-NS sector build an
${\cal N}$ = 1 supergravity multiplet. Likewise, the NS-R and R-R sector are combined 
into one multiplet of ${\cal N}$ = 1. 

Let us move on now to little strings.
One way to derive properties of little strings in type IIB is to consider a gauge instanton on
a D5 brane. 
Half of the bulk supersymmetries are unbroken by the D5-brane. Sixteen real supercharges are thus found, 
obeying an ${\cal N}$=(1,1) 
superalgebra. Of the sixteen supersymmetries, half are broken by the instanton. On the 
macroscopic string, there are
thus eight fermionic modes, four left-moving and four right-moving ones. Upon quantization of these
zero-modes, one finds 8 bosonic and 8 fermionic states. They correspond to the on-shell degrees
of freedom of one ${\cal N}$=(1,1) vectormultiplet.%
\footnote{Since the little strings of type IIB are non-chiral, we will name them type iia little strings.} 
Upon decomposition with respect to ${\cal N}$ = 1 six-dimensional supersymmetry, 
one
hypermultiplet and one vectormultiplet result. It is conceivable that a similar analysis 
in type IIA, if possible at all, would give a ${\cal N}$=(2,0) tensormultiplet, decomposing into
one hypermultiplet and one tensormultiplet of ${\cal N}$=1 supersymmetry. We remark that the
tensormultiplet has a bosonic content that consists of one scalar and one two-form potential
with self-dual field strength. 

All in all, one may wish to draw the following analogies between type ii
little strings and type II strings. In both cases, there is one universal sector, the
hypermultiplet resp. the (NS-NS + R-NS) sector. The other sector depends on whether or not the
supersymmetry in the
``bulk'' is chiral, where ``bulk'' means the fivebrane worldvolume for the little string 
and ten-dimensional spacetime for
the type II string. For the non-chiral type IIA and little type iia, the remaining
bosonic fields are odd $q$-form potentials, coupling to even $p$ D$p$- resp. d$p$-branes. As to chiral
type IIB/iib, the R-R sector has only even $q$-form potentials, of which one has a self-dual field
strength. In these theories, only odd $p$ D$p$ or d$p$-branes occur. So this analogy, however formal,
is at least remarkable. We shall visualize it in the tables at the end of this
section.

As to the various fivebranes, the type IIA NS fivebrane theory is conjectured to be 
described by chiral iib little strings, whereas iia little strings would do the job for type
IIB fivebranes.

One may then proceed and construct ``little type 0a/b strings'' by
mimicking the procedure for the bulk strings: spacetime fermions are removed and there is a 
doubled R-R spectrum compared to the superstring.
Concretely, this would imply that the massless spectrum of little type 0a strings consists of
the bosonic contents of one hypermultiplet and two vectormultiplets, in ${\cal N}$=1 language. 
This precisely matches with the spectrum on the type 0B NS fivebrane. Little type 0b strings
would then generate the bosonic content of one hypermultiplet, one selfdual tensormultiplet
and one anti-selfdual tensormultiplet. Also here, this seems to be consistent with the spectrum
of the type 0A NS fivebrane. 

\label{tables}
In the following tables, we summarize the spectra of type II versus type 0 string
theories
\begin{center}
\begin{tabular}{|l|c|c|c|c|} \hline
&\multicolumn{2}{c|}{II}&\multicolumn{2}{c|}{0}\\ \cline{2-5}
&A&B&A&B\\ \hline
Universal&\multicolumn{2}{c|}{$g_{\m\n},B_{\m\n},\Phi$}&
  \multicolumn{2}{c|}{$g_{\m\n},\,B_{\m\n},\,\Phi$}\\ 
sector& \multicolumn{2}{c|}{+ fermions}&\multicolumn {2}{c|}{}\\ \hline
Other&$C_1,\,C_3$ & $C_0,\, C_2,\,C_4^+$ & $C_1,\,C_1',\,C_3,\,C_3'$ & 
 $C_0,\,C_0',\, C_2,$\\
sector& + fermions&+ fermions&&$C_2',\,C_4^+,\,C_4^-$\\ 
&(non-chiral)&(chiral)&(non-chiral)&(non-chiral)\\ \hline
\end{tabular}
\end{center}
and of type ii versus type o little string theories:
\begin{center}
\begin{tabular}{|l|c|c|c|c|} \hline
&\multicolumn{2}{c|}{ii}&\multicolumn{2}{c|}{o}\\ \cline{2-5}
&a&b&a&b\\ \hline
Universal&\multicolumn{2}{c|}{4 scalars}&\multicolumn{2}{c|}{4 scalars}\\ 
sector& \multicolumn{2}{c|}{+ fermions}&\multicolumn {2}{c|}{}\\ \hline
Other&vector&$T_2^+$, scalar &2 vectors& $T_2^+,\,T_2^-$,\\
sector&  + fermions&+ fermions&&2 scalars\\
&(non-chiral)&(chiral)&(non-chiral)&(non-chiral)\\ \hline
\end{tabular}
\end{center}  

To summarize \sect{type0:NS5}, we have derived the massless spectra of type 0
NS5-branes. We have shown that they are consistent with anomaly considerations. We
have proposed a speculative interpretation of the spectra in terms of ``type 0
little strings''.    
%%%%%%%%%%%%%%%%%%%%%%%%%%%%%%%%%%%%%%%%%%%%%%%%%%%%%%%%%%%%%%%%%%%%%%%%%%%%%%%%
\section{Type 0B S-duality}\label{type0:dual}
In this section, we combine results from \sect{type0:D:inflow} and
\sect{type0:NS5} to comment \cite{0dual} on the type 0B S-duality conjectured in
\cite{berggab}. 

\paragraph{Strings} In type IIB string theory, the fundamental string is mapped to the D-string under S-duality.
The Green-Schwarz light-cone formulation  of the fundamental type IIB string involves eight spacetime
bosonic fields and sixteen spacetime fermionic ones. All fermions transform in 
the ${\bf 8}_s$ representation
of the transverse SO(8) rotation group. Upon quantization, their zero-modes generate the 256-fold degenerate
groundstate, transforming in the $({\bf 8}_v+{\bf 8}_c)\otimes({\bf 8}_v+
{\bf 8}_c)$ of SO(8). 

The excitations of a D1-brane can be described in string perturbation theory by quantizing the open strings
beginning and ending on it. Doing so, the NS sector gives eight massless bosons and the R sector sixteen
massless fermions.

Note that these massless excitations of both the fundamental string and the D-string can be interpreted as
Goldstone modes for broken translation invariance and broken supersymmetry.

It is believed that the IIB D-string at strong (fundamental) string coupling is the same object as a
fundamental IIB string at weak coupling. The evidence for this conjecture is largely based on supersymmetry.
For one thing, the ground state of the D-string is BPS, so that it can safely be followed to strong coupling
and compared to the ground state of the weakly coupled fundamental string. 

Type 0B string theory is not supersymmetric, which complicates the analysis of its strong coupling limit
enormously. Let us nevertheless try to proceed.

The fundamental type 0B string can be obtained from the fundamental type IIB string by performing a
$(-1)^{F_s}$ orbifold, where $F_s$ is the spacetime fermion number. In the 
untwisted sector of this
orbifold, the massless states in the $({\bf 8}_v\otimes{\bf 8}_v)+({\bf 8}_c\otimes
{\bf 8}_c)$ survive the $(-1)^{F_s}$ projection. The
light modes in the twisted sector, which has to be added for modular invariance, 
are a singlet tachyon and the extra massless R-R states in the 
${\bf 8}_s\otimes{\bf 8}_s$.

Bergman and Gaberdiel conjectured \cite{berggab} that the  fundamental type 0B 
string should be S-dual to a
bound state of an electric and a magnetic D1-brane. This proposal immediately raises two questions. First, when
one quantizes the open strings on a coincident electric-magnetic D-string pair, one 
naively seems to find twice as many
massless excitations as on a fundamental string: sixteen bosons and thirty-two fermions. This would
lead to too many states of the D-string pair compared to the fundamental string. How could the extra
degrees of freedom disappear? Second, assuming that we have found a way to get rid of the superfluous degrees
of freedom, how does the $(-1)^{F_s}$ projection appear on the D-string pair?

As to the first question, in order to count modes one should really compactify 
the system: saying that one has two continua of states rather than one continuum
does not make much sense. 
When a direction is compactified, the proposal in \cite{berggab}
involves a monodromy on one side of the duality. It turns out that this 
monodromy eliminates the redundant would-be massless fermions. 

As to the second question, the $(-1)^{F_s}$ projection might be related to a gauging 
of a $\Zbar_2$ subgroup of the gauge symmetry on the D-string pair. However, it 
looks hard to make this idea more precise.

\paragraph{Fivebranes} Type IIB NS5-branes are S-dual to D5-branes. These objects have the same light
excitations (the reduction of an ${\cal N}=1$, D=10, U(1) vector multiplet to D=6). As in the case of IIB
strings, part of the spectrum is protected by supersymmetry.

Given the story for strings, it is tempting to propose \cite{berggab} 
that the type 0B NS5-brane is S-dual to
an electric-magnetic D5-brane pair. This raises two puzzles. 

First, how do the spectra match? Is the doubled gauge spectrum of type 0B NS5-branes at weak coupling related 
to the two
gauge fields on the D5-brane pair? This is not clear, since to compare the two objects one has to take one of
them to strong coupling, where it is unclear what happens. For one thing, the doubling of the gauge field on the
type 0B NS5-brane (compared to the IIB NS5-brane) is related to the doubled R-R spectrum of 
the bulk of type 0A, as shown in
\sect{type0:NS5:spectra}. The latter doubling is not expected to persist at 
strong coupling \cite{berggab}.
Are there fermionic zero-modes on NS5-branes? We have not found any from a 
zero-mode analysis of the bulk
gravity theory, but this does not exclude that they could appear in a non-perturbative way. This would be
analogous to the (not completely understood) case of the fundamental type 0 string, where it looks impossible to
interpret the fermion zero-modes as zero-modes of bulk fields.

Second, do type 0B NS5-branes carry anomalous gravitational couplings, which are known to be carried by type 0B
D-branes \cite{BCR}? Since these couplings are related to anomalies and thus 
to topology, we expect that their presence or
absence could be invariant under continuously changing the coupling from zero to infinity. If this is the case,
answering this question could provide a test of type 0B S-duality. The rest of this section will be devoted to
analysing this question.

We first remind the reader of the familiar story in type IIB. Consider, for instance, the intersection of two
D5-branes on a string. The open strings from one brane to the other give rise to chiral fermions living on the
intersection. These chiral fermions lead to gauge and gravitational anomalies, which are cancelled by anomaly
inflow from the branes into the intersection \cite{GHM}. For this to happen, 
the D5-branes need, amongst other terms, 
a
\beq
C_2\wedge p_1
\eeq     
term in their Wess-Zumino action, where $C_2$ is the R-R two-form potential and $p_1$ the first Pontrjagin class
of the tangent bundle of the D5-brane worldvolume. We have checked the presence 
of this term by an explicit
string computation \cite{benfred}. Now, type IIB S-duality implies the presence of chiral fermions on the
intersection of two NS5-branes on a string, and thus a
\beq
B\wedge p_1
\eeq
term on the worldvolumes of each NS5-brane ($B$ is the NS-NS two-form). Indeed, the following argument%
\footnote{In fact, this is how the $C_2\wedge p_1$ coupling on D5-branes was first discovered.}
for the
presence of this term was given in \cite{bersadvaf}. Under T-duality in a
transverse direction, the 
NS5-brane is mapped to a Kaluza-Klein monopole in type IIA. This object is described by a Euclidean Taub-NUT
metric in the transverse space.  Since $p_1({\rm ETN})\neq 0$ and the $B\wedge X_8$ term discussed 
in
the previous section contains a piece proportional to $B\wedge (p_1)^2$, reducing this term 
gives rise to the required $B\wedge p_1$ term on the IIA Kaluza-Klein
monopole and thus on the IIB NS5-brane.

What about type 0B? On the one hand, it has been argued \cite{BCR} that chiral 
fermions and thus anomalies are present on the
intersection of electric and magnetic D5-branes (see 
\sect{type0:NS5:anomalies}). 
Hence, type 0 D5-branes have the $p_1$ couplings in their
Wess-Zumino actions (this has also been checked via a string computation \cite{BCR}). We expect that the number
of chiral fermion zero-modes on these intersections cannot change under a continuous increase of the string 
coupling. On the other hand, in \sect{type0:NS5:anomalies} we have argued that 
there is no $B\wedge X_8$ term in type 0A, so we
expect no  $B\wedge p_1$ coupling of type 0B NS5-branes, and thus no chiral fermions on intersections of
NS5-branes. If this argument is correct, it could be a problem for type 0B S-duality.

%Non-BPS branes
\chapter{Non-BPS branes}\label{nonbps}

This chapter deals with non-BPS D-branes. These are objects on which open
strings can end, but that preserve no supersymmetry. In some string theories,
such objects can be stable, but in ten-dimensional type II string theory they are
not. Our main aim will be to derive a Wess-Zumino action for non-BPS D-branes in
type II string theory. This will be done in \sect{nonbps:anomalous}, which is
based on our paper \cite{ournonbps}.

Before we come to that, we review how work by Sen and other people has motivated 
the study of non-BPS branes. Good references include
\cite{sen9803,sen9805,sen9808,wittenK,horava,senreview,lerdarusso}.

%In \sect{nonbps:mot} we review how
%Sen's work motivated the study of non-BPS branes.  In \sect{nonbps:anomalous} we
%propose a Wess-Zumino action for the unstable non-BPS D-branes of type II string
%theory \cite{ournonbps}. On the one hand, we argue that the action we propose
%is consistent with the interpretation of BPS D-branes as topologically
%non-trivial tachyon configurations on a non-BPS D-brane. On the other hand, we
%check the presence of the terms we propose by explicit string computations. 

\section{Motivation}\label{nonbps:mot}

In Sen's construction \cite{sen9805,sen9808,senreview}, non-BPS D-branes are 
constructed 
as kink solutions of the tachyon field (i.e., a field with negative mass squared)
on a coincident D-brane--anti-D-brane pair. In \sect{nonbps:mot:brane}, we give 
an introduction to brane-antibrane systems. In \sect{nonbps:mot:testing}, we 
outline how such systems have been used in testing string dualities (we
concentrate on one example \cite{sen9808}). Finally, in \sect{nonbps:mot:K}, we
indicate how a closely related development leads to K-theory
\cite{atiyah,karoubi} as the appropriate
mathematical framework to discuss D-brane charges \cite{wittenK}.

\subsection{Brane--anti-brane systems}\label{nonbps:mot:brane}

\begin{figure}
\begin{center}
\epsfig{file=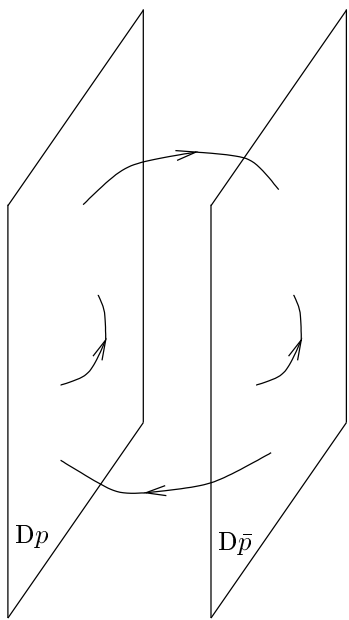}
\end{center}
\caption{The four kinds of open strings in a system with a D$p$-brane and an 
anti-D$p$-brane. The strings stretching from the brane to the antibrane, or vice
versa, have the ``wrong'' GSO-projection, as explained in the main text and in
\fig{fig:cylindernon}.}\label{fig:ppbar}
\end{figure}

Consider in a type II string theory a D$p$-brane coincident with an
anti-D$p$-brane (denoted D$\bar p$).%
\footnote{We shall call such a configuration a D$p$-D$\bar p$ pair, without
explicitly mentioning the fact that the constituents are coincident.}
The dynamics of this system is described by
the four kinds of open strings beginning and ending on D$p$ or D$\bar p$ (see
\fig{fig:ppbar}). 
If it were not for the GSO projection, one would find the following
spectrum for each of the four kinds of open strings. 
In the NS sector, the ground state is tachyonic, meaning that it
corresponds to a fluctuation of the D$p$-D$\bar p$ system with negative mass
squared. At the massless
level, one finds the reduction of a ten-dimensional vector: a vector and $9-p$
scalars. The other states have masses of the order of the string scale. The R
sector contains the reduction of a ten-dimensional Majorana spinor and a tower
of massive fermions. 

For the $pp$-strings (going from D$p$ to itself) and the
$\bar p\bar p$-strings, the GSO-projection is the ordinary one (see the end of
\sect{superstrings:strings:worldsheet}): it eliminates the tachyon,
keeps the massless vector and scalars, and restricts the massless fermions 
to a single
ten-dimensional chirality. The scalars can be interpreted as the Goldstone modes
for broken translation invariance: they describe the transverse fluctuations of
the brane or anti-brane. The sixteen fermionic states are the Goldstone fermions
corresponding to the supersymmetries broken by the ground state of the brane. 
The spectrum of the  $pp$-strings (and of the  $\bar p\bar p$-strings) is
supersymmetric.

\begin{figure}
\begin{center}
\epsfig{file=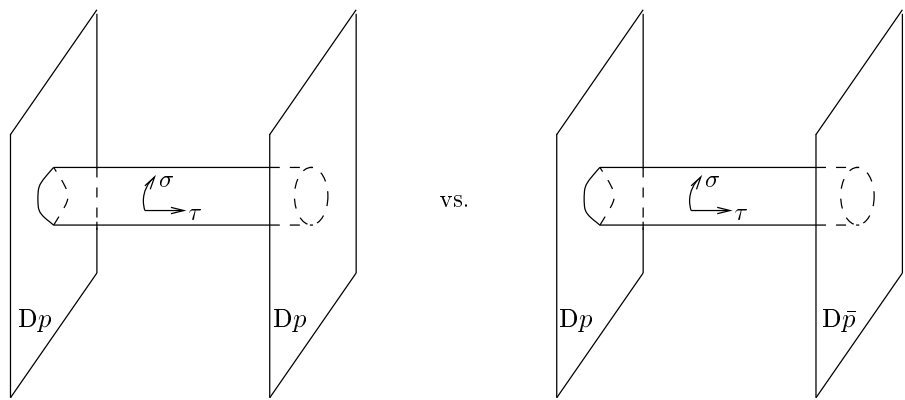}
\end{center}
\caption{In the closed string channel (see \fig{fig:cylinder0}), the difference
between both diagrams is the sign of the R-R contribution. In the open string
channel, this corresponds to the two diagrams having a different GSO-projection.}
\label{fig:cylindernon}
\end{figure}

The other (``wrong'') GSO projection is taken for the $p\bar p$- and 
$\bar pp$-strings: in
the NS sector, the tachyon is kept, and in the R sector, the projection is on the
other chirality than above. The resulting spectrum is clearly non-supersymmetric
(since the bosons and the fermions have different masses). To see why the
``wrong'' GSO-%
projection should be performed, consider the one-loop vacuum diagram of, e.g., the
$p\bar p$-strings (see \fig{fig:cylindernon}). 
The world-sheet is a cylinder, with world-sheet time running
around the loop. By world-sheet duality (i.e., interchanging world-sheet time and
space), this diagram can equivalently be regarded as a closed string tree
diagram: the two branes interact by emitting and absorbing closed string states,
such as the graviton and (fluctuations of) the appropriate R-R potential (see
\fig{fig:cylindernon}). 
In the open string
channel, the amplitude can be written as the sum of four pieces (one of which
vanishes in the cases we will be considering): the NS and R sector 
contributions,
in each case with or without $(-1)^F$ from the GSO projection inserted, where 
$(-1)^F$ is the world-sheet fermion number (see
\sect{superstrings:strings:worldsheet}). It turns out that the contributions
with  $(-1)^F$ inserted correspond, in the closed string channel, to exchange
diagrams of closed strings in the R-R sector. Knowing this, we can argue why, for
instance, the
$p\bar p$ strings have the other GSO projection than the $pp$ strings. The
difference between a brane and an anti-brane is the sign of their charges, in
particular of their R-R charges (see, for instance, \eq{RRanti}). 
Therefore, the sign of the R-R contribution to the
amplitude is different for the two cases, which in the open string channel is
interpreted as a different GSO projection.

The lowest-lying bosonic spectrum of the D$p$-D$\bar p$ system can be summarized
in the following matrix:
\beq
\left(\begin{array}{cc}{\cal A}^{(1)}&T\\ \bar T&{\cal A}^{(2)}
\end{array}\right)~,
\eeq   
where the first (second) row/column denotes the strings beginning/ending on the
brane (anti-brane) and by ${\cal A}$ we mean the full reduction of a ten-dimensional
vector to $p+1$ dimensions (so not only the vector $A$ but also the scalars). The
two tachyons (distinguished by the orientation of the strings they are a mode of)
are denoted $T$ and $\bar T$.

Now specialize to a type IIB D-string--anti-D-string system, and perform the
world-sheet parity projection $\O$ (see the end of
\sect{superstrings:strings:worldsheet})
leading to a type I
D-string--anti-D-string system. Only the $T+\bar T$ combination (to be denoted by
$T$ in the following section) of the two
tachyons survives the projection, so that we end up with one real tachyon. The
gauge fields $A^{(1)}$ and $A^{(2)}$ are projected out by $\O$, so that the U(1)
gauge symmetries on each of the branes are lost. However, a discrete $\Zbar_2$
subgroup of each U(1) survives the projection. This makes it possible to turn on
a $\Zbar_2$ Wilson line on the string and/or anti-string. Further, tadpole
cancellation implies that the vacuum must be filled with 32 D9-branes. This
leads to additional degrees of freedom on the (anti-)D-strings, corresponding to open 
strings between the (anti-)D-strings and the D9-branes.
%%%%%%%%%%%%%%%%%%%%%%%%%%%%%%%%%%%%%%%%%%%%%%%%%%%%%%%%%%%%%%%%%%%
\subsection{Testing string dualities}\label{nonbps:mot:testing}

The brane--anti-brane systems described in \sect{nonbps:mot:brane} have recently
been studied by Sen. His main motivation \cite{sen9803}
was checking string dualities at a non-BPS level. For instance,
the type I superstring has been conjectured to be dual to the SO(32) heterotic 
string
\cite{variousdimensions, PolWit}. This duality is a strong/weak duality and as
such rather difficult to test. One piece of evidence is that the low-energy
limits of both theories are equivalent \cite{variousdimensions}, but this just
follows from supersymmetry. Other evidence is provided by the fact that the
D-string of type I has the same world-sheet structure as the heterotic string
\cite{PolWit}. Non-trivial as this statement is, supersymmetry does play an
important role in it: one has to use the BPS property of the D-string in order
to be able to follow it to strong coupling. We refer to \sect{central:susy} for
more information about BPS states.

What Sen was after was a test of this and other dualities beyond the BPS level.
Such a test was already hinted at in a footnote in \cite{variousdimensions}:
`The SO(32) heterotic string has particles that transform as spinors of SO(32);
these are absent in the elementary string spectrum of Type I and would have to
arise as some sort of solitons if these two theories are equivalent.' In 1998,
Sen found this sort of soliton in the type I string theory \cite{sen9808}. 
Note that the
SO(32) spinor particles in the heterotic string spectrum do not preserve any
supersymmetry: they are non-BPS. Therefore, the identification of these states as
solitons in the type I string theory tests the heterotic/type I duality
at a non-supersymmetric level. The states
that are being compared in both theories are non-BPS but stable: by charge
conservation, the lightest SO(32) spinor state has nothing it can decay into.

It is not difficult to get a rough idea of Sen's construction
\cite{sen9805} (the complete analysis 
is much more sophisticated and will not be dealt with here; we refer to the
original paper \cite{sen9808} and the reviews \cite{senreview} and
\cite{lerdarusso}). First,
compactify one space direction of ten-dimensional Minkowski space-time to a
circle (this circle will eventually be decompactified). Consider
a type I D-string--anti-D-string pair wrapped around this compact dimension. 

\begin{figure}
\begin{center}
\epsfig{file=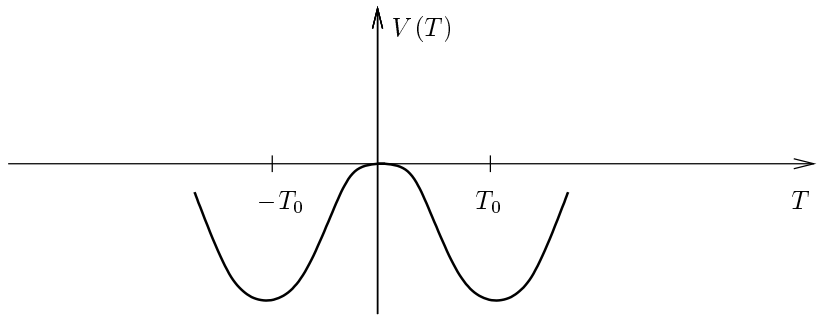}
\end{center}
\caption{Conjectured form of the potential for the tachyon on a
type I D-string--anti-D-string pair.}\label{fig:tachpot}
\end{figure}

The effective potential for the real tachyon field $T$ that survives the $\O$ 
projection 
is not known%
\footnote{However, see \cite{BSZ} for string field theory computations of this
potential.}%
, but it is known to be even and we assume that it has precisely
two minima, at $\pm T_0$ (see \fig{fig:tachpot}). 
The fact that a small fluctuation of the field $T$
around $T=0$ has negative mass squared, from which the field inherits the name
`tachyon', signals that the system is unstable. It has
been argued \cite{sen9805, BSZ}
that the system can decay into the vacuum if the tachyon
field condenses to one of the minima of its potential. 

\begin{figure}
\begin{center}
\epsfig{file=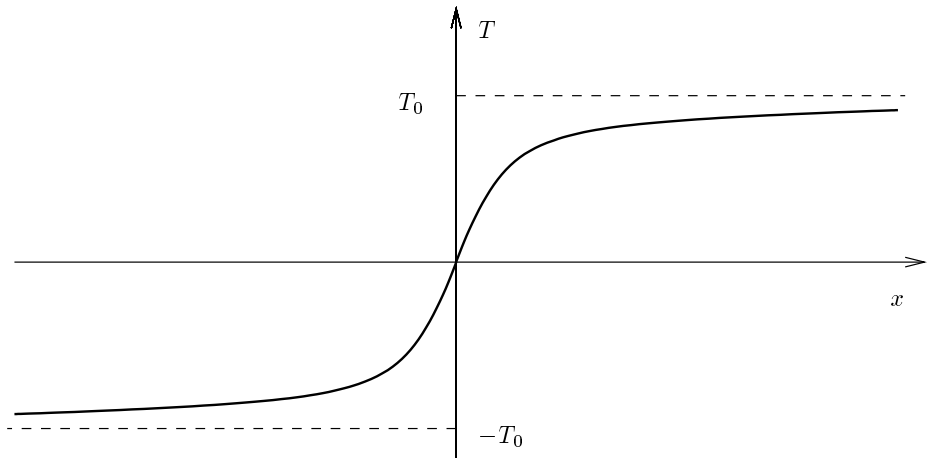,width=12.5cm}
\end{center}
\caption{Kink configuration of the tachyon on a
type I D-string--anti-D-string pair, corresponding to a non-BPS D-particle.}
\label{fig:kink}
\end{figure}

However, more interesting tachyon configurations are possible. One can
turn on a $\Zbar_2$ Wilson line on the anti-string (or the string). Doing this 
implies that the tachyon field $T$ is anti-periodic along the compactification
circle, so the trivial condensation considered above is no longer allowed.
In the decompactification limit, this amounts to considering a topological sector
in which the tachyon approaches $-T_0$ for $x\rightarrow -\infty$ (where $x$ is
the coordinate along the direction that used to be compactified) and $T_0$ for 
$x\rightarrow +\infty$ (see \fig{fig:kink}). 
Such a kink solution of the tachyon equations of motion
would have a core where the tachyon field vanishes, and where its energy is
concentrated (as the configuration resembles the vacuum far away from the core).
Hence, the type I D-string--anti-D-string system with this tachyon 
configuration can be interpreted as a localized
particle, which we call a non-BPS D0-brane. The reason why we call it a D0-brane
is that open strings can end on it. Intuitively, this is because near the core of
the kink the system resembles the original D-string--anti-D-string pair, so open
strings should be allowed to end there.

This D0-brane is a spinor
of SO(32). To see this, we go back to the  D-string--anti-D-string 
pair  wrapped around the compact dimension, with a $\Zbar_2$ Wilson line on
the anti-string (and not on the string). The open strings between the D-string
and the 32 D9-branes in the vacuum give rise to 32 fermion zero modes, whose
quantization leads to the conclusion that the  states of the D-string
transform as a spinor of SO(32). If there were no Wilson line on 
the anti-D-string, the same conclusion would hold for the anti-D-string as well,
so that the total system would not be a spinor. However, putting in the Wilson 
line 
eliminates the zero modes of the anti-D-string, so that the total system, and
consequently the non-BPS D0-brane, is a spinor. 

For future
reference, note that the topological stability of the kink solution is
guaranteed by the fact that the set of minima of the tachyon potential is
disconnected: $\pi_0(\{-T_0,T_0\})=\Zbar_2$.

The use of non-BPS branes in checking string dualities is not limited to
heterotic/type I duality. We refer the reader to \cite{senother,BGnonbps,BGhet} 
for other examples.  
%%%%%%%%%%%%%%%%%%%%%%%%%%%%%%%%%%%%%%%%%%%%%%%%%%%%%%%%%%%%%%%%%%%
\subsection{K-theory}\label{nonbps:mot:K}

This section is based on \cite{sen9808,wittenK,horava}.
Consider a D-string coincident with an anti-D-string in type IIB rather than 
type I. In
this case, there is complex tachyon $T$ living on the common world-sheet of the
D-string and the anti-D-string. We assume the vacuum manifold to be a circle
$S^1$.
Since this circle is connected, the analogue of the kink solution described in
the previous subsection is not stable: a real tachyon remains on the
resulting non-BPS D0-brane world-volume.% 
\footnote{By the way, this tachyon is odd under the 
world-sheet parity operator $\O$, which is consistent with the fact that the
type I D-string--anti-D-string configuration studied in \sect{nonbps:mot:testing}
is stable.} 
In fact, the excitations of the non-BPS D0-brane are described by open strings
{\it without} a GSO-projection.
To summarize: it is possible to construct a type
IIB D0-brane, but this object is unstable, which is signaled by a tachyonic
fluctuation in the world-volume theory.  

This unstable D0-brane can also be described by a boundary state 
\cite{BGnonbps,9903123}. In this approach, the absence of a GSO-projection on the
open strings is
reflected by the absence of a R-R component of the boundary state.
Analogous unstable D$p$-branes can be constructed for all odd (even) values of
$p$ in type IIA (IIB). 

We have just used the fact that the first homotopy 
group of $S^1$ is trivial to
argue that a kink configuration of the tachyon field is not topologically
stable. However, the circle does have a non-trivial homotopy group, namely the
first one:  $\pi_0(S^1)=\Zbar$. For a complex tachyon, this opens up the
possibility that a vortex configuration  is topologically stable. Consider, for 
instance, a D2--anti-D2
system in type IIA. The complex tachyon is charged oppositely under the gauge
fields $A^{(1)}$ and $A^{(2)}$ on the D2 and the anti-D2, respectively.
Therefore, its kinetic term takes the form
\beq
|D_\mu T|^2~,
\eeq
where
\beq
D_\mu T=(\partial_\mu-\ii A^{(1)}_\mu+\ii A^{(2)}_\mu)\,T~.
\eeq
Consider the following static, finite energy vortex configuration, in polar
coordinates $(r,\t)$:
\beq
T\simeq T_0 e^{\ii\t}~,~~A^{(1)}_\t-A^{(2)}_\t\simeq 1~~~~{\rm
as}~r\rightarrow\infty~.
\eeq
This soliton has its energy concentrated near the core (where the tachyon
vanishes) and describes a stable, finite mass particle in type IIA string
theory. 

To see which particle it is, observe that 
\beq
\oint(A^{(1)}-A^{(2)})\cdot dl=2\pi~,
\eeq
so there is one unit of magnetic flux associated with the gauge field 
$(A^{(1)}-A^{(2)})$. This implies that the particle carries (one unit of)
D0-brane charge, as can be seen as follows. The Wess-Zumino actions for the D2
and the anti-D2 contain
\beq\label{RRanti}
\frac{T_2}\k\int_{\rm D2}\hat C_3+\hat C_1\wedge 2\pi\a'dA^{(1)}\,
-\frac{T_2}\k\int_{\overline{\rm D2}}\hat C_3
+\hat C_1\wedge 2\pi\a'dA^{(2)}~,
\eeq
where the minus sign is due to the fact that the second brane is an anti-brane 
rather than a brane. If the two world-volumes coincide, this reduces to
\beq
2\pi\a'\frac{T_2}\k\int_{D2=\overline{\rm D2}}\hat C_1\wedge d(A^{(1)}-A^{(2)})~,
\eeq
so that the core of the vortex solution (where the flux is localized) 
carries charge $4\pi^2\a'T_2/\k=T_0/\k$ (see \eq{Tp})
under $C_1$. As a consequence, the soliton we are studying has every right to 
be identified with the usual, BPS D0-brane of type IIA. Thus, we have constructed
a stable type IIA D0-brane out of a D2--anti-D2 pair.

In the previous discussion, it was crucial that the ``relative'' gauge field
$(A^{(1)}-A^{(2)})$ carried magnetic flux. This flux implies that the gauge 
bundles on
the brane and the anti-brane are inequivalent. Formal differences of vector
bundles are the subject of K-theory, which is the appropriate mathematical setting
for the constructions described in this section \cite{MM,wittenK}. The mathematical
literature on K-theory includes \cite{atiyah,karoubi}. A reference that is
perhaps more easily accessible to physicists is \cite{EGH}.

Slightly generalizing the above constructions, we have the following relations
between the stable and unstable type II D-branes. First, starting from a 
D$(p+2)$--D($\overline{p+2}$) system (i.e., an unstable combination of stable
constituents) we can construct a stable, BPS D$p$-brane by considering a vortex 
configuration of the complex tachyon field. Second, an unstable D($p+1$)-brane
can be obtained as a kink configuration of the real part of the complex tachyon
of the D$(p+2)$--D($\overline{p+2}$) system. The instability of the resulting 
brane is signaled by the presence of a real tachyonic fluctuation. (For
the case $p=-1$ this tachyon can be projected out by going to type I.) Third,
a result we have not discussed yet is that a BPS D$p$-brane can be viewed as
a kink configuration of the left-over real tachyon on the unstable  
D($p+1$)-brane \cite{horava}. The latter mechanism is the subject of
\sect{nonbps:anomalous}.

Iterating the first construction in the previous paragraph (or via a more direct
procedure \cite{wittenK}), one can construct
all stable type IIB D-branes out of sufficiently many space-time-filling
D9-brane--anti-D9-brane pairs \cite{wittenK}. As shown above, in this procedure
the origin of the lower D-brane charges is clear from the appearance of
magnetic fluxes in the Wess-Zumino action. This gives a unified K-theoretic
picture of type IIB D-brane charges.

In type IIA, the situation is complicated by the fact that there are no stable,
space-time filling D-branes. However, a K-theoretic construction of all branes
is possible starting from a sufficient number of unstable type IIA D9-branes
\cite{horava}. Of
course these objects cannot be obtained as tachyon configurations on a higher-%
dimensional system, but they can be introduced by their boundary states as in
\cite{BGnonbps}. These unstable branes are non-BPS and do not carry
Ramond-Ramond charges. This raises the following question: how can 
BPS branes, which do carry R-R charge, be obtained as tachyon configurations on 
non-BPS branes?
This question is addressed in the following section.
%%%%%%%%%%%%%%%%%%%%%%%%%%%%%%%%%%%%%%%%%%%%%%%%%%%%%%%%%%%%%%%%%%%%%%%%%%%%%%%%
\section{Anomalous couplings}\label{nonbps:anomalous}

This section mainly follows our paper \cite{ournonbps}.
In \sect{superstrings:D-branes}, we have seen that type II BPS D-branes couple 
to Ramond-Ramond gauge fields 
through the Wess-Zumino action 
\beq \label{wzbps}
S_{\rm WZ}=\frac{T_p}{\kappa}\int_{p+1}\hat C\wedge \tr\, e^{2\pi\a '\,F+\hat B}
\wedge \sqrt{\hat{A}(R_T)/\hat{A}(R_N)}~.
\eeq
Here, $T_p/\kappa$ denotes the D$p$-brane tension, $C$ a formal sum of R-R 
potentials, $F$ the gauge field on  the brane and $B$ the NS-NS two-form.
The trace is over the Chan-Paton indices.
Further, $R_T$ and $R_N$ are the curvatures of the tangent and normal 
bundles of the D-brane world-volume, and $\hat{A}$ denotes the A-roof genus.

In the setup of \cite{wittenK}, where one starts with an unstable 
configuration of supersymmetric branes 
and anti-branes, the R-R couplings of the BPS D-branes corresponding to 
nontrivial tachyon configurations 
are inherited from the similar couplings of the parent branes. 

In the scenario of \cite{horava}, one starts from non-BPS D-branes, on which
certain vortex configurations of the tachyon field correspond to BPS D-branes.
Before our paper \cite{ournonbps} appeared, it was not clear how the resulting 
objects acquire the desired couplings in \eq{wzbps}. 

In this section, we argue that all type II non-BPS branes couple 
universally to Ramond-Ramond fields as given by \cite{ournonbps}
\beq \label{wznonbps}
S'_{\rm WZ}=a\int_{p+1}\hat C\wedge d\,\left\{\tr\, T \, e^{2\pi\a
'\,F}\right\}\wedge e^{\hat B}\wedge
\sqrt{\hat{A}(R_T)/\hat{A}(R_N)}~,
\eeq
where $a$ is a constant% 
\footnote{This constant will be fixed in the next paragraph by imposing that the
BPS D($p-1$)-brane we find there have the expected R-R charge. Then this action predicts the R-R charges
of the lower BPS D-branes that can be constructed from the non-BPS D$p$-brane.}
and $T$ is the real tachyon field living on the non-BPS brane. Like the gauge
field strength $F$, $T$ transforms in the adjoint representation of the gauge
group. 

One term of this action (the one describing the coupling of a non-BPS D$p$-brane 
to $C_p$) was discussed in \cite{sen9812, horava, senreview}. 
Below, we will show how, in nontrivial configurations of the tachyon field,
these non-BPS ``Wess-Zumino'' couplings induce the appropriate Wess-Zumino
action for the resulting BPS-branes. The cases D$p$ $\rightarrow$ D($p-1$) 
and D$p$ $\rightarrow$ D($p-3$) will be treated in detail. It will turn out, 
for instance,
that the \mbox{R-R} charges of the D8-branes and D6-branes one constructs from 
unstable D9-branes \cite{horava} have the expected ratio.
Moreover, we check the presence of these R-R couplings by performing 
various disc amplitudes with an open string tachyon inserted at the boundary
\cite{ournonbps}.

Before we proceed, we note that the action \eq{wznonbps} is gauge invariant. For
instance, the argument for invariance under R-R gauge transformations can be
copied from \eq{RRgaugeinv}.
%%%%%%%%%%%%%%%%%%%%%%%%%%%%%%%%%%%%%%%%%%%%%%%%%%%%%%%%%%%%%%%%%%%%%%
\paragraph{Relation to Wess-Zumino action} 

In \cite{horava}, 
Horava described how to construct BPS D($p-2k-1$)-branes as 
bound states of (sufficiently many) unstable D$p$-branes. The  D($p-2k-1$)-branes
arise as vortex configuration of a tachyon field, accompanied by 
non-trivial gauge fields.
We show now how the R-R couplings that we propose in \eq{wznonbps} account 
for the R-R couplings \eq{wzbps} that the BPS D($p-2k-1$)-branes must possess.

Consider first a single non-BPS D$p$-brane. There is a  real
tachyon field living on its world-volume. The tachyon potential 
is assumed to be such that the vacuum manifold consists of the two points 
$\{T_0,-T_0\}$.\footnote{The symbol $T_0$ should not be confused with the tension of a D0-brane, which
will never explicitly appear in this section.}
Consider a non-trivial (anti)-kink configuration $T(x)$ depending on a
single coordinate. 
The R-R coupling \eq{wznonbps} on the D$p$-brane reads in this case 
\begin{equation}
\label{m1}
a\int_{p+1}\hat C\wedge dT\wedge {\rm e}^{2\pi\alpha' F +\hat B} \wedge 
\sqrt{\hat{A}(R_T)/\hat{A}(R_N)}~.
\end{equation}
The term 
\beq\label{CpT}
a\int_{p+1}\hat C_p\wedge dT
\eeq
was suggested in  \cite{horava} and shown to be present by a 
disc computation (in an alternative formalism) in \cite{sen9812,senreview}.
It involves the topological density  $\partial_x T(x)$, which is localized at 
the core of the kink and is such that $\int dT(x) = \pm 2 T_0$. 
In the limit of zero kink size, we would have $dT(x) = 2 T_0\delta(x-x_0) dx$, 
and the above action would take the form\footnote{%
Actually, trying to follow the reduction of a D$p$-brane
to a lower-dimensional one, there is a puzzle  
concerning the gravitational part
$\sqrt{\hat{A}(R_T)/\hat{A}(R_N)}$. 
The directions along the parent brane transverse to the 
smaller brane contribute originally to $\hat A(R_T)$. It is not clear to us how they are 
reassigned to the normal bundle in the reduced action.
In fact, this problem seems also to be present for the reduction of brane-antibrane pairs
to lower-dimensional BPS branes as in \cite{wittenK}, where only the standard 
WZ actions \eq{wzbps} are involved.} 
of the usual Wess-Zumino
effective action for a BPS D($p-1$)-brane, localized in the $x$-direction at
$x_0$:
\begin{equation}
\label{m2}
2T_0a\int_{p}\hat C\wedge {\rm e}^{2\pi\alpha' F +\hat B} 
\wedge \sqrt{\hat{A}(R_T)/\hat{A}(R_N)}~.
\end{equation}
In reality, the D($p-1$)-brane will have a certain thickness in the direction of the kink.
\par
Note that the constant $a$ can be fixed in terms of $T_0$ by equating $2T_0a$ with the tension 
$T_{p-1}/\kappa$ of a BPS D($p-1$)-brane. This being done, the remainder of this paragraph provides a
non-trivial check on our couplings in \eq{wznonbps}.  
\par
As a less trivial example, 
let us start from two coincident unstable D$p$-branes. The tachyon field $T$,
transforming in the adjoint of the ${\rm U}(2)$ gauge group, 
can form a non-trivial vortex configuration in co-dimension three.
The tachyon potential is assumed to be such that the matrices $T$ minimizing
the potential have eigenvalues $(T_0,-T_0)$, so that the vacuum manifold is 
${\cal V} = {\rm U}(2)/({\rm U}(1)\times {\rm U}(1))=S^2$. The possible
stable vortex configurations  $T({\bf x})$, depending on $3$ coordinates 
$x^i$ transverse to the ($p-2$)-dimensional core of the vortex, are classified by the
non-trivial embeddings of the ``sphere at infinity'' $S^{2}_\infty$ into the
vacuum manifold, namely by $\pi_{2}({\cal V}) = \Zbar$.

Apart from  the ``center of mass'' ${\rm U}(1)$ 
subgroup, we are in the situation of the Georgi-Glashow model,
where the tachyon field $T({\bf x})= T^a({\bf x}) \sigma_a/2$ 
($\sigma_a$ being the Pauli matrices) 
%$\sigma^a/2$ are the generators in the
%two-dimensional representation of SU(2), normalized as in \eq{tnormalization}) 
transforms in the adjoint representation of  ${\rm SU}(2)$ 
and the vacuum manifold is described by $T^a T^a = 4T_0^2$.
The vortex configuration of winding number one, which is the 't Hooft-Polyakov
monopole, is of the form
\begin{equation}
\label{m3}
T({\bf x}) = f(r) \sigma_a x^a~,
\end{equation}
where $r$ is the radial distance in the three transverse directions, 
and the prefactor $f(r)$ goes to a constant for $r\to 0$ and approaches
$T_0/r$ for $r\to\infty$.

The finite energy requirement implies that $D_i T^a$ 
vanishes sufficiently fast at infinity, from which it follows that a vortex is accompanied by a 
non-trivial gauge field. For the case \eq{m3} above, the non-trivial 
part of the SU(2) gauge field has the form
\begin{equation}
\label{m4}
{\cal A}^a_i({\bf x}) = h(r) \epsilon^a{}_{ij} x^j~,
\end{equation} 
with $h(r)$ approaching a constant for $r\to 0$, while $h(r)\sim 1/r^2$ 
at infinity.

Define
\begin{equation}
\label{m5}
{\cal G}_{ij} = {T^a\over 2T_0} {\cal F}^a_{ij}~.
\end{equation}
Because of the finite energy condition, ${\cal G}$
approaches 't~Hooft's  U(1) field strength%
\footnote{'t Hooft's U(1) field strength ${\cal F}^{\rm U(1)}$ is defined by \cite{hooftpol}
\beq
{\cal F}^{\rm U(1)}_{ij}=\hat T^a {\cal F}^a_{ij}-\e^{abc}\hat T^a D_i\hat T^b D_j\hat T^c~,
\eeq
where $\hat T^a=T^a/(T^bT^b)^{1/2}$. It has the property that it reduces to 
\beq
{\cal F}^{\rm U(1)}_{ij}=\partial_i {\cal A}^3_j - \partial_j {\cal A}^3_i
\eeq
in regions where $\hat T=(0,0,1)$. A vortex configuration carries a magnetic 
charge
\beq
g = \int_{S^{2}_\infty} {\cal F}^{\rm U(1)}=4\pi n~,
\eeq
where $S^{2}_\infty$ denotes the `sphere at infinity' and $n$ is the winding 
number of the vortex configuration. Thus, the mininal magnetic charge is $4\pi$ in
our conventions, which is consistent with \eq{Fq}, \eq{gq} and \eq{convBen}.}
far away from the core of the vortex. Thus, we have 
\beq
\int_{S^{2}_\infty} {\cal G}=4\pi
\eeq
for our vortex configuration \eq{m3} of winding number one. In the approximation
that the ``magnetic charge'' (as measured by the field ${\cal G}$) is concentrated 
in one point at the center of the core, we may write
\beq\label{magncharge}
d{\cal G} =4\pi\delta^3({\bf x}).
\eeq

The  WZ action \eq{wznonbps} for the D$p$-brane can be rewritten as
\begin{equation}
\label{m6}
a\int_{p+1} C\wedge d\,\tr\{ T\, {\rm e}^{2\pi\alpha' {\cal F}}\}
\wedge{\rm e}^{2\pi\alpha' {\hat F} + B} \wedge 
\sqrt{\hat{A}(R_T)/\hat{A}(R_N)}~,
\end{equation}
where we have split the U(2) field-strength into its SU(2) part ${\cal F}$
and its U(1) part $\hat F$.%
\footnote{To avoid confusion about our notation, we note that the hat has 
nothing to do with a pullback.}
Inserting the 't Hooft-Polyakov configuration
for the tachyon and the SU(2) gauge field, we see that \eq{m6} involves 
precisely the field ${\cal G}= T^a {\cal F}^a/2T_0$. Using \eq{magncharge}, 
we get 
\begin{eqnarray}
\label{m7}
& & 2\pi\alpha' a\int_{p+1}  C\wedge d\tr\{ T\,{\cal F}\}\wedge
{\rm e}^{2\pi\alpha' {\hat F} + B} \wedge 
\sqrt{\hat{A}(R_T)/\hat{A}(R_N)}\nonumber\\
&=& 2\pi\alpha' a T_0
\int_{p+1}  C\wedge 4\pi\delta^3({\bf x})\wedge
{\rm e}^{2\pi\alpha' {\hat F} + B} \wedge 
\sqrt{\hat{A}(R_T)/\hat{A}(R_N)}~.
\end{eqnarray}
Thus, we have a distribution of D($p-3$)-brane charge localized at the 
core of the vortex. In particular, in the zero core size approximation we are
working in we recover
the R-R couplings \eq{wznonbps} of a BPS D($p-3$)-brane that supports the 
U(1) gauge field $\hat F$.

We note that \eq{m7}
and the remark after \eq{m2} lead to the expected ratio $4\pi^2\alpha'$ for the R-R charges of 
D($p-3$)-~and D($p-1$)-branes. 

%Since the minimal magnetic charge $g$ is $4\pi$ in our conventions (see
%\eq{Fq}, \eq{gq} and \eq{convBen}),
%F_{\m\n}^a=\partial_\m A_\n^a-\partial_\n A_\m^a+2\e^{abc}A_\m^bA_\n^c;
%\tr T{\cal F}=2T^a {\cal F}^a
%\eq{m7}
%and the remark after \eq{m2} lead to the expected ratio $4\pi^2\alpha'$ for the R-R charges of 
%D($p-3$)-~and D($p-1$)-branes.

The mechanism we have just described generalizes
to the construction of a BPS D$(p-2k-1)$-brane as a vortex solution of the
tachyon field on a
non-BPS D$p$-brane, described in \cite{horava}. In this case, it is 
convenient to start with $2^k$ unstable D$p$-branes. The configuration of 
vorticity one for the tachyon field, which sits in the adjoint of U$(2^k)$, is 
of the form 
\begin{equation}
\label{m8}
T({\bf x}) = f(r) \, \Gamma_i x^i~,
\end{equation}  
where $r$ is the radius in the $2k+1$ transverse dimensions $x^i$, and the
$\Gamma$-matrices in these dimensions are viewed as 
U($2^k$) elements. Eq.~(\ref{m8}) is a direct generalization of the 't Hooft-%
Polyakov case, Eq.~(\ref{m3}). Again, the finite energy requirement should imply
a non-trivial gauge field configuration, leading to a non-zero 
generalized magnetic charge $\int_{S^{2k}_\infty}\tr\{T {\cal F}^k\}$.
In such a background, the WZ action \eq{wznonbps} contains the factor
$d\,{\rm Tr}\{T {\cal F}^k\}= \rho({\bf x}) d^{2k+1}x$; the (generalized) magnetic
charge density $\rho$ is concentrated at the core of the vortex, and in the
zero-size limit reduces to a delta-function in the transverse space. Thus, we
are left with the WZ action for a D$(p-2k-1)$-brane.    
%$4\pi w/e$
%for a vortex of winding number $w$ if $e$ is the gauge coupling).
%%%%%%%%%%%%%%%%%%%%%%%%%%%%%%%%%%%%%%%%%%%%%%%%%%%%%%%%%%%%%%%%%%%%%%%%%%%%%%%%%%%%%%%%%%%%%%%%%%%%%%
\paragraph{String computation}

To compute the disc scattering amplitudes necessary to check \eq{wznonbps}, it is convenient to
conformally map the disc to the upper half plane and use the ``doubling trick'' as described, for
instance, in \cite{hashimoto} (see also the paragraph on `Closed and open
strings' in \sect{superstrings:strings:worldsheet}). 
This trick consists in replacing, e.g., $\bar X^{\mu}
(\bar z)$ by $S^\mu_{~\nu} X^\nu(\bar z)$, where $S^{\mu}_{~\nu}$ is diagonal, with entries $1$ in the 
worldvolume and $-1$ in the transverse directions, and then treating the fields depending on 
$\bar z$  as if $\bar z$ were a holomorphic variable living on the lower half plane. The fermionic 
$\psi^\mu$ fields are treated in
the same way. As to the spin fields in the \mbox{R-R} sector, for BPS D$p$-branes in type IIA 
$\bar S^{\dot\alpha}(\bar z)$ is replaced by 
$(\gamma^0\gamma^1\cdots\gamma^p)^{\dot\alpha}{}_\beta\,S^\beta(\bar z)$ (where the chirality flips
because $p$ is even). For type IIB,  $\bar S^{\alpha}(\bar z)$ is replaced by
$(\gamma^0\gamma^1\cdots\gamma^p)^{\alpha}{}_\beta\,S^\beta(\bar z)$, where now $p$ is odd. 
For the
non-BPS D$p$-branes we are studying here, $p$ is odd in IIA and even in IIB, so that there is a
chirality flip in IIB and not in IIA. The explicit computations below will be done for IIA, but the
story is, of course, completely analogous for IIB.

\begin{figure}
\begin{center}
\epsfig{file=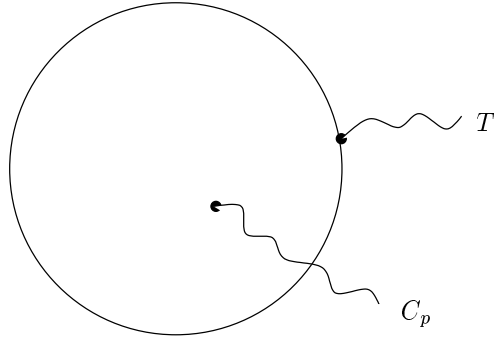}
\end{center}
\caption{Disc amplitude for the two point function of the open string tachyon and
the $p$-form R-R potential in the presence of a single non-BPS D$p$-brane. The aim of the
computation is to establish the term \eq{CpT} in the non-BPS D$p$-brane action.}
\label{fig:CT}
\end{figure}

The first amplitude we are going to compute is the two point function of one open string tachyon and
a R-R potential in the presence of a single non-BPS D$p$-brane in IIA
(see \fig{fig:CT}).\footnote{This has been done before in a
formalism in which non-BPS D-branes are constructed in an alternative way
\cite{sen9812,senreview}.}
This will establish the first term in the 
expansion
of \eq{wznonbps}. We take the R-R vertex operator in the ($-1/2,-1/2$) picture (which
exhibits the R-R field strengths% 
\footnote{
In this chapter, we denote the bispinor containing these field strengths
(see \eq{FAB}) by $H_{\a\dot\b}$ rather than $F_{AB}$. The choice for $H$ rather
than $F$ is to avoid confusion with the field strength $F$ of the gauge field on
the brane. The $A,B$ indices of \eq{FAB} are
32-component spinor indices. For type IIA, the GSO-projections mentioned below 
\eq{FAB} impose that the first index on the bispinor $F_{AB}$ should be chiral
and the second one antichiral. These chiralities are reflected by the
16-component $\a,\dot\b$ spinor indices; the presence or absence of a dot 
reflects the chirality.  
} 
rather than the potentials):
\bea
V_{\rm RR}&=&c(z)\tilde c(\bar z)e^{-\phi(z)/2}e^{-\tilde\phi(\bar z)/2}
H_{\alpha\dot\beta}\, S^\alpha(z)\, \bar S^{\dot\beta}(\bar z)\,
e^{\ii k\cdot X(z,\bar z)}\\
&\rightarrow& c(z)c(\bar z)e^{-\phi(z)/2}e^{-\phi(\bar z)/2}
H_{\alpha\dot\beta}\, S^\alpha(z)\, (\gamma^0\gamma^1\cdots\gamma^p)
^{\dot\beta}{}_{\dot\gamma}\,S^{\dot\gamma}(\bar z)\nonumber\\
&&e^{\ii k\cdot X(z)}\,
e^{\ii k\cdot S\cdot X(\bar z)}~,\nonumber
\eea
where $H_{\alpha\dot\beta}$
%\propto H_{\mu_1\ldots\mu_{p+1}} (C\gamma^{\mu_1\ldots\mu_{p+1}})_{\alpha\dot\beta}$ 
is the bispinor containing the R-R field strengths
%($C$ is the charge conjugation matrix); 
and $k$ the momentum of the R-R
potential. We do not keep track of the overall normalization,
since we are not able to directly determine the constant $a$ in \eq{wznonbps} anyway.\footnote%
{However, we are
interested in the relative normalization of this amplitude with respect to the ones with photons
inserted. The constant $a$ itself was fixed in the previous paragraph.}
The tachyon vertex operator is put in the $-1$ picture:
\beq
V_T=c(y)e^{-\phi(y)}T(k')\,e^{\ii k'\cdot X(y)}~~,
\eeq 
where $T$ and $k'$ are the tachyon polarization and momentum and $y$ is a point on the real axis. 
The three insertion points $z$, $\bar z$ and $y$ have been fixed by 
introducing ghost
fields. The contributions of the ghost, superghost and $X$ sectors combine into $(z-\bar
z)^{5/4}$. The contraction of the two spin fields in the fermionic sector gives
\beq
<S^\alpha(z)\,S^{\dot\gamma}(\bar z)>=(z-\bar z)^{-5/4}\,C^{\alpha\dot\gamma}~~,
\eeq
with $C$ the charge conjugation matrix. 
Thus, the amplitude becomes 
\beq
T\,H_{\alpha\dot\beta}
(\gamma^0\gamma^1\cdots\gamma^p)^{\dot\beta}{}_{\dot\gamma}\,C^{\alpha\dot\gamma}\times K~~,
\eeq
where $K$ is a global factor.
Tracing over the spinor indices, only the part of $H_{\alpha\dot\beta}$
proportional to $H_{\mu_1\ldots\mu_{p+1}} (C\gamma^{\mu_1\ldots\mu_{p+1}})_{\alpha\dot\beta}$
contributes, making the amplitude proportional to
$T\, H_{\mu_1\ldots\mu_{p+1}}\epsilon^{\mu_1\ldots\mu_{p+1}}$. Upon integration by parts,
this confirms the first term of~\eq{wznonbps}.

There is a kinematical subtlety in this computation. String scattering amplitudes can only be
computed for on-shell external particles. It is easy to convince oneself that, since the tachyon
carries only momentum along the brane and the momentum along the brane is conserved, the tachyon and
the R-R potential cannot be both on-shell. As a way out, one could consider branes with Euclidean
signature, for which this kinematical problem does not occur, and then extrapolate the couplings one
finds there to their Minkowski cousins. The situation is analogous to the one we
commented on in footnote~\ref{Euclidean} on page \pageref{Euclidean}.

\begin{figure}
\begin{center}
\epsfig{file=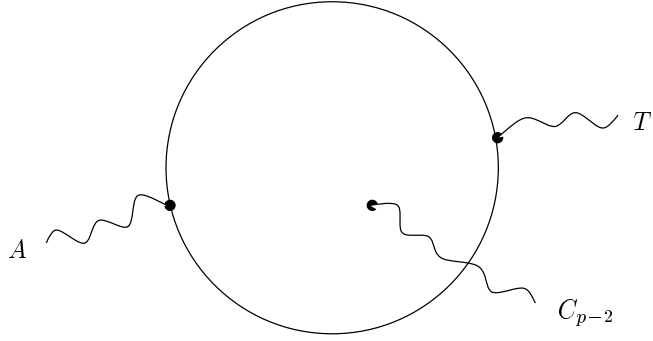}
\end{center}
\caption{Disc amplitude for the three point function of the open string tachyon, 
the ($p-2$)-form R-R potential and the Born-Infeld gauge field in the presence 
of a single non-BPS D$p$-brane.}\label{fig:CTF}
\end{figure}

To check the second term of \eq{wznonbps}, depending linearly on $F$, we add to the previous
amplitude a vertex operator for a gauge field (see \fig{fig:CTF}). 
This vertex operator is in the 0 picture:
\beq\label{photon}
V_A=A_\mu(\ii\dot X^\mu(w)+2\alpha'\,p\cdot\psi\psi^\mu(w))\,e^{\ii p\cdot X(w)}~~,
\eeq
where this time we have kept track of all normalization factors. Here, $A_\mu$ is the polarization of
the gauge field, $p$ is its momentum and $w$ is on the real axis. Only the fermionic part of the 
photon vertex operator
can lead to terms of the type we are looking for (the photon should provide two 
gamma-matrices).
We compute that part of the amplitude to lowest order 
in the photon momentum. This means that we will put $p$ equal to zero in the bosonic sector,
thus keeping only the explicit $p$ dotted with a $\psi$ in \eq{photon}. We follow the previous
computation as closely as possible by fixing again $z$, $\bar z$ and $y$, such that only $w$ needs
to be integrated over. In the limit of small photon momentum, the ghost, superghost and $X$ sector
contributions are unchanged (they multiply to $(z-\bar z)^{5/4}$). The fermionic correlator is
\beq
2\alpha' p_\nu A_\mu <S^\alpha(z)\psi^\nu\psi^\mu(w)S^{\dot\gamma}(\bar z)>=
-\ii\alpha' p_\nu A_\mu(\gamma^{\nu\mu})^{\alpha\dot\gamma} 
(w-z)^{-1}(w-\bar z)^{-1}(z-\bar z)^{-1/4}.
\eeq
The resulting integral can be done by a contour integration:
\beq\label{onedimint}
(z-\bar z)\int_{-\infty}^{+\infty}\,dw\,(w-z)^{-1}(w-\bar z)^{-1}=2\pi\ii~~,
\eeq
leading to
\beq
2\pi\alpha'\, p_\nu\,A_\mu\,T\,H_{\alpha\dot\beta}\,
(\gamma^{\nu\mu})^{\alpha\dot\gamma}(\gamma^0\gamma^1\cdots\gamma^p)^{\dot\beta}{}_{\dot\gamma}\,
\times K
\eeq
for the amplitude. This corresponds indeed to the term in \eq{wznonbps} linear in $F$. Note that the
factor $2\pi\alpha'$ multiplying $F$ in \eq{wznonbps} comes out correctly.

The generalization to multiple (low-momentum) photon insertions is straightforward. The dependence 
on the photon insertion points of the relevant part of the fermionic correlator factorizes, such
that each integration reduces to the one-dimensional integral 
in \eq{onedimint}.
One can also include Chan-Paton factors in the computations, leading to the trace in
\eq{wznonbps}. Finally, one could check the presence of the gravitational terms in \eq{wznonbps}
explicitly. Since all graviton vertex operators can be inserted in the ($0,0$) picture, the various
contractions will be identical to the ones used in
\cite{benfred,stefanski,normal}. 

Note that, from a technical point of view, the only role of the tachyon in the above computations 
is to provide its superghost part, allowing one to insert the R-R vertex operator in the $(-1/2,-1/2)$
picture, instead of the ($-3/2,-1/2$) picture. Thus, the inclusion of the tachyon proves wrong one's first 
impression that non-BPS D-branes cannot couple to the closed string R-R sector because of the 
GSO-projection. Apart from this, the above computations perfectly parallel their counterparts for
BPS D-branes.     

We mention some of the related developments in the recent literature.
First, a Wess-Zumino action for brane--anti-brane pairs has been given in
\cite{KW}.
Second, it has been checked \cite{SK,Groningen} that
the Wess-Zumino action \eq{wznonbps} for non-BPS D-branes in type II
string theory is consistent with T-duality, at least if the curvature terms are
neglected. Third, this Wess-Zumino action has been supplemented with a 
Born-Infeld action \cite{senBI, Garousi, SK, Groningen}. 
Finally, in \cite{HHK}, the non-BPS D-branes in type II
string theory have been associated to non-trivial loops in the configuration
space of type II string theory.

To summarize \chap{nonbps}, we have introduced non-BPS D-branes as kink 
solutions of the tachyon field on a coincident D-brane--anti-D-brane pair. We
have indicated how non-BPS D-branes play a role in checking string dualities and
how D-brane charges are related to K-theory. The main result in this chapter is
that non-BPS D-branes in type II string theory couple to R-R fields via the
Wess-Zumino action \eq{wznonbps} \cite{ournonbps}. This action explains how
vortex solutions of the tachyon field on the non-BPS D-brane, which were
conjectured to correspond to BPS D-branes \cite{horava}, acquire R-R charge.
We have checked the action \eq{wznonbps} by explicit string computations.

\appendix
% samenvatting
\chapter{Samenvatting}

\def\theequation{\arabic{equation}}

\section{Snaartheorie}\label{snaar}
\selectlanguage{dutch}

Een goede fysische theorie verklaart vele experimentele waarnemingen op een
consistente manier vanuit een klein aantal fundamentele principes.

De algemene relativiteitstheorie is hiervan een voorbeeld. Haar principe is dat
gravitatie veroorzaakt wordt door kromming van de ruimte-tijd. Ze verklaart
diverse fenomenen, gaande van de zwaartekracht op de aarde tot
planetenbewegingen en de afbuiging van licht door de zon. Een ander voorbeeld
van een goede fysische theorie is het standaardmodel van de
elementaire-deeltjesfysica. Het fundamentele principe is dat ijksymmetrie\"en
aan de basis liggen van de elektromagnetische en kernkrachten. Vele
voorspellingen van het standaardmodel zijn bevestigd in experimenten in
deeltjesversnellers, soms met een verbazingwekkende precisie.

We hebben gezien dat de algemene  relativiteitstheorie een goede theorie is
voor de gravitationele interactie, terwijl het standaardmodel een goede
beschrijving geeft van de sterke, zwakke en elektromagnetische interacties.
Krijgen we een goede theorie van de vier fundamentele wisselwerkingen door de
algemene relativiteitstheorie en het standaardmodel te combineren? Het probleem
is dat de algemene relativiteitstheorie een klassieke theorie is, terwijl het
standaardmodel een kwantumtheorie is. Samen geven ze geen {\it consistente}
beschrijving van de natuur. Men zou kunnen proberen de algemene
relativiteitstheorie te kwantiseren, maar alle rechtstreekse pogingen om dat te
doen zijn mislukt.

Nochtans bestaat er een consistente kwantumtheorie die gravitatie beschrijft. Ze
was `per ongeluk' ontdekt op het einde van de jaren zestig en in het begin van
de jaren zeventig, toen mensen een theorie voor de sterke interacties aan het
zoeken waren. Pogingen om bepaalde experimentele waarnemingen over sterk
interagerende deeltjes te begrijpen leidden tot het idee dat de sterke
interacties door een theorie van snaren zouden kunnen worden beschreven. De
populariteit van dit idee daalde toen bleek dat een andere theorie,
kwantumchromodynamica (`quantum chromodynamics', QCD), de sterke interacties goed
beschreef. Dit betekende echter niet het einde van snaartheorie: men had ontdekt
dat snaartheorie onder andere een massaloos spin-2 deeltje beschrijft. Zo'n
deeltje verwachtte men als drager van de gravitationele interactie in een
kwantumtheorie van gravitatie!

Ge\"\i nspireerd door het feit dat ze gravitatie bevat, opperde men het idee
dat snaartheorie% 
\footnote{We verwijzen naar Hoofdstuk~\ref{superstrings} voor een inleiding tot
snaartheorie.}
alle interacties zou kunnen verenigen. Inderdaad
ontdekte men dat snaartheorie ook de ijksymmetrie\"en van het
standaardmodel bevat. Bijgevolg verenigt snaartheorie gravitatie en de
ijksymmetrie\"en van het standaardmodel op consistente wijze! Consistentie
alleen maakt snaartheorie echter nog geen goede fysische theorie. Daarvoor is
het ook nodig dat de theorie vele experimentele waarnemingen verklaart uitgaande
van een klein aantal fundamentele principes.

Wat de experimentele waarnemingen betreft, de eerste vereiste is dat
snaartheorie de voorspellingen van de algemene relativiteitstheorie en het
standaardmodel reproduceert, tenminste wanneer de omstandigheden zodanig zijn
dat men weet dat deze theorie\"en een goede beschrijving van de natuur geven.
Snaarfenomenologie is de tak van snaartheorie die zich bezighoudt met het zoeken
naar zo'n snaarmodellen. De huidige situatie is dat het standaardmodel (nog)
niet exact gereproduceerd is, maar dat er wel `semi-realistische' modellen
bestaan die aardig in de buurt komen.

Wat de fundamentele principes van snaartheorie betreft, een aantal hiervan
moet vermoedelijk nog ontdekt worden. Nochtans zijn er verschillende
aanwijzingen over de structuur van de theorie.

Ten eerste is het duidelijk dat de structuur van perturbatieve snaartheorie veel
beperkender is dan die van kwantumveldentheorie: terwijl men in
kwantumveldentheorie veel verschillende interacties kan bedenken, laat
perturbatieve snaartheorie enkel het splitsen en samenkomen van snaren toe. Een
volledige, niet-perturbatieve beschrijving van snaartheorie ontbreekt echter
nog. Wat we weten is dat snaartheorie niet enkel een theorie van snaren is: ze
bevat ook uitgebreide (d.w.z. meerdimensionale) objecten, `branen' genoemd (de
naam is een veralgemening van `membranen'). Het is niet duidelijk of de
volledige theorie kan worden geformuleerd als een theorie van snaren.

Ten tweede speelt supersymmetrie,%
\footnote{Supersymmetrie wordt ge\"\i ntroduceerd in Sectie~\ref{central:susy}.}
een symmetrie tussen bosonen en fermionen,
een belangrijke rol in snaartheorie. Supersymmetrie zou een van de organiserende
principes van snaartheorie kunnen zijn. Dat neemt niet weg dat er recent
pogingen geweest zijn om betekenis te geven aan niet-supersymmetrische
snaartheorie\"en.

Ten slotte zijn dualiteiten%
\footnote{Zie Sectie \ref{central:duality} voor een inleiding tot dualiteiten.}
prominent aanwezig in de huidige ontwikkelingen in snaartheorie. Ze bieden de
mogelijkheid aspecten van niet-perturbatieve snaartheorie te exploreren.

Tot nog toe hebben we snaartheorie besproken als mogelijke Theorie Van Alles.
Snaartheorie kan echter ook op een andere manier een omwenteling teweegbrengen
in ons begrip van de natuur: we kunnen snaartheorie gebruiken als hulpmiddel om
iets te leren over kwantumveldentheorie, in het bijzonder over de ijktheorie\"en
van het standaardmodel. Er is bijvoorbeeld hoop dat snaartheorie van nut kan
zijn om het berucht sterke-koppelingsprobleem van QCD op te lossen. Hierop gaan
we nu iets dieper in.

Kwantumchromodynamica (QCD) is een theorie van de sterke interacties. Ze behoort
tot het standaardmodel. Bij hoge energie is de theorie zwak gekoppeld, zodat
storingstheorie ons in staat stelt voorspellingen te doen voor
verstrooiingsamplituden bij hoge energie. Die voorspellingen komen overeen met
de experimentele resultaten. Bij lage energie is de theorie echter sterk
gekoppeld, zodat storingstheorie nutteloos is. Het is niet bekend hoe
berekeningen
kunnen worden gedaan in dit regime. Nochtans zijn zulke berekeningen nodig om
fenomenen als de opsluiting van quarks te begrijpen.

Onlangs is het vermoeden gerezen dat de sterke-koppelingslimiet van bepaalde
veldentheorie\"en kan worden beschreven door een zwak gekoppelde duale
snaartheorie \cite{maldacena}. Dit verband heet de AdS/CFT correspondentie.%
\footnote{In Sectie~\ref{AdSCFT} geven we een korte maar ietwat technischer 
inleiding tot de AdS/CFT correspondentie.}   
In zijn huidige vorm is het vermoeden niet echt van toepassing op QCD, maar het
idee dat gerelateerde theorie\"en door specifieke snaartheorie\"en kunnen worden
beschreven vertoont intrigerende gelijkenissen met de oorspronkelijke
sterke-interactie motivatie voor snaartheorie!  
%%%%%%%%%%%%%%%%%%%%%%%%%%%%%%%%%%%%%%%%%%%%%%%%%%%%%%%%%%%%%%%%%%%%%%%%%%%%%%%%
\selectlanguage{english}
\section{D-branen}\label{D-braan}
\selectlanguage{dutch}
\subsection{Algemeenheden}
In de meeste recente ontwikkelingen in snaartheorie spelen {\it D-branen}%
\footnote{We verwijzen naar Sectie~\ref{superstrings:D-branes} voor meer details
over D-branen.}
een cruciale rol. D-branen maakten hun intrede in snaartheorie als hypervlakken
waarop open strings kunnen eindigen. Het is echter duidelijk geworden dat ze
dynamische objecten zijn: ze kunnen bewegen en hun vorm veranderen. D-branen
vormen een bepaalde klasse van de uitgebreide objecten die we vermeldden in
Sectie~\ref{snaar}, toen we zeiden dat snaartheorie niet enkel een theorie van
snaren is. Wat D-branen onderscheidt van andere uitgebreide objecten is dat ze,
hoewel ze zelf geen perturbatieve snaartoestanden zijn, een beschrijving hebben
in perturbatieve snaartheorie: de dynamica van een D-braan wordt beschreven 
door de open snaren die erop eindigen. Dankzij deze eigenschap zijn D-branen
veel makkelijker te bestuderen dan andere uitgebreide objecten in snaartheorie.
D-branen zijn om de volgende redenen van belang.

Ten eerste duiken ze op in vele modellen van snaarfenomenologie. De hoofdreden
is dat er ijkvelden op hun wereld-volume leven, in het bijzonder niet-abelse
ijkvelden als verscheidene D-branen samenvallen. Deze ijkvelden kunnen de rol
spelen van de ijkvelden van het standaardmodel. Van bijzonder belang voor
fenomenologische doeleinden is de observatie dat D-branen kunnen zorgen voor het
breken van supersymmetrie. Bijvoorbeeld zijn systemen met D-branen en 
anti-D-branen%
\footnote{Zie Sectie~\ref{nonbps:mot:brane} voor enkele elementaire
beschouwingen over zulke braan--anti-braan systemen.}
onlangs gebruikt om semi-realistische modellen te construeren \cite{ibanez}.%
\footnote{We verwijzen naar \cite{fredd} en verwijzingen daarin voor een
gerelateerde toepassing van D-branen: ze kunnen aangewend worden om uit
snaartheorie niet-perturbatieve resultaten af te leiden 
over supersymmetrische veldentheorie\"en die eventueel zwak gekoppeld zijn aan 
gravitatie.}

Ten tweede zijn ze nadrukkelijk aanwezig in snaardualiteiten. Vaak%
\footnote{Zie Secties~\ref{superstrings:dualities:S}, \ref{type0:dual} and
\ref{nonbps:mot:testing} voor voorbeelden.}
corresponderen D-branen in de ene perturbatieve beschrijving van snaartheorie
met fundamentele snaren in een duale beschrijving. In die zin zijn D-branen even
fundamenteel als snaren.

Ten derde zijn D-branen gerelateerd met (meerdimensionale) zwarte gaten. In
feite leveren D-branen en sommige zwarte gaten complementaire beschrijvingen 
van hetzelfde object. Welke beschrijving aangewezen is, hangt af van de
koppelingsconstante van de snaartheorie. Deze correspondentie kan aangewend
worden om het aantal toestanden te tellen van supersymmetrische (BPS)%
\footnote{In Sectie~\ref{central:susy} leggen we uit wat `BPS' betekent.}
zwarte gaten. Voor zulke zwarte gaten verandert het aantal toestanden niet
wanneer we de koppelingsconstante wijzigen, zodat hun toestanden evengoed
kunnen worden geteld in het regime waar de beschrijving als D-braan geldig is.
Deze procedure heeft geleid tot een microscopische verklaring voor de entropie
van zwarte gaten, tenminste voor bepaalde supersymmetrische zwarte gaten.

Ten vierde liggen D-branen aan de basis van de AdS/CFT correspondentie, zoals
blijkt uit Secties~\ref{AdSCFT} and \ref{type0:intro}.
%%%%%%%%%%%%%%%%%%%%%%%%%%%%%%%%%%%%%%%%%%%%%%%%%%%%%%%%%%%%%%%%%%%%%%%%%%%%%%%%
\subsection{Aspecten bestudeerd in deze thesis}\label{aspecten}
Nu we het belang van D-branen in snaartheorie aangetoond hebben, gaan we in op
de concrete aspecten die in deze thesis bestudeerd worden.

Het grootste deel van de thesis is gewijd aan D-branen. In het bijzonder zijn we
ge\"\i nteresseerd in de D-braan effectieve actie. Dat is een lage-energie
effectieve actie voor de massaloze excitaties van het D-braan, d.w.z. voor de
massaloze modes van de open snaren die de dynamica van het braan beschrijven.
Men kan zich voorstellen dat men alle massieve modes van de open snaren
`uitge\"\i ntegreerd' heeft. Het hoeft geen betoog dat zo'n effectieve,
`macroscopische' beschrijving handiger is voor berekeningen dan de volledige,
`microscopische' beschrijving in termen van open snaren.

Onze werkwijze om termen van de D-braan effectieve actie te berekenen
bestaat erin verstrooiingsamplituden van snaren in de aanwezigheid van een
D-braan te berekenen in de `microscopische' snaartheorie. Dan leggen we op dat,
tenminste bij lage energie, die amplituden gereproduceerd worden in de
effectieve beschrijving. Het formalisme dat we gebruiken om de meeste van die
amplituden te berekenen in snaartheorie is het {\it randtoestand}
(`boundary state') formalisme. In dat formalisme wordt een D-braan voorgesteld als
een bron van gesloten-snaartoestanden.

De termen in de D-braan actie waarin we vooral ge\"\i nteresseerd zijn, worden
{\it anomale koppelingen} genoemd. Zij maken deel uit van de D-braan Wess-Zumino
actie. Via het `anomalie instroom mechanisme' zorgen ze ervoor dat de ijk- en
gravitationele anomalie\"en in bepaalde configuraties van D-branen gecompenseerd
worden. Op die manier zijn ze belangrijk voor de beroemde consistentie van
snaartheorie. Onze belangrijkste bijdrage in dat kader bestaat in het expliciet
nagaan van de aanwezigheid van de anomale koppelingen in de D-braan actie. Als
een bonus vinden we extra termen in de D-braan actie. Deze extra termen zijn
nieuw.

We breiden het anomalie instroom argument en de verstrooiingsberekeningen uit
naar D-branen in de niet-supersymmetrische type 0 snaartheorie\"en. We leiden
een Wess-Zumino actie af die opmerkelijk sterk lijkt op die voor D-branen in de
supersymmetrische type II snaartheorie\"en. Zulke gelijkenissen tussen
supersymmetrische en niet-supersymmetrische situaties in snaartheorie vormen een
belangrijk thema in deze thesis.
De type 0 D-branen zijn gebruikt om de AdS/CFT correspondentie uit te breiden
naar niet-supersymmetrische veldentheorie\"en.  
We zetten onze studie van type 0 snaartheorie\"en verder door het spectrum af te
leiden van NS-vijfbranen, andere uitgebreide objecten in die theorie\"en. We
combineren onze kennis van D-branen en NS-vijfbranen om een recent
dualiteitsvoorstel te becommentari\"eren.

Om een niet-supersymmetrische D-braan veldentheorie te bekomen, hoeft men niet
van een niet-supersymmetrische snaartheorie te vertrekken. Men kan ook
niet-supersymmetrische (niet-BPS) configuraties van D-branen beschouwen in een
supersymmetrische snaartheorie. Zulke configuraties kunnen bijvoorbeeld bestaan
uit BPS D-branen en hun anti-branen. Men kan ook niet-BPS D-branen beschouwen.
Dat zijn onstabiele objecten in type II snaartheorie. Wij bestuderen die
niet-BPS D-branen en leiden er een Wess-Zumino actie voor af. Die actie legt
gedeeltelijk uit dat BPS D-branen overeenkomen met monopool-achtige
configuraties in de wereld-volume theorie van niet-BPS D-branen.

Ten slotte gebruiken we de effectieve actie van BPS D-branen in type II
snaartheorie om D-branen in de achtergrond van andere D-branen te bestuderen.
Zulke configuraties hebben vele toepassingen. Wij concentreren ons op een
toepassing in de AdS/CFT correspondentie, namelijk het analogon in snaartheorie
van een extern baryon in veldentheorie.

We hebben ervoor gekozen onze aandacht in deze thesis bijna uitsluitend toe te 
spitsen op gesloten-snaartheorie\"en in tien vlakke, niet-compacte dimensies.
Dat houdt in dat we een aantal gerelateerde onderwerpen waarover we 
gepubliceerd hebben, niet behandelen. In het bijzonder besteden we geen aandacht
aan speciale K\"ahlermeetkunde \cite{special}, anomale koppelingen van
ori\"entifold-vlakken \cite{benfred,normal}, de studie van D-branen in D-braan
achtergronden via superalgebra's \cite{CGMV}, niet-supersymmetrische
ijktheorie\"en afkomstig van type 0 orbifolds \cite{BCR} en het randtoestand
formalisme voor orbifolds \cite{BCR}. We verwijzen ge\"\i nteresseerde lezers
naar onze artikels over die onderwerpen.

Tot besluit, de voornaamste methoden die we gebruiken om D-branen te bestuderen,
zijn het berekenen van verstrooiingsamplituden en het gebruik van 
consistentie-argumenten. We berekenen de meeste van die verstrooiingsamplituden
in het randtoestand formalisme. De consistentie-argumenten houden verband
met de compensatie van ijk- en gravitationele anomalie\"en via het mechanisme
van anomalie instroom. We zijn ge\"\i nteresseerd in niet-supersymmetrische
D-braan configuraties. In het bijzonder proberen we delen van de mooie structuur
van supersymmetrische snaren en D-branen uit te breiden naar
niet-supersymmetrische situaties.
%%%%%%%%%%%%%%%%%%%%%%%%%%%%%%%%%%%%%%%%%%%%%%%%%%%%%%%%%%%%%%%%%%%%%%%%%%%%%%%%
\selectlanguage{english}
\section{Overzicht en samenvatting van de resultaten}
\selectlanguage{dutch}
In Sectie \ref{aspecten} hebben we een kort overzicht gegeven van de onderwerpen
die in deze thesis aan bod komen. Daarbij hebben we onze motivatie en methoden
benadrukt. In deze sectie schetsen we de structuur van de thesis en geven we een
iets meer gedetailleerde samenvatting van de resultaten. 

De meeste van de resultaten zijn gepubliceerd in de artikels
\cite{benfred, normal, CGMV, BCR, ournonbps, 0dual}. Het is onze bedoeling ze
in deze thesis op een coherente manier te presenteren en voldoende
achtergrondmateriaal te voorzien om ze toegankelijk(er) te maken voor
niet-specialisten. In het bijzonder denken we aan mensen die vertrouwd zijn met
kwantumveldentheorie maar geen gedetailleerde kennis bezitten van deze tak van
snaartheorie.

Enerzijds introduceren we in Hoofdstuk \ref{central} een aantal sleutelbegrippen
in een niet-snaartheoretische context, in het bijzonder in kwantumveldentheorie.
Die begrippen zijn dualiteit, supersymmetrie en anomalie instroom.  Anderzijds
proberen we in Hoofdstuk~\ref{superstrings} en Sectie~\ref{anomalous:boundary}
genoeg technische informatie over snaren, D-branen en randtoestanden aan te
reiken opdat gemotiveerde lezers in staat zouden zijn de berekeningen in
Hoofdstuk~\ref{anomalous} te begrijpen, in het bijzonder die in
Sectie~\ref{anomalous:BC}, die meer in detail gegeven worden. 
Hoofstuk~\ref{superstrings} is ook een noodzakelijke voorbereiding voor
Hoofdstukken~\ref{baryon}, \ref{type0} en \ref{nonbps}. In wat volgt geven we
voor de
verschillende hoofdstukken een overzicht en een samenvatting van de resultaten.

\subsection{Hoofdstuk~\ref{central}: Enkele sleutelbegrippen}
In Hoofdstuk~\ref{central} introduceren we drie sleutelbegrippen, meestal in een
veldentheoretische context. In Sectie~\ref{central:duality} geven we
verscheidene voorbeelden van dualiteiten. We concentreren ons vooral op
elektromagnetische dualiteit, in het bijzonder op de dualiteitsconjectuur van
Montonen en Olive \cite{montol} en enkele raadsels die daarmee verbonden waren. In
Sectie~\ref{central:susy} tonen we hoe die raadsels op een natuurlijke wijze
opgelost worden in supersymmetrische veldentheorie. We introduceren de
supersymmetrie-algebra en het begrip `BPS toestand'. In
Sectie~\ref{central:anomalies} presenteren we enkele resultaten uit de
literatuur over anomalie\"en en anomalie instroom. We concentreren ons op 
\'e\'en voorbeeld \cite{CH,naculich}, waar men het instroom mechanisme 
expliciet aan het werk ziet. Omdat het anomalie instroom mechanisme een cruciaal
begrip is in deze thesis, gaan we er nu kort op in.

Bij anomalie instroom kan men zich het volgende voorstellen \cite{CH,naculich}.
Beschouw een ijkveld dat op een niet-chirale manier aan fermionen gekoppeld is,
laat ons zeggen in een vier-dimensionale ruimte-tijd. 
Bij gebrek aan chirale ijk-koppelingen, zal de ijksymmetrie niet-anomaal zijn. 
Veronderstel nu dat de fermionen bovendien op een chirale manier koppelen aan
een scalair veld. Voor bepaalde configuraties van het scalaire veld is het
mogelijk dat bepaalde fermion modes effectief op een lager-dimensionaal 
`defect' van
de ruimte-tijd leven, bijvoorbeeld op een twee-dimensionaal oppervlak. Als zulke
fermion modes chiraal zijn in twee dimensies, kunnen ze een ijkanomalie
veroorzaken in hun effectieve twee-dimensionale theorie. Dat kan op het eerste
gezicht in tegenspraak lijken met het feit dat we vertrokken waren van een
vierdimensionale theorie zonder chirale ijk-koppelingen. 

De oplossing is dat de anomale variatie van de effectieve
twee-dimensionale actie gecompenseerd wordt door een ijkvariatie veroorzaakt
door de overblijvende fermion modes, die in de `bulk' van de ruimte-tijd leven.
Er is dus `anomalie instroom' van de vier-dimensionale ruimte-tijd naar het
twee-dimensionale `defect'. In meer fysische termen kan met het als volgt
stellen. Kijkt men enkel naar de vrijheidgraden die effectief op het `defect'
leven, dan lijkt het alsof er (voor bepaalde achtergrondconfiguraties) `uit het
niets' lading gecre\"eerd wordt op het defect. Kijkt men echter ook naar de 
vrijheidsgraden in de `bulk', dan merkt men dat die lading in werkelijkheid
gewoon vanuit de `bulk' naar het defect stroomt, en dat lading in de volledige
theorie behouden is.  

We zien hier dus dat, wanneer men de chirale vrijheidsgraden op het `defect'
kent, men bepaalde voorspellingen kan doen over de `bulk' theorie. In het
eenvoudige voorbeeld dat we besproken hebben, kan het lijken dat we hieruit
niets leren wat we nog niet wisten. 
Het instroom mechanisme is echter een krachtig hulpmiddel
in meer ingewikkelde situaties, zoals die
welke we bespreken in Secties~\ref{superstrings:inflow}, \ref{type0:D:inflow} 
en \ref{type0:NS5:anomalies}.   

\subsection{Hoofdstuk~\ref{superstrings}: Snaren en branen}
In Hoofdstuk~\ref{superstrings} voeren we supersnaren en D-branen in. We doen
een inspanning om op een coherente manier de begrippen in te voeren die relevant
zijn voor deze thesis, in het bijzonder voor de berekeningen in
Hoofdstuk~\ref{anomalous}. 

In Sectie~\ref{superstrings:strings:worldsheet} voeren we snaren in vertrekkend
van hun wereld-oppervlak beschrijving. We bestuderen onder andere het verschil 
tussen open en gesloten snaren, de verschillende sectoren (bijvoorbeeld R en
NS voor open snaren), 
de spectra van snaarexcitaties en de GSO-projectie. We bespreken
ook enkele meer technische aspecten, zoals spoken en BRST-kwantisatie. Deze
technische aspecten zijn uiteraard van belang, maar om het grootste deel van
onze expliciete berekeningen te volgen, is het niet nodig ze te kennen. De lezer
kan er dus voor kiezen alles over spoken en BRST-operatoren over te slaan. In
Sectie~\ref{superstrings:strings:spacetime} bespreken we de
lage-energie effectieve beschrijving van snaartheorie als een
supergravitatietheorie in een tien-dimensionale ruimte-tijd. De wisselwerking
tussen de wereld-oppervlak- en ruimte-tijd-beschrijvingen van snaartheorie is
een van de hoofdthema's in deze thesis. Sectie~\ref{superstrings:background}
gaat over supersnaren in achtergrondvelden.

In Sectie~\ref{superstrings:scattering} verzamelen we een aantal hulpmiddelen om
verstrooiingsamplituden te berekenen in snaartheorie. We introduceren het
belangrijke begrip `vertex-operator'. Een groot deel van deze sectie is gewijd
an technische aspecten, zoals (super)spoken, beelden (`pictures') en
BRST-invariantie. Deze aspecten kunnen weer overgeslagen worden door lezers die
er niet bijzonder in ge\"\i nteresseerd zijn. Ze zijn vooral nodig om onze
latere keuzes van vertex-operatoren te rechtvaardigen. In  
Sectie~\ref{superstrings:scattering:closed} geven we een aantal van die
vertex-operatoren.

In Sectie~\ref{superstrings:D-branes} introduceren we D-branen en hun effectieve
actie. In het bijzonder tonen we een aantal anomale termen van de D-braan 
Wess-Zumino actie 
\beq
S_{\rm WZ}=\frac{T_p}{\kappa}\int_{p+1} C\wedge e^{2\pi\a '\,F+B}\wedge
\sqrt{\hat{A}(R_T)/\hat{A}(R_N)}
\eeq
in detail.% 
\footnote{We verwijzen naar Sectie~\ref{superstrings:D-branes} voor de notatie.}
De inhoud van deze sectie is van cruciaal belang om
de thesis te begrijpen. Sectie~\ref{superstrings:inflow} gaat over het anomalie
instroom argument in de context van D-branen en NS-vijfbranen. We geven weinig
technische details, maar de idee\"en nemen een centrale plaats in in de thesis.
Bijvoorbeeld zullen we anomalie instroom argumenten gebruiken in 
Secties~\ref{type0:D:inflow} en \ref{type0:NS5:anomalies}.
In Sectie~\ref{superstrings:dualities} bestuderen we drie dualiteiten in
snaartheorie. De discussie van type IIB S-dualiteit is nuttig als voorbereiding
op Sectie~\ref{type0:dual}, waar we een veralgemening ervan kritisch bestuderen.

\subsection{Hoofdstuk~\ref{anomalous}: Anomale koppelingen uit snaarberekeningen}
In dit hoofdstuk gaan we de aanwezigheid na van sommige van de anomale D-braan
koppelingen in (\ref{WZ}). Dat doen we door verstrooiingsamplituden 
van gesloten-snaartoestanden in de aanwezigheid van een D-braan expliciet te 
berekenen in snaartheorie. Daartoe gebruiken we het randtoestand formalisme. In
Sectie~\ref{anomalous:boundary} voeren we de randtoestand in en geven we een
verzameling preciese regels over hoe hij moet worden gebruikt in 
verstrooiingsamplituden met een sfeer als wereld-oppervlak.%
\footnote{We doen hier echter geen poging deze regels echt af te leiden.} 
Deze sectie bevat niet veel origineel werk, al is, voor zover wij weten, onze 
verzameling regels niet in de literatuur te vinden.

In Sectie~\ref{anomalous:overview} geven we een overzicht van de tests waaraan
de Wess-Zumino actie (\ref{WZ}) onderworpen is. Het blijkt dat er
onrechtstreekse argumenten zijn voor de aanwezigheid van alle termen in
(\ref{WZ}), maar dat voor de termen met NS-NS velden alle rechtstreekse tests
ontbraken tot ons artikel \cite{benfred} verscheen. De indirecte argumenten
gebruiken expliciet de consistentie van snaartheorie, in het bijzonder de
afwezigheid van ijk- en gravitationele anomalie\"en. Directe tests van (\ref{WZ})
gaan dus na of snaartheorie met D-branen inderdaad een consistente theorie is.

In Sectie~\ref{anomalous:BC} testen we de aanwezigheid van de term
\beq
\frac{T_p}{\kappa}\int_{p+1}\hat C_{p-1}\wedge \hat{B}
\eeq
in de D-braan Wess-Zumino actie. Dankzij het voorbereidende werk in 
Hoofdstuk~\ref{superstrings} en Sectie~\ref{anomalous:boundary} kunnen we de
berekeningen in detail tonen. We vinden de gezochte term inderdaad.
Sectie~\ref{anomalous:BC} is een uitgebreide
versie van een hoofdstuk van ons artikel \cite{benfred}.

In Sectie~\ref{anomalous:RRC} bevestigen we de aanwezigheid 
(met de juiste co\"effici\"ent) van de termen
\beq
\frac{T_p}{\kappa}\int_{p+1}\hat C_{p-3}\wedge \frac{(4\pi^2\a ')^2}{384\pi^2}
(\tr R_T^2-\tr R_N^2)~,
\eeq 
waarin viervormen geconstrueerd uit krommings-tweevormen een rol spelen. Deze
berekeningen werden eerst uitgevoerd in onze artikels \cite{benfred,normal}.

In Sectie~\ref{anomalous:D:non}, die gebaseerd is op ons artikel \cite{normal}, 
beschouwen we de termen
\beq
\frac{T_p}{\kappa}\int_{p+1}\hat C_{p-7}\wedge 
\left(\frac{(4\pi^2\a ')^4}{294912\pi^4}(\tr R_T^2-\tr R_N^2)^2+
\frac{(4\pi^2\a ')^4}{184320\pi^4} (\tr R_T^4-\tr R_N^4)\right)~.
\eeq 
Hierin staat een achtvorm opgebouwd uit krommings-tweevormen. Gebruik makend van
de resultaten van \cite{stefanski}, kunnen we de aanwezigheid van al deze termen
bevestigen, inclusief de co\"effici\"enten. 
Bovendien vinden we, als bonus, extra D-braan koppelingen
\cite{normal}. Die extra
koppelingen, die nieuw zijn, zijn niet-anomaal, wat meteen verklaart waarom ze
niet eerder gevonden waren door anomalie instroom argumenten. 

\subsection{Hoofdstuk~\ref{baryon}: Baryon vertex in AdS/CFT}
In dit hoofdstuk bespreken we een toepassing van de effectieve D-braan actie: we
tonen hoe ze gebruikt kan worden om snaren eindigend op  D-branen in
D-braan achtergronden te bestuderen. We beschouwen onder andere
een D5-braan gewikkeld rond de vijf-sfeer $S^5$ van een $AdS_5\times S^5$ 
ruimte. $AdS_5\times S^5$ is de structuur van de ruimte-tijd dichtbij de 
horizon van D3-branen. We zijn dus een D5-braan in de `near-horizon' meetkunde
van D3-branen aan het bestuderen. Deze configuratie speelt een rol in de AdS/CFT
correspondentie \cite{maldacena}: een D5-braan gewikkeld rond $S^5$ is het
analogon in snaartheorie van een baryon in de corresponderende veldentheorie
\cite{baryons}.

In Sectie~\ref{baryon:intro} geven we een korte inleiding tot de AdS/CFT
correspondentie, de baryon vertex en de methode die we gebruiken om die baryon
vertex te bestuderen: de BPS methode. Het vervolg van   Hoofdstuk~\ref{baryon}
is gebaseerd op ons artikel \cite{CGMV}. In Sectie~\ref{baryon:review} leggen we
de BPS methode uit in een iets eenvoudigere situatie dan die welke ons zal
interesseren. In Sectie~\ref{baryon:ham} bestuderen we de baryon vertex
configuratie en een gerelateerde configuratie in ons formalisme. We reproduceren
de BPS-vergelijking (\ref{BPScondition}) van \cite{Imamura,Callan}. Een mooi
aspect van onze methode is dat we resultaten kunnen afleiden die nauw
verbonden zijn met supersymmetrie, en dit zonder expliciet fermionen te
beschouwen.%
\footnote{Dit is ook het geval voor de BPS-ongelijkheid voor magnetische
monopolen, zie Secties~\ref{central:duality} en \ref{central:susy}.}
Een ander voordeel van onze methode is dat ze de mogelijkheid opent de
BPS-vergelijking te herinterpreteren in termen van superalgebra's \cite{CGMV}.
In deze thesis bespreken we die herinterpretatie echter niet.

\subsection{Hoofdstuk~\ref{type0}: Type 0 snaartheorie}
Dit hoofdstuk is gewijd aan type 0 snaartheorie. Vergeleken met type II snaren
hebben type 0 snaren een tachyon, dubbel zoveel R-R velden en geen ruimte-tijd
fermionen in hun perturbatieve spectrum. De tachyonische fluctuatie signaleert 
dat we rond een onstabiel `vacuum' aan het expanderen zijn.
Wegens de problemen die de aanwezigheid van zo'n tachyonische mode met zich 
meebrengt, werden type 0 snaartheorie\"en pas anderhalf jaar geleden populair.
Toen werd opgemerkt \cite{polya,9811035} dat het tachyon niet noodzakelijk een
probleem hoeft te zijn wanneer men, zoals in de AdS/CFT correspondentie, de
effectieve veldentheorie\"en op het wereld-volume van D-branen wil bestuderen.

In Sectie~\ref{type0:intro} bespreken we deze motivatie om type 0 snaartheorie
te bestuderen. Type 0 snaren worden ge\"\i ntroduceerd in
Sectie~\ref{type0:strings}. In Sectie~\ref{type0:D} voeren we type 0 D-branen
in. Zoals kon worden verwacht op basis van het verdubbelde  het R-R spectrum
van type 0 snaren ten opzichte van type II snaren, zijn er ook tweemaal zoveel
verschillende D-branen in type 0. Zoals in ons artikel \cite{BCR} leiden we de
Wess-Zumino actie (\ref{WZaction}) af voor type 0 D-branen:
\beq
S_{\rm WZ}=\frac{T_p}{\kappa}\int_{p+1}(C)_\pm\wedge 
{\rm e}^{2\pi\a '\,F+B}\wedge
\sqrt{\hat{A}(R_T)/\hat{A}(R_N)}~.
\eeq
De $\pm$ geeft het verschil aan tussen `elektrische' and `magnetische' branen. 
Om deze actie af te leiden, maken we gebruik van een anomalie instroom argument.
De aanwezigheid van de verschillende termen kan ook rechtstreeks getest worden
zoals we deden voor type II D-branen.

We merken op dat niet alleen de Wess-Zumino actie voor type 0 D-branen zelf,
maar ook de argumenten waarmee ze afgeleid en getest werd, zeer sterk lijken op
wat we kennen van de supersymmetrische type II D-branen. Hoewel van
(ruimte-tijd) supersymmetrie geen sprake is voor type 0 D-branen, lijken ze in
sommige opzichten dus sterk op supersymmetrische type II D-branen.  

Sectie~\ref{type0:NS5} is gebaseerd op ons artikel \cite{0dual}, waar we
NS-vijfbranen in type 0 snaartheorie\"en bestuderen. We leiden hun massaloze
spectra af en vinden dat die spectra niet-chiraal en puur bosonisch zijn voor
zowel type 0A als type 0B. We stellen een speculatieve interpretatie van de type
0 NS-vijfbraan spectra voor in termen van `type 0 kleine snaren'.

In Sectie~\ref{type0:NS5:anomalies} hebben we gezien
dat type IIA NS-vijfbranen een chiraal spectrum hebben. Dit chiraal spectrum
geeft aanleiding tot een gravitationele anomalie, die gecompenseerd wordt door
anomalie instroom van de `bulk' van de ruimte-tijd naar het vijfbraan.  
We berekenen dat voor zowel type 0A als type 0B de analoge `bulk' term afwezig
is. Er is dus geen instroom van de bulk naar de vijfbranen, wat consistent is
met onze bevinding dat die vijfbranen een niet-chiraal spectrum hebben en dus 
geen gravitationele anomalie \cite{0dual}.

In \cite{berggab} zijn een aantal dualiteiten van type 0 snaartheorie\"en
voorgesteld. Een ervan is S-dualiteit van type 0B snaartheorie.
In Sectie~\ref{type0:dual} combineren we onze studies van type 0 D-branen en
NS-vijfbranen om enkele vragen op te werpen over type 0B S-dualiteit.

\subsection{Hoofdstuk~\ref{nonbps}: Niet-BPS branen}
In dit hoofdstuk bestuderen we niet-BPS D-branen. Dat zijn objecten waarop open
snaren kunnen eindigen, maar die geen supersymmetrie bewaren. In sommige
snaartheorie\"en kunnen zulke objecten stabiel  zijn, maar in tien-dimensionale
type II snaartheorie zijn ze het niet. In Sectie~\ref{nonbps:mot} geven we aan
hoe werk van Sen het onderzoek naar niet-BPS D-branen op gang bracht. Het
belangrijkste resultaat van Hoofdstuk~\ref{nonbps} staat in 
Sectie~\ref{nonbps:anomalous}, die gebaseerd is op ons artikel \cite{ournonbps}.
Daar stellen we voor de onstabiele niet-BPS D-branen van type II snaartheorie
de Wess-Zumino actie (\ref{wznonbps}) voor
\cite{ournonbps}:
\beq 
S'_{\rm WZ}=a\int_{p+1}\hat C\wedge d\,\left\{\tr\, T \, e^{2\pi\a
'\,F}\right\}\wedge e^{\hat B}\wedge
\sqrt{\hat{A}(R_T)/\hat{A}(R_N)}~,
\eeq
waar $T$ het tachyonveld is dat de instabiliteit van het niet-BPS D-braan
signaleert.
Enerzijds argumenteren we \cite{ournonbps}
dat die actie consistent is met de interpretatie \cite{horava} van
BPS D-branen als topologisch niet-triviale tachyon-configuraties op een niet-BPS
D-braan. Anderzijds testen we de aanwezigheid van de verschillende termen door
expliciete berekeningen in snaartheorie \cite{ournonbps}.

\subsection{Besluit}
De meest opvallende resultaten in deze thesis zijn waarschijnlijk
\begin{itemize}
\item
de bevestiging van de aanwezigheid van verschillende anomale termen in de 
D-braan Wess-Zumino actie door expliciete berekeningen in snaartheorie
(Hoofdstuk~\ref{anomalous}),

\item
het vinden van nieuwe, niet-anomale termen in de D-braan Wess-Zumino actie
(Sectie~\ref{anomalous:D:non}),

\item
het uitbreiden van de BPS-methode naar D-branen in D-braan achtergronden
(Hoofdstuk~\ref{baryon}),

\item
de Wess-Zumino actie (\ref{WZaction}) voor type 0 D-branen  
(Sectie~\ref{type0:D:inflow}),

\item
de type 0 NS-vijfbraan spectra  (Sectie~\ref{type0:NS5}) 

\item
en de Wess-Zumino actie (\ref{wznonbps}) voor niet-BPS D-branen in type II 
snaartheorie (Hoofdstuk~\ref{nonbps}).
\end{itemize} 

\selectlanguage{english}

% bibliography

\end{document}